\documentclass[preprint,preprintnumbers,nofootinbib,tightenlines,12pt,superscriptaddress]{revtex4}

\usepackage{graphicx}
\usepackage{amssymb}
\usepackage{amsmath,mathrsfs,verbatim}
\usepackage{times}
\usepackage{latexsym}
\usepackage{hyperref}
\usepackage{color}
\def\beq{\begin{equation}}
\def\eeq{\end{equation}}
\def\bey{\begin{eqnarray}}
\def\eey{\end{eqnarray}}

\def\lsim{\mathrel{\raise.3ex\hbox{$<$\kern-.75em\lower1ex\hbox{$\sim$}}}}
\def\gsim{\mathrel{\raise.3ex\hbox{$>$\kern-.75em\lower1ex\hbox{$\sim$}}}}
\def\Msun{{\rm \, M}_\odot}

\def\kms{\, {\rm km/s} }

\newcommand{\be}{\begin{equation}}
\newcommand{\ee}{\end{equation}}

\newcommand{\mx}{m_\chi}
\newcommand{\ax}{\alpha_\chi}
\newcommand{\mphi}{m_\phi}

\newcommand{\vrel}{v_{\rm rel}}
\newcommand{\HI}{H{\small{\sc I}} }

\begin{document}

\title{Dark Matter Self-interactions and Small Scale Structure}

\author{Sean Tulin}
\email{stulin@yorku.ca}
\affiliation{Department of Physics and Astronomy, York University, Toronto, Ontario M3J 1P3, Canada}

\author{Hai-Bo Yu}
\email{haiboyu@ucr.edu}
\affiliation{Department of Physics and Astronomy, University of California, Riverside, California 92521, USA}

\date{\today}

\begin{abstract}

We review theories of dark matter (DM) beyond the collisionless paradigm, known as self-interacting dark matter (SIDM), and their observable implications for astrophysical structure in the Universe.  Self-interactions are motivated, in part, due to the potential to explain long-standing (and more recent) small scale structure observations that are in tension with collisionless cold DM (CDM) predictions.  Simple particle physics models for SIDM can provide a universal explanation for these observations across a wide range of mass scales spanning dwarf galaxies, low and high surface brightness spiral galaxies, and clusters of galaxies.  At the same time, SIDM leaves intact the success of $\Lambda$CDM cosmology on large scales.  This report covers the following topics: {\it (1) small scale structure issues}, including the core-cusp problem, the diversity problem for rotation curves, the missing satellites problem, and the too-big-to-fail problem, as well as recent progress in hydrodynamical simulations of galaxy formation; {\it (2) N-body simulations for SIDM}, including implications for density profiles, halo shapes, substructure, and the interplay between baryons and self-interactions; {\it (3) semi-analytic Jeans-based methods} that provide a complementary approach for connecting particle models with observations; {\it (4) merging systems}, such as cluster mergers (e.g., the Bullet Cluster) and minor infalls, along with recent simulation results for mergers; {\it (5) particle physics models}, including light mediator models and composite DM models; and {\it (6) complementary probes for SIDM}, including indirect and direct detection experiments, particle collider searches, and cosmological observations.  We provide a summary and critical look for all current constraints on DM self-interactions and an outline for future directions.

\end{abstract}


\maketitle

\tableofcontents

\newpage 

\section{Introduction}

\subsection{The dark matter puzzle}

It was long ago pointed out by Oort that the distributions of mass and light in galaxies have little resemblance to one another~\cite{1940ApJ....91..273O}.  At the time, observations of galaxy NGC 3115 found a rotation curve rising linearly with radius---indicating a constant mass density---despite its luminosity falling by over an order of magnitude.   Oort derived a mass-to-light ratio that increased with distance up to $\sim 250$ in solar units\footnote{Modern distance estimates revise Oort's value down to $\sim 40$.} at the outermost radius.  He concluded that the luminous disk of NGC 3115 is ``imbedded in a more or less homogeneous mass of great density''~\cite{1940ApJ....91..273O}. The same phenomenon was observed in M31 by Babcock a year earlier~\cite{1939LicOB..19...41B}.  These observations stood in contrast to the Milky Way (MW), where, by studying vertical dynamics of nearby stars, the mass-to-light ratio {\it within the disk} was known to be only $\sim 2$ at the solar radius \cite{Opik1915,1922ApJ....55..302K,1922MNRAS..82..122J,1932BAN.....6..249O}.  Earlier pioneering observations of the Coma cluster by Zwicky~\cite{1933AcHPh...6..110Z} and Virgo cluster by Smith~\cite{1936ApJ....83...23S} inferred similarly striking discrepencies between mass and light on much larger scales.

These issues were not fully appreciated for several decades, when astronomers were able to study galactic rotational velocities at much larger distances.  While many rotation curves exhibited a linearly increasing velocity at small distances, they were expected to turn over eventually and fall as $r^{-1/2}$ according to Kepler's laws~\cite{1963ARA&A...1..179K}.  Instead, optical observations by Rubin and Ford of M31~\cite{1970ApJ...159..379R}---soon after extended to larger radii using 21-cm radio observations~\cite{1975ApJ...201..327R,1973A&A....26..483R}---revealed a circular velocity that did not fall off, but remained approximately constant.  Many other spiral galaxies were found to exhibit the same behavior~\cite{1970ApJ...160..811F,1973A&A....26..483R,1980ApJ...238..471R,1981AJ.....86.1825B,Sofue:2000jx}, indicating that most of the mass in galaxies is found in massive, nonluminous halos extending far beyond the spatial extent of luminous stars and gas~\cite{1974ApJ...193L...1O,1974Natur.250..309E}.  Massive spherical dark halos could also explain the apparent stability of bulgeless spiral galaxies~\cite{1973ApJ...186..467O}, which by themselves are unstable to the formation of bars~\cite{1964ApJ...139.1217T,1971ApJ...168..343H}.

The large amount of dark matter (DM) required by observations pointed toward its nonbaryonic nature.\footnote{In this context, the term ``baryons'' represents protons, neutrons, and electrons that constitute normal atomic matter.}  The total mass density found in halos was estimated to be around $20\%$ of the critical density~\cite{1974ApJ...193L...1O,1974Natur.250..309E}, in remarkable agreement with present cosmological values~\cite{Ade:2015xua}.  If composed of baryons, the cosmological baryon density would be in tension with upper limits inferred from nucleosynthesis arguments~\cite{1973ApJ...179..343W}, as well as being difficult to hide from astrophysical observations~\cite{1983PhLB..126...28H}.  Therefore, the ``missing mass'' puzzle in galaxies, as well as clusters, suggested the existence of a new dominant form of matter, such as an elementary particle leftover from the Big Bang.

At first, the neutrino appeared to be a promising DM candidate within the Standard Model (SM) of elementary particles~\cite{Gershtein:1966gg,Cowsik:1972gh,Szalay:1976ef}.  A relic thermal bath of neutrinos, produced alongside the cosmic microwave background (CMB), could yield the required mass density if the neutrino mass was $\mathcal{O}(10 \; {\rm eV})$.  However, because neutrinos are ``hot'' DM---they decouple from photons and electrons around the nucleosynthesis epoch while still relativistic---free-streaming erases density fluctuations below supercluster scales~(see, e.g., Ref.~\cite{Primack:2000iq}).  Numerical simulations have shown that top-down structure formation, where superclusters form first in the neutrino-dominated Universe and subsequently fragment to produce galaxies, is incompatible with galaxy clustering constraints~\cite{1983ApJ...274L...1W,1984MNRAS.209P..27W,1984Natur.310..637H}.

\begin{figure}
\includegraphics[scale=0.5]{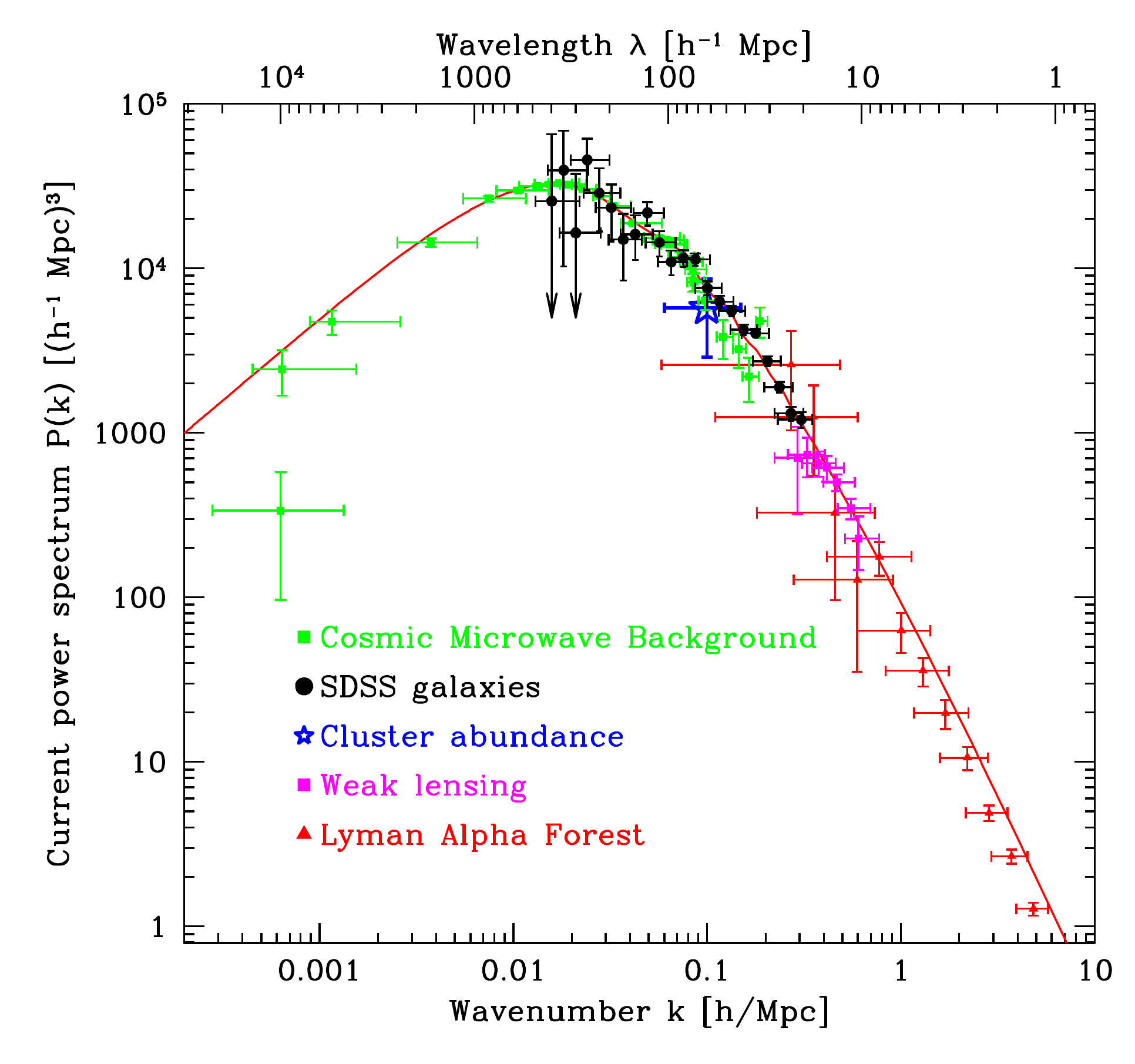}
\caption{ \it Matter power spectrum inferred through cosmological measurements.  Red line shows the best fit for $\Lambda$CDM cosmology for a simplified five-parameter model, assuming a flat spatial geometry and a scale-invariant primordial spectrum.  Reprinted from Ref.~\cite{Tegmark:2003uf}.  See therein and Ref.~\cite{Tegmark:2003ud} for further information.
}
\label{fig:power}
\end{figure}

Cosmological data has converged upon the $\Lambda$CDM paradigm as the standard model of cosmology (e.g., Ref.~\cite{Bahcall:1999xn}).  Of the total mass-energy content of the Universe, approximately $26\%$ is cold dark matter (CDM) and $5\%$ is baryonic matter (while the remainder is consistent with a cosmological constant $\Lambda$), with a nearly scale-invariant spectrum of primordial fluctuations~\cite{Ade:2015xua}.  In this picture, structure in the Universe forms as primordial overdensities collapse under gravity.  Since CDM, acting as a pressureless fluid, is more dominant and collapses more readily than baryonic matter, it provides the gravitational potential underlying the distribution of visible matter in the Universe.  The observed matter power spectrum, as obtained from a variety of cosmological probes, is in remarkable agreement with $\Lambda$CDM cosmology, shown in Fig.~\ref{fig:power}. In addition, the $\Lambda$CDM model also explains many important aspects of galaxy formation~\cite{Springel:2006vs,TrujilloGomez:2010yh}. 

Despite this success, all evidence to date for DM comes from its gravitational influence in the Universe.  With no viable DM candidate within SM, the underlying theory for DM remains unknown.  Many new particle physics theories proposed to address shortcomings of the SM simultaneously predict new particles that can be a DM candidate.  Examples include weakly-interacting massive particles (WIMPs) motivated by the hierarchy problem, such as neutralinos in the supersymmetric models~\cite{Goldberg:1983nd,Jungman:1995df} and Kaluza-Klein states in extra dimensional models~\cite{Servant:2002aq,Cheng:2002ej}, as well as extremely light axion particles~\cite{Preskill:1982cy} associated with the solution to the strong CP problem in QCD~\cite{Peccei:1977hh}.  The comic abundance of these new particles can be naturally close to the DM abundance inferred from the cosmological observations (e.g., Ref.~\cite{Lee:1977ua}).  This coincidence has motivated decades of efforts to discover the particle physics realization of DM through experimental searches for new physics beyond the SM (e.g., see~\cite{Bertone:2004pz,Feng:2010gw,Arrenberg:2013rzp} and references therein).

On large scales, the structure of the Universe is consistent with DM particles that are cold, collisionless, and interact with each other and SM particles purely via gravity.  WIMPs and axions have interactions with the SM that are potentially large enough to be detectable in the laboratory, while on astrophysical scales these interactions are negligible and these candidates behave as CDM.  On the other hand, other particle physics candidates for DM may have interactions with the SM are too feeble to observe directly.  Here, observational tests of structure and departures from the CDM paradigm play a complementary role in probing the particle physics of DM {\it independently} of its interactions with SM particles (as had been done for hot DM candidates).

\subsection{Crisis on small scales}

$\Lambda$CDM is an extremely successful model for the large scale structure of the Universe, corresponding to distances greater than $\mathcal{O}({\rm Mpc})$ today (see Fig.~\ref{fig:power}).  On smaller scales, structure formation becomes strongly nonlinear and N-body simulations have become the standard tool to explore this regime.  Cosmological DM-only simulations have provided several predictions for the structure and abundance of CDM halos and their substructure. However, it remains unclear whether these predictions are borne out in nature. 

Since the 1990s, four main discrepancies between CDM predictions and observations have come to light. 
\begin{itemize}
\item {\it Core-cusp problem:} High-resolution simulations show that the mass density profile for CDM halos increases toward the center, scaling approximately as $\rho_{\rm dm} \propto r^{-1}$ in the central region~\cite{Dubinski:1991bm,Navarro:1995iw,Navarro:1996gj}.  However, many observed rotation curves of disk galaxies prefer a constant ``cored'' density profile $\rho_{\rm dm} \propto r^0$~\cite{Flores:1994gz,Moore:1994yx,Moore:1999gc}, indicated by linearly rising circular velocity in the inner regions. The issue is most prevalent for dwarf and low surface brightness (LSB) galaxies~\cite{1995ApJ...447L..25B,McGaugh:1998tq,2000AJ....120.3027C,vandenBosch:2000rza,Borriello:2000rv,deBlok:2001mf,deBlok:2001fe,Marchesini:2002vm,Gentile:2005de,Gentile:2006hv,KuziodeNaray:2006wh,KuziodeNaray:2007qi,Salucci:2007tm}, which, being highly DM-dominated, are appealing environments to test CDM predictions.
\item {\it Diversity problem:} Cosmological structure formation is predicted to be a self-similar process with a remarkably little scatter in density profiles for halos of a given mass~\cite{Navarro:1996gj,Bullock:1999he}.  However, disk galaxies with the same maximal circular velocity exhibit a much larger scatter in their interiors~\cite{Oman:2015xda} and inferred core densities vary by a factor of ${\cal O}(10)$~\cite{deNaray:2009xj}. 
\item {\it Missing satellites problem:} CDM halos are rich with substructure, since they grow via hierarchical mergers of smaller halos that survive the merger process~\cite{Kauffmann:1993gv}.  Observationally, however, the number of small galaxies in the Local Group are far fewer than the number of predicted subhalos.  In the MW, simulations predict $\mathcal{O}(100-1000)$ subhalos large enough to host galaxies, while only 10 dwarf spheroidal galaxies had been discovered when this issue was first raised~\cite{Moore:1999nt,Klypin:1999uc}.  Nearby galaxies in the field exhibit a similar underabundance of small galaxies compared to the velocity function inferred through simulations~\cite{Zavala:2009ms,Zwaan:2009dz,TrujilloGomez:2010yh}.
\item {\it Too-big-to-fail problem (TBTF):} In recent years, much attention has been paid to the most luminous satellites in the MW, which are expected to inhabit the most massive suhalos in CDM simulations. However, it has been shown that these subhalos are too dense in the central regions to be consistent with stellar dynamics of the brightest dwarf spheroidals~\cite{BoylanKolchin:2011de, BoylanKolchin:2011dk}. The origin of the name stems from the expectation that such massive subhalos are too big to fail in forming stars and should host observable galaxies. Studies of dwarf galaxies in Andromeda~\cite{Tollerud:2014zha} and the Local Group field~\cite{Garrison-Kimmel:2014vqa} have found similar discrepancies.
\end{itemize}
It must be emphasized, however, that these issues originally gained prominence by comparing observations to theoretical predictions from {\it DM-only} simulations.  Hence, there has been extensive debate in the literature whether these small scale issues can be alleviated or solved in the $\Lambda$CDM framework once dissipative baryonic processes, such as gas cooling, star formation, and supernova feedback, are included in simulations~\cite{Navarro:1996bv,Governato:2009bg}. We review these topics in detail in \S\ref{sec:astro}. 

A more intriguing possibility is that the CDM paradigm may break down on galactic scales. One early attempt to solve these issues supposes that DM particles are warm, instead of cold, meaning that they were quasi-relativistic during kinetic decoupling from the thermal bath in the early Universe~\cite{Colombi:1995ze,Bode:2000gq}. Compared to CDM, warm DM predicts a damped linear power spectrum due to free-streaming, resulting in a suppression of the number of substructures. Warm DM halos are also typically less concentrated because they form later than CDM ones.  Recent high-resolution simulations show that warm DM may provide a solution to the missing satellites and too-big-to-fail problems~\cite{Lovell:2011rd,Lovell:2013ola,Horiuchi:2015qri}.  However, the favored mass range of thermal warm DM is in strong tension with Lyman-$\alpha$ forest observations~\cite{Irsic:2017ixq, Viel:2013apy}\footnote{The Lyman-$\alpha$ constraints may be weakened due to uncertainties in the evolution of intergalactic medium~\cite{Kulkarni:2015fga,Garzilli:2015iwa,Cherry:2017dwu}.} and the abundance of high redshift galaxies~\cite{Menci:2016eui}.  Also, while warm DM halos have constant density cores set by the phase space density limit, the core sizes are far too small to solve the core-cusp problem given Lyman-$\alpha$ constraints~\cite{Maccio:2012qf}.

\subsection{Self-Interacting dark matter}

\begin{figure}
\includegraphics[scale=0.423]{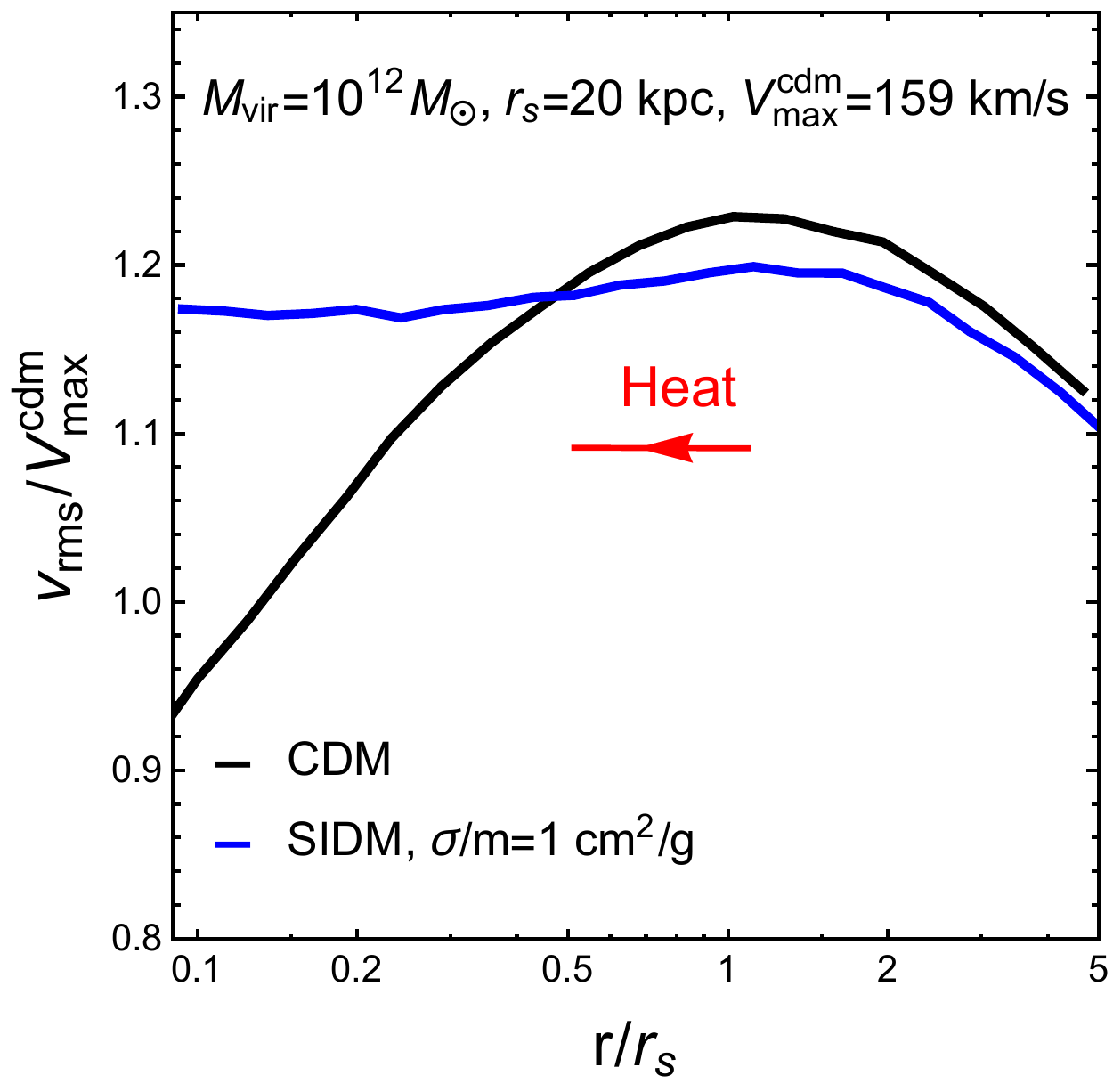}
\includegraphics[scale=0.415]{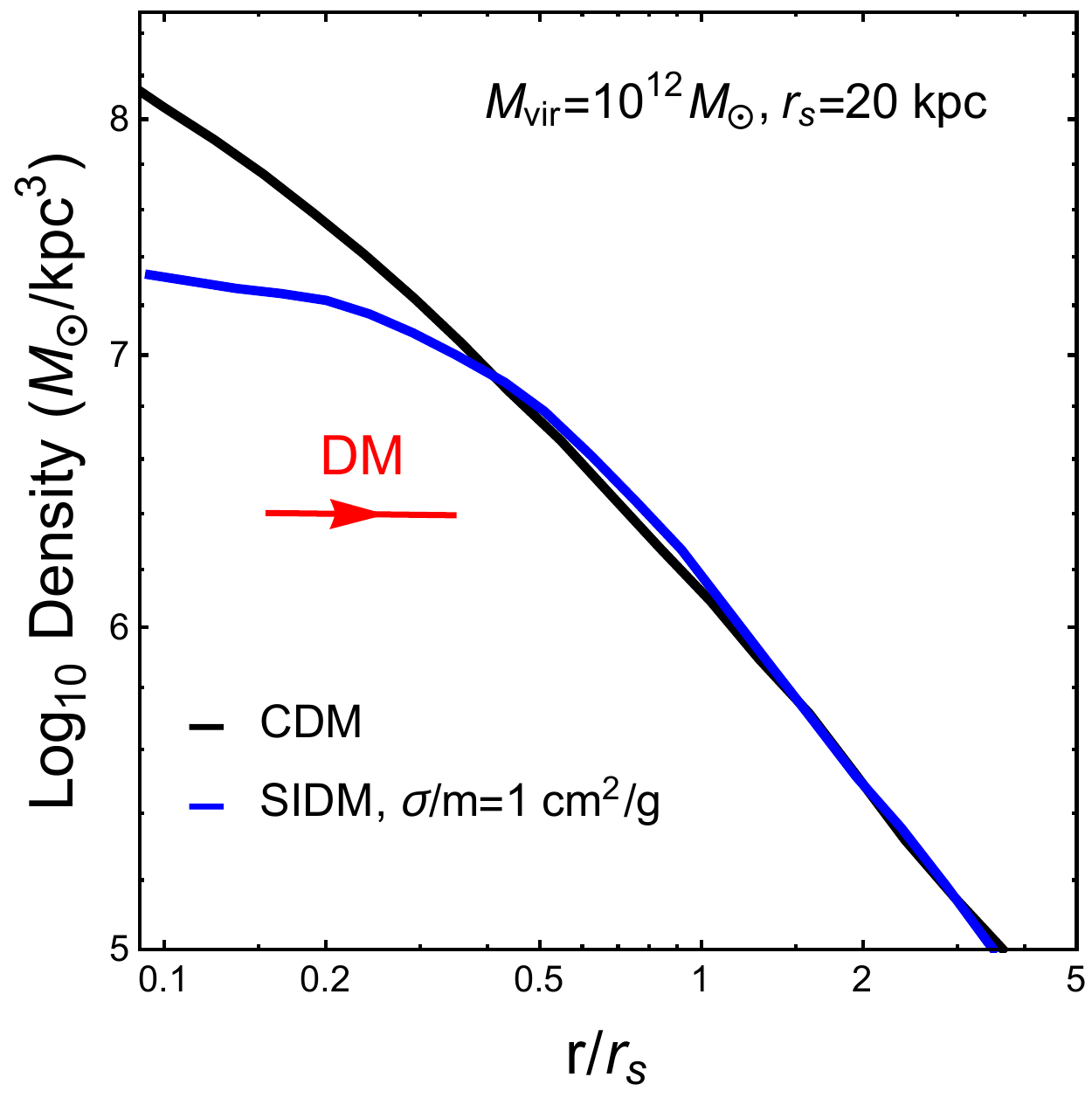}
\includegraphics[scale=0.42]{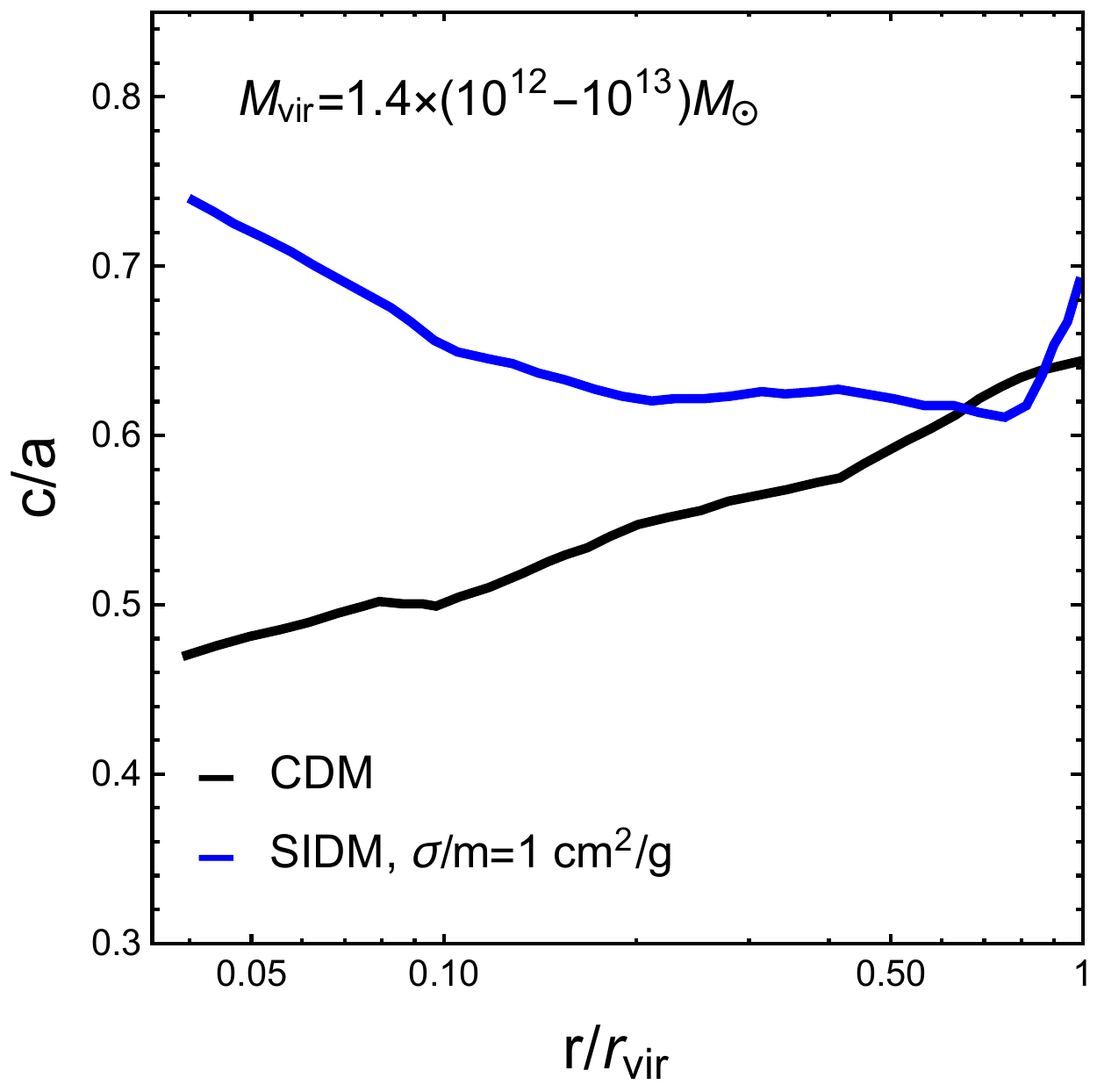}
\caption{\it Left: Density profiles (left), dispersion profiles (center), and median halo shapes (right) for SIDM with $\sigma/m=1\;{\rm cm^2/g}$ and its CDM counterpart. DM self-interactions cause heat transfer from the hot outer region to the cold inner region of a CDM halo and kinetically thermalize the inner halo, leading to a shallower density profile and a more spherical halo shape. Simulation data from Ref.~\cite{Rocha:2012jg,Peter:2012jh}.}
\label{fig:diff}
\end{figure}

Another promising alternative to collisionless CDM is self-interacting dark matter (SIDM), proposed by Spergel \& Steinhardt to solve the core-cusp and missing satellites problems~\cite{Spergel:1999mh}.  In this scenario, DM particles scatter elastically with each other through $2 \to 2$ interactions. Self-interactions lead to radical deviations from CDM predictions for the inner halo structure, shown in Fig.~\ref{fig:diff}.  We summarize the expectations for SIDM halos (blue) compared to CDM halos (black) as follows:
\begin{itemize}
\item {\it Isothermal velocity dispersion:} Although a CDM halo is a virilized object, the DM velocity dispersion, indicating the ``temperature" of DM particles, is not a constant and decreases towards the center in the inner halo.  Self-interactions transport heat from the hotter outer to the cooler inner region of a DM halo, thermalizing the inner halo and driving the velocity dispersion to be uniform with radius (Fig.~\ref{fig:diff}, left panel).  The velocity distribution function for SIDM becomes more Maxwell-Boltzmann compared to CDM~\cite{Vogelsberger:2012sa}.
\item {\it Reduced central density:} Hierarchical structure formation leads to a universal density profile for CDM halos~\cite{Navarro:1995iw,Navarro:1996gj}.  In the presence of collisions, the central density is reduced as low-entropy particles are heated within the dense inner halo, turning a cusp into a core (Fig.~\ref{fig:diff}, center panel).  
\item {\it Spherical halo shape:} While CDM halos are triaxial~\cite{Dubinski:1991bm}, collisions isotropize DM particle velocities and tend to erase ellipticity.  The minor-to-major axis ratio $c/a$ is closer to unity toward the center of SIDM halos compared to CDM halos (Fig.~\ref{fig:diff}, right panel).
\end{itemize}
Since the scattering rate is proportional to the DM density, SIDM halos have the same structure as CDM halos at sufficiently large radii where the collision rate is negligible.

The local collision rate is given by
\beq
R_{\rm scat} = \sigma v_{\rm rel} \rho_{\rm dm}   /m \approx 0.1 \; {\rm Gyr}^{-1} \times  \Big( \frac{\rho_{\rm dm}}{0.1 \; \Msun/{\rm pc}^3} \Big) \Big( \frac{v_{\rm rel}}{50 \; {\rm km/s}} \Big) \Big( \frac{\sigma /m }{1 \; {\rm cm^2/g}} \Big)   \, ,  \label{eq:rate}
\eeq
where $m$ is the DM particle mass, while $\sigma, v_{\rm rel}$ are the cross section and relative velocity, respectively, for scattering.  Within the central region of a typical dwarf galaxy, we have $\rho_{\rm dm} \sim 0.1 \; \Msun/{\rm pc}^3$ and $v_{\rm rel} \sim 50 \; {\rm km/s}$~\cite{Oh:2010ea}.  Therefore, the cross section per unit mass must be at least
\beq
\sigma/m \sim 1 \; {\rm cm^2/g} \approx 2 \times 10^{-24} \; {\rm cm^2/GeV} \label{eq:sigmam}
\eeq
to have an effect on the halo, corresponding to at least one scattering per particle over 10 Gyr galactic timescales.  For $\sigma/m\sim1\; {\rm cm^2/g}$, the mean free path of DM particles is larger than the core radii (Knudsen number larger than unity) and heat conduction is effective in the inner halo.  Provided $\sigma/m$ is not dramatically larger than this value, $R_{\rm scat}$ is negligible during the early Universe when structure forms.\footnote{Although self-interactions may be active in the {\it very} early Universe, long before matter-radiation equality, they rapidly fall out of equilibrium due to the Hubble dilution and redshifting of DM particles.}  Therefore SIDM retains the success of large-scale structure formation from $\Lambda$CDM, affecting structure at late times and only on small scales in the dense inner regions of halos.

Cored density profiles lead to shallower rotation curves for dwarf and LSB spiral galaxies at small radii, in accord with observations, while cores in satellite galaxies ameliorate the too-big-to-fail problem by reducing the predicted stellar line-of-sight velocity dispersion for the largest subhalos.   If $\sigma/m$ is fixed as in Eq.~\eqref{eq:sigmam}, the effect of self-interactions on the halo scales approximately in a self-similar fashion, and larger halos, such as those for massive elliptical galaxies and clusters, may be impacted by self-interactions at proportionally larger radii.

In principle, self-interactions can also affect substructure, reducing the subhalo mass function to solve the missing satellites problem.  SIDM subhalos are prone to tidal disruption, being less concentrated than their CDM counterparts, as well as evaporation due to ram pressure stripping from the host halo~\cite{Spergel:1999mh}.  However, this mechanism requires a too-large cross section that is excluded for reasons we discuss in \S\ref{sec:sidmhalos}.

The effect of self-interactions on DM halos has been borne out through N-body simulations.  Soon after Spergel \& Steinhardt's proposal, a first wave of SIDM simulations was performed~\cite{Moore:2000fp,Yoshida:2000bx,Burkert:2000di,Kochanek:2000pi,Yoshida:2000uw,Dave:2000ar,Colin:2002nk}.  In conjunction with these simulations, a number of constraints on SIDM emerged, the most stringent of which were strong lensing measurements of the ellipticity and central density of cluster MS2137-23~\cite{MiraldaEscude:2000qt,Meneghetti:2000gm}.  These studies limited the cross section to be below $0.1 \; {\rm cm^2/g}$, which excludes self-interactions at a level to explain small scale issues in galaxies.  The Bullet Cluster provides additional evidence for the collisionless nature of DM, requiring $\sigma/m \lesssim 1\; {\rm cm^2/g}$~\cite{Randall:2007ph}.  Although ad hoc velocity dependencies were put forth to evade these constraints~\cite{Yoshida:2000uw,Firmani:2000ce,Firmani:2000qe,Colin:2002nk}, SIDM largely fell into disfavor in light of these difficulties.

More recently, new N-body simulations for SIDM, with dramatically higher resolution and halo statistics, have revived DM self-interactions~\cite{Vogelsberger:2012ku,Rocha:2012jg,Peter:2012jh,Zavala:2012us,Elbert:2014bma,Vogelsberger:2014pda,Fry:2015rta,Dooley:2016ajo}.  Many of the constraints from larger scales are much weaker than previously thought~\cite{Rocha:2012jg,Peter:2012jh}.  In particular, constraints based on ellipticity have been overestimated: self-interactions do not erase triaxiality as effectively as previously supposed, and moreover, an observed ellipticity has contributions along the line-of-sight from regions outside the core where the halo remains triaxial~\cite{Peter:2012jh}.  The conclusion is that $\sigma/m\sim 0.5 - 1 \; {\rm cm^2/g}$ can solve the core-cusp and TBTF issues on small scales, while remaining approximately consistent with other astrophysical constraints on larger scales~\cite{Rocha:2012jg,Peter:2012jh,Zavala:2012us,Elbert:2014bma}.  However, more recent studies based on stacked merging clusters~\cite{Harvey:2015hha} and stellar kinematics within cluster cores~\cite{Kaplinghat:2015aga} suggest some tension with these values.  Viable SIDM models are preferred to have a scattering cross section with a mild velocity-dependence from dwarf to cluster scales.  We discuss these issues in further detail in \S\ref{sec:sidmhalos}.

On the theory side, a new semi-analytical SIDM halo model has been developed based on the Jeans equation~\cite{Kaplinghat:2013xca,Kaplinghat:2015aga,Kamada:2016euw}.  It can reproduce the simulation results for SIDM profiles within 10--20\% while being much cheaper computationally.  Discussed in \S\ref{sec:jeans}, this approach provides insight for understanding the baryonic influence on SIDM halo properties~\cite{Kaplinghat:2013xca}, testing SIDM models from dwarf to cluster scales~\cite{Kaplinghat:2015aga}, and addressing the diversity in rotation curves~\cite{Kamada:2016euw}. 

Furthermore, there has been important progress in particle physics models for SIDM. Both numerical and analytical methods have been developed to accurately calculate the cross section for SIDM models involving the Yukawa~\cite{Feng:2009hw,Tulin:2013teo,Schutz:2014nka} or atomic interactions~\cite{Cline:2013pca,Boddy:2016bbu}.  These studies make it possible to map astrophysical constraints on $\sigma/m$ to the particle model parameters, such as the DM and mediator masses and coupling constant.  In addition, an effective theory approach has been proposed in parametrizing SIDM models with a set of variables that are directly correlated with astrophysical observations~\cite{Cyr-Racine:2015ihg,Vogelsberger:2015gpr}. 

\begin{table}[t] 
\begin{tabular}{l|c|c|l|l} 
\hline
{\bf Positive observations} & $\sigma/m$ & $v_{\rm rel}$ & \multicolumn{1}{|c|}{Observation} & Refs. \\
\hline
\hline
Cores in spiral galaxies & $\gtrsim 1 \; {\rm cm^2/g} $ & $30-200 \;  {\rm km/s}$ & Rotation curves &  \cite{Dave:2000ar,Kaplinghat:2015aga} \\
 (dwarf/LSB galaxies) & & & \\
\hline
Too-big-to-fail problem & & & \\
Milky Way & $\gtrsim 0.6 \; {\rm cm^2/g}$ &  $50 \; {\rm km/s}$ & Stellar dispersion  & \cite{Zavala:2012us} \\
Local Group & $\gtrsim 0.5 \; {\rm cm^2/g}$ &  $50 \; {\rm km/s}$ & Stellar dispersion  &  \cite{Elbert:2014bma}  \\
\hline
Cores in clusters & $\sim 0.1 \; {\rm cm^2/g}$ & $1500 \; {\rm km/s}$ & Stellar dispersion, lensing & \cite{Kaplinghat:2015aga,Elbert:2016dbb} \\
\hline
{\it Abell 3827 subhalo merger} & $\sim 1.5 \; {\rm cm^2/g}$ & $1500 \; {\rm km/s}$ & DM-galaxy offset & \cite{Kahlhoefer:2015vua} \\
\hline
{\it Abell 520 cluster merger} & $\sim 1 \; {\rm cm^2/g}$ & $2000-3000 \; {\rm km/s}$ & DM-galaxy offset  & \cite{Jee:2014hja,Kahlhoefer:2013dca,Sepp:2016tfs} \\
\hline
\hline
\multicolumn{5}{l}{\bf Constraints } \\
\hline
\hline
Halo shapes/ellipticity & $\lesssim 1 \; {\rm cm^2/g}$ & $1300 \; {\rm km/s}$ & Cluster lensing surveys & \cite{Peter:2012jh}\\
\hline
Substructure mergers & $\lesssim 2 \; {\rm cm^2/g}$ & $\sim 500-4000 \; {\rm km/s}$ & DM-galaxy offset & \cite{Harvey:2015hha,Wittman:2017gxn} \\
\hline
Merging clusters & $\lesssim {\rm few} \; {\rm cm^2/g}$ & $2000-4000 \; {\rm km/s}$ & Post-merger halo survival & Table~\ref{tab:mergers}\\
& & & (Scattering depth $\tau < 1$) & \\
{\it Bullet Cluster} & $\lesssim 0.7 \; {\rm cm^2/g}$ & $4000 \; {\rm km/s}$ & Mass-to-light ratio & \cite{Randall:2007ph}\\
\hline
\end{tabular}
\caption{\label{tab:constraints} Summary of positive observations and constraints on self-interaction cross section per DM mass. Italicized observations are based on {\it single individual systems}, while the rest are derived from sets of multiple systems.  Limits quoted, which assume constant $\sigma/m$, may be interpreted as a function of collisional velocity $v_{\rm rel}$ provided $\sigma/m$ is not steeply velocity-dependent. References noted here are limited to those containing quoted self-interaction cross section values.  Further references, including original studies of observations, are cited in the corresponding sections below.
}
\end{table}

In Table~\ref{tab:constraints}, we summarize the present status of astrophysical observations related to SIDM .  The positive observations indicate discrepancies with CDM-only simulations and the required cross section assuming self-interactions are responsible for solving each issue.  Dwarf and LSB galaxies favor cross sections of at least $1 \; {\rm cm^2/g}$ to produce large enough core radii in these systems, which is also consistent with alleviating the too-big-to-fail problem among MW satellites and field dwarfs of the Local Group.  The cross section need not be particularly fine-tuned. Values as large as $50 \; {\rm cm^2/g}$ provide consistent density profiles on dwarf scales~\cite{Elbert:2014bma}.  (The upper limit on $\sigma/m$ on dwarf scales due to the onset of gravothermal collapse, which we discuss in \S\ref{sec:sidmhalos}, is not well-known.)  

Lastly, for completeness, let us mention that SIDM, in its original conception, was introduced much earlier by Carlson, Machacek \& Hall~\cite{Carlson:1992fn} to modify large scale structure formation.  In their scenario, DM particles (e.g., glueballs of a nonabelian dark sector) have $3 \to 2$ (or $4 \to 2$) number-changing interactions by which they annihilate and eventually freeze-out to set the relic density, as well as $2 \to 2$ elastic self-interactions that maintain kinetic equilibrium.  Unlike standard freeze-out~\cite{Scherrer:1985zt}, DM particles retain entropy as their number is depleted and therefore cool more slowly than the visible sector.  When this theory was proposed, a mixture of both cold plus hot DM within a flat, matter-dominated Universe ($\Lambda=0$) was a viable and theoretically appealing alternative to $\Lambda$CDM~\cite{Davis:1992ui,Klypin:1992sf,Primack:1997av}. Although this original SIDM was proposed as a hybrid between hot plus cold DM, it provided too much small scale power suppression in Lyman-$\alpha$ observations relative to larger scales and was found to be excluded~\cite{deLaix:1995vi}.\footnote{We emphasize that this exclusion is for a {\it matter}-dominated cosmology.  Recently, this SIDM scenario has been revived within an observationally consistent (i.e., $\Lambda$-dominated) cosmology~\cite{Hochberg:2014dra}, discussed in \S\ref{sec:models}.}


\subsection{From astrophysics to particle theory}

The figure of merit for self-interactions, $\sigma/m$, depends on the underlying DM particle physics model.  WIMPs have self-interactions mediated through the weak force, Higgs boson, or other heavy states.  Since WIMP interactions and masses are set by the weak scale, yielding $\sigma/m \sim 10^{-38} \; {\rm cm^2/GeV}$, they effectively behave as collisionless CDM.  If self-interactions indeed explain the small scale issues, then DM cannot be a usual WIMP.

\begin{figure}
\includegraphics[scale=.8]{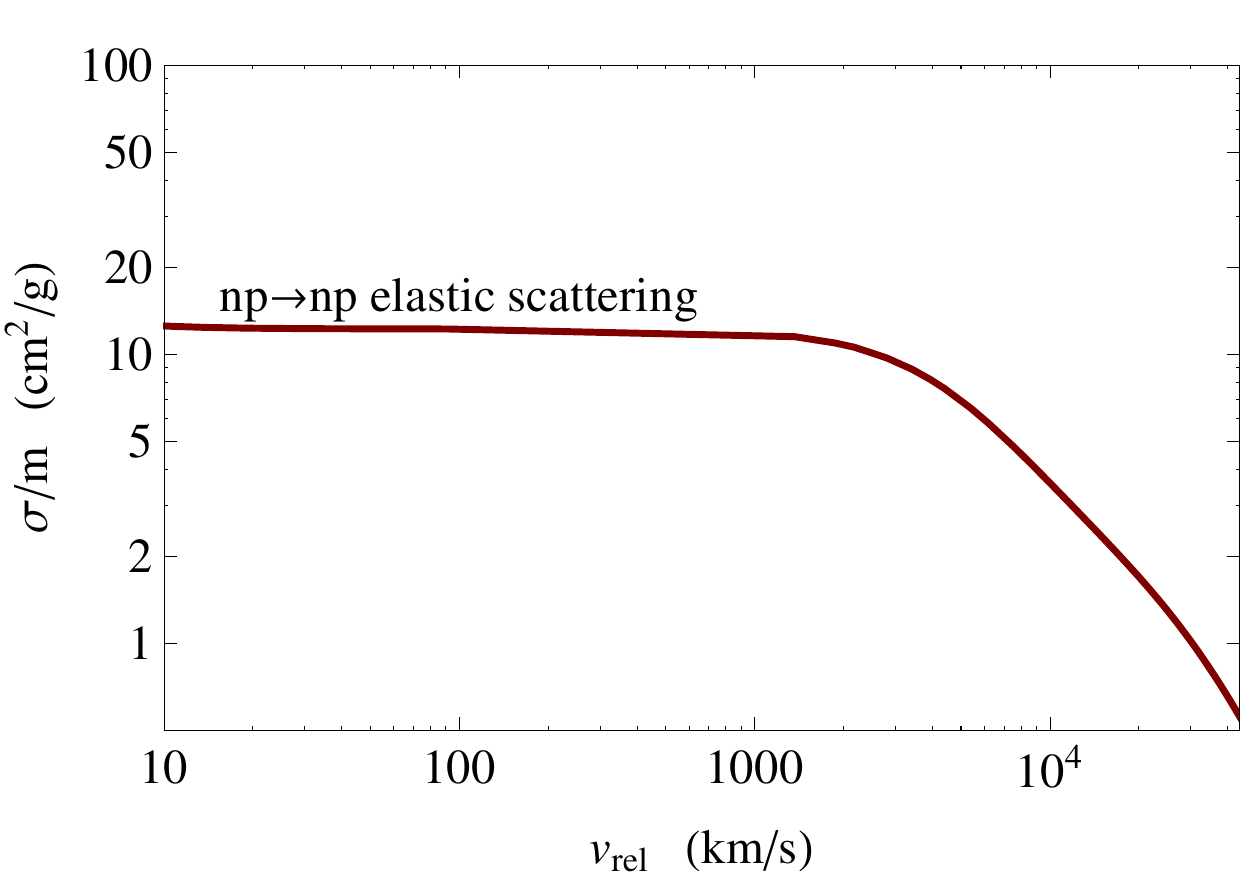}
\caption{ \it As an analogy to SIDM, we show $\sigma/m$ for ``nucleon self-interactions'' (neutron-proton elastic scattering) as a function of scattering velocity $v_{\rm rel}$.  Nuclear interactions are similar in magnitude to what is required for DM self-interactions to explain small scale structure issues.   
}
\label{fig:npscat}
\end{figure}

What sort of particle theory can have a large enough $\sigma/m$?  An analogy is provided by nuclear interactions, mediated by pion exchange.  In Fig.~\ref{fig:npscat}, we show $n$-$p$ elastic scattering data for $n$ kinetic energies 0.5 eV $-$ 10 MeV~\cite{CHADWICK20112887}, expressed in terms of $v_{\rm rel}$ and $\sigma/m$ (here, $m$ is the nucleon mass).  The required cross section for SIDM is comparable in magnitude to nuclear cross sections for visible matter.  The lesson here is that $1 \; {\rm cm^2/g}$ or larger can be achieved if the interaction scale lies below $\sim1$ GeV.  However, unlike nuclear scattering, the theory of self-interactions need not be strongly-coupled, nor does the DM mass need to be below 1 GeV.  For example, self-interactions can be a weakly-coupled dark force~\cite{Ackerman:2008gi,Feng:2009mn,Feng:2009hw,Buckley:2009in,Loeb:2010gj,Tulin:2012wi,Tulin:2013teo}, with the mediator particle denoted by $\phi$.  A perturbative calculation gives (in the limit $v_{\rm rel} = 0$)
\beq
\sigma/m = \frac{4 \pi \alpha^{\prime 2} m}{m_\phi^4}  \approx 1 \; {\rm cm^2/g} \times 
\Big(\frac{\alpha^\prime }{0.01} \Big)^2 \Big(\frac{m}{10  \; {\rm GeV}} \Big) \Big(\frac{m_\phi }{40 \; {\rm MeV} } \Big)^{-4} \label{eq:born}
\eeq
where $\alpha^\prime$ is the DM analog of the electromagnetic fine structure constant, $\alpha_{\rm em} \approx 1/137$.  Self-interactions that are electomagnetic strength (or weaker) are sufficient, as are weak-scale DM masses, provided the mediator mass $m_\phi$ is light enough. Note that Eq.~(\ref{eq:born}) is only valid in the weakly-coupled Born limit, $\alpha^\prime m/\mphi\ll1$, and there are important corrections outside of this regime~\cite{Tulin:2013teo}.

Another important point is that $\sigma$ need not be constant in velocity.  DM particles in larger mass halos have larger characteristic velocities compared to those in smaller halos.  Therefore, observations from dwarf to cluster scales provide complementary handles for constraining the velocity-dependence of $\sigma$.  Along these lines, Refs.~\cite{Yoshida:2000uw,Firmani:2000ce,Firmani:2000qe,Colin:2002nk} advocated a cross section of the form $\sigma \propto 1/v_{\rm rel}$ in order to evade cluster constraints and fit a constant central density across all halo scales.  We caution that such a dependence is not motivated by any theoretic framework for SIDM.  

On the other hand, many well-motivated particle physics scenarios predict velocity-dependent cross sections that are not described by simple power laws.  Scattering through a dark force can be a contact-type interaction at small velocity, with constant $\sigma$; yet once the momentum transfer is larger than the mediator mass, scattering is described by a Rutherford-like cross section that falls with higher velocity.  These models naturally predict a larger $\sigma/m$ on dwarf scales compared to clusters, depending on the model parameters.  The velocity-dependence may be qualitatively similar to $n$-$p$ scattering, shown in Fig.~\ref{fig:npscat}, or it can have a more complicated form~\cite{Tulin:2012wi,Tulin:2013teo}.  We discuss models for SIDM and their expected velocity-dependence in \S\ref{sec:models}.  

The particle physics of SIDM is not just about self-interactions.  While searching for their effect on structure is the only probe if DM is {\it completely} decoupled from the SM, this is unlikely from an effective field theory point of view.  For example, a dark force can interact with the SM through hypercharge kinetic mixing~\cite{Holdom:1985ag} if it is a vector or the Higgs portal operator~\cite{Patt:2006fw} if it is a scalar.  Such couplings allow mediator particles thermally produced in the early Universe to decay, avoiding overclosure~\cite{Kaplinghat:2013yxa}.  At the same time, these operators provide a window to probing SIDM in direct and indirect detection experiments, as well as collider searches.  Similar to the WIMP paradigm, SIDM motivates a multifaceted program of study combining complementary data from both astrophysical and terrestrial measurements.  Though the coupling between the dark and visible sectors must be much weaker than for WIMPs, this can be compensated by the fact that mediator states must be much lighter than the weak scale.  We discuss in \S\ref{sec:comp} these complementary searches for DM (and other dark sector states) within the context of the SIDM paradigm.



\section{Astrophysical observations}
\label{sec:astro}

The kinematics of visible matter is a tracer of the gravitational potential in galaxies and clusters, allowing the underlying DM halo mass distribution to be inferred.  Observations along these lines have pointed to the breakdown of the collisionless CDM paradigm on small scales.   In this section, we review these astrophysical observations and discuss possible solutions from baryonic physics and other systematic effects.  For other recent reviews of these issues, we direct the reader to Refs.~\cite{deBlok:2009sp,2010arXiv1009.4505B,Weinberg:2013aya,DelPopolo:2016emo}.


The issues discussed here may share complementary solutions.  For example, mechanisms that generate cored density profiles may help reconcile the too-big-to-fail problem by reducing central densities of MW subhalos, as well as accommodating the diversity of galactic rotation curves.  On the other hand, the too-big-to-fail problem may share a common resolution with the missing satellites problem if the overall subhalo mass function is reduced.  

\subsection{Core-Cusp Problem}

Collisionless CDM-only simulations predict ``cuspy'' DM density profiles for which the logarithmic slope, defined by $\alpha = d\ln \rho_{\rm dm}/d\ln r$, tends to $\alpha \sim -1$ at small radii~\cite{Dubinski:1991bm,Navarro:1995iw,Navarro:1996gj,Moore:1999gc}.  Such halos are well-described by the Navarro-Frenk-White (NFW) profile~\cite{Navarro:1995iw,Navarro:1996gj},
\begin{eqnarray}
\rho_{\rm NFW}(r)=\frac{\rho_s}{(r/r_s)(1+r/r_s)^2} \, ,
\end{eqnarray}
where $r$ is the radial coordinate and $\rho_s$ and $r_s$ are characteristic density and scale radius of the halo, respectively.\footnote{High resolution simulations have found that collisionless CDM profiles become shallower than $\alpha \sim -1$ at small radii~\cite{Navarro:2008kc,Stadel:2008pn}, in better agreement with the Einasto profile~\cite{1965TrAlm...5...87E}.  However, the enclosed mass profile is slightly larger at small radii compared to NFW fits~\cite{Navarro:2008kc}, further exacerbating the issues discussed here.}  On the other hand, as we discuss below, many observations do not find evidence for the steep inner density slope predicted for collisionless CDM, preferring ``cored'' profiles with inner slopes $\alpha \sim 0$ that are systematically shallower~\cite{Flores:1994gz,Moore:1994yx}.  This discrepancy is known as the ``core-cusp problem.''  

A related issue, known as the ``mass deficit problem,'' emerges when observed halos are viewed within a cosmological context~\cite{McGaugh:1998tq,deBlok:2001rgg,McGaugh:2006vv,Oman:2015xda}.  
The mass-concentration relation predicted from cosmological CDM simulations implies a tight correlation between $\rho_s$ and $r_s$ (see, e.g., Refs.~\cite{Ludlow:2013vxa,Dutton:2014xda,Rodriguez-Puebla:2016ofw}).  Since cosmological NFW halos are essentially a single-parameter profile (up to scatter), determination of the halo at large radius fixes the halo at small radius.  However, many observed systems have less DM mass at small radii compared to these expectations.  Alternatively, if NFW halos are fit to data at {\it both} large and small radii, the preferred profiles tend to be less concentrated than expected cosmologically.

\vspace{2mm}

\underline{ Rotation curves:} Late-type dwarf and LSB galaxies are ideal laboratories for halo structure.  These systems are DM dominated down to small radii (or over all radii) and environmental disturbances are minimized.  Flores \& Primack~\cite{Flores:1994gz} and Moore~\cite{Moore:1994yx} first recognized the core-cusp issue based on \HI rotation curves for several dwarfs, which, according to observations, are well described by cored profiles~\cite{1988ApJ...332L..33C,1990AJ.....99..547L,1990AJ....100..648J}.\footnote{In these early studies, the Hernquist profile~\cite{Hernquist:1990be} was used to model collisionless CDM halos.  It has the same behavior as the NFW profile in the inner regions, $\rho_{\rm dm}\propto r^{-1}$, but falls as $r^{-4}$ at large radii.  On the other hand, a variety of cored profiles have been adopted in the literature, including the nonsingular isothermal profile~\cite{1926ics..book.....E}, Burkert profile~\cite{1995ApJ...447L..25B}, and pseudo-isothermal profile $\rho_{\rm dm}(r) = \rho_0 (1+r^2/r_c^2)^{-1}$, where $\rho_0$ and $r_c$ are the core density and radius, respectively~\cite{1985ApJ...295..305V}.  }

Indeed, rotation curve measurements for dwarfs and LSBs have been a long-standing challenge to the $\Lambda$CDM paradigm~\cite{1995ApJ...447L..25B,McGaugh:1998tq,2000AJ....120.3027C,vandenBosch:2000rza,Borriello:2000rv,deBlok:2001mf,deBlok:2001fe,Marchesini:2002vm,Gentile:2005de,Gentile:2006hv,KuziodeNaray:2006wh,KuziodeNaray:2007qi,Salucci:2007tm}.
For axisymmetric disk galaxies, circular velocity can be decomposed into three terms
\beq \label{eq:vcirc}
V_{\rm circ}(r)=\sqrt{V_{\rm halo}(r)^2+ \Upsilon_* V_{\rm star}(r)^2+V_{\rm gas}(r)^2}  \, ,
\eeq
representing the contributions to the rotation curve from the DM halo, stars, and gas, respectively.  The baryonic contributions to the rotation curve are modeled from the respective surface luminosities of stars and gas.  However, the overall normalization between stellar mass and light remains notoriously uncertain: stellar mass is dominated by smaller and dimmer stars, while luminosity is dominated by more massive and brighter stars.  Estimates for the stellar mass-to-light ratio---denoted by $\Upsilon_*$ in Eq.~\eqref{eq:vcirc}---rely on stellar population synthesis models and assumptions for the initial mass function, with uncertainties at the factor-of-two level~\cite{Conroy:2013if}.  Modulo this uncertainty, the DM profile can be fit to observations.  For a spherical halo, the DM contribution to the rotation curve is $V_{\rm halo}(r)=\sqrt{G M_{\rm halo}(r)/r}$, where $G$ is Newton's constant and $M_{\rm halo}(r)$ is the DM mass enclosed within $r$.  

\begin{figure}
\includegraphics[scale=0.6]{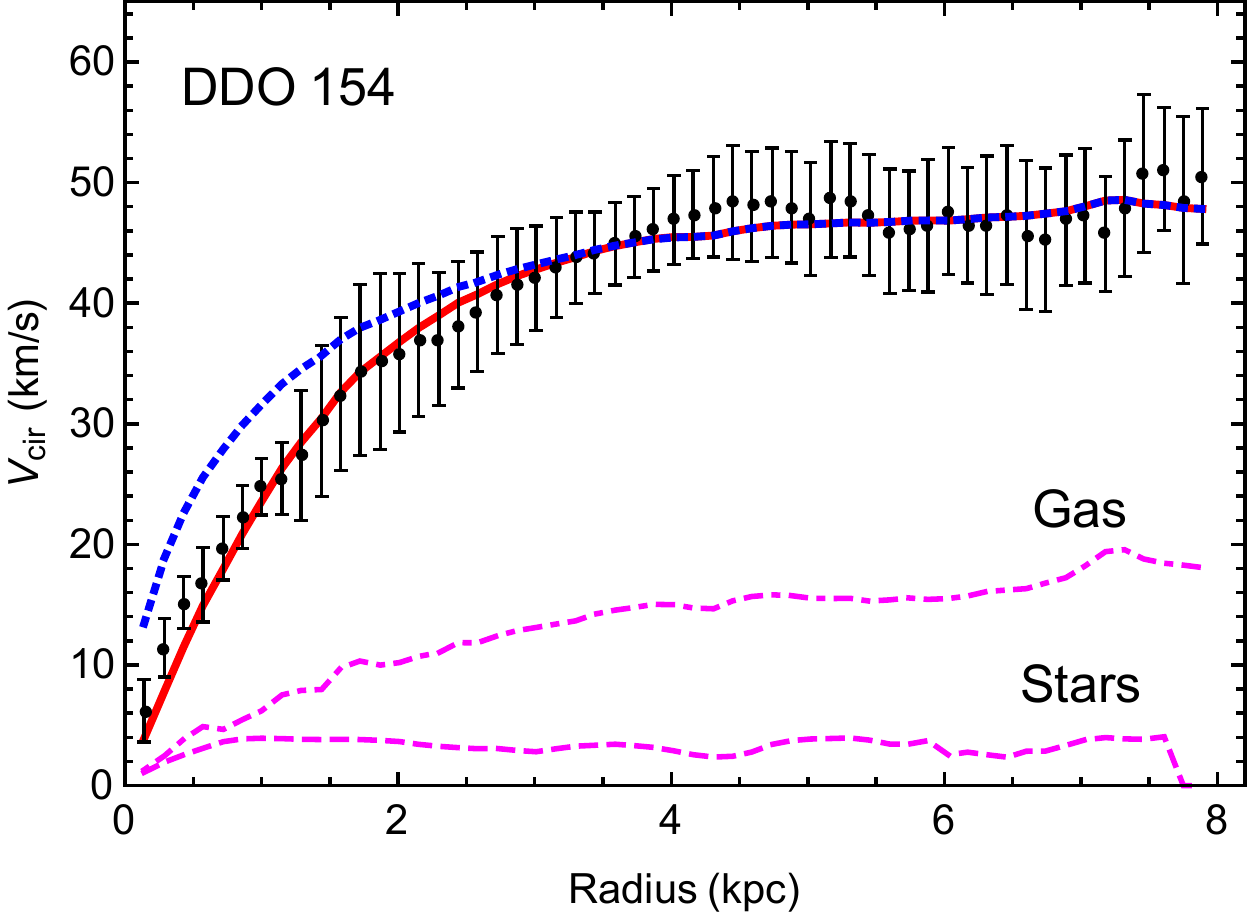}\;\;\;\;\;\;\includegraphics[scale=0.61]{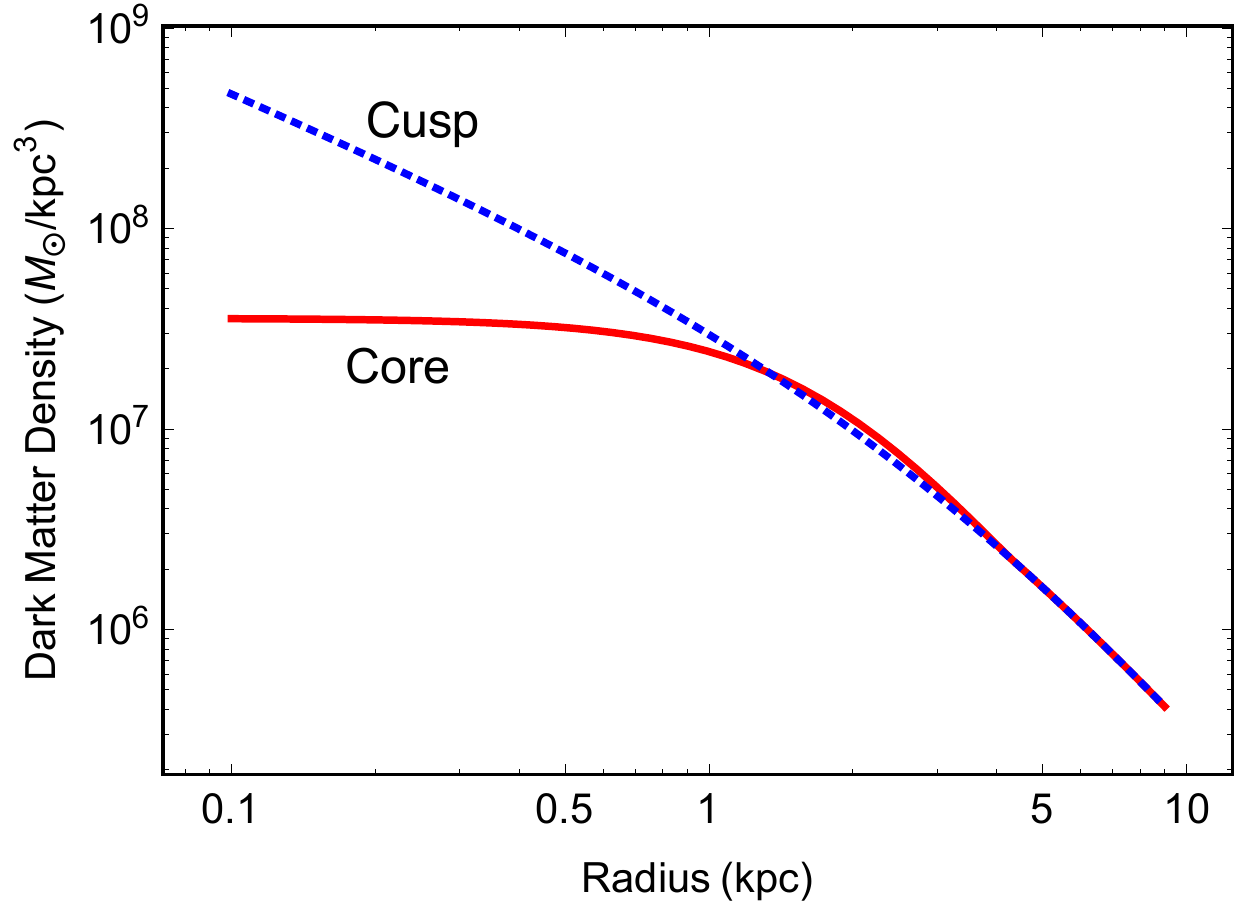}
\caption{\it Left: Observed rotation curve of dwarf galaxy DDO 154 (black data points)~\cite{Oh:2015xoa} compared to models with an NFW profile (dotted blue) and cored profile (solid red).  Stellar (gas) contributions indicated by pink (dot-)dashed lines.
Right: Corresponding DM density profiles adopted in the fits.  NFW halo parameters are $r_s\approx3.4~{\rm kpc}$ and $\rho_s\approx1.5\times10^{7} \, \Msun/{\rm kpc^3}$, while the cored density profile is generated using an analytical SIDM halo model developed in~\cite{Kaplinghat:2015aga,Kamada:2016euw}.
}
\label{fig:ddo154}
\end{figure}

Fig.~\ref{fig:ddo154} illustrates these issues for dwarf galaxy DDO 154. The left panel shows the measured \HI rotation curve~\cite{Oh:2015xoa} compared to fits with cuspy (NFW) and cored profiles, which are shown in the right panel.  The NFW halo has been chosen to fit the asymptotic velocity at large radii and match the median cosmological relation between $\rho_s$ and $r_s$~\cite{Kamada:2016euw}.  However, this profile overpredicts $V_{\rm circ}$ in the inner region.  This discrepancy is a symptom of too much mass for $r\lesssim2~{\rm kpc}$, while the data favors a shallower cored profile with less enclosed mass.  An NFW profile with alternative parameters can provide an equally good fit as the cored profile, but the required concentration is significantly smaller than preferred cosmologically~\cite{Oh:2015xoa}. We note not all dwarf galaxies require a density core, as we we will discuss later.

Recent high-resolution surveys of nearby dwarf galaxies have given further weight to this discrepancy.  The \HI Near Galaxy Survey (THINGS) presented rotation curves for seven nearby dwarfs, finding a mean inner slope $\alpha=-0.29\pm0.07$~\cite{Oh:2010ea}, while a similar analysis by LITTLE THINGS for 26 dwarfs found $\alpha=-0.32\pm0.24$~\cite{Oh:2015xoa}.  These results stand in contrast to $\alpha \sim -1$ predicted for CDM.  

However, this discrepancy may simply highlight the inadequacy of DM-only simulations to infer the properties of real galaxies containing both DM and baryons.  One proposal along these lines is that supernova-driven outflows can potentially impact the DM halo gravitationally, softening cusps~\cite{Navarro:1996bv,Oh:2010mc}, which we discuss in further detail in \S\ref{sec:astro} E.  Alternatively, the inner mass density in dwarf galaxies may be systematically underestimated if gas pressure---due to turbulence in the interstellar medium---provides radial support to the disk~\cite{Dalcanton:2010bp,2017MNRAS.466...63P}.  In this case, the observed circular velocity will be smaller than needed to balance the gravitational acceleration, as per Eq.~\eqref{eq:vcirc}, and purported cores may simply be an observational artifact.

In light of these uncertainties, LSB galaxies have become an attractive testing ground for DM halo structure.  A variety of observables---low metallicities and star formation rates, high gas fractions and mass-to-light ratios, young stellar populations---all point to these galaxies being highly DM-dominated and having had a quiescent evolution~\cite{deBlok:1997zlw}.  Moreover, LSBs typically have larger circular velocities and therefore deeper potential wells compared to dwarfs.  Hence, the effects of baryon feedback and pressure support are expected to be less pronounced.

Rotation curve studies find that cored DM profiles are a better fit for LSBs compared to cuspy profiles~\cite{McGaugh:1998tq,deBlok:2001mf,deBlok:2001fe,KuziodeNaray:2006wh,KuziodeNaray:2007qi}. In some cases, NFW profiles can give reasonable fits, but the required halo concentrations are systematically lower than the mean value predicted cosmologically.  Although early \HI and long-slit H$\alpha$ observations carried concerns that systematic effects---limited resolution (beam-smearing), slit misalignment, halo triaxiality and noncircular motions---may create cores artificially, these issues have largely been put to rest with the advent of high-resolution \HI and optical velocity fields (see Ref.~\cite{deBlok:2009sp} and references therein).  Whether or not baryonic feedback can provide the solution remains actively debated~\cite{deNaray:2011hy,Oman:2015xda,Katz:2016hyb,Pace:2016oim}.  Cored DM profiles have been further inferred for more luminous spiral galaxies as well~\cite{Gentile:2004tb,Salucci:2007tm,Donato:2009ab}.  

Although observational challenges remain in interpreting the very inner regions of galaxies, other studies have shown that small scale issues persist out to larger halo radii as well.  McGaugh et al.~\cite{McGaugh:2006vv} examined rotation curves restricted to $r\geq 1$ kpc for 60 disk galaxies spanning a large range of mass and types.  Under plausible assumptions for $\Upsilon_*$, the inferred halo densities are systematically lower than predicted for CDM.  Gentile et al.~\cite{Gentile:2007sb} analyzed rotation curves of 37 spiral galaxies.  For each galaxy, an NFW halo is identified by matching the total enclosed DM mass with the observed one at the last measured rotation curve point.  The resulting NFW halos are overdense in the central regions, but less dense in the outer regions, compared to the DM distribution inferred observationally.   These studies support the picture that cosmological NFW halos tend to be too concentrated to fit observed rotation curves.

\vspace{2mm}

\underline{Milky Way satellites:} Stellar kinematics for dwarf spheroidal (dSph) galaxies around the MW support the existence of cored DM density profiles.  These DM-dominated galaxies---unlike disk galaxies discussed above---are gas-poor, dispersion-supported systems in which the stellar kinematics are dominated by random motions.  Since available observations only encode half of the full 6D phase space information (i.e., two spatial dimensions in the plane of the sky and one velocity dimension along the line of sight), there is a well-known degeneracy between halo mass and stellar velocity anisotropy, given by the parameter
\beq
\beta_{\rm aniso} = 1 - \sigma^2_t/\sigma^2_r \, ,
\eeq
where $\sigma^2_{r,t}$ denotes the radial ($r$) and tangential ($t$) stellar velocity dispersions, which follows from the equilibrium Jeans equation.  Fortunately, however, for a wide range of halo models and anisotropies, the halo mass can be robustly estimated at the half-light radius, the radius enclosing half of the luminosity of the galaxy, with little scatter~\cite{Walker:2009zp,Wolf:2009tu}.

Further studies have exploited the fact that Sculptor and Fornax dSphs have two chemo-dynamically distinct stellar subpopulations (metal-rich and metal-poor)~\cite{Tolstoy:2004vu}.  Both populations trace the same underlying gravitational potential---but at different radii---effectively constraining the halo mass at two distinct radii and allowing the DM halo profile to be determined.  The analyses by Battaglia et al~\cite{Battaglia:2008jz}, based on the Jeans equation, and by Amorisco \& Evans~\cite{2012MNRAS.419..184A}, based on modeling stellar distribution functions, find that a cored DM profile provides a better fit to the velocity dispersion profiles of Sculptor compared to NFW.  A third method by Walker \& Pe\~{n}arrubia~\cite{Walker:2011zu} reached a similar conclusion for both Sculptor and Fornax.  Their method used a simple mass estimator to measure the slope of the mass profile, excluding a NFW profile at high significance.\footnote{Strigari et al.~\cite{Strigari:2014yea} performed a different analysis for the Sculptor galaxy and argued that an NFW profile is consistent with observations (see also Ref.~\cite{Breddels:2013qqh}). The discrepancy could be due to different assumptions about the anisotropy of the stellar velocity dispersions.}  With the recent identification of three distinct stellar subpopulations in Fornax, Amorisco et al.~\cite{2013MNRAS.429L..89A} further showed that the galaxy resides in a DM halo with a $\sim 1~{\rm kpc}$ constant density core, while an NFW halo is incompatible with observations unless its concentration is unrealistically low.

Observations of kinematically cold stellar substructures in dSphs lend indirect support to the presence of DM cores.  Kleyna et al.~\cite{Kleyna:2003zt} detected a stellar clump with low velocity dispersion in the Ursa Minor dSph and argued that its survival against tidal disruption is more likely in a cored---rather than cuspy---host halo. Walker et al.~\cite{Walker:2006qr} found a similar clump in the Sextans dSph, also implying a cored halo~\cite{Lora:2013fla}.  Lastly, the wide spatial distribution of five globular clusters in the Fornax dSph again favors a cored host halo~\cite{SanchezSalcedo:2006fa,Goerdt:2006rw,Read:2006fq,Cole:2012ns}, as dynamical friction within a cuspy halo would cause these clusters to sink to the center in less than a Hubble time. 

\vspace{2mm}

\underline{Galaxy clusters:} There is evidence that the core-cusp problem may be present for massive galaxy clusters as well.  Sand et al.~\cite{Sand:2002cz,Sand:2003bp} have advocated that the inner mass profile in relaxed clusters can be determined using stellar kinematics of the brightest cluster galaxy (BCG) that resides there.  In combination with strong lensing data, they find a mean inner slope $\alpha \approx - 0.5$ for the three best-resolved clusters in their sample, shallower than expected for CDM, albeit with considerable scatter between clusters~\cite{Sand:2003bp}.  Further studies by Newman et al.~\cite{Newman:2009qm,Newman:2011ip,Newman:2012nv,Newman:2012nw} have supported these findings.  A joint analysis of stellar BCG kinematics and lensing data for seven clusters with masses $\sim (0.4-2) \times 10^{15} \, \Msun$ has found that, while the {\it total} mass density profiles are consistent with NFW, the density profiles for DM {\it only} become softer than NFW for $r \lesssim$ 30 kpc~\cite{Newman:2012nv,Newman:2012nw}.  The halos are described equally well by a shallower inner slope $\alpha \approx -0.5$ or a cored profile with core radius $\sim 10-20$ kpc.  

Baryon dynamics from active galactic nuclei (AGN) can be important, similar to the effect of supernovae in smaller halos.  Hydrodynamical simulations of clusters by Martizzi et al.~\cite{Martizzi:2011aa,Martizzi:2012ci} have found that repeated outflows from gas accretion and ejection by AGN can produce a cored cluster halo.  On the other hand, weaker feedback prescriptions in simulations by Schaller et al.~\cite{Schaller:2014gwa}---argued to produce a more realistic stellar mass function for cluster galaxies---do not produce cored DM profiles and yield {\it total} mass profiles that are somewhat steeper than NFW, seemingly in tension with the Newman et al.~results~\cite{Newman:2012nv,Newman:2012nw}.  

Other systematic effects may also be relevant for reconciling this descrepency.  BCGs are dispersion-supported galaxies dominated by the stellar density in their interiors and there is a systematic degeneracy between $\Upsilon_*$ and $\beta_{\rm aniso}$ that is important when disentangling the DM and baryonic densities.  Schaller et al.~\cite{Schaller:2014gwa} argue that radially-biased $\beta_{\rm aniso} \approx 0.2-0.3$, as suggested by their simulations, can push their results into closer agreement with Refs.~\cite{Newman:2012nv,Newman:2012nw}.  On the other hand, Newman et al.~\cite{Newman:2009qm,Newman:2012nw} find that radially biased anisotropies only serve to make DM density profiles shallower in their fits.  In any case, the comparison is not strictly apples-to-apples since the Newman clusters~\cite{Newman:2012nv,Newman:2012nw} are up to a factor of five larger than the largest simulated ones by Schaller et al.~\cite{Schaller:2014gwa}.

\subsection{Diversity Problem}

In $\Lambda$CDM, hierarchical structure formation produces self-similar halos well-described by NFW profiles.  Since the halo parameters (e.g.,~$\rho_s$ and $r_s$) are strongly correlated, there is only one parameter specifying a halo.  For example, once the maximum circular velocity $V_{\rm max}$ (or any other halo parameter) is fixed, the halo density profile is completely determined at all radii including the inner density cusp (up to the scatter).  On the other hand, the inner rotation curves of observed galaxies exhibit considerable diversity.  Galaxies of the same $V_{\rm max}$ can have significant variation in their central densities.   Any mechanism to explain the core-cusp issue must also accommodate this apparent diversity.

\begin{figure}
\includegraphics[trim={0 3cm 0 4cm},clip,scale=0.35]{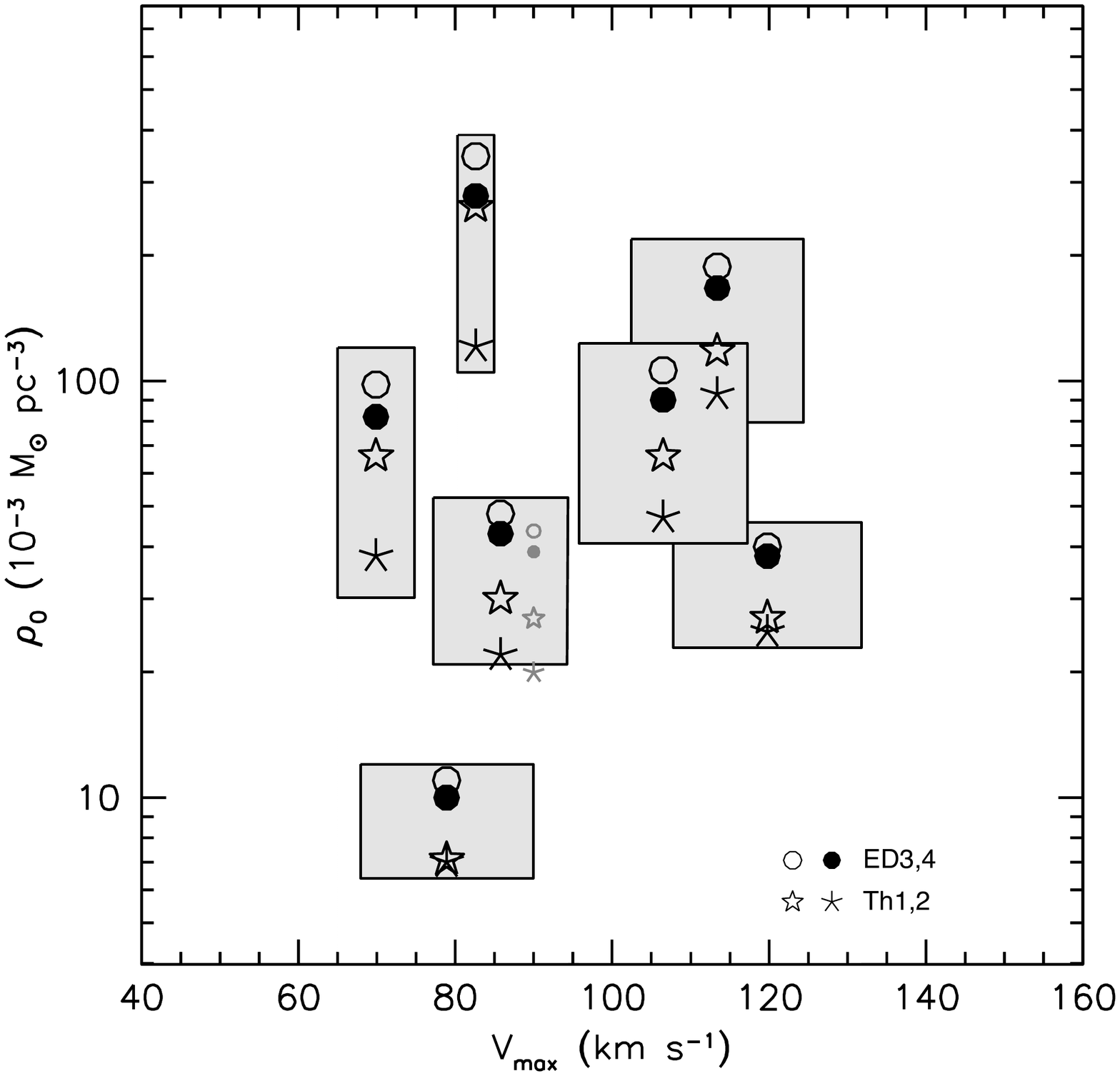}\;\;\;\;\;\includegraphics[scale=0.58]{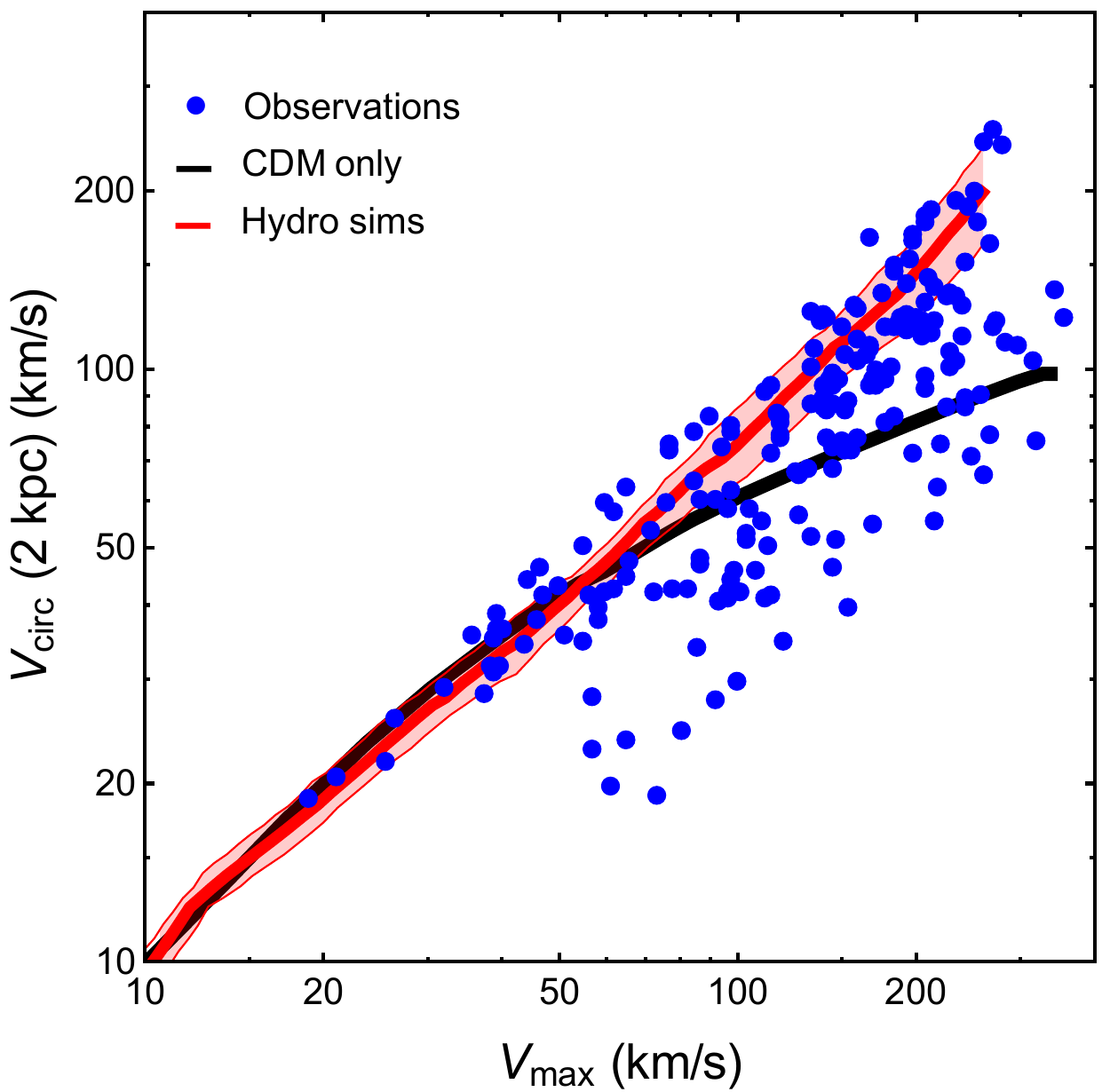}
\caption{\it Left: Inferred central core density $\rho_0$ as a function of the maximum observed rotation velocity of seven low-surface-brightness galaxies. Each symbol represents a different model for the cored DM halo density profile. For a given model, $\rho_0$ is not a constant for a fixed $V_{\rm max}$. The small gray symbols indicate the results when a non-zero stellar mass-to-light ratio is assumed.  Reprinted from Ref.~\cite{deNaray:2009xj}.
Right: The total (mean) rotation speed measured at $2~{\rm kpc}$ versus the maximum rotation speed for observed galaxies. Solid black line indicates CDM-only prediction expected for NFW haloes of average concentration. Thick red line shows the mean relation predicted in the cosmological hydrodynamical simulations~\cite{Oman:2015xda}, and the shaded areas show the standard deviation. Data compiled in~\cite{Oman:2015xda}.}
\label{fig:diversity}
\end{figure}

To illustrate this issue, Kuzio de Naray et al.~\cite{deNaray:2009xj} fitted seven LSB galaxies with four different cored halo models, including, e.g., a pseudoisothermal profile $\rho_{\rm dm}(r) = \rho_0 (1+r^2/r_c^2)^{-1}$.  Fig.~\ref{fig:diversity} (left) shows the central DM density $\rho_0$---inferred by the inner slope of $V_{\rm circ}(r)$---versus $V_{\rm max}$ for these galaxies.  Within the sample, there is no clear correlation between the inner ($\rho_0$) and outer ($V_{\rm max}$) parts of the halo.  Moreover, the spread in $\rho_0$ can be large for galaxies with similar $V_{\rm max}$, up to a factor of ${\cal O}(10)$ when $V_{\rm max}\sim80~{\rm km/s}$.  The result is independent of the choice of halo model and mass-to-light ratio.

Instead of fitting to a specific halo profile, Oman et al.~\cite{Oman:2015xda} parametrized the diversity of rotation curves more directly by comparing $V_{\rm circ}(2~{\rm kpc})$ versus $V_{\rm max}$, which represent the inner and outer halos, respectively.  Fig.~\ref{fig:diversity} (right) shows the scatter in these velocities for observed galaxies (blue points) compared to the correlation expected from CDM-only halos (solid line) and CDM halos with baryons (red band).  For $V_{\rm max}$ in the range of $50\textup{--}300~{\rm km/s}$, the spread in $V_{\rm circ}(2~{\rm kpc})$ is a factor of $\sim3$ for a given $V_{\rm max}$.  For example, when $V_{\rm max}\sim70~{\rm km/s}$, CDM (only) predicts $V_{\rm circ}(2~{\rm kpc})\sim50~{\rm km/s}$ (solid line), but observed galaxies span from $V_{\rm circ}(2~{\rm kpc})\sim20~{\rm km/s}$ to $70~{\rm km/s}$.  Galaxies at the low end of this range suffer from the mass deficit problem discussed above.  For these outliers, including the baryonic contribution will make the comparison worse.  On the other hand, galaxies at the upper end of this range could be consistent with CDM predictions once the baryonic contribution is included~\cite{Oman:2015xda}.  However, the spread in the baryon distribution plays a less significant role in generating the scatter in $V_{\rm circ}(2~{\rm kpc})$ for CDM halos since the enclosed DM mass in the cusp tends to dominate over the baryon mass.

\subsection{Missing Satellites Problem}

\begin{figure}
\includegraphics[trim={0 3cm 0 4cm},clip,scale=0.35]{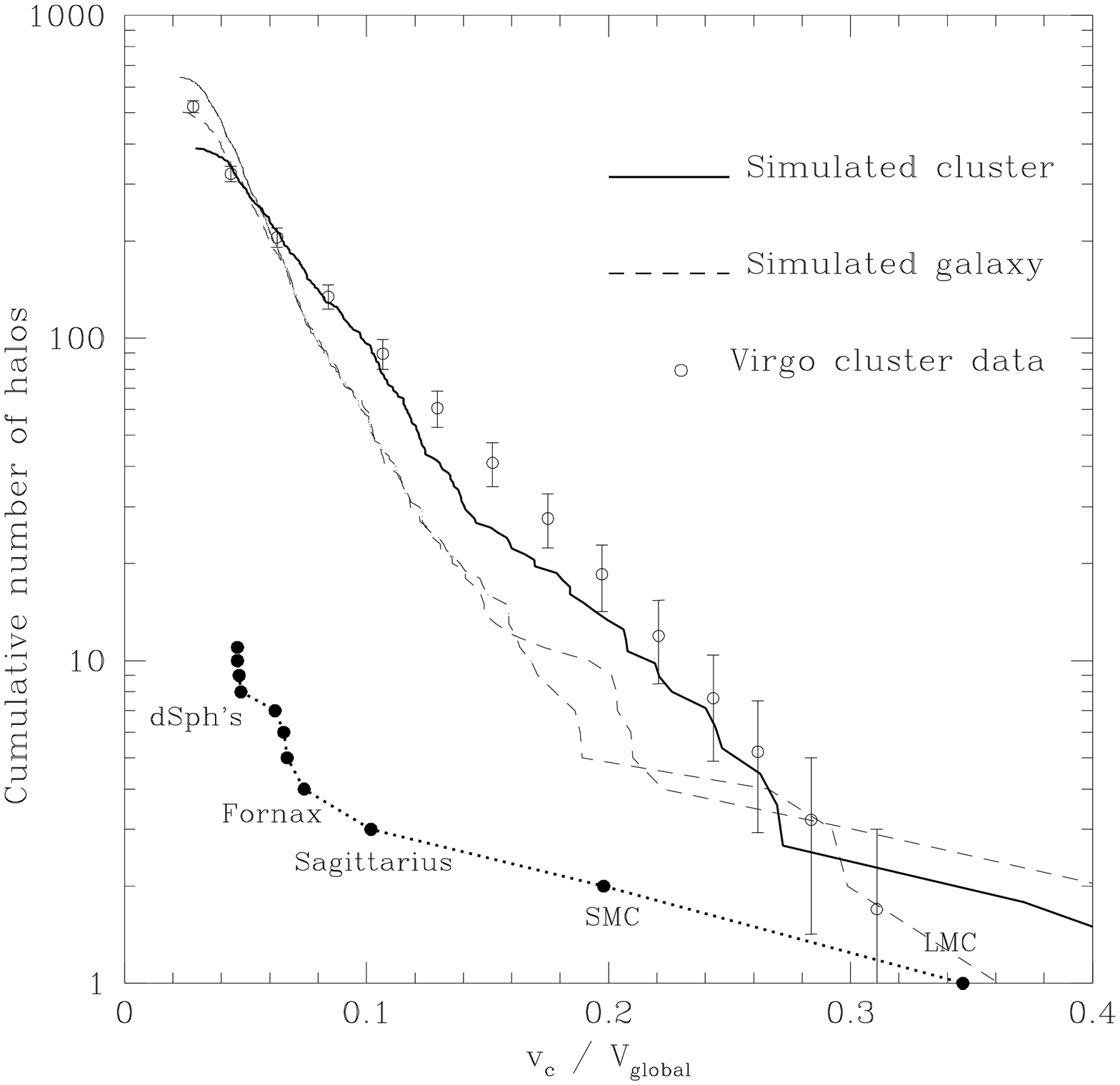}\;\;\;\includegraphics[scale=0.5]{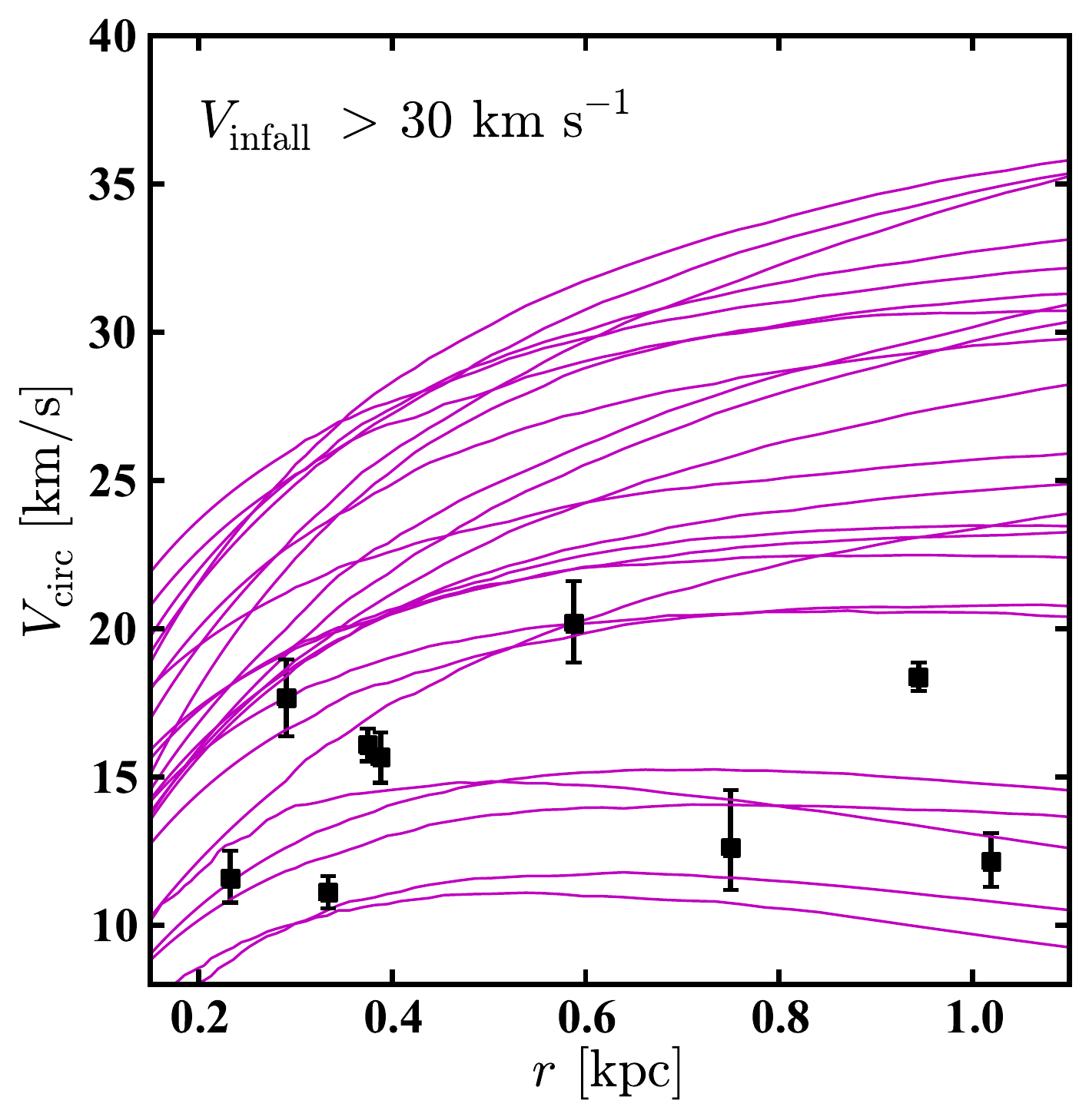}
\caption{\it Left: Abundance of subhalos within the MW (dashed) and Virgo cluster (solid) in $\Lambda$CDM simulations, compared with the distribution of observed MW satellites (filled circles) and galaxies in the Virgo cluster (open circles). Reprinted from Ref.~\cite{Moore:1999nt}.
Right: Circular velocity profiles for MW subhalos with $V_{\rm max}>10~{\rm km s^{-1}}$ predicted from CDM simulations (purple lines).  Each data point corresponds to $V_{circ}$ evaluated at the half-light radius for nine brightest MW dwarf spheroidal galaxies. Reprinted from Ref.~\cite{BoylanKolchin:2011dk}.}
\label{fig:tbtf}
\end{figure}

The hierarchical nature of structure formation predicts that CDM halos should contain large numbers of subhalos~\cite{Kauffmann:1993gv}. Numerical simulations predict that a MW-sized halo has a subhalo mass function that diverges at low masses as $dN/dM_{\rm halo}\propto M^{-1.9}_{\rm halo}$~\cite{Springel:2008cc}.  Consequently, the MW should have several hundred subhalos with $V_{\rm max} \sim 10-30~{\rm km/s}$ within its virial radius, each in principle hosting a galaxy~\cite{Klypin:1999uc,Moore:1999nt}.  However, only 11 dwarf satellite galaxies were known in the MW when the problem was originally raised in 1999, shown in Fig.~\ref{fig:tbtf} (left).  The mismatch also exists between the abundance of observed satellites in the Local Group and that predicted in simulations. This conflict is referred as the ``missing satellites problem.''  We note that a similar descrepancy does not appear for galactic-scale substructure in galaxy clusters (shown in Fig.~\ref{fig:tbtf} (left) for the Virgo cluster).

One possibility is that these subhalos exist but are invisible because of the low baryon content. For low-mass subhalos, baryonic processes may play an important role for suppressing star formation. For instance, the ultraviolet photoionizing background can inhibit gas collapse into DM halos by heating the gas and reducing the gas cooling rate, which could suppress galaxy formation in halos with circular velocities less than $30~{\rm km/s}$~\cite{Thoul:1996by,Bullock:2000wn}. In addition, after the initial star formation episode, supernova-driven winds could push the remaining gas out of the shallow potential wells of these low mass halos~\cite{Dekek:1986gu}. 

The discovery of many faint new satellites in the Sloan Digital Sky Survey has suggested that as many as a factor of $\sim5-20$ more dwarf galaxies could be still undiscovered due to faintness, luminosity bias, and limited sky coverage~\cite{Tollerud:2008ze,Walsh:2008qn,Bullock:2009gv}.  More recently, seventeen new candidate satellites have been found in the Dark Energy Survey~\cite{Bechtol:2015cbp,Drlica-Wagner:2015ufc}.  Given these considerations, the dearth of MW subhalos may not be as severe as thought originally.  

A similar abundance problem has arisen for dwarf galaxies in the field of the Local Volume.  The velocity function---the number of galaxies as a function of their \HI line widths---provides a useful metric for comparing to CDM predictions since \HI gas typically extends out to large distances to probe $V_{\rm max}$ for the halo~\cite{1989MNRAS.237.1127C,1993ApJ...413...59S}.  While in accord with observations for larger galaxies, the velocity function for CDM overpredicts the number smaller galaxies with $V_{\rm max} \lesssim 80 \; \kms$~\cite{Zavala:2009ms,Zwaan:2009dz,TrujilloGomez:2010yh}.  For example, Klypin et al.~\cite{Klypin:2014ira} find $\sim 200$ nearby galaxies within 10 Mpc with $V_{\rm max} \sim 30-50 \; \kms$, while CDM predicts $\mathcal{O}(1000)$.  Unlike the satellites, which are considerably smaller and fainter, these galaxies are relatively bright dwarf irregulars where observations are essentially complete within this volume.  

One explanation for this missing dwarf problem is that \HI line widths may be biased tracers for $V_{\rm max}$.  \HI measurements for many dwarf galaxies are limited to the rising part of the rotation curve and therefore do not sample the full gravitational potential of the DM halo.  Whether this bias can reconcile the observed velocity function with CDM~\cite{Brook:2015eva,Maccio:2016egb,Brooks:2017rfe}, or if the discrepancy still persists~\cite{Trujillo-Gomez:2016pix}, remains to settled.

\subsection{Too-Big-to-Fail Problem}

\underline{Satellites within the Local Group:} Boylan-Kolchin et al.~\cite{BoylanKolchin:2011de,BoylanKolchin:2011dk} showed that the population of the MW's brightest dSph galaxies, which are DM dominated at all radii, exhibit another type of discrepancy with CDM predictions.  Since these satellites have the largest stellar velocities and luminosities, they are expected to live in the most massive MW subhalos.  However, the most massive subhalos predicted by CDM-only simulations have central densities too large to host the observed satellites.  As shown in Fig.~\ref{fig:tbtf} (right), simulations predict ${\cal O}(10)$ subhalos with $V_{\rm max} > 30~{\rm km/s}$, whereas the bright MW dSphs have stellar dispersions corresponding to CDM subhalos with $12\lesssim V_{\rm max} \lesssim25~{\rm km/s}$.  It is puzzling that the most massive subhalos should be missing luminous counterparts since their deep potential wells make it unlikely that photoionizing feedback can inhibit gas accretion and suppress galaxy formation.  Hence, these substructures should be too big to fail to form stars. 

Several proposed mechanisms may address the TBTF problem without invoking DM physics. First, the MW halo mass may be underestimated.  The TBTF discrepancy is based on simulated MW-like halos with masses in the range $1-2 \times 10^{12} \, \Msun$~\cite{BoylanKolchin:2011de,BoylanKolchin:2011dk}.  However, since the number of subhalos scales with host halo mass, the apparent lack of massive subhalos might be accommodated if the MW halo mass is around $5 \times 10^{11}~\Msun$, although Boylan-Kolchin et al.~argue against this possibility (see also Ref.~\cite{Wang:2015ala} for a summary of different estimates.)  Second, the MW may have less massive subhalos due to scatter from the stochastic nature of structure formation~\cite{Purcell:2012kd}.  However, a recent analysis finds that there is only $\sim1\%$ chance that MW-sized host halos have a subhalo population in statistical agreement with that of the MW~\cite{Jiang:2015vra}.  In addition, similar discrepancies also exist for the brightest dwarf galaxies in Andromeda~\cite{Tollerud:2014zha} and the Local Group field~\cite{Garrison-Kimmel:2014vqa}, which further disfavor these explanations. Baryonic physics---including environmental effects from the MW disk---may also play a role, discussed in the next section.

\vspace{2mm}

\underline{Field galaxies:} A similar TBTF problem also arises for dwarf galaxies in the field~\cite{Ferrero:2011au,Papastergis:2014aba,Garrison-Kimmel:2014vqa,Brook:2014hda,Klypin:2014ira}.  According to abundance matching, galaxies are expected to populate DM halos in a one-to-one relationship that is monotonic with mass (i.e., larger galaxies in larger halos).  However, the galaxy stellar mass function, inferred by the Sloan Digital Sky Survey~\cite{Baldry:2008ru,Li:2009ce}, is shallower at low mass than the halo mass function in $\Lambda$CDM, which suggests that galaxy formation becomes inefficient for halos below $\sim 10^{10} \, \Msun$~\cite{Guo:2009fn}.  Hence, it is expected that most faint dwarf galaxies populate a narrow range of DM halos with $V_{\rm max} \sim 30 \; {\rm km/s}$, while halos with $V_{\rm max} \lesssim 30 \; {\rm km/s}$ would have no galactic counterpart~\cite{Ferrero:2011au}.  

Ferrero et al.~\cite{Ferrero:2011au} find that rotation curves for faint dwarf galaxies do not support these conclusions.  A large fraction of faint galaxies in their sample appear to inhabit CDM halos with masses below $\sim 10^{10} \, \Msun$, which would imply that other more massive halos---which should be too big to fail---lack galaxies.  On the other hand, optical galaxy counts may simply have missed a large fraction of low surface brightness objects.

\begin{figure}
\includegraphics[trim={2cm 3cm 2cm 4cm},clip,scale=0.6]{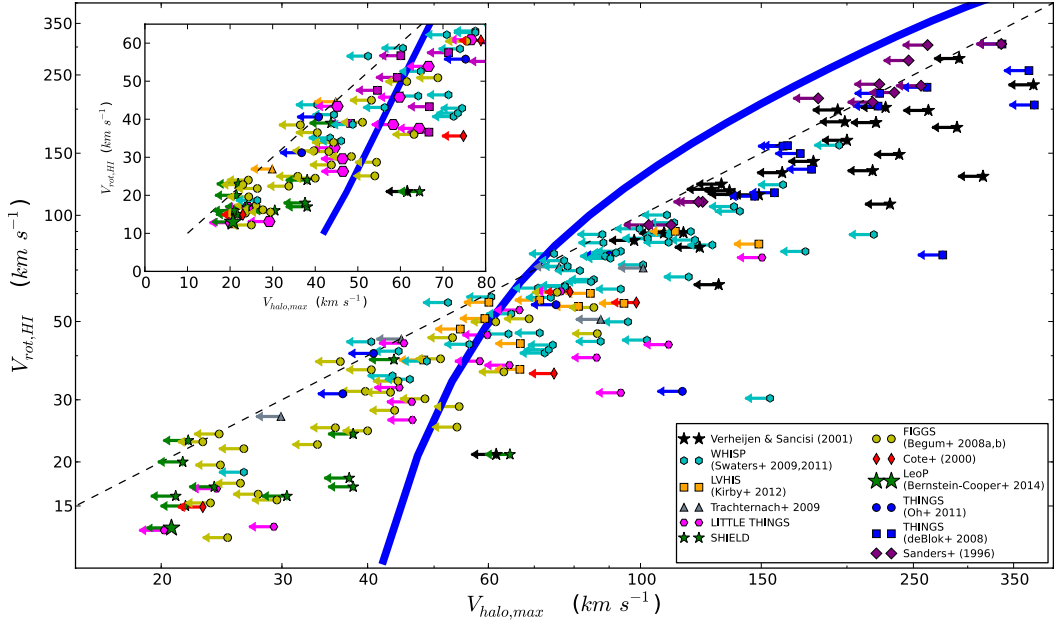}
\caption{\it The $V_{\rm rot, \HI}\textup{--}V_{\rm halo, max}$ relation for galaxies. The colored points represent observed galaxies, and they are drawn as upper limits, as the baryonic contribution to the rotation curve is neglected. The blue line is inferred from abundance matching in $\Lambda$CDM cosmology. Reprinted from Ref.~\cite{Papastergis:2014aba}.}
\label{fig:tbtffield}
\end{figure}

More recently, Papastergis et al.~\cite{Papastergis:2014aba} have confirmed the findings of Ref.~\cite{Ferrero:2011au} by utilizing data from the Arecibo Legacy Fast ALFA 21-cm survey.  Since the majority of faint galaxies are gas-rich late-type dwarfs, these observations provide a more complete census without a bias against low stellar luminosity objects.  Fig.~\ref{fig:tbtffield} shows their main results, illustrating the TBTF problem in the field.  
\begin{itemize}
\item The blue curve is the prediction of abundance matching between observed \HI rotational velocities ($V_{\rm rot,\HI}$) and predicted maximum rotational velocities for $\Lambda$CDM halos ($V_{\rm halo, max}$).  
\item The various symbols indicate the observed correlation for a sample of 194 nearby galaxies.  For each galaxy, $V_{\rm rot,\HI}$ is the rotational velocity at the outermost measured radius and $V_{\rm halo, max}$ is determined by matching a CDM halo rotation curve at that point.  
\item Since baryonic contributions to the rotation curve are neglected, each symbol is regarded as an {\it upper limit} on $V_{\rm halo, max}$. Hence, the symbols are indicated by leftward arrows.
\end{itemize}
Galaxies to the right of the blue line can potentially be reconciled with abundance matching predictions since baryons can make up the difference.  On the other hand, the lowest mass galaxies to the left of the blue line seem inconsistent with these arguments.  Abundance matching predicts they should inhabit halos with $V_{\rm halo,max} \sim 40 \; {\rm km/s}$, yet this is incompatible with their measured \HI circular velocities.  

\vspace{2mm}

\underline{Common ground with the core-cusp problem:} The essence of the TBTF problem is that low-mass galaxies have gas or stellar velocities that are too small to be consistent with the CDM halos they are predicted to inhabit.  At face value, the issue is reminiscent of the core-cusp/mass deficit problems for rotation curves and other observations.  Thus, one way to resolve the TBTF problem is if these galaxies have reduced central densities compared to CDM halos.  By generating cored profiles in low-mass halos, self-interactions may resolve this issue for the dwarf galaxies in the MW~\cite{Vogelsberger:2012ku,Zavala:2012us}, Local Group~\cite{Elbert:2014bma}, and the field~\cite{Schneider:2016ayw}.

\subsection{Baryon feedback}

It has been more than 20 years, since the small scale ``crisis" was first posed in the 1990s. Since then, there has been extensive discussion on whether all issues can be resolved within the CDM paradigm once baryonic feedback processes---gas cooling, star formation, supernovae, and active galactic nuclei---are accounted for.  Here we give a brief review of feedback on galactic scales.

\vspace{2mm}

\underline{Cores in galaxies:}  As galaxies form, gas sinks into the inner halo to produce stars. The deepening gravitational potential of baryons causes the central density and velocity dispersion of DM to increase through adiabatic contraction~\cite{Blumenthal:1985qy}.  Although at first sight this makes the core-cusp problem worse, the situation can be very different due to non-adiabatic feedback from supernova-driven gas outflows~\cite{Navarro:1996bv,Governato:2009bg}.  If a sizable fraction of baryons is suddenly removed from the inner halo, DM particles migrate out to larger orbits.  In this way, repeated bursts of star formation and outflows (followed by reaccretion) can reduce the central halo density through a purely {\it gravitational} interaction between DM and baryons.  Non-adiabaticity is essential for the mechanism to work.  This is achieved by assuming a ``bursty'' star formation history with $\mathcal{O}(10)$ variation in the star formation rate over time scales comparable to the dynamical time scale of the galaxy~\cite{Teyssier:2012ie}.  (See Refs.~\cite{Pontzen:2014lma,Pontzen:2011ty} for a pedagogical explanation.)

Governato et al.~\cite{Governato:2009bg} argued that bursty star formation in dwarf galaxies could be connected to another long-standing puzzle in galaxy formation: bulgeless disk galaxies.  If strong outflows are necessary to remove low angular momentum gas to prevent bulge formation in certain galaxies, they may also induce DM cores.  High-resolution hydrodynamical simulations show that supernova feedback can generate $\mathcal{O}({\rm kpc)}$ cores in CDM halos for dwarf galaxies~\cite{Governato:2009bg}.  Notably, the shallow inner slope of the DM distribution inferred from THINGS and LITTLE THINGS dwarf galaxies could be consistent with CDM halos once feedback is included (except DDO 210)~\cite{Oh:2010mc,Governato:2012fa,Oh:2015xoa} (see, e.g., Fig. 6 of~\cite{Oh:2015xoa}). 

\begin{figure}
\includegraphics[trim={0cm 0cm 0 1cm},clip,scale=0.5]{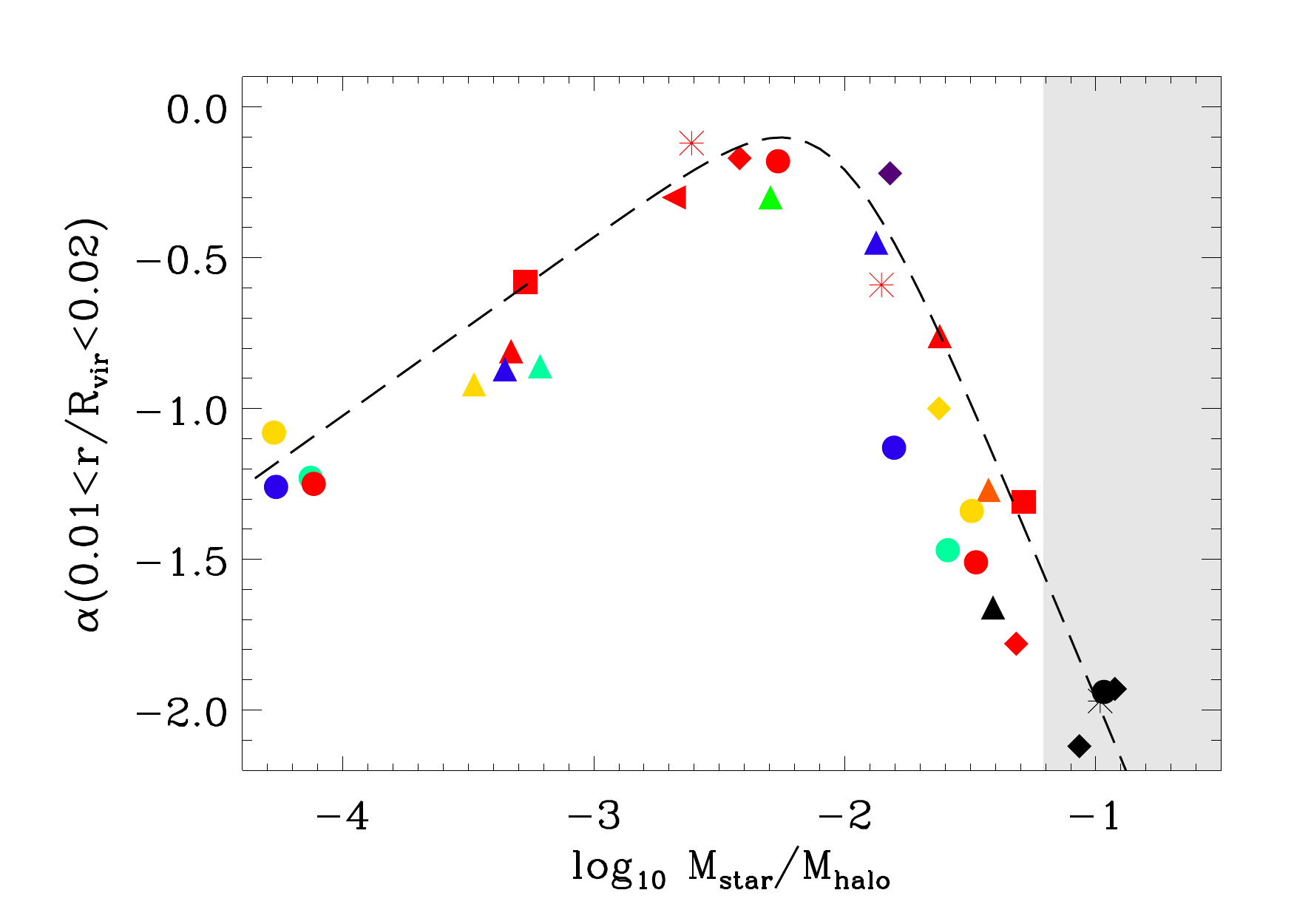}\;\;\includegraphics[scale=0.4]{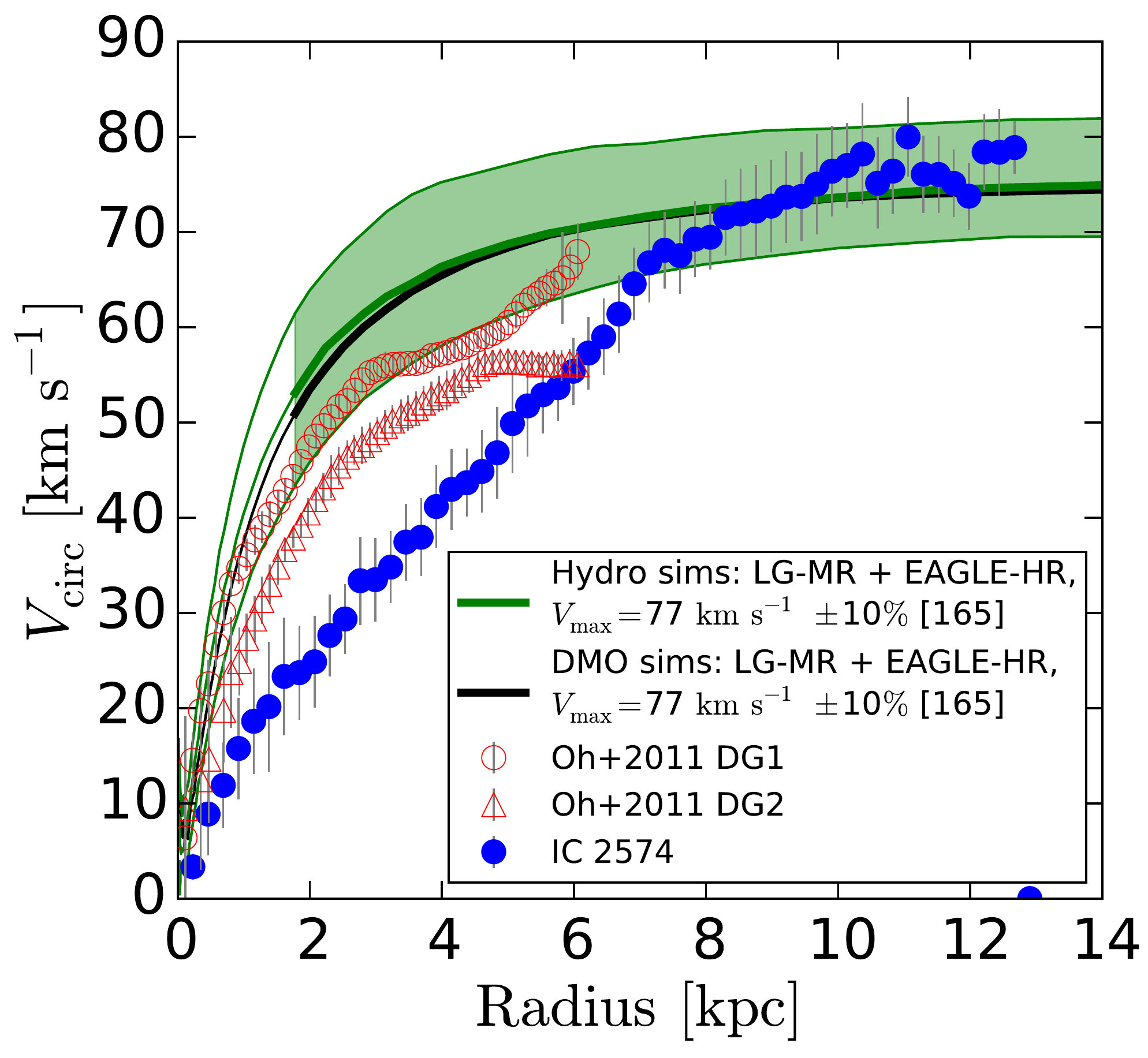}
\caption{\it Left: Slope of DM density profile vs the ratio of the stellar mass to the halo mass, predicted in $\Lambda$CDM hydrodynamical simulations with strong stellar feedback. Different colors denote the feedback schemes, and different simulated galaxies are represented with symbols. The best fitted function is indicated with the dashed curve. Reprinted from Ref.~\cite{DiCintio:2013qxa}.
Right: Observed rotation curves of dwarf galaxy IC 2574 (blue), compared to simulated ones from Oman et al.~\cite{Oman:2015xda} (green band) and Oh et al.~\cite{Oh:2010mc} (open circles and triangles). Reprinted from Ref.~\cite{Oman:2015xda}.}
\label{fig:baryon}
\end{figure}

Further studies by Di Cintio et al.~\cite{DiCintio:2013qxa,DiCintio:2014xia} investigated the effect of feedback as a function of halo mass.  Their sample includes 31 galaxies from hydrodynamical simulations spanning halo mass $10^{10}\textup{--}10^{12}\Msun$.  The corresponding stellar masses are fixed by abundance matching, with stellar-to-halo mass ratio in the range $M_{\rm star}/M_{\rm halo}\sim10^{-4}\textup{--}10^{-1}$.  Fig.~\ref{fig:baryon} (left) shows how the DM inner density slope $\alpha$ (measured at $1\textup{--}2\%$ of the virial radius) depends strongly on $M_{\rm star}/M_{\rm halo}$.  Di Cintio et al.~\cite{DiCintio:2013qxa} conclude the following points:
\begin{itemize}
\item Maximal flattening of the inner DM halo occurs for $M_{\rm star}/M_{\rm halo}\sim 0.5\%$.  According to abundance matching, this corresponds to stellar mass $\sim 3\times10^8 \, \Msun$, halo mass $\sim6\times10^{10}\Msun$, and asymptotic circular velocity $\sim50~{\rm km/s}$.  The THINGS dwarfs are right in the ``sweet spot'' for supernova feedback to be maximally effective.
\item Supernova feedback is less effective in larger halos, which (by abundance matching) correspond to larger values of $M_{\rm star}/M_{\rm halo}$.  Despite more star formation available to drive feedback, the deeper potential well for the halo suppresses the effect.  For these galaxies, adiabatic contraction dominates and halos are cuspy.
\item When the ratio is less than $10^{-4}$, there is too little star formation for feedback to affect the inner DM density.  This implies that halos hosting galaxies with $M_{\rm star}\lesssim 3\times10^6\Msun$ remain cuspy, consistent with earlier findings~\cite{Governato:2012fa}
\end{itemize}
These results have been further confirmed by the Numerical Investigation of Hundred Astrophysical Objects (NIHAO) simulations~\cite{Tollet:2015gqa}, where the same stellar feedback model was used for a larger halo sample. Other simulations with similar feedback prescriptions have also confirmed core formation in halos~\cite{2013MNRAS.429.3068T,Madau:2014ija}.

Cosmological simulations with Feedback In Realistic Environments (FIRE)~\cite{Hopkins:2013vha} have emphasized that core formation in DM halos is tightly linked with star formation history~\cite{Onorbe:2015ija,Chan:2015tna}.  O\~{n}orbe et al.~\cite{Onorbe:2015ija} targeted low mass galaxies relevant for the TBTF problem and showed that kpc-sized cores may arise through feedback.  However, cores form only when star formation remains active after most of the halo growth has occurred ($z \lesssim 2$).  Otherwise, even if a core forms at early times, subsequent halo mergers may erase it if feedback has ceased.  Lower mass galaxies, such as ultra-faint dwarfs, have too little star formation to affect the DM halo.

Chan et al.~\cite{Chan:2015tna} explored a wider mass range using the FIRE simulations, $M_{\rm star}\approx10^4\textup{--}10^{11}\Msun$ and $M_{\rm halo}\approx10^9\textup{--}10^{12} \Msun$. The DM density profiles become shallow for $M_{\rm halo}\approx10^{10}\textup{--}10^{11}\Msun$ due to strong feedback, with the maximal effect at $M_{\rm halo}\sim10^{11}\Msun$. The result is broadly consistent with Governato et al.~\cite{Governato:2012fa} and Di Cintio et al.~\cite{DiCintio:2013qxa}. Chan et al.~also found that large cores can form only if bursty star formation occurs at a late epoch when cusp-building mergers have stopped, as pointed out in~\cite{Onorbe:2015ija}.

Although baryonic feedback may reconcile galactic rotation curves with the CDM paradigm (see, e.g., Refs.~\cite{Brook:2015qja,Katz:2016hyb}), questions yet remain:
\begin{itemize}
\item {\it Can feedback generate ultra-low density cores?} Feedback-created cores are limited to the inner $\sim 1$ kpc where the star formation rate is high.   However, some galaxies, such as dwarf IC 2574, are extreme outliers with huge core sizes and mass deficits that are far beyond what feedback can do.  Fig.~\ref{fig:baryon} (right) shows the rotation curve for IC 2574, together with simulated ones with similar values of $V_{\rm max}$.  The observed core size is $\sim 6$ kpc~\cite{Oh:2010ea}---set by the region over where $V_{\rm circ}(r)$ rises linearly with radius---which is too large to be created with bursty feedback (open symbols)~\cite{Oh:2010mc} or smoother implementations (green band)~\cite{Oman:2015xda}.  Fig.~\ref{fig:diversity} (right) shows that many galaxies have inner mass deficits out to 2 kpc and it is challenging for feedback to remove enough DM from their central regions to be consistent with observations.
\item {\it Does feedback explain the full spectrum of observed rotation curves?} Galaxies exhibit a broad spread in rotation curves, encompassing both cored and cuspy profiles~\cite{Oman:2015xda}.   While some systems have large cores, there are galaxies with the same $V_{\rm max}$ which are consistent with a cuspy CDM halo.  For fixed $V_{\rm max}$, the spread in the velocity at 2 kpc is a factor of $\sim 3$, as shown in Fig.~\ref{fig:diversity} (right).  It is unclear whether feedback prescriptions can account for such a scatter.
\item {\it Is bursty feedback required?} Since star formation is far below the resolution of simulations, baryon dynamics depend on how feedback is modeled, especially the density threshold for star formation.\footnote{In various simulations, the gas density thresholds for star formation are taken to be $\gtrsim10\textup{--}100$ (Brooks et al.~\cite{Brooks:2012ah}), $10\textup{--}100$ (FIRE, Chan et al.~\cite{Chan:2015tna}), $1000$ (FIRE-2, Wetzel et al.~\cite{Wetzel:2016wro}), and $0.1\textup{--}1~\rm{atoms/cm^3}$ (Oman et al.~\cite{Oman:2015xda}).}  Large thresholds lead to bursty star formation since energy injection to the dense medium causes strong outflows, which disturb the potential violently to generate DM cores~\cite{Governato:2009bg,Governato:2012fa}.  However, Oman et al.~\cite{Oman:2015xda} have analyzed galaxies from the EAGLE and LOCAL GROUP simulation projects---which adopt a far smaller threshold---and find that feedback is negligible. These simulations have star formation occuring throughout the gaseous disk---without sudden fluctuations in the gravitational potential needed to produce DM cores---and yet produce galaxies that are consistent with other observational constraints.
\item {\it What is the epoch of core formation?} Detailed comparisons between hydrodynamical simulations with bursty feedback would be useful to understand what systematic differences may be present (if any).  One potential point of divergence is the epoch when star formation bursts need to occur to yield cores.  Pontzen \& Governato~\cite{Pontzen:2011ty} argue that DM cores may form due to feedback at an early time $2 < z < 4$.  However, Refs.~\cite{Onorbe:2015ija,Chan:2015tna} find that cores remain stable only if outflows remain active at later times, $z \lesssim 2$, once halo growth has slowed. This question is important for connecting observational tracers of starbursts in galaxies to feedback-driven core formation (e.g., Ref.~\cite{2012ApJ...744...44W,Kauffmann:2014cda}).
\end{itemize}

\vspace{2mm}

\underline{Substructure:} A number of studies have found that baryon dynamics may solve the missing satellites and TBTF problem within $\Lambda$CDM, independently of whether star formation is bursty or not.  Sawala et al.~\cite{Sawala:2015cdf} performed a suite of cosmological hydrodynamical simulations of 12 volumes selected to match the Local Group.  These simulations adopt a smooth star formation history, as in Ref.~\cite{Oman:2015xda}, and do not yield cored profiles.  Regardless, the number of satellite galaxies is reduced significantly in MW and M31-like halos (within 300 kpc), as well as in the broader Local Group (within 2 Mpc), in better accord with observations compared to the expectation from DM-only simulations.  Supernova feedback and reionization deplete baryons in low mass halos and only a subset of them can form galaxies.  In addition, the most massive subhalos in MW-like galaxies have an inner mass deficit due to ram pressure stripping and a suppressed halo growth rate due to the baryon loss. These massive subhalos could be consistent with kinematical observations of the MW dSphs, solving the TBTF problem in the MW, within measurement uncertainties~\cite{Fattahi:2016nld}.  

Earlier studies by Zolotov et al.~\cite{Zolotov:2012xd} and Brooks et al.~\cite{Brooks:2012vi,Brooks:2012ah} reached a similar conclusion in simulations with a bursty star formation scheme, as in Ref.~\cite{Governato:2012fa} (see Ref.~\cite{Stinson:2006cp} for details).  Strong feedback produces cored profiles in subhalos, which in turn enhances tidal effects from the stellar disk of host galaxies (see also~\cite{Penarrubia:2010jk}).  More recently, FIRE simulations in the Local Group environment by Wetzel et al.~\cite{Wetzel:2016wro} also showed that the population of satellite galaxies with $M_{\rm star}\gtrsim10^5  \Msun$ does not suffer from the missing satellites and TBTF problems.  

\vspace{2mm}

\underline{Too-big-to-fail in the field:} Papastergis \& Shankar~\cite{2016A&A...591A..58P} argued that the TBTF problem for field dwarf galaxies cannot be solved in $\Lambda$CDM even with baryonic feedback effects. They adapted the halo velocity function proposed in Sawala et al.~\cite{Sawala:2015cdf}, which includes the effects of baryon depletion and reionization feedback, and the modified CDM halo density profile due to supernova feedback from Di Cintio et al.~\cite{DiCintio:2013qxa}. While reionization effectively suppresses star formation for halos with $V_{\rm max}\lesssim20~{\rm km/s}$, the TBTF problem here concerns halos with $V_{\rm max}\approx25\textup{--}45~{\rm km/s}$, which are too massive to be affected significantly.  In addition, cored profiles may help alleviate the tension for galaxies whose stellar kinematics are measured only in the very inner region, but not those with $V_{\rm circ}$ observed at large radii.  

More recently, Verbeke et al.~\cite{Verbeke:2017rfd} argued that the TBTF problem in the field may be resolved by noncircular motions in \HI gas.  According to their simulations, H{\small{\sc I}} kinematics is an imperfect tracer for mass in small galaxies due to turbulence (as pointed out in Refs.~\cite{Dalcanton:2010bp,2017MNRAS.466...63P}).  In particular, galaxies with $V_{\rm rot,\HI} \lesssim 30$ km/s (see Fig.~\ref{fig:tbtffield}) may be consistent with living in larger halos than would otherwise be inferred without properly accounting for this systematic effect.

\vspace{2mm}

\underline{Summary:} The predictions of $\Lambda$CDM cosmology can be modified on small scales when dissipative baryon physics is included in simulations.  Strong baryonic feedback from supernova explosions and stellar winds disturb the gravitational potential violently, resulting in a shallow halo density profile.  Reionization can suppress galaxy formation in low mass halos and strong tides from host's stellar disk can destroy satellite halos. These effects can also be combined in shaping galaxies. For example, if feedback induces cores in low mass halos, galaxy formation is more effectively suppressed by reionization and the halo is more vulnerable to tidal disruption.  In simulations, the significance of these effects depends on the specific feedback models, in particular, the gas density threshold for star formation.  It seems that strong feedback---leading to the core formation and significant suppression of star formation in low mass halos---is {\it not} required for reproducing general properties of observed galaxies in simulations (see, e.g., Refs.~\cite{Oman:2015xda,Sawala:2015cdf,Sales:2016dmm}). Thus, it remains unclear to what degree baryon dynamics affect halo properties in reality.  Moreover, cored profiles generated by feedback, as proposed in Refs.~\cite{DiCintio:2013qxa,DiCintio:2014xia}, may be inconsistent with the correlations predicted in $\Lambda$CDM cosmology, namely the mass-concentration and $M_{\rm star}\textup{--}M_{\rm halo}$ abundance matching relations~\cite{Pace:2016oim} (however, see Ref.~\cite{Katz:2016hyb}).  Thus, it remains an intriguing possibility that small scale issues may imply the breakdown of the CDM paradigm on galactic scales, as we will focus on in the rest of this article.


\section{N-body simulations and SIDM halo properties} 
\label{sec:sidmhalos}

N-body simulations have been the primary tools for understanding the effect of self-interactions on structure~\cite{Moore:2000fp,Yoshida:2000bx,Burkert:2000di,Kochanek:2000pi,Yoshida:2000uw,Dave:2000ar,
Colin:2002nk,Vogelsberger:2012ku,Rocha:2012jg,Peter:2012jh,Zavala:2012us,Elbert:2014bma,Vogelsberger:2014pda,Fry:2015rta,Dooley:2016ajo}.
Early SIDM simulations used smoothed particle hydrodynamics to model DM collisions~\cite{Moore:2000fp,Yoshida:2000bx}.  This approach treats SIDM as an ideal gas described by fluid equations, which are valid in the optically-thick regime where the mean free path $\lambda_{\rm mfp}$ is much smaller than typical galactic length scales.  Due to efficient thermalization, SIDM halos in this context form a singular isothermal profile, $\rho_{\rm dm} \propto r^{-2}$, which is steeper than collisionless CDM halos and exacerbates the core-cusp issue rather than solving it~\cite{Moore:2000fp,Yoshida:2000bx}.\footnote{The fluid equations approach has been used as a semi-analytical framework for modeling SIDM halos under simplifying assumptions (spherical symmetry and self-similar evolution)~\cite{Balberg:2002ue,Ahn:2004xt,Koda:2011yb}.  In fact, this model agrees well with N-body simulations for isolated halos, even for the optically-thin regime that is beyond the validity of the fluid approximation; however, the model is not yet able to reproduce N-body results for cosmological halos due to departure from self-similar evolution~\cite{Koda:2011yb}.}

Consequently, most SIDM simulations have focused the more promising case of self-interactions in the optically-thin regime, with cross sections spanning $0.1 - 50 \; {\rm cm^2/g}$
~\cite{Burkert:2000di,Kochanek:2000pi,Yoshida:2000uw,Dave:2000ar,
Colin:2002nk,Vogelsberger:2012ku,Rocha:2012jg,Peter:2012jh,Zavala:2012us,Elbert:2014bma,Vogelsberger:2014pda,Fry:2015rta,Dooley:2016ajo}.  
In this case, $\lambda_{\rm mfp}$ is larger than the typical $\mathcal{O}({\rm kpc})$ core radius over which self-interactions are active.  Here we discuss these simulations and their implications for astrophysical observables.  

The majority of simulations assume a contact-type interaction where scattering is isotropic and velocity-independent, described by a fixed $\sigma/m$.  These studies have converged on $\sigma/m \approx 0.5 - 1 \; {\rm cm^2/g}$ to solve the core-cusp and TBTF issues on small scales, while remaining approximately consistent with other astrophysical constraints on larger scales, such as ellipticity measurements.\footnote{This value is consistent with Spergel \& Steinhardt's original estimate, $\sigma/m \sim 0.45 - 450 \; {\rm cm^2/g}$, based on having $\lambda_{\rm mfp} = (\rho_{\rm dm} \sigma/m)^{-1}$ in the range $1 \; {\rm kpc} - 1 \; {\rm Mpc}$ with DM density $\rho_{\rm dm} \approx 0.4\; {\rm GeV/cm^3}$ as in the local solar neighborhood~\cite{Spergel:1999mh}.}  However, more recent studies based on massive clusters disfavor the constant cross section solution, prefering somewhat smaller values $\sigma/m \approx 0.1\; {\rm cm^2/g}$ on these scales~\cite{Kaplinghat:2015aga,Elbert:2016dbb}.  

Before we discuss these topics in greater detail, we emphasize a few key points:
\begin{itemize}
\item Self-interaction cross sections are generically velocity-dependent, as predicted in many particle models for SIDM (see \S\ref{sec:models}). Therefore, constraints on $\sigma/m$ must be interpreted {\it as a function of} halo mass, since DM particles in more massive halos will have larger typical velocities for scattering.  (``Halo mass'' refers to the virial mass, for which we note that different studies have adopted somewhat different conventions in defining.)  
\item The effect from self-interactions on a halo is not a monotonic function of $\sigma/m$, which can be understood as follows.  If DM is collisionless, its velocity dispersion in the halo is peaked near the scale radius $r_s$, while particles in both the center and outskirts of the halo are colder.  Once collisions begin to occur, only the inner region of the halo is in thermal contact.  The center of the halo forms a core that grows larger as more heat flows inward due to the positive temperature gradient.  Eventually, once thermal contact is reached with the outskirts of the halo where the temperature gradient is negative, heat is lost outward and gravothermal collapse of the core ensues~\cite{LyndenBell:1968yw,Colin:2002nk}.  
\item Our discussion mostly focuses on DM-only simulations and is subject to the usual caveat that baryons have been ignored.  While baryonic feedback may explain away the issues that SIDM aims to address, this depends strongly on the prescription for feedback adopted.  Without a more definitive understanding, making quantitative statements about the evidence for SIDM remains limited by these systematics.  At the end of this section, we discuss recent simulations for SIDM including baryonic dynamics~\cite{Vogelsberger:2014pda,Fry:2015rta}.
\item The present section focuses on the quasi-equilibrium structure of SIDM halos.  We discuss simulations of merging halos, such as Bullet Cluster-like systems, in a later section (\S\ref{sec:merging}).
\end{itemize}


\subsection{Implementing self-interactions in simulations}

In N-body simulations, DM particles are represented by ``macroparticles'' of mass $m_p \gg \Msun$, each representing a phase space patch covering a vast number of individual DM particles.  Self-interactions, assumed to be a short-range force on galactic scales, are treated using a Monte Carlo approach~\cite{Burkert:2000di}.  
Two macroparticles scatter if a random number between $[0,1]$ is less than the local scattering probability within a given simulation time step $\Delta t$, which must be small enough to avoid multiple scatterings.  Outgoing trajectories preserve energy and linear momentum\footnote{Angular momentum is not conserved for individual scatterings, due to finite separations between macroparticles, but any nonconservation is expected to average to zero over the halo since separation vectors are randomly oriented~\cite{Kochanek:2000pi}.}, with a scattering angle chosen randomly, under the assumption that scattering is isotropic in the center-of-mass frame.  

Different simulations have adopted different prescriptions for determining the scattering probability.  Many earlier studies used a background density method~\cite{Burkert:2000di,Kochanek:2000pi,Yoshida:2000uw,Colin:2002nk}.  In this approach, the probability for an individual macroparticle $i$ to scatter is
\beq  \label{eq:Pi}
P_i = \rho_{{\rm dm}}^{(i)} v_{{\rm rel}}^{(i)} (\sigma/m) \Delta t \, ,
\eeq
where $\rho_{{\rm dm}}^{(i)}$ is the mean local background density, spatially averaged over its nearest neighbors, and $v_{\rm rel}^{(i)}$ is the relative velocity between $i$ and one or more of its neighbors.  If a scattering occurs, particle $i$ is paired up with a nearby recoiling partner $j$.

An alternative approach treats scattering as a pair-wise process between macroparticles $i,j$ with velocities $\mathbf v_{i,j}$~\cite{Dave:2000ar,Rocha:2012jg}.  The probability is
\beq \label{eq:Pij}
P_{ij} = \rho_{ij} |\mathbf v_i - \mathbf v_j| (\sigma/m) \Delta t \, ,
\eeq
where $\rho_{ij}$ represents the target density from $j$ to be scattered by $i$.  Rocha et al.~\cite{Rocha:2012jg} provide an insightful derivation of Eq.~\eqref{eq:Pij} starting from the Boltzmann collision term.  Each macroparticle $i$, centered at $\mathbf r_i$, is coarse-grained over a finite spatial patch using a cubic spline kernel $W(|\mathbf r - \mathbf r_i|,h)$ with smoothing length $h$~\cite{1985A&A...149..135M}.\footnote{The smoothing kernel is defined as $W(r,h) = h^{-3} w(r/h)$, where 
\beq
w(x) = \frac{8}{\pi} \left\{ \begin{array}{cc} 
1 - 6 x^2 + 6 x^3 , & 0 \le x \le \tfrac{1}{2} \\ 
2(1-x)^3 , & \tfrac{1}{2} < x \le 1 \\ 
 0  , &  x > 1 \end{array} \right. .
\eeq
$W$ is normalized as $\int d^3 r \, W(|\mathbf r - \mathbf r_i|,h) = 1$ and has dimensions of number density.}  The resulting collision rate between patches $i,j$ is determined by the overlap integral 
\beq \label{rhoij}
\rho_{ij} = m_p \int d^3 r \, W(|\mathbf r - \mathbf r_i|,h) \, W(|\mathbf r - \mathbf r_j|,h) \, .
\eeq
Numerical convergence studies show that $h$ must not be too small compared to the mean particle spacing $a = (m_p/\rho_{\rm dm})^{1/3}$, requiring $h/a \gtrsim 0.2$~\cite{Rocha:2012jg}.  Simulations in Ref.~\cite{Rocha:2012jg} have fixed $h$ such that $0.2 \lesssim h/a \lesssim 1$ throughout the inner halo where the self-interaction rate is large.  However, self-interactions are artificially quenched in the outer halo where $h/a < 0.1$, although scattering is not expected to be relevant here due to simple rate estimates.

Alternatively, Vogelsberger et al.~\cite{Vogelsberger:2012ku} follow a hybrid between these approaches.  First, $P_{ij}$ is computed in Eq.~\eqref{eq:Pij} with $\rho_{ij} = m_p W(|\mathbf r_i - \mathbf r_j|,h)$.  Here, $\rho_{ij}$ can be interpretted as the ``background'' density from $j$ at the position of $i$ (which follows from Eq.~\eqref{rhoij} by setting the smoothing kernel for $i$ to be a delta function).  The smoothing length is also taken to be much larger than the Ref.~\cite{Rocha:2012jg} approach, with $h$ adjusted dynamically such that $\sim 40$ particles are in range of $i$.  Since $h \sim \sqrt[3]{40}\times a$, scatterings are less localized.  The total probability for $i$ to scatter is determined by $P_i  = \tfrac{1}{2} \sum_j P_{ij}$, and if a scattering occurs, the scattering partner $j$ is chosen with probabilty weighted by $P_{ij}$.  It is unknown what differences, if any, may arise between these various methods (although both Refs.~\cite{Rocha:2012jg,Vogelsberger:2012ku} are in mutual agreement with the Jeans modeling approach~\cite{Kaplinghat:2015aga} within 10--20\%). 

Next, we turn to convergence issues.  Since self-interactions affect mainly the innermost radii in the halo, simulations must have sufficient resolution to robustly model these dynamics.  It is well-known that coarse-graining introduces an artificial relaxation process due to two-body gravitational scattering that can alter the inner halo profile, with a timescale that scales with particle number, $t_{\rm relax} \propto N/\log N$~\cite{1987gady.book.....B}. Power et al.~\cite{Power:2002sw} showed that this process is the main limit for resolving the innermost halo structure for collisionless CDM.  The innermost radius of convergence, termed the Power radius, occurs where there are too few enclosed particles, $N(r)$, such that $t_{\rm relax}(r)$ becomes shorter than the Hubble time. This radius $r$ is determined by the condition $N(r)/\log N(r) \approx 0.3  \sqrt{ \bar{\rho}_{\rm dm}(r)/\rho_{\rm crit} }$, where $\bar{\rho}_{\rm dm}(r)/\rho_{\rm crit}$ is the enclosed mean DM density contrast relative to the critical density~\cite{Power:2002sw}.  

In fact, SIDM halos are more robust and better converged below the Power radius compared to their CDM counterparts.  This is sensible since the rate for self-interactions is larger---by construction, the goal is to have more than one self-interaction per Hubble time---than the rate for gravitational scattering~\cite{Vogelsberger:2012ku,Elbert:2014bma}.

At the same time, the cosmological environment at the outermost radii is also important for the SIDM halo.  Its thermal evolution is affected by heating from mergers and infall, which slows the gravothermal collapse of cores.  Hence, simulations of isolated halos are not sufficient for SIDM, as we discuss below.

\subsection{Halo density profiles}  

Cosmological simulations of SIDM halos show a mass deficit at small radii compared to collisionless CDM halos, provided self-interactions are in the optically-thin regime~\cite{Dave:2000ar,Yoshida:2000uw}.  Here we discuss these results and the implications for $\sigma/m$.

\vspace{2mm}

\underline{Galactic scales:} Dav\'{e} et al.~\cite{Dave:2000ar} performed the first cosmological SIDM simulations targeting dwarf galaxies ($M_{\rm halo} \sim 10^{10} \, \Msun$).  Their simulation volume contained several dwarf galaxies, resolved down to $\mathcal{O}({\rm kpc})$ scales, simulated with $\sigma/m \approx 0.5$ and $5 \; {\rm cm^2/g}$.  Their results preferred $5 \; {\rm cm^2/g}$ in order to produce galaxies with core densities $\rho_0 \approx 0.02 \; \Msun/ {\rm pc}^3$, broadly consistent with observed galaxies~\cite{Firmani:2000ce}, while $0.5 \; {\rm cm^2/g}$ yielded densities a few times larger but still viable.  
Moreover, for their largest simulated halo ($M_{\rm halo} \sim 6 \times 10^{11} \; \Msun$), no evidence of core collapse was seen with cross sections as large as $\sigma/m \approx 50 \; {\rm cm^2/g}$.

\begin{figure}
\includegraphics[scale=0.25]{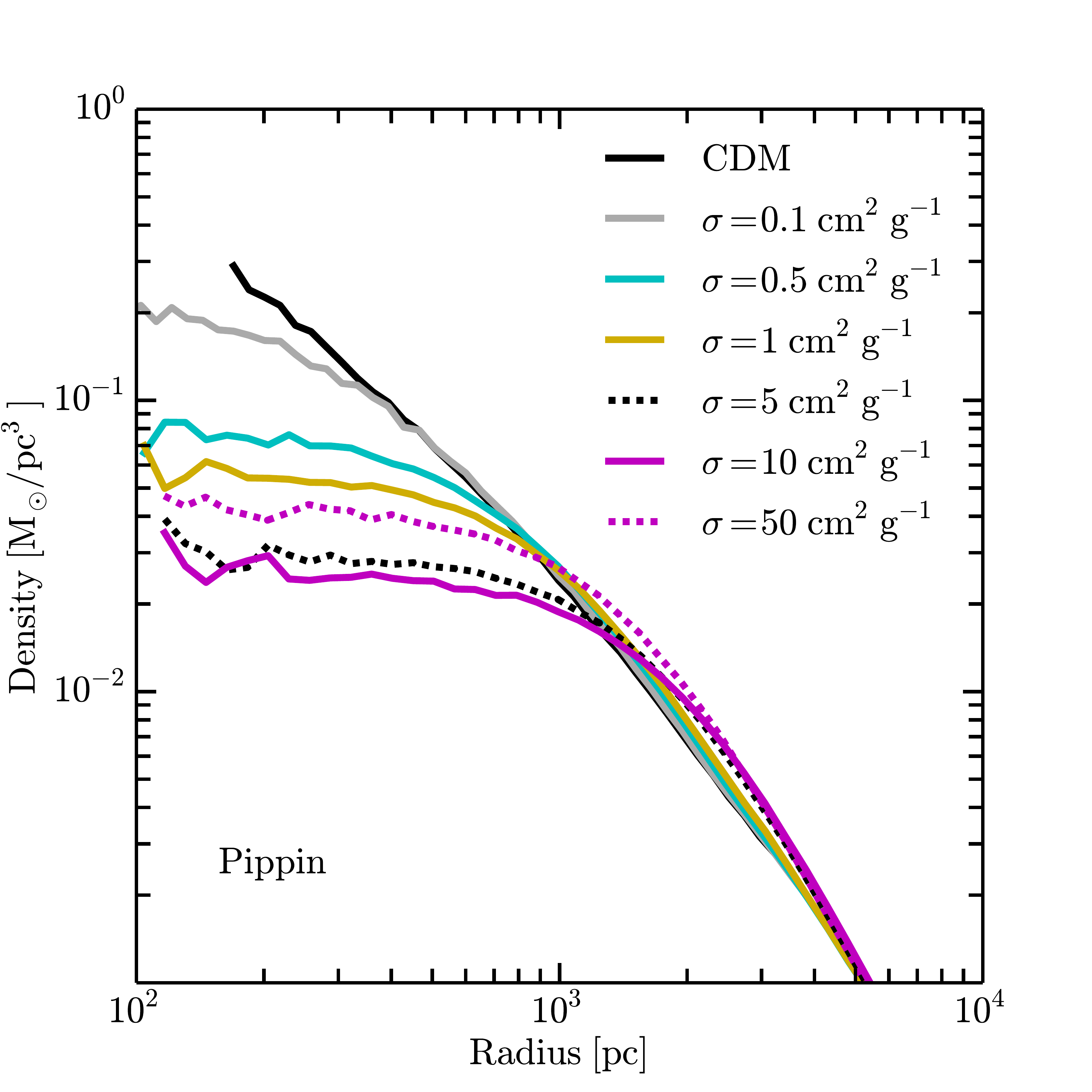}\includegraphics[scale=0.25]{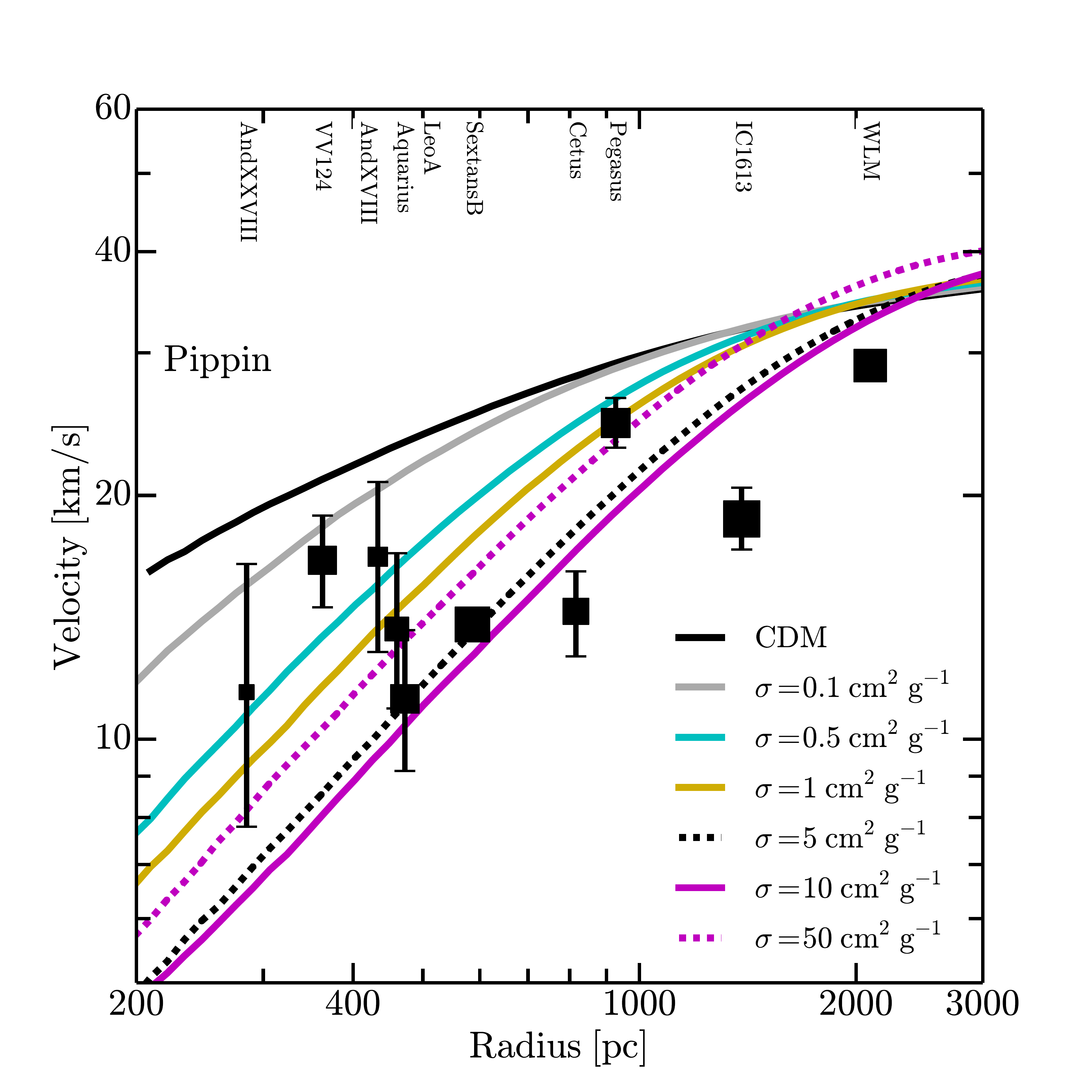}
\caption{ \it Left: Density profiles for halo with mass $\sim 10^{10} \; \Msun$ (dubbed ``Pippin'') from DM-only simulations with varying values of $\sigma/m$.  Right: Rotation curves for Pippin halo with $\sigma/m \gtrsim 0.5 \; {\rm cm^2/g}$ are broadly consistent with measured stellar velocities (evaluated at their half-light radii) for field dwarf galaxies of the Local Group.  Reprinted from Ref.~\cite{Elbert:2014bma}.
}
\label{fig:pippin}
\end{figure}

More recently, Elbert et al.~\cite{Elbert:2014bma} simulated two (cosmological) dwarf galaxies, resolved down to $\mathcal{O}(100 \; {\rm pc})$ scales and for many values of $\sigma/m$.  Fig.~\ref{fig:pippin} (left) shows their results for one dwarf (``Pippin'') and demonstrates that central density is not a monotonic function of $\sigma/m$.  For $\sigma/m \lesssim 10 \; {\rm cm^2/g}$, self-interactions are predominantly in the core-growth regime (heat flowing in), and the central density decreases with increasing $\sigma/m$.  However, a larger self-interaction rate, $\sigma/m = 50 \; {\rm cm^2/g}$, leads to an {\it increasing} central density, indicating this halo has entered core collapse.  Nevertheless, core collapse is mild.  Density profiles with $\sigma/m = 0.5 - 50 \; {\rm cm^2/g}$, spanning two orders of magnitude, vary in their central densities by only a factor of $\sim 3$.  Comparing with data for field dwarfs in the Local Group, Fig.~\ref{fig:pippin} (right) shows that predicted SIDM rotation curves for $0.5 - 50 \; {\rm cm^2/g}$ are consistent with the velocities and half-light radii inferred from several observed galaxies.  This illustrates not only how SIDM affects both the core-cusp and TBTF problems simultaneously, but that $\sigma/m$ need not be fine-tuned to address these issues.

The conclusion from these studies is that $\sigma/m \gtrsim 0.5 \; {\rm cm^2/g}$ can produce $\mathcal{O}({\rm kpc})$ cores needed to resolve dwarf-scale anomalies~\cite{Elbert:2014bma}.  However, the upper limit on $\sigma/m$ at these scales---due to core collapse producing a too-cuspy profile---remains unknown.

\vspace{2mm}

\underline{Cluster scales:} Next, we turn to clusters ($M_{\rm halo} \sim 10^{14} - 10^{15} \, \Msun)$.  The first cosmological simulations at these scales were performed by Yoshida et al.~\cite{Yoshida:2000uw}, which studied a single $10^{15} \, \Msun$ halo for $\sigma/m = 0.1$, $1$, and $10 \; {\rm cm^2/g}$.  More recently, Rocha et al.~\cite{Rocha:2012jg} performed simulations targeting similar scales, but over much larger cosmological volume, for $\sigma/m = 0.1$ and $1 \; {\rm cm^2/g}$.  The best-resolved halos in their volume span $10^{12} - 10^{14} \, \Msun$.  For $1 \; {\rm cm^2/g}$, the central density profiles are clearly resolved for the Yoshida halo and for $\sim 50$ Rocha halos.  On cluster scales, SIDM halos have $\mathcal{O}(100-200 \; {\rm kpc})$ radius cores and central densities $\rho_0 \sim {\rm few} \times 10^{-3} \; \Msun/{\rm pc}^3$.  For $\sigma/m = 0.1 \; {\rm cm^2/g}$, the simulations lack sufficient resolution to fully resolve the cored inner halo, though $\mathcal{O}(30 \; {\rm kpc})$ radius cores seem a reasonable estimate.  For $\sigma/m = 10 \; {\rm cm^2/g}$, the Yoshida halo has a similar density profile compared to $1 \; {\rm cm^2/g}$, although the former is considerably more spherical (ellipticity is discussed below).  

It is important to note that SIDM halos exhibit variability in their structure.  Within the Rocha et al.~\cite{Rocha:2012jg} halo sample, SIDM halos, with fixed  $\sigma/m = 1 \; {\rm cm^2/g}$ and fixed $V_{\rm max}$, show an order-of-magnitude scatter in their central densities.  The dwarf halo samples from Dav\'{e} et al.~\cite{Dave:2000ar} show a similar scatter in central density, albeit with lower resolution.  This variation reflects the different mass assembly histories for different halos.

Strong gravitational lensing data has been used to constrain the core size and density in clusters relevant for SIDM~\cite{Firmani:2000ce,Firmani:2000qe,Wyithe:2000si,Meneghetti:2000gm}.  Mass modeling of cluster CL0024+1654 found a cored DM profile with radius $\sim 50$ kpc~\cite{Tyson:1998vp}---quite similar to the Yoshida $0.1 \; {\rm cm^2/g}$ halo---which was interpretted as evidence for self-interactions at these scales~\cite{Firmani:2000ce,Firmani:2000qe}.  However, interpretation for this particular cluster is complicated by the fact that it has undergone a recent merger along the line of sight (see Ref.~\cite{Umetsu:2009hk} and references therein).  

Meneghetti et al.~\cite{Meneghetti:2000gm} placed the strongest constraint on cluster cores by examining the ability of SIDM halos (specifically, the halo from Yoshida et al.~\cite{Yoshida:2000uw,Yoshida:2000bx}) to produce ``extreme'' strong lensing arcs, i.e., radial arcs or giant tangential arcs.  Giant tangential arcs, with length-to-width ratio $l/w \ge 10$, are present in many lensing observations~\cite{1993MNRAS.262..187W}, but simulated SIDM halos with $\sigma/m = 1$ or $10 \; {\rm cm^2/g}$ lack sufficient surface mass density, limiting arcs to $l/w \lesssim 3.5$.  Observations of radial arcs pose a more severe constraint.  Since they do not occur even for the SIDM halo with $0.1 \; {\rm cm^2/g}$ (as well as for larger $\sigma/m$), Meneghetti et al.~conclude $\sigma/m < 0.1 \; {\rm cm^2/g}$ on cluster scales.  However, there are several caveats to keep in mind (as acknowledged in \cite{Meneghetti:2000gm}).  First, due to the aforementioned variation in SIDM density profiles, constraints based on a single simulated halo require caution.  Second, and more importantly, the simulated SIDM halo does not include the baryonic density from a central galaxy.  Recent studies of massive clusters by Newman et al.~\cite{Newman:2012nw,Newman:2012nv}---two of which exhibit radial arcs (MS2137-23 and A383)---demonstrate that an $\mathcal{O}(10 \; {\rm kpc})$ radius DM core is consistent with lensing data provided the baryonic mass is included.  The important point is that the {\it total} density profile is well-described by an NFW profile, which is well-known to permit radial arcs~\cite{1996A&A...313..697B}, even though the DM density by itself may have a cored profile.  Thus, we conclude that cluster cores are consistent with $0.1 \; {\rm cm^2/g}$, but $1 \; {\rm cm^2/g}$ is excluded.  (We make these statements more precise in \S\ref{sec:jeans}).

\vspace{2mm}

\underline{Isolated vs cosmological simulations:} Lastly, we note that simulations of {\it isolated} SIDM halos can evolve much differently than cosmological halos.  Kochanek \& White~\cite{Kochanek:2000pi} found that cores in isolated halos can be short-lived, collapsing very soon after formation and evolving toward a steeper $\rho \propto r^{-2}$ profile.  For example, an isolated halo similar to the Pippen halo with $50 \; {\rm cm^2/g}$ is expected to form a core and recollapse all within $\sim 1 \; {\rm Gyr}$, in contrast to results from Elbert et al.~\cite{Elbert:2014bma}.  On the other hand, isolated halos with smaller cross sections can have cores that persist over a Hubble time.  This points toward the importance of cosmological infall for mitigating gravothermal collapse---especially for larger values of $\sigma/m$---since the influx of high entropy DM particles slows energy loss from the core.  In fact, it is the occasional violent major merger, rather than the smoother continuous infall of smaller clumps, which is predominantly responsible for ``resetting the clock'' for an otherwise collapsing core~\cite{Yoshida:2000uw,Colin:2002nk}.  This further supports the fact that SIDM halos are expected to have scatter in their structure that reflects the stochastic halo assembly process~\cite{Colin:2002nk,Brinckmann:2017uve}.  

In addition, core collapse for isolated halos is dependent on the initial density profile, which must be set by hand as an initial condition for non-cosmological simulations.  For example, SIDM cores evolving from an initial Hernquist profile, as assumed in Ref.~\cite{Kochanek:2000pi}, collapse twice as quickly relative to an initial NFW profile~\cite{Koda:2011yb}.

\subsection{Halo shapes: ellipticity}

Self-interactions are expected to make DM halos more spherical compared to triaxial collisionless CDM halos, at least in the inner regions where the scattering rate is largest.  In fact, halo shape observations of elliptical galaxies and clusters have provided some of the most stringent constraints on self-interactions that exist in the literature.  These limits are based on the assumption that one scattering per particle is sufficient to substantially affect the ellipticity of a halo.  Early simulations by Dav\'{e} et al.~\cite{Dave:2000ar} supported this conclusion.  For their largest and best-resolved halos ($\sim 10^{11-12}\, \Msun$), SIDM halos yield a minor-to-major axis ratio $c/a \gtrsim 0.85$ at radii where at least one scattering has occurred, while CDM halos have $c/a \approx 0.6-0.7$ at these radii.  However, these conclusions have been revisited in light of recent simulations with larger halo masses and statistics, as we now discuss.

  \vspace{2mm}

\underline{Cluster ellipticity:}  One of the strongest quoted limits on SIDM is due strong lensing measurements of cluster MS2137-23, based on its apparent ellipticity at distances down to $\sim 70$ kpc~\cite{1993ApJ...407...33M}.  Using these observations and assuming the halo should be spherical where $R_{\rm scat}^{-1} \lesssim 5$ Gyr, Miralda-Escud\'{e} obtained a limit $\sigma/m \lesssim 0.02 \; {\rm cm^2/g}$ on cluster scales~\cite{MiraldaEscude:2000qt}, which utterly rules out a contact-type interaction for SIDM.  However, it is crucial to clarify the extent to which these arguments are borne out in N-body simulations.

More recently, Peter et al.~\cite{Peter:2012jh} revisted these issues in detail.  Their study (companion to Rocha et al.~\cite{Rocha:2012jg}, discussed above) involves a sample of $\sim 550$ simulated SIDM halos in the range $10^{11} - 10^{14} \, \Msun$ with $\sigma/m = 0.03$, $0.1$, and $1 \; {\rm cm^2/g}$.  They conclude that the Miralda-Escud\'{e} limit at $0.02 \; {\rm cm^2/g}$ is far overestimated, due to several reasons:
\begin{itemize}
\item For $0.03 \; {\rm cm^2/g}$ (close to the Miralda-Escud\'{e} limit), ellipticities for SIDM and CDM halos were found to be virtually identical, at least down the inner $\mathcal{O}(10 - 20 \; {\rm kpc})$ as limited by resolution.  Although MS2137-23 is a factor of four larger than the largest Peter et al.~halo, large deviations from CDM should be visible in this halo at $\sim 35$ kpc, which matches the same scattering rate in MS2137-23 at $\sim 70$ kpc, if Miralda-Escud\'{e}'s argument is correct. 
\item Even for larger cross sections, SIDM halos are not spherical at the radius where one or two scatterings have occured, but rather remain somewhat elliptical with a median value of $c/a \approx 0.7$ for halos in the range $10^{13} - 10^{14} \, \Msun$.   Brinckmann et al.~\cite{Brinckmann:2017uve} have largely corroborated these results for larger halo masses $\sim 10^{15} \, \Msun$ comparable to MS2137-23.
\item Differing mass assembly histories lead to scatter in $c/a$ at the $\sim 10-20 \%$ for individual halos, which complicates drawing a limit based on a single system.  
\item Lensing observables are sensitive to the projected mass density along the line of sight.  Thus, the surface density at small projected radius includes contributions from the outer halo, which may remain elliptical since self-interactions are not efficient.
\end{itemize}
Peter et al.~conclude that even $1 \; {\rm cm^2/g}$ is not excluded by strong lensing for MS2137-23.\footnote{This conclusion is further supported by the Jeans analysis in Sec.~\ref{sec:jeans} below, which combines lensing and stellar kinematics data to show MS2137-23 is consistent with $\sigma/m \approx 0.1 \; {\rm cm^2/g}$. See Fig.~\ref{fig:clusters} (right).}  

Statistical studies of ensembles of lensing observations, as opposed to single measurements, offer a more powerful probe of self-interactions.  To illustrate the potential of this approach, Peter et al.~\cite{Peter:2012jh} consider a subset of five clusters from the Local Cluster Substructure Survey (LoCuSS)~\cite{Richard:2009yd}, which are chosen based on having a parametrically similar surface density profile compared to their five most massive simulated SIDM halos.  The distribution of ellipticities for SIDM halos with $0.1 \; {\rm cm^2/g}$ is consistent with the LoCuSS observations, while those with $1 \; {\rm cm^2/g}$ are not.  Peter et al.~conclude that these data tentatively suggest $\sigma/m \lesssim 1 \; {\rm cm^2/g}$, albeit with caution due to the limited statistics, unknown selection bias, and lack of baryons included in the SIDM lens modeling.  Indeed, for $0.1 \; {\rm cm^2/g}$, noticible differences in ellipticity between SIDM and CDM halos become apparent at only small radii where the stellar density becomes important~\cite{Peter:2012jh,Brinckmann:2017uve}.

\vspace{2mm}

\underline{Elliptical galaxies:} These massive systems offer the opportunity to constrain self-interactions in halos on $10^{12} - 10^{13} \, \Msun$ scales.  Constraints on SIDM have focused on the isolated elliptical galaxy NGC 720~\cite{Feng:2009hw,Feng:2009mn,Lin:2011gj}, which has $M_{\rm halo} \sim 7 \times 10^{12} \, \Msun$~\cite{Humphrey:2006rv}, based on X-ray shape measurements from the {\it Chandra} telescope~\cite{Buote:2002wd}. Buote et al.~\cite{Buote:2002wd} have shown that the X-ray isophotes remain elliptical at least down to projected radii $\sim 5$ kpc. (At smaller radii, shape measurements suffer from systematic uncertainties due to point source subtraction.)  Since the X-ray emissivity scales as $\rho_{\rm gas}^2$, where the gas density $\rho_{\rm gas}$ has a fairly steep radial dependence, emission along the line of sight is strongly weighted toward physical radii near the projected radius (in contrast with lensing measurements, as described above).  To determine the shape of the DM density, 
Buote et al.~treat the gas as a single isothermal component in hydrostatic equilibrium and model the total mass density with a spheroidal profile that is either NFW, Hernquist~\cite{1990ApJ...356..359H}, or isothermal ($\rho_{\rm dm} \propto r^{-2}$).  While the isothermal model provides the best $\chi^2$ fit, all models (whether prolate or oblate) produce a similar ellipticity $\epsilon \approx 0.4$.  

Constraints from NGC 720 have been based on the assumption that one self-interaction per particle within the inner 5 kpc over 10 Gyr is enough sphericalize the inner halo~\cite{Feng:2009hw,Feng:2009mn,Lin:2011gj}.  This yields a stringent constraint, $\sigma/m \lesssim 0.01 \; {\rm cm^2/g}$, assuming a mean DM density $\rho_{\rm dm} \approx 0.1 \; \Msun/{\rm pc}^3$ and relative velocity $\rm v_{\rm rel} \approx 540$ km/s~\cite{Humphrey:2006rv,Feng:2009hw}.  

However, the simulations by Peter et al.~\cite{Peter:2012jh} show such constraints to be substantially overestimated as well.  As above, SIDM halos retain ellipticity even where $R_{\rm scat} \sim 0.1 \; {\rm Gyr}^{-1}$.  Based on their sample of simulated SIDM halos with mass $3-10 \times 10^{12} \, \Msun$, the distribution of ellipticities for $0.1 \; {\rm cm^2/g}$ is perfectly consistent with $\epsilon \approx 0.4$, and even $1 \; {\rm cm^2/g}$ halos are marginally allowed.  On the other hand, the mean central densities for SIDM halos with $1 \; {\rm cm^2/g}$ are typically too small compared to NGC 720 (by a factor of a few), while $0.1 \; {\rm cm^2/g}$ provides better agreement.  But before any robust limit can be made, it is essential to consider the gravitational influence of baryons on the DM density~\cite{Kaplinghat:2013xca}.  If baryons dominate the total mass density within the inner $\sim 5 - 10$ kpc (as in the case for several elliptical galaxies that have received detailed mass modeling~\cite{Humphrey:2006rv}), the SIDM density profile can be steeper and more elliptical than expected without baryons.

\subsection{Substructure}

In the collisionless CDM paradigm, hierarchical structure formation leads to an abundance of substructure within halos~\cite{Kauffmann:1993gv}.  Observationally, these ``halos within halos'' manifest as dwarf galaxies around larger galaxies, or galaxies within groups or clusters.  Self-interactions tend to erase substructure, provided the scattering rate is sufficiently large, through two effects~\cite{Spergel:1999mh}.  First, self-interactions within subhalos lead to density profiles that are less concentrated and more prone to tidal disruption.  Second, self-interactions lead to evaporation of subhalos via ram pressure stripping as they pass through their host halo, since the former have a much lower velocity dispersion compared to the latter.  Suppressing the subhalo mass function on MW scales is relevant for addressing the missing satellites problem~\cite{Moore:1999nt,Klypin:1999uc}, while reducing the central density profiles of subhalos is relevant for addressing the TBTF problem~\cite{BoylanKolchin:2011de,BoylanKolchin:2011dk}.  However, substructure on cluster scales must be preserved~\cite{Moore:1999nt,Gnedin:2000ea}.

\begin{figure}
\includegraphics[scale=0.26]{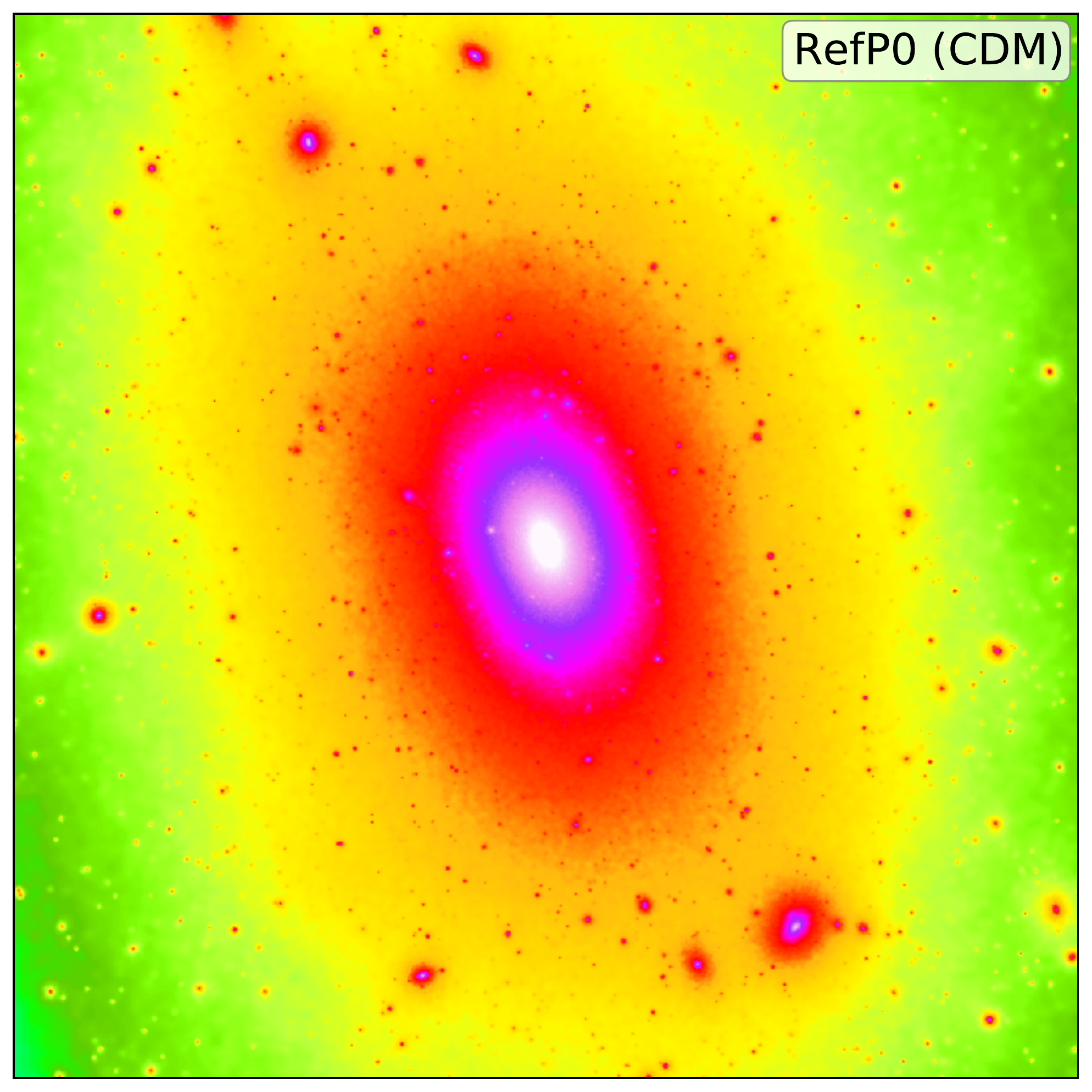}
\includegraphics[scale=0.26]{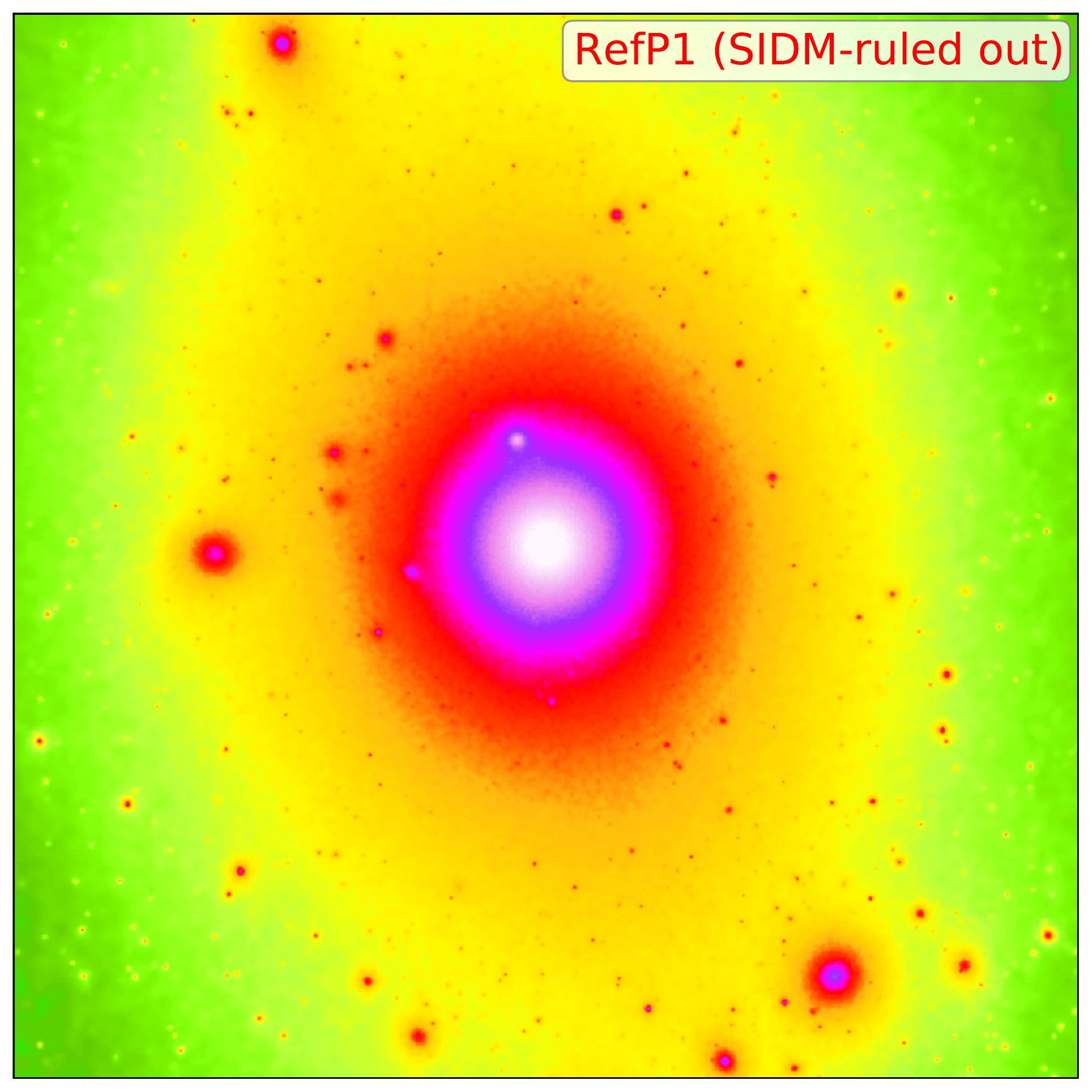}
\includegraphics[scale=0.26]{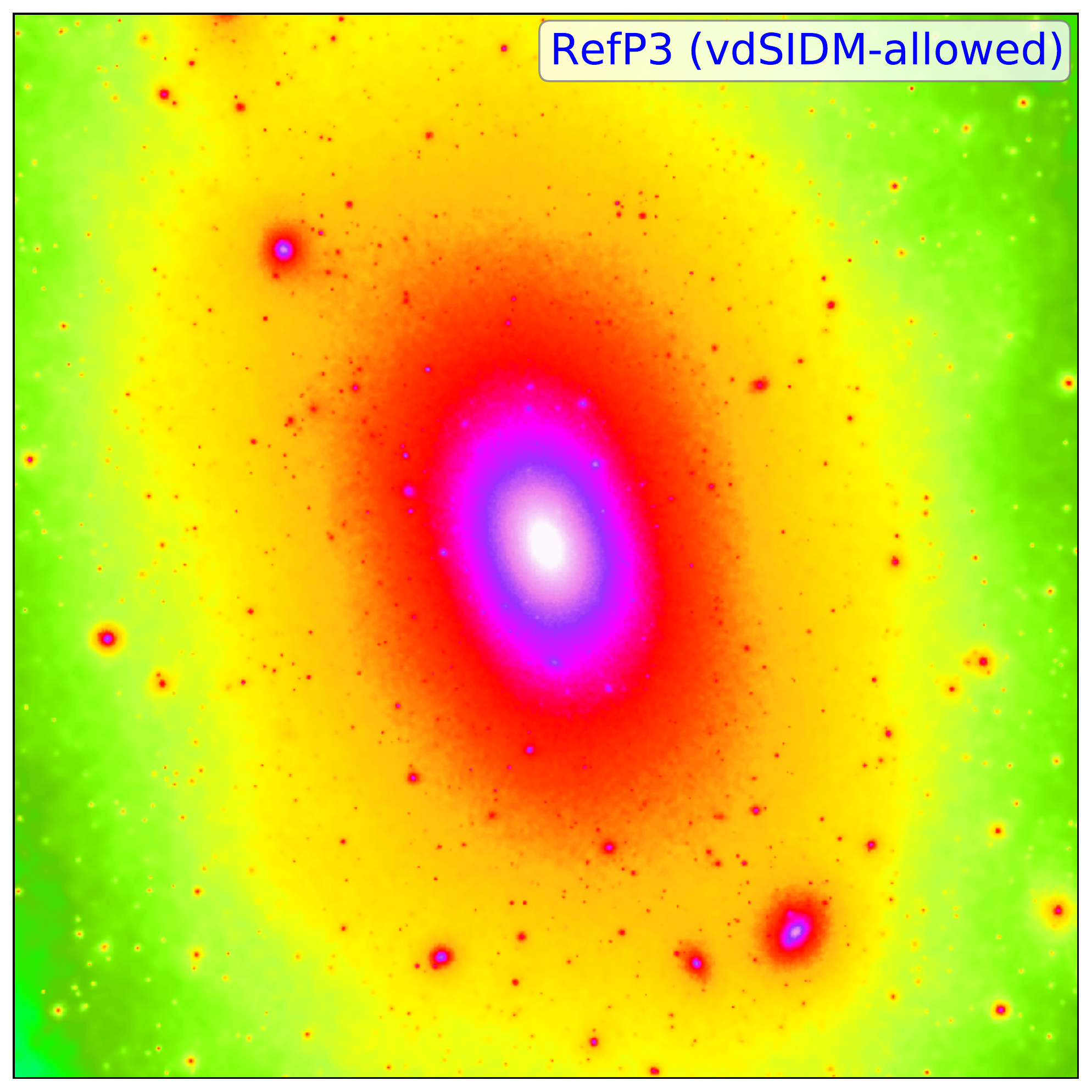}
\caption{ \it MW-like DM halo and its substructure for three scenarios: collisionless CDM (left); SIDM with a large, constant cross section $\sigma/m = 10 \; {\rm cm^2/g}$ (center); and SIDM with velocity-dependent scattering (right).  Velocity-dependent SIDM model has $\sigma/m = 35 \; {\rm cm^2/g}$ at $v= 10 \; {\rm km/s}$ (dwarf scales) and $\sigma/m \lesssim 10^{-2} \; {\rm cm^2/g}$ for $v\gtrsim 200 \; {\rm km/s}$ (MW and larger scales).  Each panel shows projected densities for a $(270 \; {\rm kpc})^3$ cube.  Reprinted from Ref.~\cite{Vogelsberger:2012ku}.
}
\label{fig:MWhalo}
\end{figure}

\vspace{2mm}

\underline{Milky Way substructure:}  Early simulations by Dav\'{e} et al.~\cite{Dave:2000ar} reported modest reductions ($30-50\%$) in the subhalo mass function below $10^9 \, \Msun$ for $\sigma/m \approx 0.5$ and $5 \; {\rm cm^2/g}$, which could help, but not alleviate, the missing satellites problem.  However, more recent, higher resolution SIDM simulations have reached a more pessimistic conclusion~\cite{Colin:2002nk,Vogelsberger:2012ku,Rocha:2012jg,Zavala:2012us,Dooley:2016ajo}.  Fig.~\ref{fig:MWhalo} shows the projected densities for a MW-like halo from Ref.~\cite{Vogelsberger:2012ku} for different DM models.  
For constant cross sections of $\sigma/m = 1 \; {\rm cm^2/g}$ or less, the subhalo mass function remains unchanged compared to collisionless CDM~\cite{Zavala:2012us}, which is illustrated in Fig.~\ref{fig:MWhalo} (left).  For a larger cross section of $\sigma/m = 10 \; {\rm cm^2/g}$, the subhalo mass function could be reduced by $\mathcal{O}(30\%)$ for subhalos below $10^{8.5} \, \Msun$, while the number of larger subhalos remains unaffected.  This scenario corresponds to Fig.~\ref{fig:MWhalo} (center).  While it is evident that substructure is suppressed---particularly for smaller halos located nearer to the center of the host halo---even this modest reduction in substructure comes at a price of making the host halo spherical out to distances $\sim {\rm 50 \; kpc}$.  This scenario is excluded according to the ellipticity constraints discussed above (albeit for halos a factor of a few more massive than the MW).  

Alternatively, self-interactions may be velocity-dependent, which effectively allows the cross section to vary across halo mass scales.  Fig.~\ref{fig:MWhalo} (right) shows a MW-like SIDM halo with velocity-dependent self-interactions~\cite{Vogelsberger:2012ku}.  The cross section chosen, motivated by classical scattering from a Yukawa potential~\cite{Feng:2009hw,Loeb:2010gj}, is sizable on dwarf scales but suppressed on MW and larger scales.  It is clear from Fig.~\ref{fig:MWhalo} (right) that the subhalo mass function is indistinguishable from collisionless CDM.  Even though the scattering rate is large on dwarf scales, this does not translate into evaporation of substructure since the relative velocity between the subhalo and its host halo is set by the velocity dispersion of the latter, for which the cross section is suppressed. 

Despite being one of its original motivations~\cite{Spergel:1999mh}, the conclusion is that SIDM {\it cannot} solve the missing satellites problem.\footnote{This conclusion only applies to the {\it minimal} SIDM scenario where the only interaction is elastic scattering between DM particles. Nonminimal variations of SIDM can have a substantial impact on substructure (see \S\ref{sec:comp}).}  Having a constant cross section around $0.5\; {\rm cm^2/g}$ is insufficient to erase MW substructure~\cite{Colin:2002nk,D'Onghia:2002hm}.  Reducing substructure requires a much larger cross section on MW scales, which is excluded based on halo shape constraints, while merely having a large cross section on dwarf scales is insufficient to erase substructure~\cite{Vogelsberger:2012ku,Zavala:2012us}.  However, allowed SIDM models can still impact the stellar mass function for satellite galaxies~\cite{Dooley:2016ajo}.  SIDM subhalos, while negligibly impacted by DM evaporation, have cored density profiles, resulting in a stellar density that is less tightly bound and more prone to tidal stripping.

\begin{figure}
\includegraphics[scale=0.24]{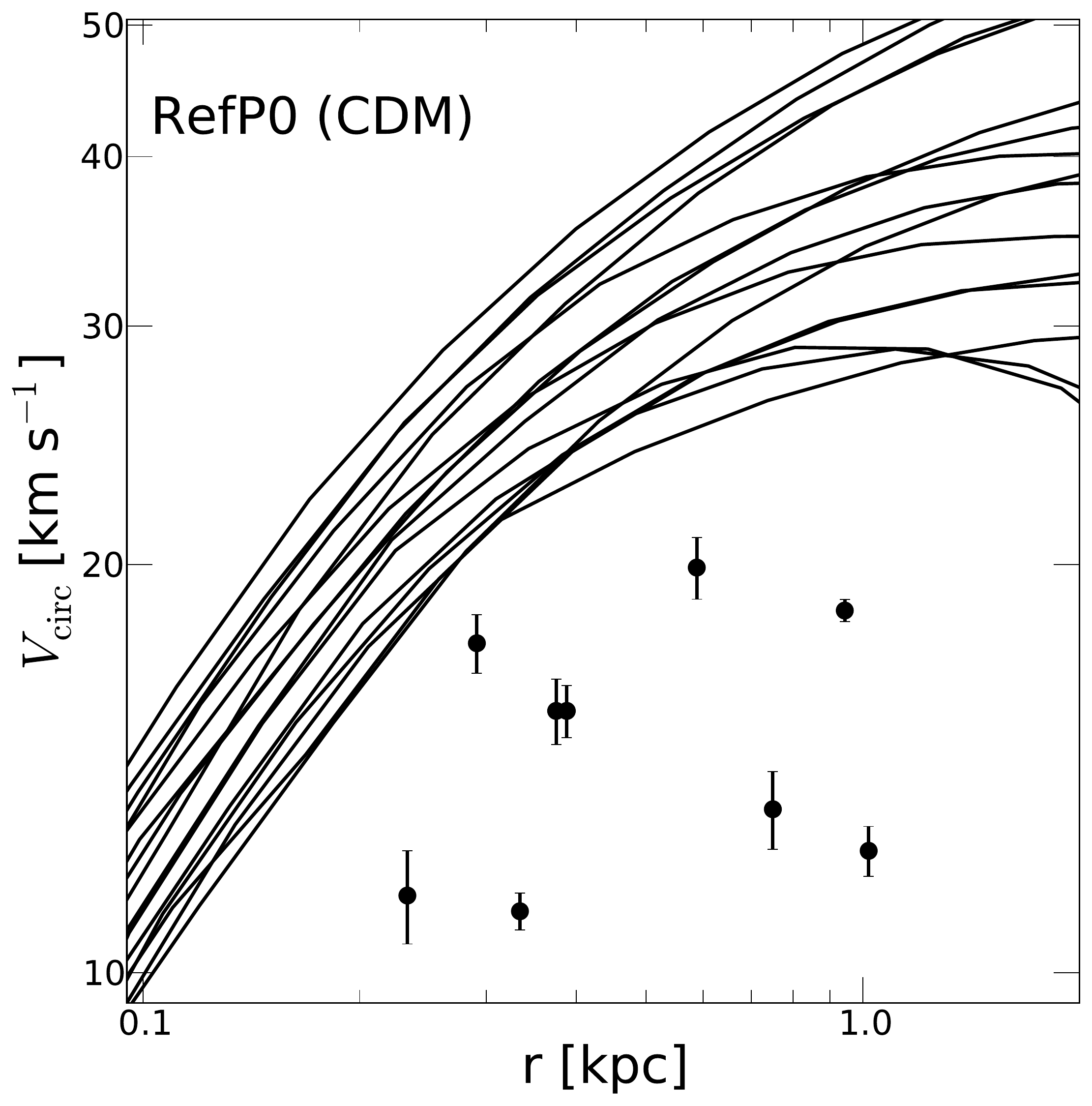}
\includegraphics[scale=0.24]{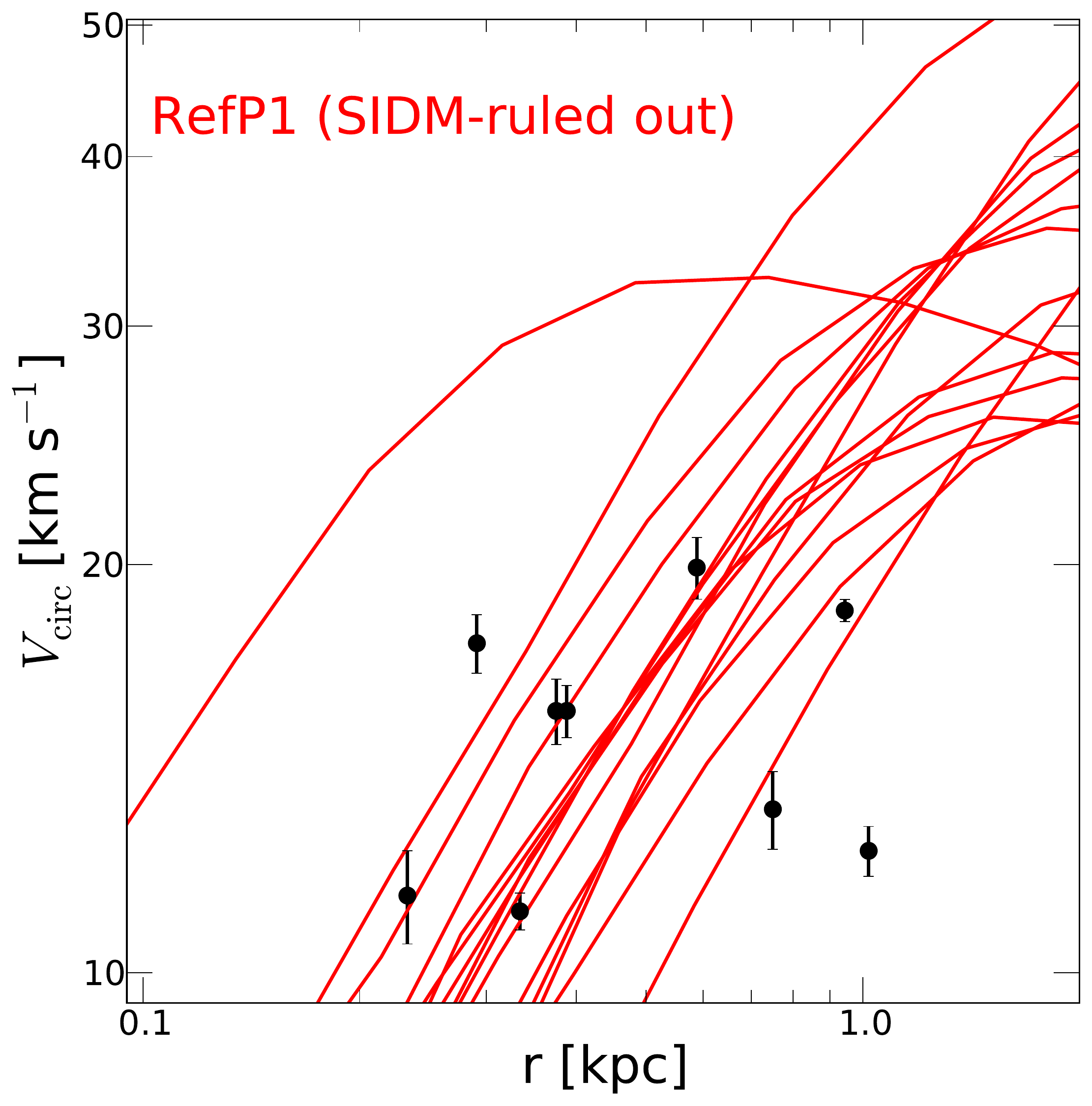}
\includegraphics[scale=0.24]{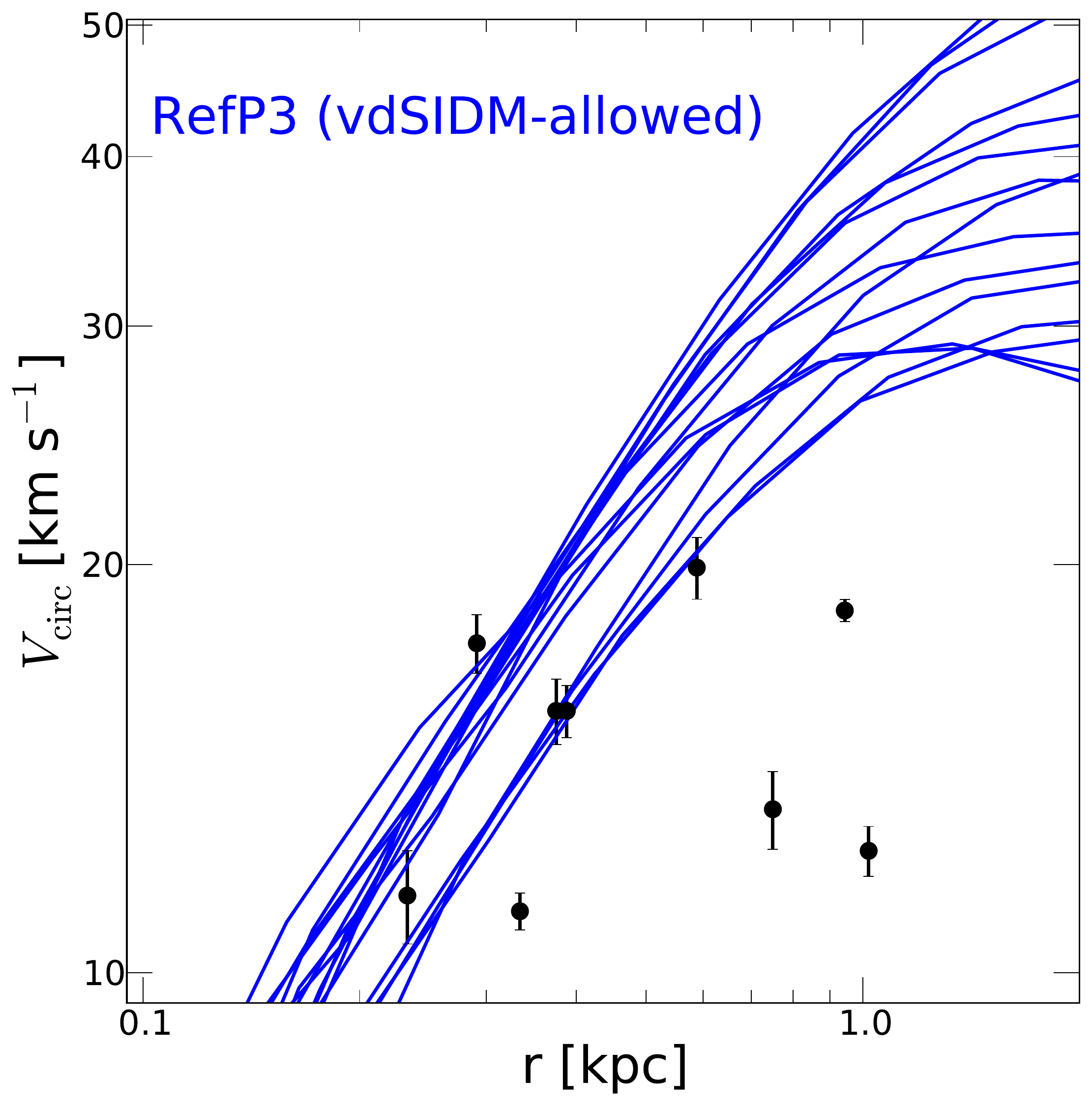}
\caption{ \it Circular velocity profiles for 15 most massive subhalos for three scenarios shown in Fig.~\ref{fig:MWhalo}: collisionless CDM (left); SIDM with $10 \; {\rm cm^2/g}$ (center); and SIDM with velocity-dependent scattering (right).  Data points show the inferred circular velocities at the half-light radii for MW dwarf spheroidal galaxies. Reprinted from Ref.~\cite{Vogelsberger:2012ku}.
}
\label{fig:MWhaloTBTF}
\end{figure}

Next, we turn to the TBTF problem.  For collisionless CDM, the most massive subhalos within a simulated MW-like halo are too dense, with too-large predicted stellar velocity profiles, to match observed velocity dispersions for the MW dSphs~\cite{BoylanKolchin:2011dk}. Despite having little effect on the abundance or total mass of subhalos, self-interactions can reduce their central densities and thereby reduce their velocity profiles in accord with observations, as shown in Fig.~\ref{fig:MWhaloTBTF}.  Self-interactions at the level of $1\; {\rm cm^2/g}$ (or the velocity-dependent models considered therein) yield a reduced central density of $\rho_0 \sim 0.1 \; \Msun/{\rm pc}^3$~\cite{Vogelsberger:2012ku,Zavala:2012us}.  These SIDM scenarios are consistent with the observed stellar kinematics for the MW dSphs~\cite{Strigari:2008ib,Wolf:2009tu}, while for $ 0.1\; {\rm cm^2/g}$ self-interactions are insufficient~\cite{Zavala:2012us}.

\vspace{2mm}

\underline{Cluster-scale substructure:}  In contrast to MW scales, there is no ``missing galaxies'' problem for clusters.  Moore et al.~\cite{Moore:1999nt} showed that the abundance of substructure within the Virgo cluster is well-described by simulations of a collisionless CDM halo with mass $5 \times 10^{14} \, \Msun$.  However, this effect does not provide a strong constraint on self-interactions.  SIDM simulations with $1 \; {\rm cm^2/g}$ find only a modest effect from subhalo evaporation on cluster scales, while observables related to the central cluster densities and core sizes are in principle much more stringent~\cite{Rocha:2012jg}.  

Using analytic estimates, Gnedin \& Ostriker~\cite{Gnedin:2000ea} excluded $\sigma/m$ between $0.3 - 10^4 \; {\rm cm^2/g}$ based on the effect of DM evaporation on elliptical galaxies in cluster halos.  As an elliptical galaxy passes through its host cluster halo, stars in the elliptical galaxy expand adiabatically as DM mass is lost through self-interactions, while the luminosity remains unchanged.  This causes a shift in the fundamental plane for elliptical galaxies found in clusters and in the field, since the latter do not experience evaporation, and no significant environmental dependence is observed~\cite{Kochanek:1999rj}.\footnote{More recent studies have found a weak dependence to the fundamental plane intercept dependent on the local galaxy density.  However, this dependence correlates with the stellar mass-to-light ratio, while the total dynamical mass (which would be affected by evaporation in denser environments) shows no such correlation~\cite{LaBarbera:2010jt}.}  However, based on simulation results for $1 \; {\rm cm^2/g}$, Gnedin \& Ostriker's constraint may be overestimated, although detailed comparisons between N-body simulations and the fundamental plane relation are lacking~\cite{Rocha:2012jg}.

\subsection{SIDM simulations with baryons}
\label{sec:sidmbar}

While DM-only simulations have played a crucial role toward understanding the impact of self-interactions on DM halos, the next step is to incorporate self-interactions within hydrodynamic simulations that include the dynamics of baryons.  Feedback processes remain the leading ``vanilla'' explanation for small scale structure issues apparent in CDM-only simulations, despite considerable debate over precisely how they are implemented.  Hydrodynamic simulations with SIDM may provide guidance for how to disentangle and distinguish the effects of self-interactions and feedback.  Additionally, it is important to understand how the predictions of SIDM are influenced by the presence of baryons, as well as conversely how baryonic tracers for DM are influenced by self-interactions.

Recently, Vogelsberger et al.~\cite{Vogelsberger:2014pda} and Fry et al.~\cite{Fry:2015rta} have performed the first N-body simulations including both self-interactions and baryons, both targeting dwarf scales.  However, both sets of simulations have implemented star formation and feedback to opposite effect.  In Ref.~\cite{Fry:2015rta}, galaxies have bursty star formation histories (provided enough baryons are present) marked by episodic supernovae-driven gas outflows over times much shorter than the dynamical time scale of the galaxy.  This process---motivated in part by explaining the formation of bulgeless dwarf galaxies by expelling low angular momentum gas that would otherwise form a stellar bulge---leads to nonadiabatic changes in the gravitational potential that can significantly impact the central DM distribution~\cite{Governato:2009bg,Pontzen:2011ty,Governato:2012fa}.  On the other hand, Ref.~\cite{Vogelsberger:2014pda} has implemented a smoother prescription for star formation and feedback, that, while successfully suppressing bulge formation, provides a much smaller influence on the DM halo.

Vogelsberger et al.~\cite{Vogelsberger:2014pda} present results for two dwarf galaxies, simulated for a range of self-interaction cross sections.  These galaxies are DM-dominated at all radii and are comparable to observed THINGS dwarfs~\cite{Oh:2010ea} in terms of their stellar mass and maximum circular velocity, $V_{\rm max} \sim 50 - 80 \; {\rm km/s}$.  Whether DM is self-interacting or collisionless, baryonic feedback provides a negligible impact on the DM halo, and the DM central density is primarily influenced by the effect of self-interactions as in SIDM-only simulations.  However, the presence of an $\mathcal{O}({\rm kpc})$ core in SIDM halos can affect the baryon component compared to a more concentrated DM profile, resulting in an $\mathcal{O}(30\%)$ reduction in star formation and reducing the central densities of stars and gas.  In particular, the stellar distribution inherits a core radius that correlates with the core radius for DM, which may provide a useful observational handle for SIDM.

Fry et al.~\cite{Fry:2015rta} have simulated somewhat smaller systems, including three dwarf galaxies with $V_{\rm max} \sim 30-60 \; {\rm km/s}$---comparable to field dwarfs of the Local Group~\cite{Wolf:2009tu}---as well as making predictions for even smaller dwarfs that are below current observational sensitivities.  For the larger dwarfs, baryonic feedback by itself produces reduced central densities and $\mathcal{O}({\rm kpc})$ cores for collisionless CDM, upon which the further inclusion of self-interactions with $2 \; {\rm cm^2/g}$ has only marginal effect.  For smaller dwarfs with $V_{\rm max} \lesssim 20 \; {\rm km/s}$, baryonic feedback is insufficient to form cores due to a lack of star formation.  At the same time, self-interactions with $2 \; {\rm cm^2/g}$ are unable to form cores larger than 500 pc in these smaller halos (consistent with simple rate arguments).  

It is evident that the conclusions of Refs.~\cite{Vogelsberger:2014pda,Fry:2015rta} are quite different due to the feedback prescriptions adopted therein.  This underscores the role of baryonic physics as a systematic uncertainty in the small scale structure puzzle, as well the importance of taking new approaches to test the consistency of feedback models with collisionless CDM (e.g., Refs.~\cite{Katz:2016hyb,Pace:2016oim,2016ApJ...820..131E,2017ApJ...835..193E}).


\section{Jeans approach to relaxed SIDM halos}
\label{sec:jeans}
\subsection{Isothermal solutions to the Jeans equations}
Despite the importance of N-body simulations, these methods are limited by their intensive computational nature.  Even minimal particle models for SIDM exhibit rich dynamics for elastic scattering~\cite{Tulin:2012wi,Tulin:2013teo}, and it is not feasible to explore the full range of possibilities with simulations.  To complement these studies, there is a useful semi-analytic method based on the Jeans equation for understanding SIDM halo profiles in relaxed systems~\cite{Rocha:2012jg,Kaplinghat:2013xca,Kaplinghat:2015aga}.  This approach is well-suited to the intermediate cross section regime, $\sigma/m \sim 1\; {\rm cm^2/g}$, where DM is neither fully collisionless nor collisional throughout an entire halo.  The Jeans method can be fit directly to observations for individual systems, from dwarf galaxies to massive clusters, including the gravitational effect on the halo from the observed baryonic mass distribution.  Therefore, this method provides a bridge between simulations, astrophysical observations, and SIDM particle models.  

The Jeans equation may be derived from the collisional Boltzmann equation (see, e.g.,~\cite{1987gady.book.....B})
\beq \label{eq:boltzmanneq}
\frac{ \partial f}{\partial t} + \mathbf{v} \cdot \boldsymbol{\nabla} f -  \boldsymbol{\nabla} \Phi_{\rm tot} \cdot \frac{\partial f}{\partial \mathbf{v}} = \mathscr{C}[f]
\eeq
where $f(\mathbf r, \mathbf v, t)$ is the DM distribution function, $\Phi_{\rm tot}=\Phi_{\rm dm} + \Phi_b$ is the total gravitational potential for both DM and baryons, and $\mathscr{C}$ is the collision term from self-interactions.  In the inner halo, the collision term drives the distribution function toward kinetic equilibrium, $f \propto \exp( - \tfrac{1}{2} |{\mathbf v}|^2/\sigma_0^2 )$, where $\sigma_0$ is the isotropic one-dimensional velocity dispersion.  Taking the first moment of Eq.~\eqref{eq:boltzmanneq} and searching for quasi-equilibrium solutions in which time derivatives can be neglected, we have the (time-independent) Jeans equation
\beq \label{eq:Jeans}
\boldsymbol{\nabla} (\sigma^2_0 \rho_{\rm dm}) = - \rho_{\rm dm} \boldsymbol{\nabla}\Phi_{\rm tot} \, .
\eeq
This is simply the condition for hydrostatic equilibrium for an ideal gas of pressure $p_{\rm dm} = \sigma_0^2 \rho_{\rm dm}$.  The velocity anisotropy is assumed to vanish since collisions isotropize trajectories.  Moreover, N-body simulations have shown that DM particles are approximately isothermal within the inner halo, such that $\sigma_0$ can be taken as a constant.\footnote{The velocity dispersion is expected to increase (decrease) with halo radius for heat transfer flowing in (out), corresponding the core growth (collapse) phase of the SIDM halo, depending on the value of $\sigma/m$.  However, for a range of $\sigma/m$, N-body simulations have shown that the spatial variation in $\sigma_0$ is only $\mathcal{O}(10\%)$ within the inner halo~\cite{Elbert:2014bma}.}  Combining with Poisson's equation, we have
\beq \label{eq:Jeans2}
\sigma_0^2 \boldsymbol{\nabla}^2 \ln \rho_{\rm dm} = - 4 \pi G(\rho_{\rm dm} + \rho_b) \, ,
\eeq
where $\rho_b$ is the baryon mass density and $G$ is Newton's constant.

On the other hand, DM is effectively collisionless in the outer halo due to the reduced particle number density.  The delineation between the inner and outer halo occurs at radius $r_1$ where, on average, one collision per particle has occurred over the age of the halo, $t_{\rm age} \sim 5 - 10 \; {\rm Gyr}$.  This condition is $R(r_1) t_{\rm age} \sim 1$, where the scattering rate is given in Eq.~\eqref{eq:rate}.  The full density profile $\rho_{\rm dm}$ is taken to be a hybrid of collisional and collisionless (NFW) profiles, matched together at $r=r_1$:
\beq \label{eq:match}
\rho_{\rm dm}(r) = \left\{ \begin{array}{cc} \rho_{\rm iso}(r) \, ,   & r < r_1 \\ \rho_{\rm NFW}(r) \, ,& r > r_1 \end{array} \right. \, .
\eeq
Here, $\rho_{\rm iso}$ is the isothermal density profile defined as the solution to Eq.~\eqref{eq:Jeans2}. To match the two regions, the density profiles and enclosed mass are assumed to be equal at $r_1$. The physical picture is that self-interactions simply rearrange and thermalize the DM particle distribution in the inner halo, while leaving the outer halo as it would be in the absence of collisions. 

With this simple halo model in mind, we turn to several important questions:  Do self-interactions provide a consistent solution to the core-cusp problem in {\it all} astrophysical systems?  What are the particle physics implications for observations spanning widely different halo mass scales, from dwarf galaxies to clusters?  And lastly, does the Jeans approach for SIDM halos agree with N-body simulations?

\begin{figure}
\includegraphics[scale=0.63]{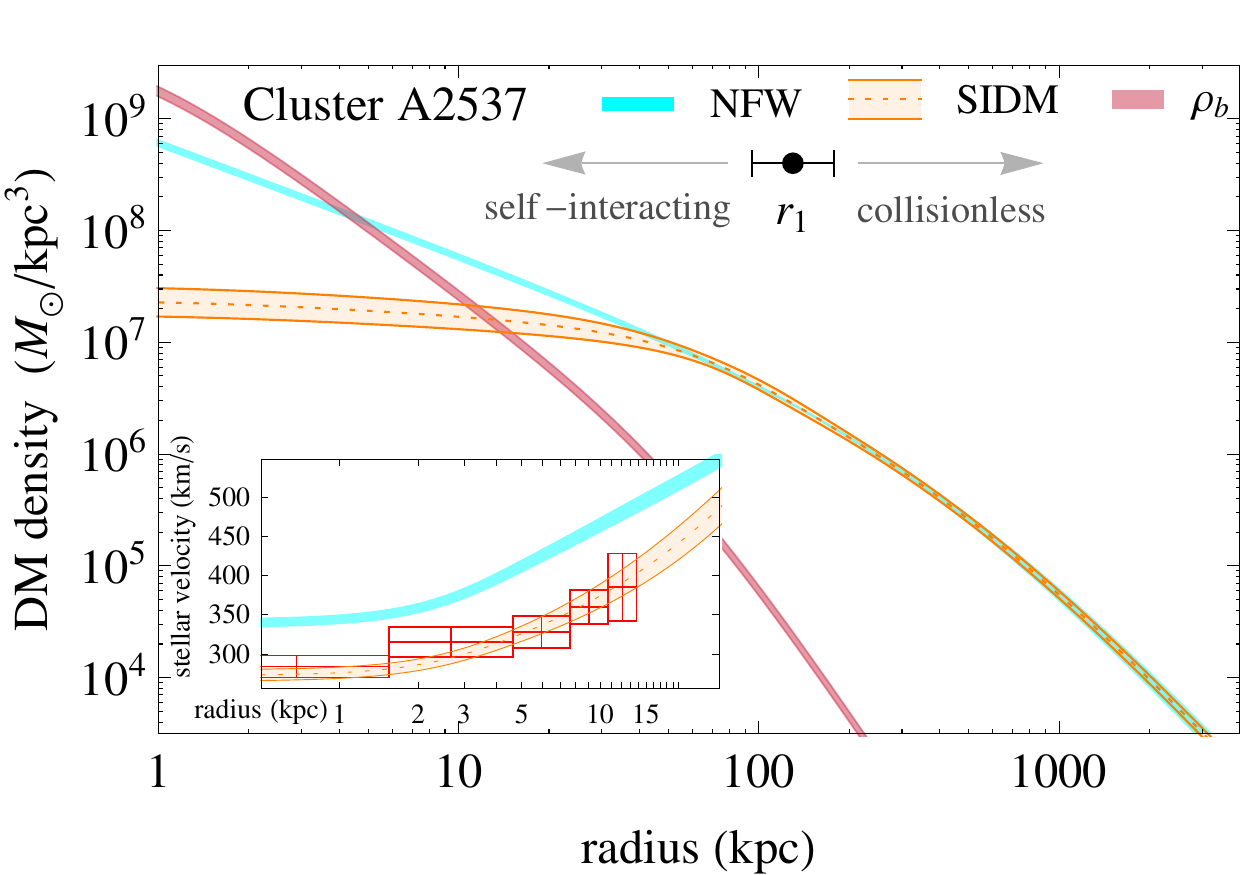}\;\;\includegraphics[scale=0.63]{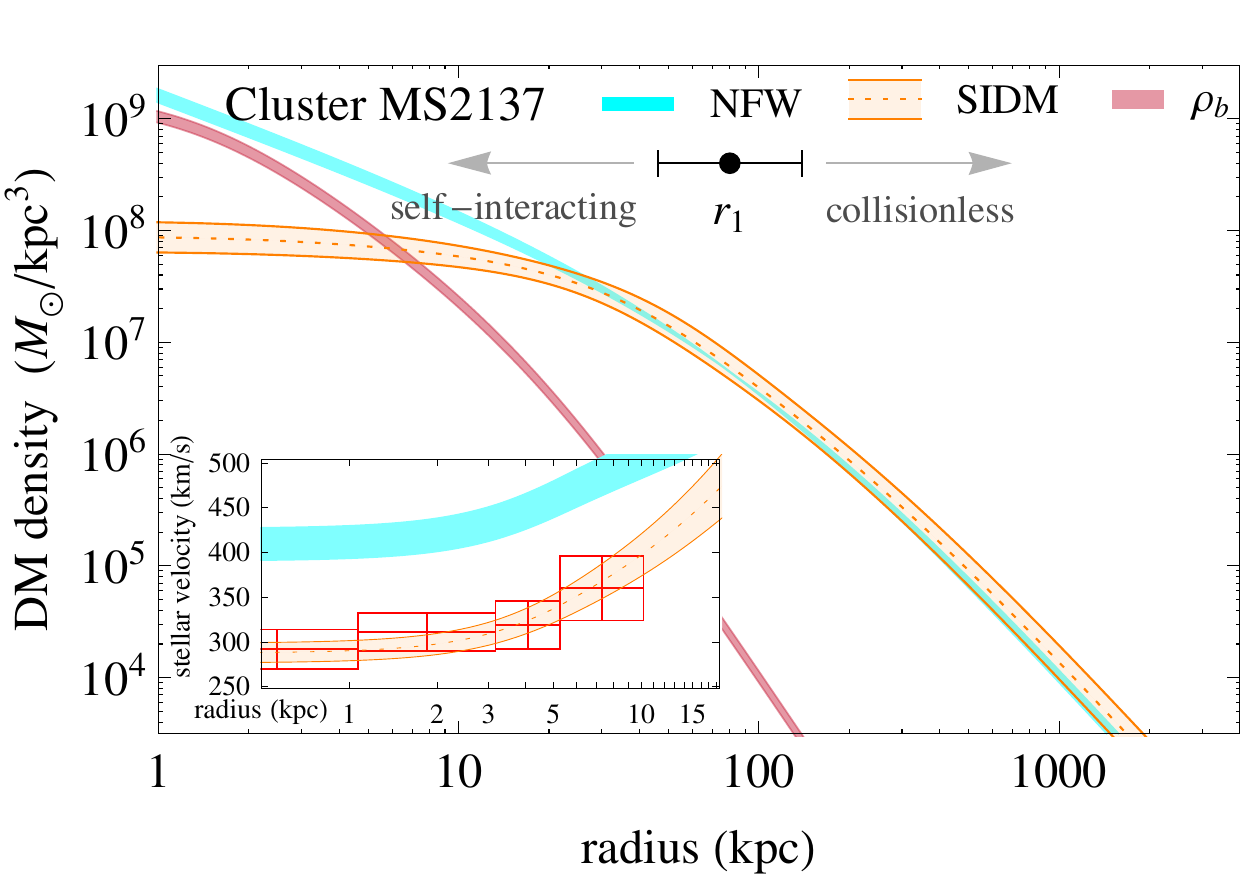}
\caption{ \it DM density profiles obtained for clusters A2537 (left) and MS2137 (right) using the Jeans method as per Eq.~\eqref{eq:Jeans2}.  Orange band shows the full SIDM profile from fitting stellar velocity kinematics at small radii (red data points in inset) and lensing data at large radii (not shown).  Red band shows the stellar density profile.  Cyan band shows the collisionless profile obtained by fitting only lensing data, which overpredicts stellar velocities in the central halo (see inset).  Both clusters are consistent with $\sigma/m \approx 0.1 \; {\rm cm^2/g}$.  Reprinted from Ref.~\cite{Kaplinghat:2015aga}.
}
\label{fig:clusters}
\end{figure}

To address these questions, Kaplinghat, Tulin \& Yu~\cite{Kaplinghat:2015aga} applied the Jeans method to an astrophysical data set spanning DM halo masses in the range $\sim 10^9 - 10^{15} \; \Msun$.  These data include rotation curves from twelve DM-dominated galaxies, including those from THINGS~\cite{2011AJ....141..193O} and LSB galaxies from Kuzio de Naray et al.~\cite{2008ApJ...676..920K}, combined with kinematic and lensing studies from six massive clusters by Newman et al.~\cite{Newman:2012nw,Newman:2012nv}, all of which exhibit cored profiles.  For each system, the cross section is obtained by
\beq
\langle \sigma v_{\rm rel} \rangle /m = (\rho_{\rm dm}(r_1) t_{\rm age})^{-1} \, 
\eeq
following Eq.~\eqref{eq:rate}.  Here, the quantity on the left-hand side represents the velocity-weighted cross section per unit mass, statistically averaged over velocity, while the $\rho_{\rm dm}(r_1)$ on right-hand side is obtained by fitting Eq.~\eqref{eq:match} to astrophysical data for each system.  Since more massive halos correlate with higher average relative velocities $\langle v_{\rm rel} \rangle$ for DM particles, the range of halos provides an important probe of the velocity-dependence of self-interactions.  Analogous to tuning the beam energy in a particle collider, the energy-dependence of scattering is crucially important for probing the underlying particle physics of SIDM.  

To illustrate the Jeans method, Fig.~\ref{fig:clusters} shows results for two clusters from Ref.~\cite{Kaplinghat:2015aga}.  The full SIDM profile has been fit to the stellar velocity dispersions for the brightest central galaxy at small radii ($\lesssim 10$ kpc) and strong and weak lensing data at larger radii ($\gtrsim 10$ kpc).  These data prefer cluster profiles with cores.  A cuspy (NFW) profile fit {\it only} from lensing data does not agree with stellar data in the central halo (see inset).  By matching the collisional and collisionless regions of the halo together to determine $r_1$, the preferred cross section for these clusters is $\sigma/m \approx 0.1 \; {\rm cm^2/g}$.  These conclusions may be weakened if stellar anisotropies are far more significant that assumed~\cite{Schaller:2014gwa} or if AGN feedback is relevant~\cite{Martizzi:2011aa}, in which case $\sigma/m \lesssim 0.1 \; {\rm cm^2/g}$.

\begin{figure}
\includegraphics[scale=0.63]{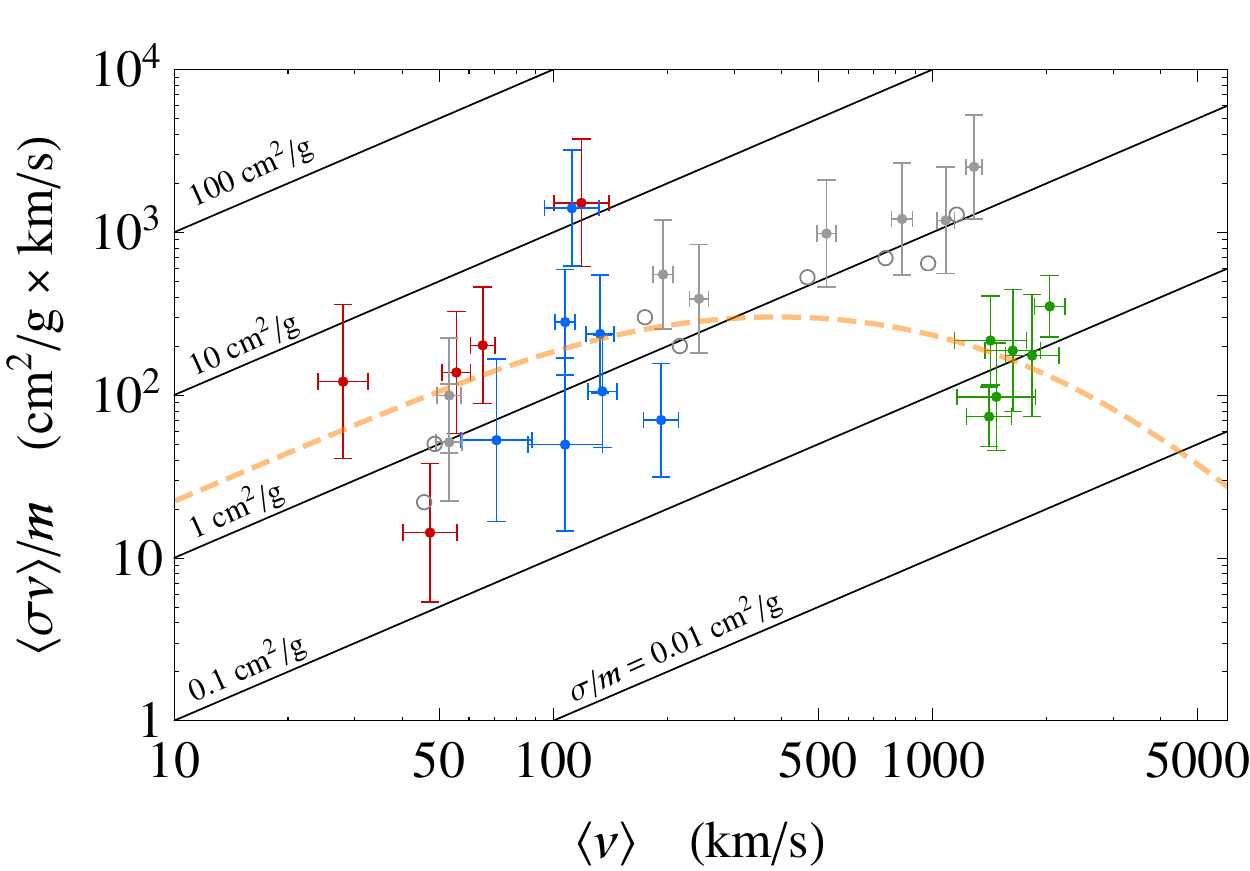}\;\;\includegraphics[scale=0.37]{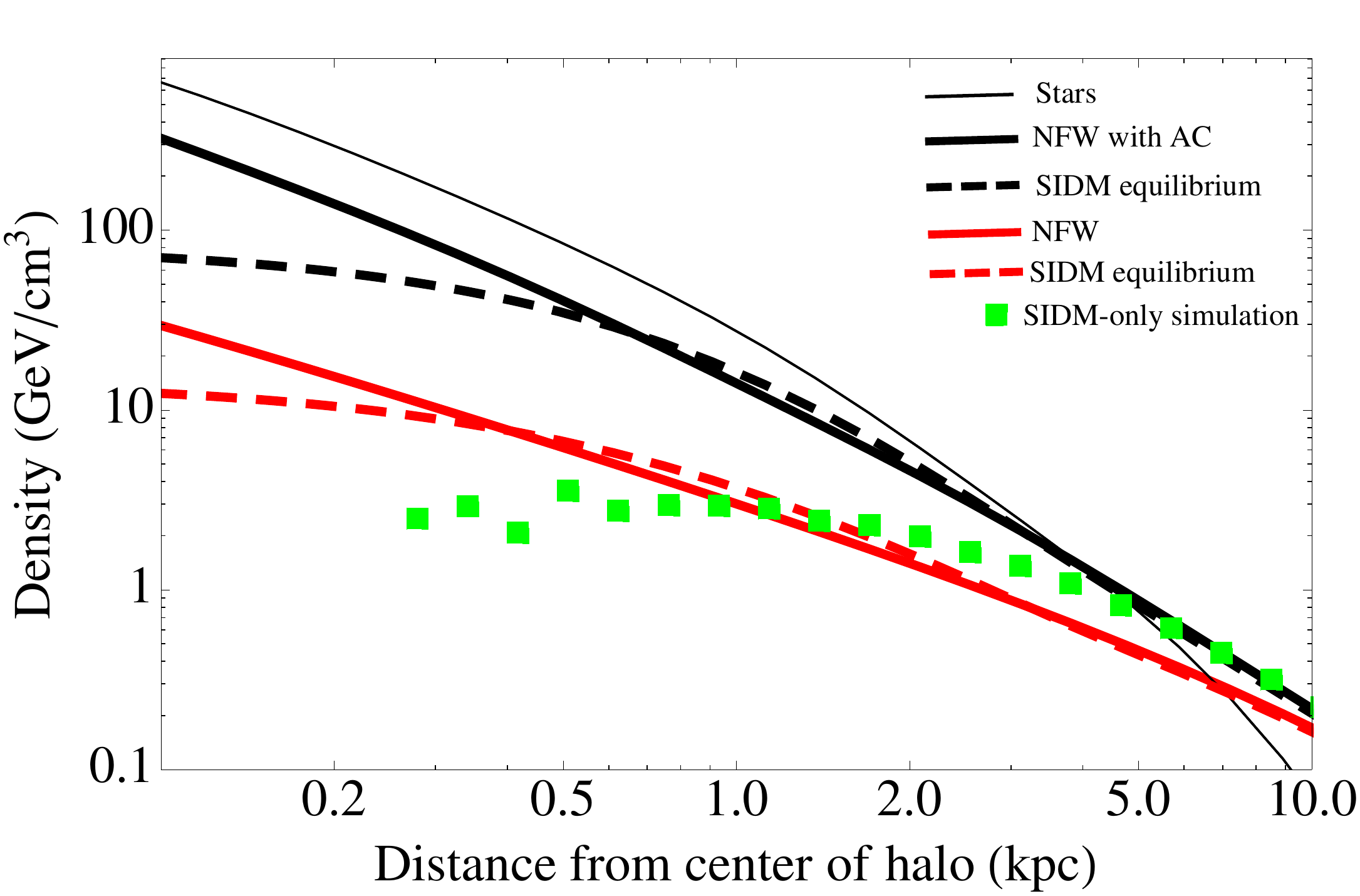}
\caption{\it Left: Velocity-weighted self-interaction cross section per unit mass as a function of average relative particle velocity in a halo.  Data points from astrophysical observations correspond to THINGS dwarf galaxies (red), LSB galaxies (blue), and clusters (green).  Diagonal lines show constant values of $\sigma/m$.  Gray points are fits to mock data from SIDM simulations, with fixed $\sigma/m = 1 \; {\rm cm^2/g}$, as a test of the Jeans method to reproduce the input cross section.  Reprinted from Ref.~\cite{Kaplinghat:2015aga}; see therein for further details.  Right: Comparison of DM density profiles for simulated SIDM-only halo (green dots) to SIDM halo with baryons (dashed curves), either with (black) or without (red) adiabatic contraction from stellar disk, where $\sigma/m\approx0.5\;{\rm cm^2/g}$.  The SIDM profile with baryons is virtually identical to the collisionless DM profile (NFW) except for the innermost $\sim 0.5$ kpc.  Reprinted from Ref.~\cite{Kaplinghat:2013xca}.
}
\label{fig:jeans}
\end{figure}

Jointly analyzing both galaxies and clusters, Fig.~\ref{fig:jeans} (left) illustrates how $\langle \sigma v_{\rm rel} \rangle /m$ depends on the average collision velocity $\langle v_{\rm rel}\rangle $, assuming self-interactions are responsible for the observed cores in these systems.  While galaxy-scale observations favor $\sigma/m \approx 2  \; {\rm cm^2/g}$, data from clusters prefers a much smaller cross section, $\sigma/m \approx 0.1 \; {\rm cm^2/g}$~\cite{Kaplinghat:2015aga}.  Taken at face value, these data imply that SIDM can provide a consistent solution to the core-cusp problem, provided self-interactions are relatively suppressed in clusters compared to dwarf galaxies.  Such a behavior is well-motivated from a particle physics perspective, as discussed below.  The data given here may be fit by a massive dark photon model (dashed orange curve in Fig.~\ref{fig:jeans}).  Lastly, to verify the validity of the Jeans approach, Ref.~\cite{Kaplinghat:2015aga} analyzed mock rotation curves produced from eight SIDM halos in a similar mass range from N-body simulations~\cite{Rocha:2012jg,Elbert:2014bma}, reproducing the input cross section value $1 \; {\rm cm^2/g}$ in those simulations.

Next, we turn to another important question: what is the interplay between self-interactions and baryons?  The first simulations including both baryons with feedback processes {\it and} self-interactions for DM have only recently been performed (see \S\ref{sec:sidmbar}), mainly targeting dwarf halos~\cite{Vogelsberger:2014pda,Fry:2015rta}.  The Jeans method provides complementary insights into the effect of baryons on SIDM, especially in systems like the MW or larger that have a significant baryon fraction~\cite{Kaplinghat:2013xca}.  

Although the Jeans approach is limited to quasi-equilibrium solutions (hence the dynamics of feedback is ignored), the static gravitational potential from baryons can dramatically change the predictions for observations compared to SIDM-only simulations.  The baryon density enters through $\rho_b$ in Eq.~\eqref{eq:Jeans2}, modifying the solution for $\rho_{\rm dm}$ from the usual cored isothermal profile.  If $\rho_b$ dominates over $\rho_{\rm dm}$ in the inner halo, the core radius shrinks substantially compared to the SIDM-only halo without baryons.  This effect is shown in Fig.~\ref{fig:jeans} (right) for a MW-like halo and a self-interaction cross section of $\sigma/m \approx 0.5 \; {\rm cm^2/g}$.  While the SIDM-only halo has a core of size $\sim 5$ kpc, the core size is reduced by an order of magnitude due to the baryonic potential from stars.  Except for the innermost $\sim 0.5$ kpc, the density profiles for both collisionless and collisional DM are virtually identical. Since $r_1\approx8.5\; {\rm kpc}$ in this case, where $\rho_b$ and $\rho_{\rm dm}$ are comparable, adiabatic contraction~\cite{Blumenthal:1985qy} may modify the initial NFW profile matched to the inner isothermal profile, as indicated in Fig.~\ref{fig:jeans} (right). However, it is negligible for the SIDM fits of the clusters shown in Fig.~\ref{fig:clusters}, because the contraction effect is very mild and it occurs $r\lesssim r_1$~\cite{Newman:2012nw}.

The baryon density may affect the shape of the DM halo as well.  While SIDM-only simulations predict halos that are spherical within the core radius, this conclusion changes once baryons are present.  For a halo like the MW, the gravitational potential of the baryonic disk causes the SIDM halo to become oblate, with aspect ratio $\sim 0.6$, out to $\sim 5$ kpc, the would-be core radius for SIDM without baryons~\cite{Kaplinghat:2013xca}.

To summarize, recent studies using the Jeans method have challenged the conventional SIDM paradigm in several important ways.  
\begin{itemize}
\item Astrophysical data from galaxies to clusters disfavor a constant self-interaction cross section.  The velocity-dependence of scattering is important for constraining the particle physics underlying SIDM.
\item Expectations for SIDM halo profiles can change dramatically from SIDM-only halos if there is a sizable baryonic component.  The central regions of halos need not have large spherical cores with slope $d \log \rho_{\rm dm}/d \log r \approx 0$.  Constraints based on halo shapes~\cite{Peter:2012jh,Feng:2009hw,Feng:2009mn} need to be re-evaluated in light of the role of baryons in those systems.
\end{itemize}

In modeling the SIDM halo properties using the Jeans method, we have neglected dynamical effects, such as baryon feedback or environmental interactions, but we expect the result is robust to the galaxy formation history.  Provided galaxy formation dynamics occurs interior to $r_1$, self-interactions will re-equilibrate the halo in response to any variations in the baryon distribution.  For $\sigma/m\sim1~{\rm cm^2/g}$, $r_1$ is close to the scale radius $r_s$, which is well outside typical radius for the stellar disk or bulge.  It is hard to imagine that feedback processes would change the halo beyond $r_1$.  Thus, the final SIDM density profile should be close to its equilibrium prediction, which depends on the baryon distribution and not the formation history.

\subsection{Diverse rotation curves for SIDM}

\begin{figure}
\includegraphics[scale=0.75]{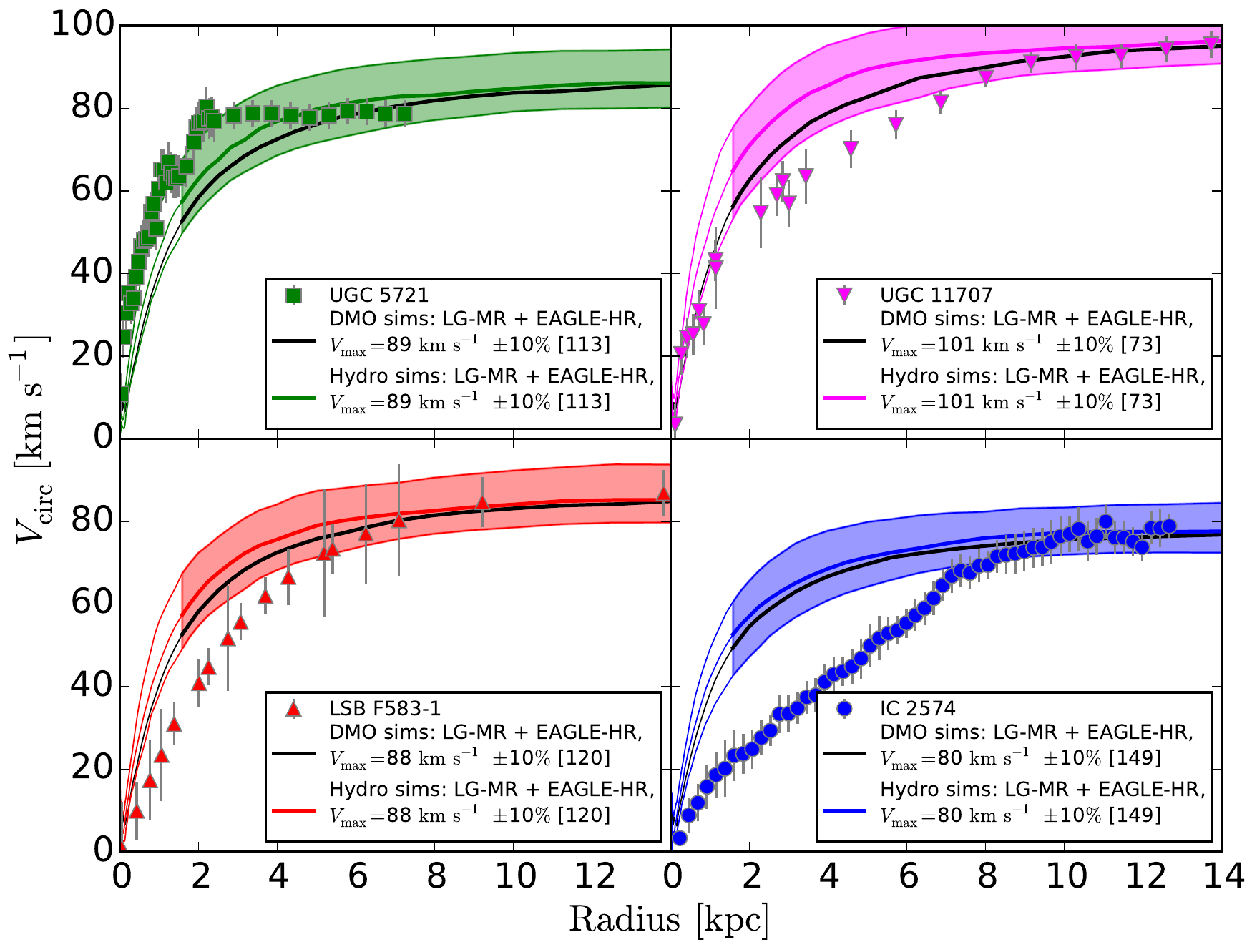}
\caption{\it Observed rotation curves of four disk galaxies (UGC 5721, UGC 11707, F583-1, and IC 2574), all with asymptotic velocity $\sim 80~{\rm km/s}$ but showing extreme diversity in the inner region. Colored band shows expected range in rotation curves from CDM hydrodynamical simulations.  Reprinted from Ref.~\cite{Oman:2015xda}.
}
\label{fig:diversityexample}
\end{figure}

The observed diversity of rotation curves poses a challenge for feedback solutions to the core-cusp problem.  The issue is made clear in recent simulations by Oman et al.~\cite{Oman:2015xda}, shown in in Fig.~\ref{fig:diversityexample}.  Each panel shows the rotation curve (data points) for one of four galaxies.  All four galaxies have comparable $V_{\rm max}\sim 80 \; {\rm km/s}$, and therefore, within $\Lambda$CDM cosmology, are expected to have similar rotation curves.  The colored bands show the scatter in rotation curves from hydrodynamical CDM simulations for similarly sized halos.  Not only do the observations lie outside the expected range, they do so in different ways: IC 2574 has a slowly rising velocity indicating a large shallow core, while UGC 5721 has a steeply rising velocity due to a cuspy profile.  Although this tension is somewhat dependent on the feedback prescription---Ref.~\cite{Oman:2015xda} adopt a smooth prescription---even bursty star formation cannot explain a core as large as IC 2574 (see Fig.~\ref{fig:baryon}).

On the other hand, DM self-interactions can explain this diversity.  There are two effects at play.  First, halos have intrinsic scatter due to variation in their mass assembly history.  For CDM halos, this is reflected in the scatter in the mass-concentration relation.  Since self-interactions rearrange the distribution of the halo interior to $r_1$, this scatter is reflected in the core size and central density for SIDM halos~\cite{Kaplinghat:2015aga,Kamada:2016euw}.   However, this cannot be the whole story since the observed scatter is larger than can be explained by assembly history alone.

Second, the baryon distribution plays an important role in SIDM halos and thermalization can lead to very different halo profiles depending on whether the inner halo is dominated by DM or baryons.  If DM-dominated, the density profile is $\rho_{\rm dm} \propto \exp(-\Phi_{\rm dm}/\sigma_0^2)$, the solution to which is the nonsingular isothermal profile~\cite{1926ics..book.....E}, while if baryon-dominated, the halo profile is $\rho_{\rm dm} \propto \exp(-\Phi_{b}/\sigma_0^2)$ and is largely set by the baryon density.  In the latter case, the DM core size is controlled by the baryonic scale radius~\cite{Kaplinghat:2013xca}.  Late-type galaxies span both cases and therefore scatter in the baryon profile translates into diverse DM profiles~\cite{Kamada:2016euw}.

To investigate these effects, Kamada et al.~\cite{Kamada:2016euw} used the Jeans approach to model rotation curves for 30 spiral galaxies with asymptotic velocities $25 -300 \; {\rm km/s}$ and fixed $\sigma/m = 3 \; {\rm cm^2/g}$. The solution to Eq.~\eqref{eq:Jeans2} provides the density profile in the inner isothermal region.  The baryon density is assumed to be a flat exponential disk
\beq
\rho_b(R,z) =\Sigma_b \Upsilon_*  e^{-R/R_d}\, \delta(z) \, ,
\eeq
where $\Sigma_b$ is the central surface density, $\Upsilon_*$ is the stellar mass-to-light ratio, $R_d$ is the disk scale radius, and $(R,z)$ are cylindrical coordinates.  In the outer halo, the density profile is matched onto an NFW profile at $r_1$, according to Eq.~(\ref{eq:match}), that is assumed to lie on the CDM mass-concentration relation within scatter~\cite{Dutton:2014xda}.  In practice, the virial mass $M_{200}$ is fixed to match the asymptotic circular velocity, while $\Upsilon_*$ and the concentration parameter $c_{200}$ are determined to fit the inner rotation curve~\cite{Kamada:2016euw}.  According to the Jeans model, the isothermal halo parameters $\rho_0, \sigma_0$ are determined once the outer NFW halo is fixed.

\begin{figure}
\includegraphics[scale=0.6]{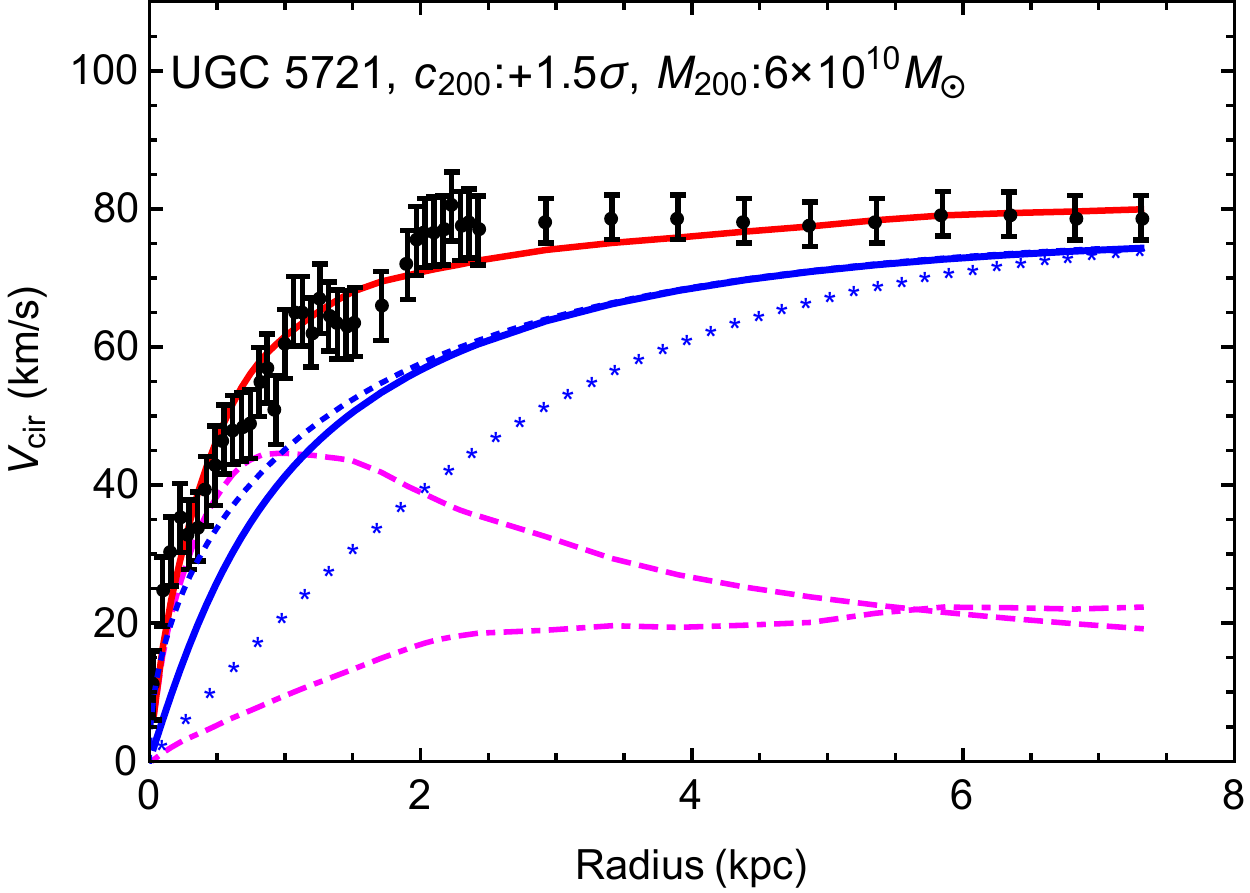}
\includegraphics[scale=0.59]{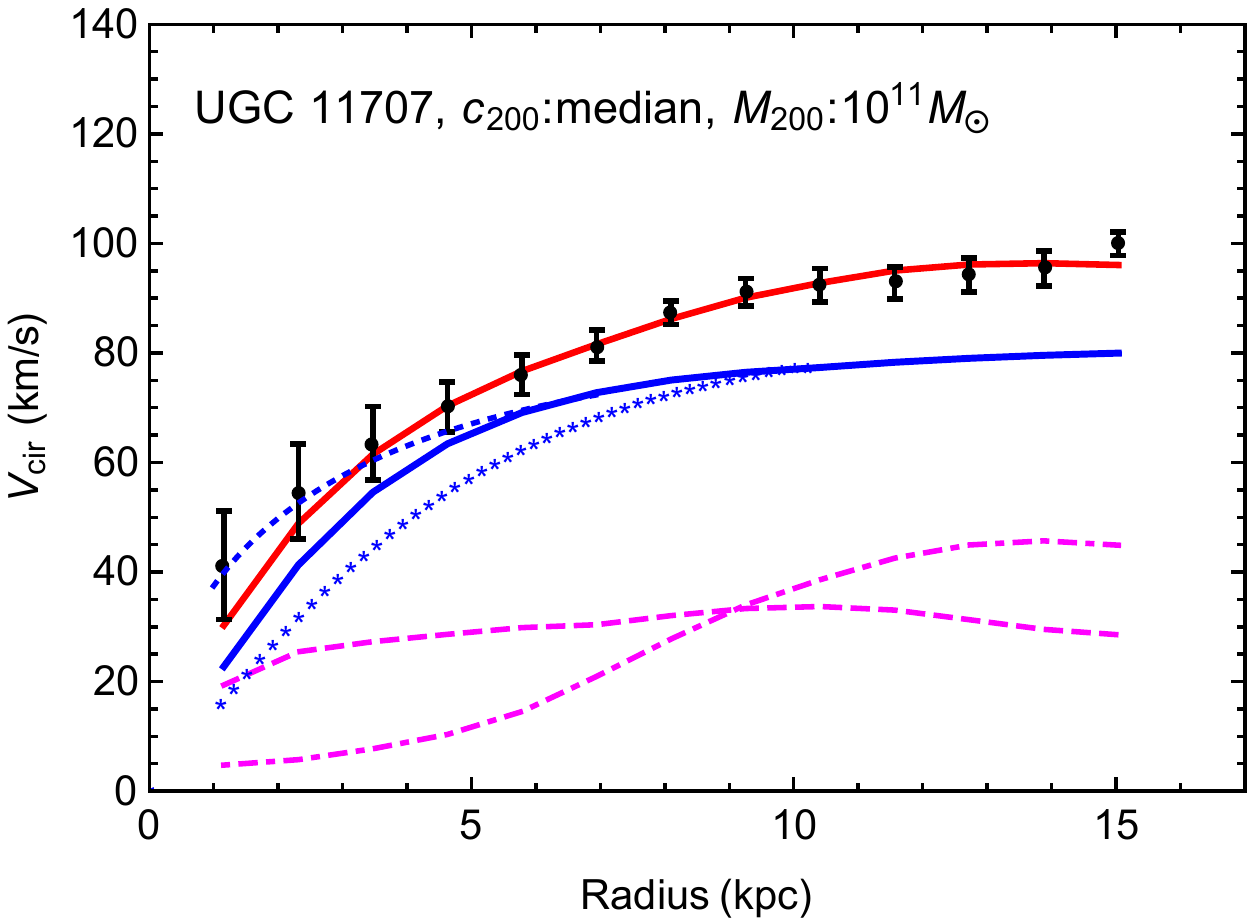}
\includegraphics[scale=0.6]{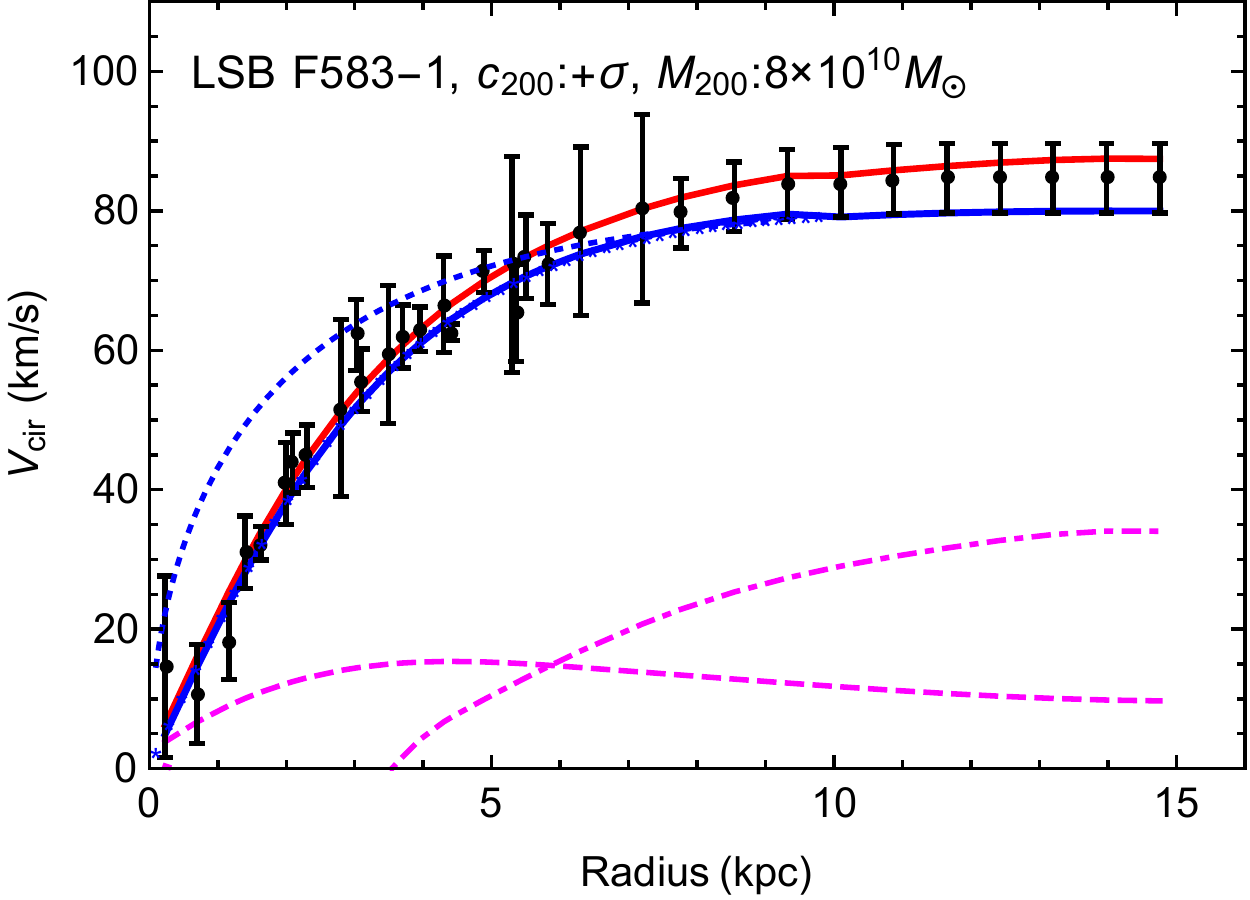}
\includegraphics[scale=0.6]{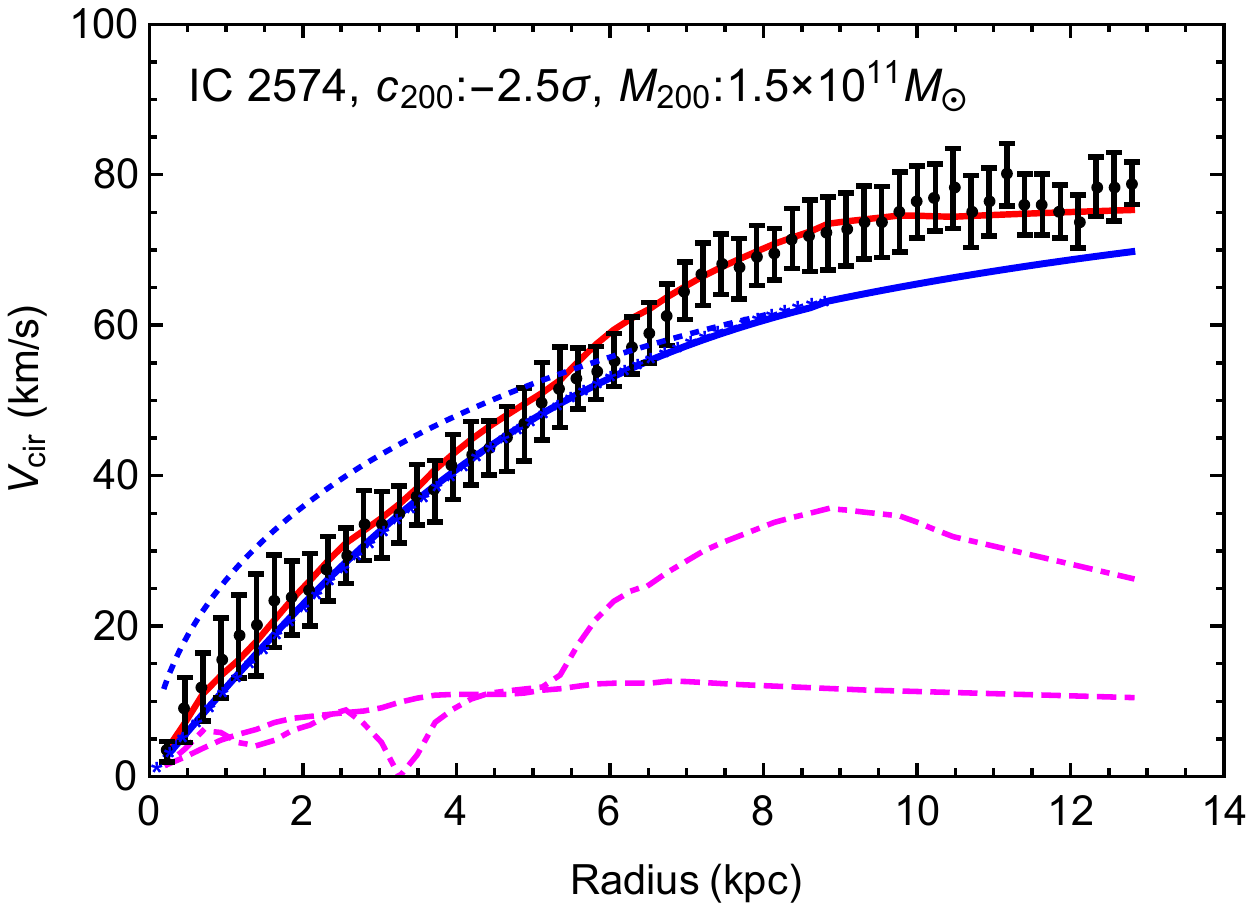}
\caption{\it Observed rotation curves of four disk galaxies (same as in Fig.~\ref{fig:diversityexample}).  Solid red lines indicate the total rotation curves for SIDM with $\sigma/m=3~{\rm cm^2/g}$, which includes the halo (solid blue), stars (magenta dashed), and gas (magenta dot-dashed) contributions.  Corresponding CDM halos (dashed blue) and the SIDM halos neglecting the influence of the
baryons (blue stars) are also shown.  Concentration parameter $c_{200}$ is indicated in terms of standard deviations with respect to median cosmological concentration for a given virial mass $M_{200}$.  Reprinted from Ref.~\cite{Kamada:2016euw}.
}
\label{fig:diversitySIDM}
\end{figure}

Fig.~\ref{fig:diversitySIDM} shows how SIDM can accommodate rotation curve diversity.  The four galaxies shown here are the same as those in Fig.~\ref{fig:diversityexample}.  The solid red curves show total rotation curves that have been fit to these galaxies.  It is quite remarkable that cosmological halos with fixed $\sigma/m$ can accommodate such disparate galaxies.  For UGC 5721, the combination of a high concentration and dense baryon distribution yields a steep DM profile (solid blue) that is closer to NFW (dashed blue) than to a typical cored SIDM profile without baryons (stars).  On the other hand, for IC 2574, a shallower core for SIDM can result from a low concentration $\sim 2.5 \sigma$ below the median plus a diffuse baryon density.  Despite having the same $V_{\rm max}$, SIDM can account for their wildly different inner velocities, $V_{\rm circ}({\rm 2 \; kpc}) \approx 70 \; {\rm km/s}$ and $25 \; {\rm km/s}$, respectively.  These galaxies represent extremes of the scatter shown in Fig.~\ref{fig:diversity} (right).  Other galaxies in Ref.~\cite{Kamada:2016euw}, such as those shown in Fig.~\ref{fig:diversityexample}, are intermediate to these extremes and can also be accommodated.

\begin{figure}
\includegraphics[scale=0.8]{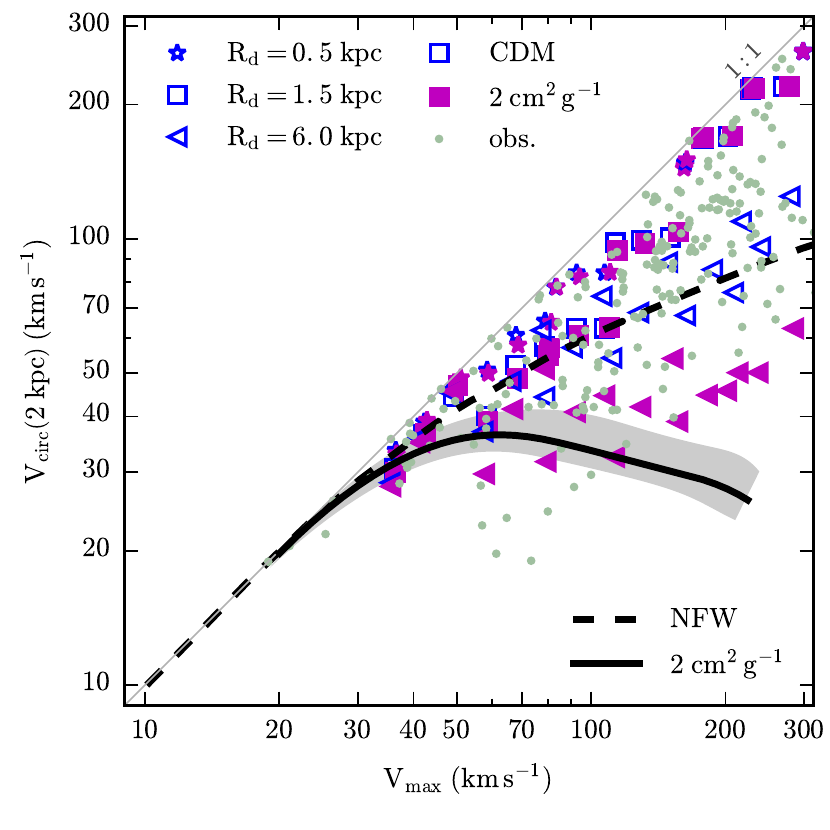}\;\;\;\;\;\includegraphics[scale=0.65]{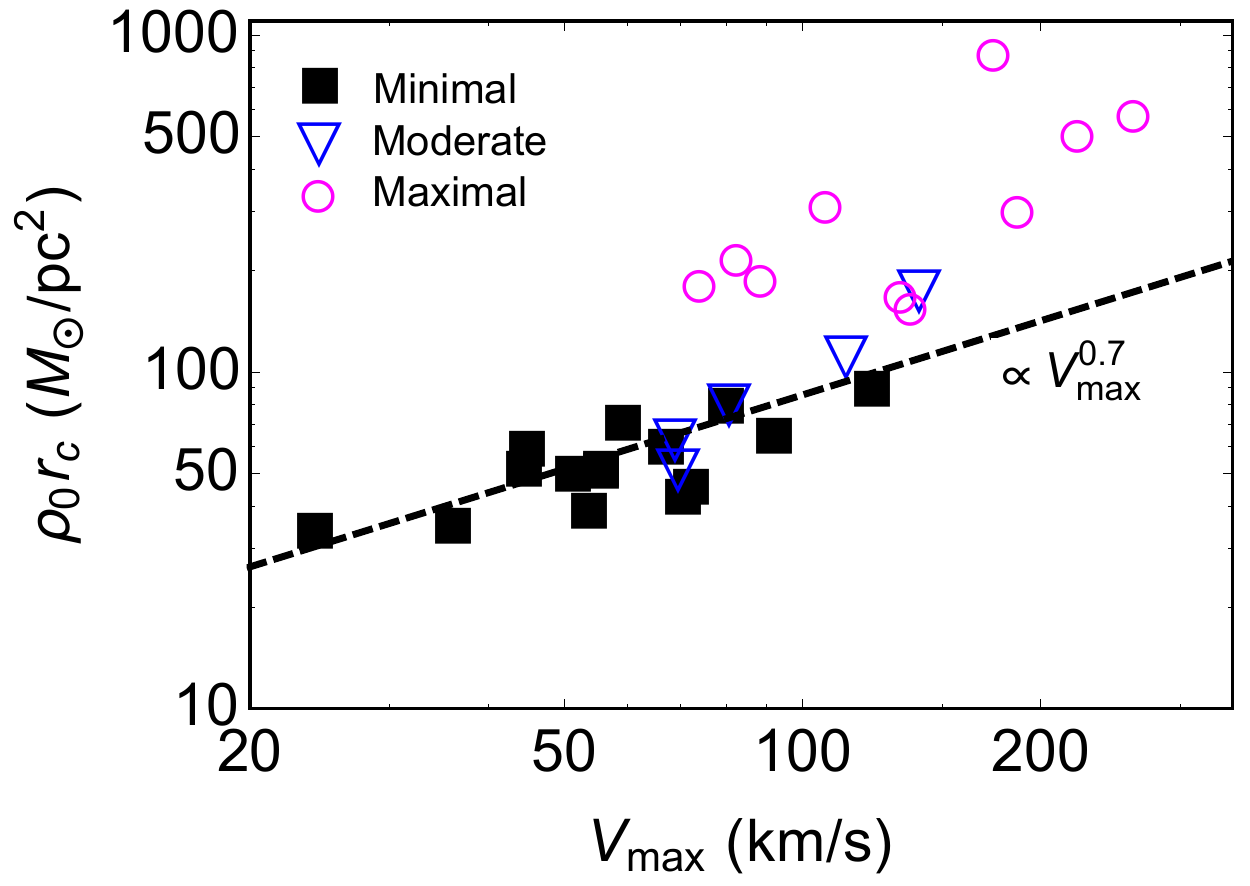}
\caption{\it Left: Scatter in $V_{\rm circ}(2~{\rm kpc})$ vs $V_{\rm max}$ for SIDM (solid magenta) and CDM (open blue) halos for different baryonic disk scale radius $R_d$. The dashed (solid) black line is the predicted relation for CDM-only (SIDM-only) halos (i.e., without baryons).  The gray band shows the scatter for SIDM-only halos without baryons (the corresponding scatter for CDM-only halos has similar width and is not shown).  See text for further details.  Reprinted from Ref.~\cite{Creasey:2016jaq}. Right: Surface density $\rho_0 r_c$ for 30 disk galaxies modeled with SIDM halos, subdivided into those where the effect of baryons is minimal, moderate, or maximal.  The minimal sample obeys $\rho_0 r_c \propto V^{0.7}_{\rm max}$ (dashed line), a reflection of the mass-concentration relation. Reprinted from Ref.~\cite{Kamada:2016euw}.
}
\label{fig:diversitySIDM2}
\end{figure}

To further investigate these effects, Creasey et al.~\cite{Creasey:2016jaq} performed a set of controlled simulations for isolated disk galaxies. In these simulations, the density profile of the initial halo is set up with a Hernquist profile, while the baryonic component is modeled as a fixed disk potential, and the disk mass is the sum of a stellar component and a gaseous one. The stellar mass is set by the abundance matching relation ($M_{\rm star}\textup{--}M_{200}$) in~\cite{Behroozi:2012iw} and the gas mass computed from the $M_{\rm gas}\textup{--}M_{200}$ relation proposed in~\cite{Huang:2012hk}.  Each halo---initialized to be in equilibrium with respect to the collisionless Boltzmann equation---is simulated with 18 different permutations: collisionless or $\sigma/m = 2 \; {\rm cm^2/g}$; concentration parameter at its median value or $\pm 2\sigma$ extreme values; and baryonic disk scale lengths $R_d=0.5$, $1.5$, and $6.0~{\rm kpc}$, representing ultra-compact, compact, and extended baryon distributions for observed galaxies, respectively.  As shown in Fig.~\ref{fig:diversitySIDM2} (left), variations in the baryon distribution induce {\it greater} diversity in SIDM halos (solid magenta symbols) compared to CDM halos (open blue symbols).  The latter has too little scatter to accommodate the diversity of observed rotation curves (green points).  

The reason SIDM halos have greater diversity is as follows~\cite{Creasey:2016jaq}.  Cored SIDM profiles have lower DM densities in their inner regions and the potential contribution of baryons is more important than for CDM, creating a larger range for $V_{\rm circ}({\rm 2\; kpc})$. For the compact disks ($R_d=0.5$ and $1.5~{\rm kpc}$), the predictions from SIDM and CDM are not very different, since the disk is the dominant contribution to $V_{\rm circ}({\rm 2\; kpc})$.  For an extended disk, $R_d=6.0~{\rm kpc}$, $V_{\rm circ}({\rm 2\; kpc})$ in SIDM is lower than predicted in CDM and this difference becomes more prominent when $V_{\rm max}$ increases. Lastly, we note that there is a gap between compact ($R_d=1.5~{\rm kpc}$) and extended ($6~{\rm kpc}$) bayon disks in Fig.~\ref{fig:diversitySIDM2} (left).  Intermediate baryon disks, which have not been included in these simulations, would populate this region.


\subsection{Scaling relations}
\label{sec:scaling}

In hierarchical structure formation, CDM halo parameters have a tight correlation known as the mass-concentration relation, which emerges from cosmological N-body simulations (see, e.g., Ref.~\cite{Dutton:2014xda}).  Since DM self-interactions only change the inner halo, while preserving the structure of the collisionless outer halo, we expect that the parameters of the inner halo---namely, the core density $\rho_0$ and core radius $r_c$---are also correlated for a given value of $\sigma/m$.

The right panel of Fig.~\ref{fig:diversitySIDM2} shows the surface density of the DM halo, given by the product $\rho_0 r_c$, as a function of $V_{\rm max}$.  Here, $\rho_0$ is the central density obtained from SIDM fits to 30 galaxies in Ref.~\cite{Kamada:2016euw} and $r_c$ is the radius at which the density is $\rho_0/2$.  In the absence of baryons, the Jeans model predicts $\rho_0 r_c\propto V^{0.7}_{\rm max}$ (dashed line), which can be traced to the mass-concentration relation~\cite{Boyarsky:2009rb,Boyarsky:2009af,Lin:2015fza}.  The ``minimal'' sample in Fig.~\ref{fig:diversitySIDM2} corresponds to galaxies where the stellar disk has negligible effect and is consistent with this relation.   For the ``moderate" and ``maximal" samples, there is a large effect on the SIDM halo from baryons and these galaxies show a deviation from the simple scaling relation.

A number of other studies have shown that the DM halo surface density is nearly a constant~\cite{Spano:2007nt,Donato:2009ab,Salucci:2011ee,Kormendy:2014ova,Burkert:2015vla,Hayashi:2015maa,Karukes:2016eiz}.  These works adopt pseudo-isothermal or Burkert profiles to fit rotation curves (and stellar dispersions of dwarf spheroidals).\footnote{In the case of a Burkert profile, the DM density at the core radius is one quarter of its central value.}  Although these studies do not impose the halo mass-concentration relation in their fits, the inferred DM surface density $\rho_c r_c \sim100  \Msun {\rm pc^{-2}}$ is consistent with the result shown here.



\section{Halo mergers}
\label{sec:merging}

Merging systems are another avenue for exploring the collisional nature of DM.  
These systems include merging clusters, with the most famous being the Bullet Cluster~\cite{Tucker:1998tp,Markevitch:2001ri,Markevitch:2003at,Clowe:2006eq}, as well as minor mergers from substructure (galactic or group halos) infalling into clusters~\cite{Furlanetto:2001tw,Natarajan:2002cw,Massey:2010nd}.  These systems probe the {\it non-equilibrium} dynamics of SIDM acting over timescales $\lesssim 1$ Gyr, which is complementary to the observables discussed above that test the quasi-equilibrium distribution of SIDM reached over $\sim$ 10 Gyr timescales.  

Major mergers between clusters are characterized by three components: 
\begin{itemize}
\item Galaxies: Since galaxies have negligible cross section for scattering, they act as collisionless test particles. (For minor mergers involving galactic-scale substructure, the stellar density plays the same role.)
\item Intracluster medium (ICM) gas: Ram pressure during the merger causes ICM gas to get shocked and dissociated from any collisionless components, namely galaxies and (potentially) DM.  
\item DM: The total mass density, which is predominantly DM, is inferred through strong and weak gravitational lensing. 
\end{itemize}
Although the initial state of the merger is not observed, X-ray observations of the shocked ICM, along with dynamical timing arguments~\cite{Dawson:2012fx} and synchotron emission associated with turbulence (radio halos) or shocks (radio relics)~\cite{Ensslin:1997kw,Skillman:2012np}, allow reconstruction of the three-dimensional geometry and velocity of the merger.  If DM is collisionless, it should remain coincident with the galaxies after the merger, while if it is strongly self-interacting, it behaves qualitatively more like collisional gas.  In the intermediate case in which the optical depth for self-scattering is nonzero but less than unity, self-interactions are expected to induce offsets between DM and galactic components due to collisional drag incurred between SIDM halos during pericenter passage~\cite{Markevitch:2003at}.  The magnitudes of such offsets have been addressed in recent numerical studies~\cite{Kahlhoefer:2013dca,Robertson:2016xjh,Kim:2016ujt,Robertson:2016qef}, which we discuss below.

The same argument can be applied to minor mergers as well~\cite{Massey:2010nd,Harvey:2013tfa}.  Since substructure infall is the dominant mechanism for structure growth, these events are far more ubiquitous than cluster mergers.  However, the expected offsets are much smaller: for collisionless DM, the expected gas-DM offsets are $\sim 20$ kpc (versus $\gtrsim 100$ kpc for major mergers) and clearly sensitivity below this scale is required to test SIDM~\cite{Harvey:2013bfa,Harvey:2013tfa}.  With the exception of rare systems with strong lensing data (such as Abell 3827), substructure merger studies must rely on weak lensing maps and require a stacked analysis of many clusters.  Harvey et al.~\cite{Harvey:2013tfa} proposed the ``bulleticity ratio'' $\beta = d_{\rm SD}/d_{\rm SG}$ as a suitable quantity to be averaged over many systems.  Here, $d_{\rm SG}$ ($d_{\rm SD}$) is the projected distance between the stellar and gas (DM) peaks.  The virtue of $\beta$ is that the orientation direction of the merger cancels in the ratio and that it has a relatively simple dependence on $\sigma/m$, under certain assumptions, as we discuss below.  In subsequent work, Harvey et al.~\cite{Harvey:2015hha} determined the first constraints on self-interactions using a statistical analysis of $\beta$ from 30 merger systems.  Their study is consistent with $\beta = 0$, as expected for collisionless DM, resulting in the strongest merger-derived limit to date (however, see Ref.~\cite{Wittman:2017gxn}).

\subsection{Observations}

\begin{figure}
\includegraphics[scale=0.41]{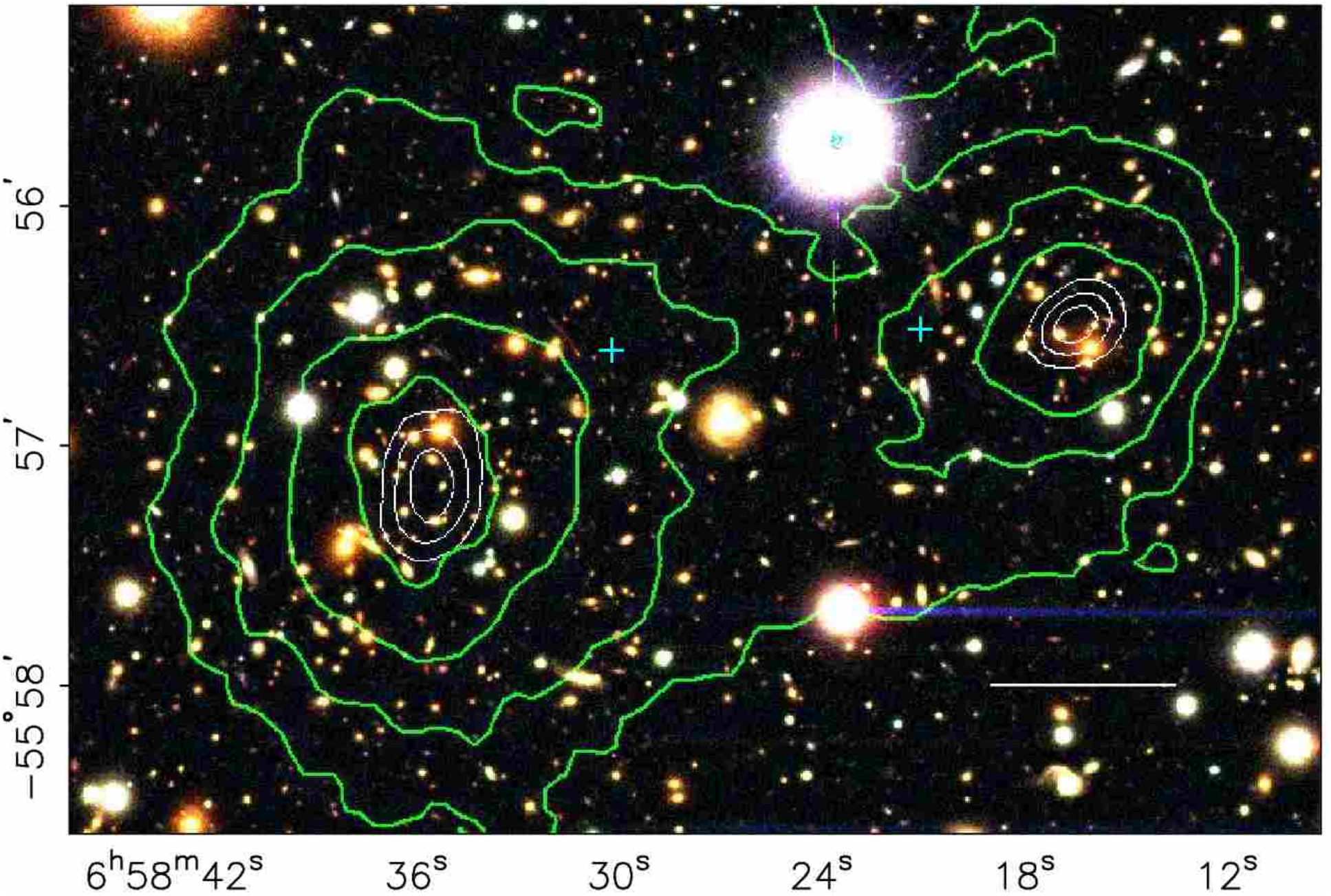}\;\;\includegraphics[scale=0.41]{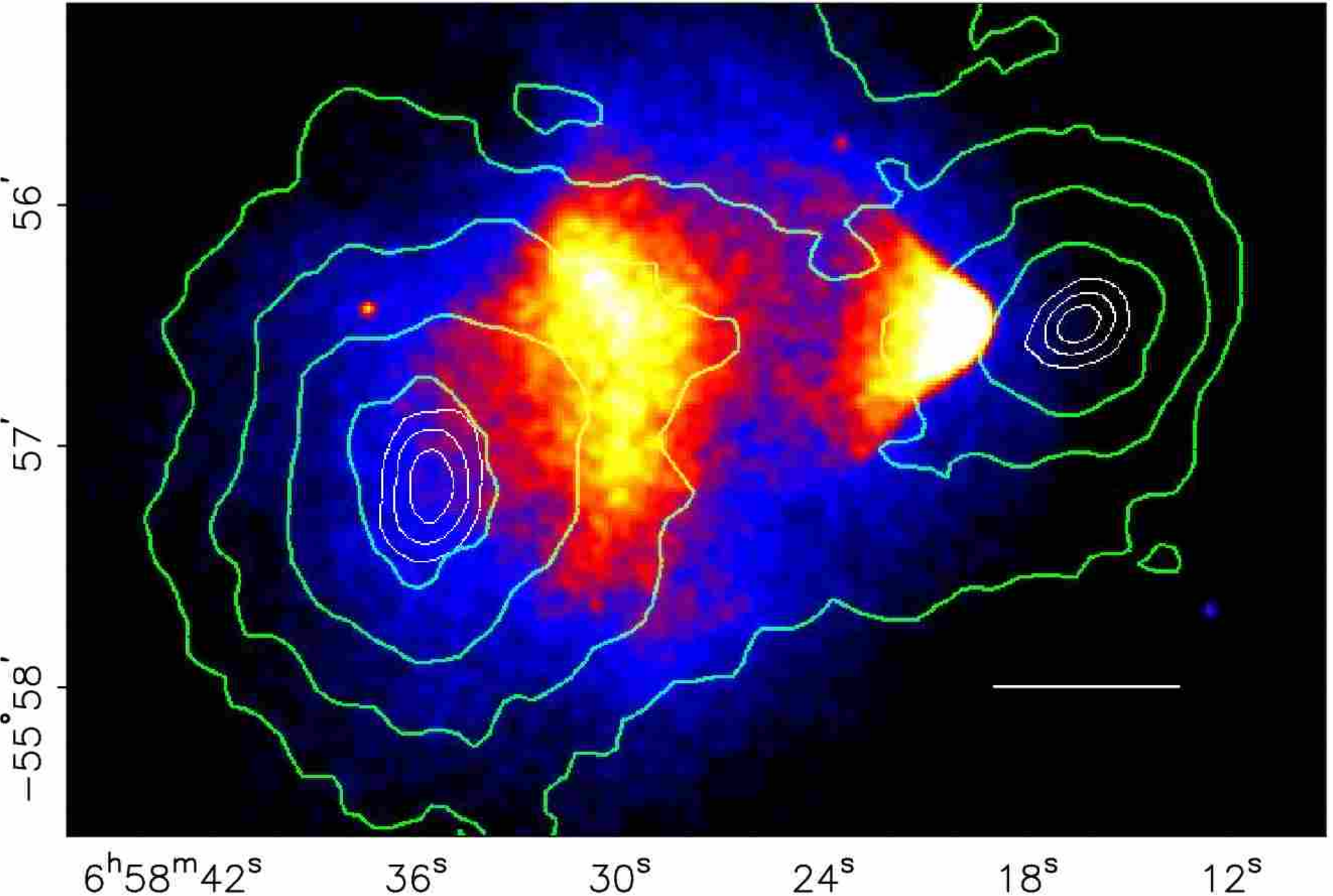}
\caption{\it Optical (left) and X-ray (right) observations of the Bullet Cluster showing the distribution of galaxies and hot gas, respectively.  Projected mass density reconstructed from weak lensing is superimposed on both panels (green contours); thin white lines indicate locations of the density peaks for the Bullet and main cluster (at 68.3\%, 95.5\%, and 99.7\% CL), separated by 720 kpc.  White bar indicates 200 kpc.  Reprinted from Ref.~\cite{Clowe:2006eq}.
}
\label{fig:bullet}
\end{figure}

The Bullet Cluster remains the archetypal dissociative merger.  In this system, a smaller subcluster (the ``Bullet'') has passed through a much larger main subcluster, evidenced by the dramatic bow shock left in the wake of the merger, shown in Fig.~\ref{fig:bullet}.  From the orientation and sharp brightness edges of the shock, it is clear that the merger was a nearly head-on collision in the plane of the sky~\cite{Markevitch:2001ri}.  The centroids for the Bullet's total mass and galactic distributions are offset by $25 \pm 29$ kpc, implying that DM behaves as collisionless CDM within uncertainties~\cite{Randall:2007ph}.  Several hydrodynamical simulations have been performed with CDM plus gas in order to reconstruct the merger~\cite{Springel:2007tu,Mastropietro:2007kr,Lage:2013yxa}.  Most recently, Lage \& Farrar~\cite{Lage:2013yxa} have simulated over a thousand Bullet Cluster realizations, including gas and magnetic fields, which were confronted against lensing, X-ray, and radio observations, as well as CMB measurements of the Sunyaev-Zel'dovich effect.  In their best-fit model, the main and Bullet subclusters are consistent with triaxial NFW profiles of $2\times 10^{15}$ and $2\times 10^{14} \, \Msun$, respectively, and infall velocity consistent with $\Lambda$CDM cosmology~\cite{Lage:2013yxa,Lage:2014yxa,Kraljic:2014soa}.  Constraints on self-interactions must be interpretted within the context of the large relative velocity between the subclusters, corresponding to $v_{\rm rel} \sim 4000$ km/s at centroid crossing~\cite{Springel:2007tu,Lage:2014yxa}.  

After the Bullet Cluster's discovery~\cite{Tucker:1998tp}, several other dissociative cluster mergers have been found.  Those for which direct constraints on self-interactions have been quoted include DLSCL J0916.2+2951 (Musket Ball Cluster)~\cite{Dawson:2011kf,2013PhDT.......211D}, MACS J0025.4-1222 (Baby Bullet)~\cite{Bradac:2008eu}, CIZA J2242.8+5301 (Sausage Cluster)~\cite{vanWeeren:2011bz,Dawson:2014jca,Jee:2014mca}, and ACT-CL J0102-4915 (El Gordo)~\cite{Menanteau:2011xy,Jee:2013gey,Ng:2014gta}, Abell 520 (Train Wreck Cluster)~\cite{Markevitch:2004qk,Mahdavi:2007yp}, and Abell 2744 (Pandora's Cluster)~\cite{Kempner:2003qk,Owers:2010uj,Merten:2011wj}.  
Among substructure mergers, Abell 3827 has received considerable attention for its implications for SIDM~\cite{Carrasco:2010jh,Williams:2011pm,Massey:2015dkw,Kahlhoefer:2015vua}. 

\vspace{2mm}

\underline{Musket Ball Cluster:} This system is a binary merger between roughly equal mass subclusters ($2-3 \times 10^{14} \, \Msun$) along a collision axis inclined $\sim 45^\circ$ to the plane of the sky~\cite{Dawson:2011kf,Dawson:2012fx}.  A feature of this merger is the large physical separation $\sim$ 1.3 Mpc between the subclusters, which implies that this system has evolved post-collision $2-5$ times longer than the Bullet Cluster~\cite{Dawson:2011kf}.  Unfortunately this does not directly translate into an enhanced sensitivity to self-interactions since, according to SIDM simulations, galaxy-DM offsets do not continue to grow with time after halo passage~\cite{Kahlhoefer:2013dca,Robertson:2016xjh,Kim:2016ujt}.  One subcluster in this system appears to have an offset $\sim 130$ kpc between the weak lensing mass peak and its corresponding galactic centroid, with the former trailing the latter along the merger axis~\cite{2013PhDT.......211D}.  However, the statistical significance of this offset is not high (less than $2 \sigma$) and does not exclude collisionless CDM~\cite{Dawson:2011kf,2013PhDT.......211D}.

\vspace{2mm}

\underline{Baby Bullet Cluster:} Two subclusters with equal mass ($2.5 \times 10^{14} \, \Msun$) have undergone a merger oriented in the plane of the sky.  The mass lensing peaks for both subclusters are concident with their respective galactic luminosity centroids, implying that DM appears collisionless, while the peak of the gas density is offset from both subclusters and is located between them~\cite{Bradac:2008eu}.

\vspace{2mm}

\underline{Sausage Cluster:} This system is dominated by two massive subclusters ($10^{15} \, \Msun$) that have undergone a merger in the plane of the sky, as evidenced by the high polarization fraction of the prominent radio relics that trace the shock fronts~\cite{vanWeeren:2011bz,Dawson:2014jca,Jee:2014mca}.  Weak lensing observations reveal two distinct mass peaks separated by $\sim 1$ Mpc, while the X-ray emission region is mainly located between the mass peaks and is highly elongated along the merger axis~\cite{Dawson:2014jca,Jee:2014mca}.  Interestingly, the mass peaks for both subclusters lag behind their respective galactic luminosity centroids by $\sim 200$ kpc~\cite{Jee:2014mca}.  However, the statistical significance of these offsets is not high (below $2\sigma$) and becomes lower if number density (instead of luminosity) is used to weight the galactic centroids.  Moreover, a number of systematic effects need to be clarified before this may be claimed as an effect due to SIDM, e.g., the total mass peaks getting skewed closer to the cluster center due to the gas density~\cite{Jee:2014mca}.

\vspace{2mm}

\underline{El Gordo:} This system is another binary merger between massive subclusters ($10^{15} \, \Msun$)~\cite{Menanteau:2011xy,Jee:2013gey}, with a merger axis inclined $\sim 20^\circ$ with respect to the plane of the sky, inferred from the polarization fraction of the radio relics~\cite{Ng:2014gta}.  El Gordo, one of the most massive clusters found to date at redshifts $z > 0.6$, has a prominent ``bullet'' of cool gas (similar to the Bullet Cluster) followed by twin-tailed wake~\cite{Menanteau:2011xy}.  The mass and luminosity peaks for each subcluster are coincident with one another, as expected for collisionless CDM.  However, the gas bullet does not lag its corresponding mass peak as expected if the two subclusters are receding from one another after first pericenter passage~\cite{Jee:2013gey}.  This may indicate that El Gordo lies on a return trajectory post apocenter~\cite{Ng:2014gta}.

\vspace{2mm}

\underline{Abell 520:} For this system, the interpretation of collisionless CDM is less clear.  Like the Bullet Cluster, this system has a prominant bow shock indicating a recent, high-velocity merger~\cite{Markevitch:2004qk}.  However, the ICM gas has a much more complicated morphology~\cite{Wang:2016kto}, while weak lensing data reveals at least five distinct mass peaks~\cite{Mahdavi:2007yp}.  Hence, Abell 520 is no mere binary collision, but rather a ``cosmic train wreck'' of multiple simultaneous mergers~\cite{Mahdavi:2007yp,Okabe:2007af}.  Of particular interest for SIDM is the observation of a dark core in Abell 520, i.e., a peak in the lensing mass map without a significant galactic counterpart~\cite{Mahdavi:2007yp,Okabe:2007af,Girardi:2008jp,Jee:2012sr}.  However, the dark core was not confirmed in other lensing studies~\cite{Clowe:2012am}.  More recently, Jee et al.~\cite{Jee:2014hja} revisited this issue with the combined datasets used in Refs.~\cite{Jee:2012sr,Clowe:2012am}, re-confirming the presence of the dark core albeit at a shifted location.\footnote{The discrepancy with Ref.~\cite{Clowe:2012am} was reconciled by systematic differences between lensing datasets, including source image coverage and differences in charge transfer inefficiencies in the CCDs used for these observations, which can skew galaxy ellipticities used for weak lensing studies (see, e.g., Ref.~\cite{Rhodes:2010yg}).}  Since the dark core lies coincident with the gas density, this suggests that DM (or a fraction thereof) may have self-interacted during the merger.  Alternatively, the dark core may be a narrow DM filament oriented along the line of sight or a subcluster with an extreme mass-to-light ratio~\cite{Mahdavi:2007yp}.

\vspace{2mm}

\underline{Abell 2744:} This complex system appears to be a simultaneous merger between at least four clusters~\cite{Owers:2010uj,Merten:2011wj}.  Previous analyses revealed a Pandora's box of perplexing features in this system, such as a clump of X-ray-emitting gas {\it leading} a dark core~\cite{Owers:2010uj,Merten:2011wj,Medezinski:2015jta}.  With more recent observations, Jauzac et al.~\cite{Jauzac:2016tjc} found eight mass peaks in Abell 2744 with masses $\sim 10^{14}\; \Msun$.  With higher resolution mass maps, the purported dark core was fully resolved into two separate peaks coincident with galaxies, while the gas clump was seen to be associated with a previously unresolved DM halo in the process of merging~\cite{Jauzac:2016tjc}.  Aside from being a remarkable coincidence of mergers~\cite{Schwinn:2016xts}, Abell 2744 is consistent with CDM.  Constraints on self-interactions have been obtained for two of the larger subclusters, which have undergone a Bullet Cluster-like merger, albeit significantly inclined along the line of sight~\cite{Merten:2011wj,Jauzac:2016tjc}.  For both subclusters, the total mass and galactic distributions remain coincident within uncertainties, while the gas is stripped, implying that DM appears collisionless.  

\vspace{2mm}

\underline{Abell 3827:} This remarkable cluster is a nearby late-stage merger of four elliptical galaxies, all located within $\sim 10$ kpc of the cluster core~\cite{Carrasco:2010jh}.  The fortutious proximity of strong lensing images near these substructures allows for a detailed reconstruction of the mass sub-peaks.  Interestingly, the best resolved sub-peak shows a significant offset from its stellar counterpart, perhaps explained by drag due to self-interactions~\cite{Williams:2011pm,Massey:2015dkw,Taylor:2017ipx}.  The size of the offset, $1.6 \pm 0.5$ kpc~\cite{Massey:2015dkw}, is inconsistent with offsets seen in hydrodynamical simulations for collisionless CDM at $>99\%$ C.L (statistics only)~\cite{Schaller:2015tia}.\footnote{After this report was completed, a new study was released for Abell 3827 that erased this discrepancy with CDM~\cite{Massey:2017cwf}.  Improvements in foreground subtraction and image identification allow for an improved lensing mass map.  The position of the sub-peak is now only $0.54^{+0.22}_{-0.23} \; {\rm kpc}$ offset from its stellar density.}

\subsection{Self-interactions in merging clusters}

To place constraints on self-interactions, merger studies have relied on three approaches: scattering depth, mass loss, and offsets.  For clarity, we focus our discussion on the Bullet Cluster, to which all three methods have been applied.  Table~\ref{tab:mergers} summarizes constraints on self-interactions for all mergers quoted in the literature.  However, as we now discuss, not all limits are equally robust.

\begin{table}[t] 
\begin{tabular}{l|c|l|l} 
\hline
Cluster & $\sigma/m$ &  \multicolumn{1}{|c|}{Method used} & Ref. \\
\hline
\hline
Bullet Cluster  & $< 3 \; {\rm cm^2/g} $ & Scattering depth $(\Sigma_{\rm dm} \approx 0.3 \; {\rm cm^2/g})$ & \cite{Markevitch:2003at} \\
 (1E 0657-558) & $< 0.7 \; {\rm cm^2/g} $ & Mass loss $< 23\%$ & \cite{Randall:2007ph} \\
 & $< 1.25 \; {\rm cm^2/g} $ & DM-galaxy offset $25 \pm 29$ kpc & \cite{Randall:2007ph} \\
\hline
Abell 520 & $3.8\pm 1.1 \; {\rm cm^2/g}$ & Scattering depth $(\Sigma_{\rm dm} \approx 0.07 \; {\rm cm^2/g})$  & \cite{Mahdavi:2007yp} \\
 & $0.94\pm 0.06 \; {\rm cm^2/g}$ & Scattering depth $(\Sigma_{\rm dm} \approx 0.14 \; {\rm cm^2/g})$ & \cite{Jee:2014hja} \\
\hline
Abell 2744 & $< 1.28 \; {\rm cm^2/g}$ & Offset & \cite{Jauzac:2016tjc} \\
& $< 3 \; {\rm cm^2/g}$ & Scattering depth $(\Sigma_{\rm dm} \approx 0.3 \; {\rm cm^2/g})$ & \cite{Merten:2011wj} \\
\hline
Musket Ball Cluster & $< 7 \; {\rm cm^2/g} $ &  Scattering depth $(\Sigma_{\rm dm} \approx 0.15 \; {\rm cm^2/g})$ & \cite{Dawson:2011kf} \\
(DLSCL J0916.2+2951) & & & \\
\hline
Baby Bullet & $< 4 \; {\rm cm^2/g}$ & Scattering depth $(\Sigma_{\rm dm} \approx 0.25 \; {\rm cm^2/g})$ & ~\cite{Bradac:2008eu} \\
(MACS J0025.4-1222) & & & \\
\hline
Abell 3827 & $\sim 1.5 \; {\rm cm^2/g}$ & Offset & \cite{Kahlhoefer:2015vua} \\
\hline
\end{tabular}
\caption{\label{tab:mergers} Summary of merging cluster constraints on SIDM.  All values for $\sigma/m$ are upper limits except for Abell 520 and 3827. For Abell 520, two quoted values of $\sigma/m$ are obtained under different assumptions in which $\tau = \Sigma_{\rm dm} \sigma/m \approx 0.25$~\cite{Mahdavi:2007yp} or $0.13$~\cite{Jee:2014hja}. }
\end{table}

The first (and most conservative) approach is based on the optical depth argument.  For the Bullet Cluster, self-interactions must not be optically thick since the Bullet halo has survived the merger.  The scattering depth $\tau$ must satisfy 
\beq \label{eq:tau}
\tau = \Sigma_{\rm dm} \sigma/m  < 1 \, .
\eeq
Here, $\Sigma_{\rm dm} \approx 0.3 \; {\rm g/cm^2}$ is the peak projected mass density along the line-of-sight, which is assumed to be the same as the column density of the larger main halo along the trajectory of the Bullet~\cite{Markevitch:2003at}.  For $\tau > 1$, the Bullet halo would have interacted more like a fluid, experiencing similar drag and stripping as the gas, which implies $\sigma/m \lesssim 3 \; {\rm cm^2/g}$.  

The second method for constraining SIDM is based on mass loss~\cite{Markevitch:2003at}.  For the Bullet Cluster, both subclusters have similar mass-to-light ratios, which is consistent with general expectations for clusters (e.g., see Ref.~\cite{Carlberg:1995ra}).  Assuming both subclusters began with equal mass-to-light ratios prior to merger, present observations require that the Bullet halo could not have lost more than $\sim 23\%$ of its initial DM mass within its innermost 150 kpc at 68\% CL~\cite{Randall:2007ph}.  SIDM simulations by Randall et al.~\cite{Randall:2007ph} show that $\sigma/m < 0.7 \; {\rm cm^2/g}$ is required to satisfy this constraint.  While this is the strongest quoted bound from the Bullet Cluster, it relies on theoretical priors for the unobserved initial condition of the subclusters, as well as the merger itself not substantially affecting the star formation rate, which may not be the case (see, e.g., Ref.~\cite{Sobral:2015bua}).

The third method for probing self-interactions is based on offsets.  Markevitch et al.~\cite{Markevitch:2003at} proposed that self-interactions can lead to an effective drag force for the Bullet halo as it traverses the main halo, with DM particles losing momentum through scattering.  This causes an apparent offset between the galactic and DM centroids, provided the drag force is sufficient to exceed the gravitational restoring force attracting the components together.  From the measured null offset in the Bullet subcluster, Randall et al.~\cite{Randall:2007ph} obtained a limit $\sigma/m < 1.25 \; {\rm cm^2/g}$ at 68\% CL according to their simulations.  

In principle, offsets provide more robust constraints on SIDM free from theoretical priors for mass-to-light ratios and with sensitivity to smaller $\sigma/m$ outside the optically thick regime.  Consequently, this method has received much theoretical attention, including N-body simulations~\cite{Randall:2007ph,Robertson:2016xjh,Kim:2016ujt,Robertson:2016qef}, analytic methods based on an effective drag force description for self-interactions~\cite{Harvey:2013tfa}, and hybrid approaches~\cite{Kahlhoefer:2013dca}.  

The offset effect depends on the type of self-interaction assumed~\cite{Kahlhoefer:2013dca,Robertson:2016qef}.  Due to the strong directionality inherent for mergers, it is important to distinguish between isotropic hard-sphere scattering and long-range interactions where scattering is forward-peaked (a la Rutherford scattering).\footnote{Long-range does not necessarily mean {\it macroscopic} over galactic distances, but rather long-range with respect to the de Broglie wavelength of DM particles.  As discussed in \S\ref{sec:models}, SIDM models of this type can have self-interactions through light force mediators with MeV-scale masses, which are clearly {\it microscopic} compared to galactic scales.}  The drag force description generally applies only for long-range interactions, in which the accumulation of many small-angle collisions retards the Bullet halo as it passes through the main halo.  

On the other hand, the drag force description can break down for hard-sphere scattering, especially for large mass ratio systems like the Bullet Cluster where the merger velocity exceeds the escape velocity of the smaller halo.  Large-angle collisions tend to eject particles from the Bullet halo, causing a centroid shift due to the expulsive tail of backward-going particles. (The tail also exerts a gravitational pull that slows down the remaining Bullet halo.  However, since this effect is gravitational, it is the same for the galactic component and does not lead to a net offset~\cite{Kahlhoefer:2013dca}.)  For mergers with $\mathcal{O}(1)$ mass ratios, offsets can arise through a combination of both drag force and expulsive tail effects~\cite{Kim:2016ujt}.

For both types of interactions, the Bullet Cluster offset constraint from Randall et al.~\cite{Randall:2007ph} appears to be overestimated~\cite{Kahlhoefer:2013dca,Robertson:2016xjh}.  For hard sphere scattering, Randall et al.~\cite{Randall:2007ph} find a 54 kpc offset for $\sigma/m = 1.25 \; {\rm cm^2/g}$, which is excluded at 68\% CL.  However, recent simulations by Robertson et al.~\cite{Robertson:2016xjh} find a much smaller offset $\sim 20$ kpc for $2 \; {\rm cm^2/g}$.  The discrepancy arises because DM halo positions within simulations are highly sensitive to the method for how they are measured~\cite{Kahlhoefer:2013dca,Robertson:2016xjh}.  Ref.~\cite{Randall:2007ph} used the ``shrinking circles'' method: the position of the Bullet halo is given as the mean position of all particles within a circle of a given radius, which is then repeated with iteratively smaller circles (down to 200 kpc radius) each centered at the previous mean position.  While this type of method is often used for locating density peaks in N-body simulations~\cite{Power:2002sw}, the centroid shift in the Bullet halo is largely due to the expulsive tail, not due to a shift in the density peak~\cite{Kahlhoefer:2013dca}.  The shrinking circles method appears less robust for this situation, generating far larger offsets and with sensitivity to the initial circle chosen, compared to more observationally-motivated methods for determining positions based on parametric fits to the projected density or lensing shear map~\cite{Robertson:2016xjh}.

Somewhat larger offsets can arise for anisotropic scattering, e.g., forward-peaked long-range interactions.  Recently, Robertson et al.~\cite{Robertson:2016qef} have performed the first N-body simulations with anisotropic self-interactions.  Comparing models with forward-peaked and isotropic angular dependencies that both give rise to the same core radius in isolated SIDM halos, the former case produces a $\sim 50\%$ larger galaxy-DM offset in a Bullet Cluster-like merger compared to the latter case.  However, this enhancement is reduced if the anisotropic cross section is associated with a Rutherford-like velocity dependence that falls with $v_{\rm rel}$~\cite{Robertson:2016qef}, as is the case in light mediator models~\cite{Ackerman:2008gi,Feng:2009mn,Buckley:2009in,Feng:2009hw,Loeb:2010gj,Tulin:2012wi,Tulin:2013teo}.  Since smaller $v_{\rm rel}$ corresponds to pairs of DM particles that move {\it opposite} to their subhalos' bulk velocity during core passage, the remaining unscattered particles will be moving faster than the bulk velocity, hence reducing any offset~\cite{Robertson:2016qef}.

In any case, it is unclear whether {\it any} SIDM model can generate offsets larger than $\sim 50 - 100$ kpc needed to reach present observational sensitivities.  Such values require $\sigma/m$ so large that the scattering depth approaches unity, in which case self-interactions enter the optically thick regime and merging subhalos would coalesce on impact~\cite{Kim:2016ujt}.  Alternatively, Kim et al.~\cite{Kim:2016ujt} have proposed searching for misalignments between the DM centroids and the brightest cluster galaxy (BCG) in merger remnants.  Due to the cores in SIDM halos, BCG misalignment and oscillation induced during the merger can persist for Gyrs, even after the cluster has otherwise relaxed.  Sensitivities down to $0.1 \; {\rm cm^2/g}$ may be achievable, although further simulations (particularly, including gas physics) are required for more definitive statements.

\subsection{Self-interactions in minor mergers}

The most conservative constraint is based on the survival of substructure in clusters~\cite{Gnedin:2000ea}.  As discussed in \S\ref{sec:sidmhalos}, the subhalo mass function in clusters is not substantially affected by self-interactions at $1 \; {\rm cm^2/g}$, at least outside the central part of the cluster~\cite{Rocha:2012jg}.\footnote{For Abell 3827, we estimate a conservative constraint based on scattering depth.  Taking a Hernquist profile for the main halo~\cite{Kahlhoefer:2015vua}, we find $\Sigma_{\rm dm} \approx 0.3 \; {\rm g/cm^2}$ for radial infall of a subhalo to 15 kpc, implying $\sigma/m \lesssim 3 \; {\rm cm^2/g}$.}  Consequently, recent substructure merger studies have focused on offsets.

Notably, Harvey et al.~\cite{Harvey:2015hha} performed a stacked analysis for the bulleticity ratio $\beta$, using observations for 72 substructures in 30 systems, including both major and minor mergers.  If self-interactions cause DM subhalos to lag their stars/galaxies, the parallel component $\beta_\parallel$ along the merger axis will be nonzero and positive.  On the other hand, the orthogonal component $\beta_\perp$ is expected to be zero and can be used as a measure of systematics.  Harvey et al.~\cite{Harvey:2015hha} find $\beta_\parallel = - 0.04 \pm 0.07$ (and $\beta_\perp = -0.06 \pm 0.07$), consistent with collisionless CDM.  This result is interpreted as a constraint on SIDM using an analytic relation 
\be \label{eq:betaformula}
\beta_\parallel = 1 - e^{-\frac{\sigma/m}{\sigma_*}}
\ee
where $\sigma_*$ is the characteristic cross section per unit mass when the halo becomes optically thick~\cite{Harvey:2013tfa}.  Eq.~\eqref{eq:betaformula} is derived assuming {\it (i)} a drag force description of self-interactions, which is valid for forward-peaked cross sections, and {\it (ii)} that the offsets between stars, gas, and DM are all small ($\lesssim 30$ kpc), which is in line with collisionless CDM simulations for {\it minor} mergers~\cite{Harvey:2013tfa}.  The resulting limit is $\sigma/m < 0.47 \; {\rm cm^2/g}$ at $95\%$ C.L.~\cite{Harvey:2015hha}.

Recently, Wittman et al.~\cite{Wittman:2017gxn} raised serious concerns about this result.  It is important to note that the highest weighted systems in the Harvey et al.~\cite{Harvey:2015hha} dataset are those with the largest stellar-gas offsets, i.e., the {\it major} mergers between clusters.\footnote{To put the stringent limit of Ref.~\cite{Harvey:2015hha} in context, we note that their sample is heavily weighted ($67\%$) by five dissociative mergers whose constraints are listed in Table~\ref{tab:mergers}~\cite{Wittman:2017gxn}.  Moreover, the two highest weighted systems, Abell 520 ($21\%$) and Abell 2744 ($16\%$), have the most complicated morphologies.}  Since these systems have far larger offsets than 30 kpc, Eq.~\eqref{eq:betaformula} is simply not valid and limits must be interpreted within the context of simulations.  Setting aside one's qualms with Eq.~\eqref{eq:betaformula}, Wittman et al.~\cite{Wittman:2017gxn} also found that many of the stellar-DM offsets in Ref.~\cite{Harvey:2015hha}, which utilize only single-band imaging for their weak lensing maps and galaxy identification, are inconsistent with values from other more comprehensive studies, which have obtained more accurate positions using strong lensing and multiband imaging data.  Accounting for these new offsets, as well as other issues, the constraint is relaxed to $\sigma/m \lesssim 2 \; {\rm cm^2/g}$~\cite{Wittman:2017gxn}.

Lastly, we turn to Abell 3827, a unique cluster in which strong lensing observations reveal a significant DM-stellar offset for one of its infalling elliptical galaxies~\cite{Williams:2011pm,Massey:2015dkw,Taylor:2017ipx}.  According to simulations, this offset is consistent with a self-interaction cross section $\sim 1.5 \; {\rm cm^2/g}$ for a contact interaction or $\sim 3 \; {\rm cm^2/g}$ for a drag force interaction~\cite{Kahlhoefer:2015vua}.  Taken at face value, however, such large cross sections appear to be inconsistent with constraints from cluster core sizes~\cite{Kaplinghat:2015aga}.


\section{Particle physics models}
\label{sec:models}

SIDM provides a compelling solution to the long-standing issues in galactic systems, while keeping all the success of CDM on larger scales. The preferred value of the DM self-scattering cross section is $\sigma/m\sim1~{\rm cm^2/g}$ in galaxies, which is much larger than the weak-scale cross section expected for a usual WIMP DM candidate. In this section, we discuss particle physics models for SIDM and show that astrophysical observations over different scales can provide complementary information on the particle nature of DM self-interactions.

\subsection{What cross section is relevant?}

Before we delve into particle physics models, we note that there are subtleties in mapping the actual observational constraints to the particle physics parameters for a given model. Since numerical simulations are the primary tools in studying the effect of self-interactions on DM structure formation, we would like to map the particle parameters to the simulation results, i.e., the preferred values of $\sigma/m$ in the scattering probability defined in Eqs.~(\ref{eq:Pi}) and~(\ref{eq:Pij}). If the DM scattering process is isotropic, such as in the $s$-wave limit, the cross section is independent of the scattering angle, the mapping is straightforward. However, if DM self-scattering is mediated by a long-range Coulomb-like interaction, it becomes difficult. In this long-range interaction limit, the total scattering cross section can be enhanced dramatically due to the singularity in the forward scattering direction, but it does not capture all relevant physics in this case. Since the small-angle forward scatterings actually play little role in conducting heat and changing the inner halo properties accordingly, the enhancement is spurious. 

On the simulation side, it is time-consuming to simulate a large $\sigma/m$, enhanced by the forward scatterings, because a small time step is required.  In addition, simulations need to take into account the angular distribution of outgoing DM particles.  Hence, it is useful to consider a proxy that properly captures important physics and is also computationally cheap. A commonly-used prescription is to assume scattering is isotropic and that the scattering probability is proportional to the so-called transfer cross section, defined as~\cite{Mohapatra:2001sx,Feng:2009hw,Buckley:2009in,Ibe:2009mk,Loeb:2010gj,Tulin:2012wi,Tulin:2013teo,Schutz:2014nka}
\beq
\sigma_T=2\pi\int^\pi_0\frac{d\sigma}{d\Omega}\, (1-\cos\theta)\sin\theta\, d\theta,
\label{eq:transfer}
\eeq
where $\theta$ is the scattering angle in the center of mass frame. Since the longitudinal momentum transfer is $\Delta p=\mx v_{\rm rel}(1-\cos\theta)/2$, $\sigma_T$ estimates the average forward momentum lost in collisions. The transfer cross section has been used to model the long-range interactions in simulations~\cite{Vogelsberger:2012ku,Zavala:2012us,Vogelsberger:2015gpr}.\footnote{Tulin, Yu \& Zurek~\cite{Tulin:2013teo} showed that the DM scattering is almost isotropic in dwarf galaxies with the model parameters adopted in Refs.~\cite{Vogelsberger:2012ku,Zavala:2012us}.}

Tulin, Yu \& Zurek~\cite{Tulin:2013teo} (see also~\cite{Cline:2013pca,Boddy:2016bbu}) suggested that the viscosity (conductivity) cross section is a better proxy in capturing the DM self-scattering effects,  
\beq
\sigma_V=2\pi\int^\pi_0\frac{d\sigma}{d\Omega}\, (1-\cos^2\theta) \sin\theta \, d\theta.
\label{eq:viscosity}
\eeq
It emphasizes scattering in the perpendicular direction ($\theta=\pi/2$) and measures the rate of energy equalization~\cite{Schultz:2008abc}. Since both forward and backward scatterings lead to little energy exchange, they do not contribute to thermalization of the inner halo. In addition, $\sigma_V$ regulates scatterings in both forward and backward directions at the same time and it can be unambiguously calculated even when the quantum interference between $t$- and $u$ channels becomes relevant for identical particles.

The difference between $\sigma_T$ and $\sigma_V$ in interpreting the observational constraints is not significant. In the $s$-wave limit, scattering is isotropic and $\sigma_T=\sigma_{\rm tot}=(3/2)\sigma_V$, where $\sigma_{\rm tot}$ is the total scattering cross section. In the case with higher partial waves, $\sigma_T$ differs from $\sigma_V$ by an ${\cal O}(1)$ factor for Yukawa interactions between distinguishable particles~\cite{Tulin:2013teo}. For identical particles, $\sigma_T$ may have a backward-scattering enhancement and $\sigma_V$ must be used, though it differs from the distinguishable case only by an ${\cal O}(1)$ factor.  Given systematic uncertainties in astrophysical observations, we expect both $\sigma_T$ and $\sigma_V$ provide a good measure for DM self-interactions, and both of them are widely used in the literature. 

Recently, Robertson et al.~\cite{Robertson:2016qef} performed the first SIDM simulations with a full treatment of anisotropic scattering.  
Their results for core sizes in isolated cluster halos show that the simplified prescription of taking $\sigma_T$ with isotropic scattering agrees with a full angular-dependent treatment within $20\%$.\footnote{Kahlhoefer et al.~\cite{Kahlhoefer:2013dca} advocated an alternative definition of the transfer cross section, $\sigma'_T=\int(1-|\cos\theta|)(d\sigma/d \Omega)d\Omega$. In the $s$-wave case, $\sigma'_T=\sigma_{T}/2$, while $\sigma'_T\approx\sigma_T$ if large-angle scattering can be neglected. Robertson et al.~\cite{Robertson:2016qef} found that $\sigma'_T$ works better than $\sigma_T$ in reproducing the full simulation results. The comparison is made in a cluster-like halo with the Hernquist DM density profile as the initial condition. Robertson et al.~\cite{Robertson:2016qef} also showed that it is important to take into account full angular dependence in simulating merging clusters because there is a preferred direction when clusters collide~\cite{Kahlhoefer:2013dca}.} It will be interesting to further compare how well $\sigma_T$ and $\sigma_V$ serve as effective parametrizations for anisotropic scattering models across different halo mass scales within a cosmological environment.

\subsection{Self-coupled scalar}

The simplest SIDM model consists of a real scalar field $\varphi$, which is a suitable DM candidate provided it is cosmologically stable~\cite{Bento:2000ah,McDonald:2001vt}.  Self-interactions arise through the quartic coupling
\beq
{\cal L_{\rm int}}= - \frac{\lambda}{4!} \varphi^4 \, 
\eeq
and the self-scattering cross section is $\sigma(\varphi\varphi\to\varphi\varphi)= \lambda^2/(128 \pi m_\varphi^2)$, where $\lambda$ is the self-coupling constant and $m_\varphi$ is the DM mass in this model.  Writing
\beq
m_\varphi \approx 8~{\rm MeV} \times \lambda^{2/3} \left(\frac{1 \; {\rm cm^2/g}}{\sigma/m_\varphi}\right)^{1/3} \, , 
\eeq
it is clear that $m_\varphi$ must lie near the MeV scale for $\lambda$ of order unity and $\sigma/m\sim1~{\rm cm^2/g}$, as required to create cores in dwarf galaxies. 

In the early Universe, thermal production and pair annihilation to SM fermions can achieve the correct DM relic abundance for $\varphi$ if, e.g., it couples to the Higgs boson~\cite{Bento:2000ah,Burgess:2000yq}.  However, this scenario is subject to strong constraints from precision Higgs decay measurements~\cite{Khachatryan:2016whc} since it contributes to invisible Higgs decay.  Alternatively, the relic abundance can be set by ``cannibalization''~\cite{Carlson:1992fn,Hochberg:2014kqa}.  If the Lagrangian has a $\varphi^3$ term, then the self-annihilation process $\varphi\varphi\varphi\to\varphi\varphi$ can occur in the early Universe and effectively deplete the $\varphi$ abundance, resulting in the correct relic density~\cite{Hochberg:2014kqa}. The DM candidate $\varphi$ could be a composite state, such as glueballs of non-Abelian gauge fields in the hidden sector~\cite{Carlson:1992fn}. In this case, both $\varphi^3$ and $\varphi^4$ terms are present naturally.

The major shortcoming of this simple model is that it predicts a constant cross section over all scales. As we discussed in~\S\ref{sec:jeans}, $\sigma/m\sim1~{\rm cm^2/g}$ is favored in dwarf galaxies, but it must be $\sim 0.1~{\rm cm^2/g}$ (or less) on cluster scales~\cite{Kaplinghat:2015aga}, otherwise DM self-interactions create a ${\cal O}(100)~{\rm kpc}$ density core in galaxy clusters, too large to be consistent with observations~\cite{Newman:2012nw}. Nevertheless, this model does illustrate a key ingredient in building a successful SIDM model: there must exist a low mass scale much below the weak scale.

\subsection{Light mediator models}
\label{subsec:yukawa}

Self-interactions can arise through the exchange of additional weakly-coupled states that interact with DM particles.  Perhaps the best-motivated setup along these lines is if DM is charged under a spontaneously broken $U(1)$ gauge symmetry.  In this case, DM stability is naturally explained by charge conservation.  As a by-product, gauge boson exchange mediates self-interactions, analogous to Rutherford scattering~\cite{Feng:2009hw,Buckley:2009in,Ibe:2009mk,Loeb:2010gj,Tulin:2012wi,Tulin:2013teo,Schutz:2014nka}.  Historically, this type of model had been motivated earlier by purported anomalies in indirect detection observations, such as the 511 keV $\gamma$-rays from the galactic bulge~\cite{Boehm:2003bt,Boehm:2003hm}, the PAMELA positron excess~\cite{ArkaniHamed:2008qn,Pospelov:2008jd,Pospelov:2007mp}, and hidden sector DM model building (e.g., Refs.~\cite{Chen:2006ni,Feldman:2006wd,Feldman:2007nf,Feng:2008ya,Feng:2008mu,Chen:2009iua}).  We also consider self-scattering mediated by scalar bosons as well.  

The model is described by the following interaction
\beq
{\cal L_{\rm int}}=\bigg\{
\begin{array}{c l}
g_\chi\bar{\chi}\gamma^\mu\chi\phi_\mu & \text{(vector mediator)}\\
g_\chi\bar{\chi}\chi\phi & \text{(scalar mediator)} \, ,
\end{array}
\eeq
where $\chi$ is the DM particle (assumed to be a fermion), $\phi$ is the mediator, and $g_\chi$ is the coupling constant. In the non-relativistic limit, self-interactions are described by the Yukawa potential
\beq
V(r)=\pm\frac{\alpha_\chi}{r}e^{-\mphi r} .
\label{eq:yukawa}
\eeq
The model parameters are the dark fine structure constant $\alpha_\chi\equiv  g^2_\chi/4\pi$, the mediator mass $m_\phi$, and the DM mass $m_\chi$.  For a vector mediator, the potential is attractive ($-$) for $\chi \bar\chi$ scattering and repulsive ($+$) for $\chi \chi$ and $\bar\chi \bar\chi$ scattering, while for a scalar mediator, the potential is purely attractive.  More detailed models~\cite{Boddy:2014yra,Ko:2014nha,Kang:2015aqa,Kainulainen:2015sva,Wang:2016lvj,Ma:2016tpf,Kitahara:2016zyb,Ma:2017ucp}, as well as potentials arising from other types of mediators~\cite{Bellazzini:2013foa}, have been considered as well.

In general, this framework admits a rich cosmological phenomenology, similar to electromagnetism in the visible sector.  If the mediator is sufficiently light, bound state formation and dark recombination~\cite{Wise:2014jva,Petraki:2014uza,Petraki:2015hla,Petraki:2016cnz,Cirelli:2016rnw}, delayed kinetic decoupling~\cite{Feng:2009mn,Aarssen:2012fx,Chu:2014lja,Buckley:2014hja,Tang:2016mot,Agrawal:2017rvu}, dark acoustic oscillations~\cite{Feng:2009mn,CyrRacine:2012fz,Pearce:2013ola,Buckley:2014hja}, and dissipation could be relevant~\cite{Fan:2013yva,Fan:2013tia}.  

Here, we follow Refs.~\cite{Tulin:2012wi,Tulin:2013teo} and restrict our attention to the effect of elastic DM self-scattering.  In the perturbative limit ($\ax\mx/\mphi\ll1$), the Born differential cross section is
\beq
\frac{d\sigma}{d\Omega}=\frac{\ax^2\mx^2}{\left[\mx^2\vrel^2(1-\cos\theta)/2+\mphi^2\right]^2}.
\eeq
In the limit of $\mphi\gg\mx\vrel$, scattering is a contact interaction and the cross section is independent of $\vrel$, as given in Eq.~\eqref{eq:born}.  In the opposite limit when $\mphi\ll\mx\vrel$, it scales as $1/\vrel^4$ \`{a} la Rutherford scattering~\cite{Feng:2009mn,Ackerman:2008gi,Agrawal:2016quu}.  Neither limit provides the mildly velocity-dependent cross section favored by observations if self-interactions are to solve small scale structure issues on galactic and cluster scales~\cite{Kaplinghat:2015aga}.  However, a small but finite mediator mass can provide the right velocity dependence.  This requires the transition between contact and Rutherford limits to occur around $v_{\rm rel} \sim 300$ km/s (between dwarf and cluster scales), implying $m_\phi/m_\chi \sim v_{\rm rel}/c \sim 10^{-3}$.
For instance, for DM with 10 GeV mass, the required mediator mass is $\mphi\sim10~{\rm MeV}$ and the corresponding range of the force is $\sim 20~{\rm fm}$.  It is remarkable that a dark force with femtoscale range can affect the dynamics of galaxies. 

Exploring the full parameter space of this model requires calculating $d\sigma/d\Omega$ beyond the pertubative limit, where multiple scattering---ladder diagrams in Feynman diagramatic language---become important.  Buckley \& Fox~\cite{Buckley:2009in} and Tulin, Yu \& Zurek~\cite{Tulin:2012wi,Tulin:2013teo} developed a numerical procedure using a nonrelativistic partial wave analysis to study the nonperturbative regime.  In some regimes, analytical expressions are also available. Feng, Kaplinghat \& Yu~\cite{Feng:2009hw} introduced formulae for the transfer cross section in the Born regime and also in the classical regime, the latter based on an empirical formula originally developed for plasma collisions~\cite{Khrapak:2003kjw,Khrapak:2014xqa}.\footnote{In the plasma medium, the photon obtains an effective mass due to the Debye screening effect and ion-ion (electron-ion) interactions can be modeled with a repulsive (attractive) Yukawa potential. Feng, Kaplinghat \& Yu~\cite{Feng:2009hw} first applied the semi-analytical fitting formula from~\cite{Khrapak:2003kjw} to study DM self-interactions. Vogelsberger et al.~\cite{Vogelsberger:2012ku} performed the first Yukawa SIDM simulations, based on the plasma formula introduced in~\cite{Feng:2009hw} and its reparametrized form suggested by Loeb \& Weiner~\cite{Loeb:2010gj}.} Ref.~\cite{Tulin:2013teo} also found an analytical expression for $s$-wave scattering, valid in the nonperturbative regime, using the Hulth\'{e}n potential as an approximation, while Braaten \& Hammer~\cite{Braaten:2013tza} studied $s$-wave resonant scattering on more general grounds.  See Ref.~\cite{Tulin:2013teo} for a map of the full parameter space and summary of numerical methods and analytical formulae.

\begin{figure}
\includegraphics[scale=0.63]{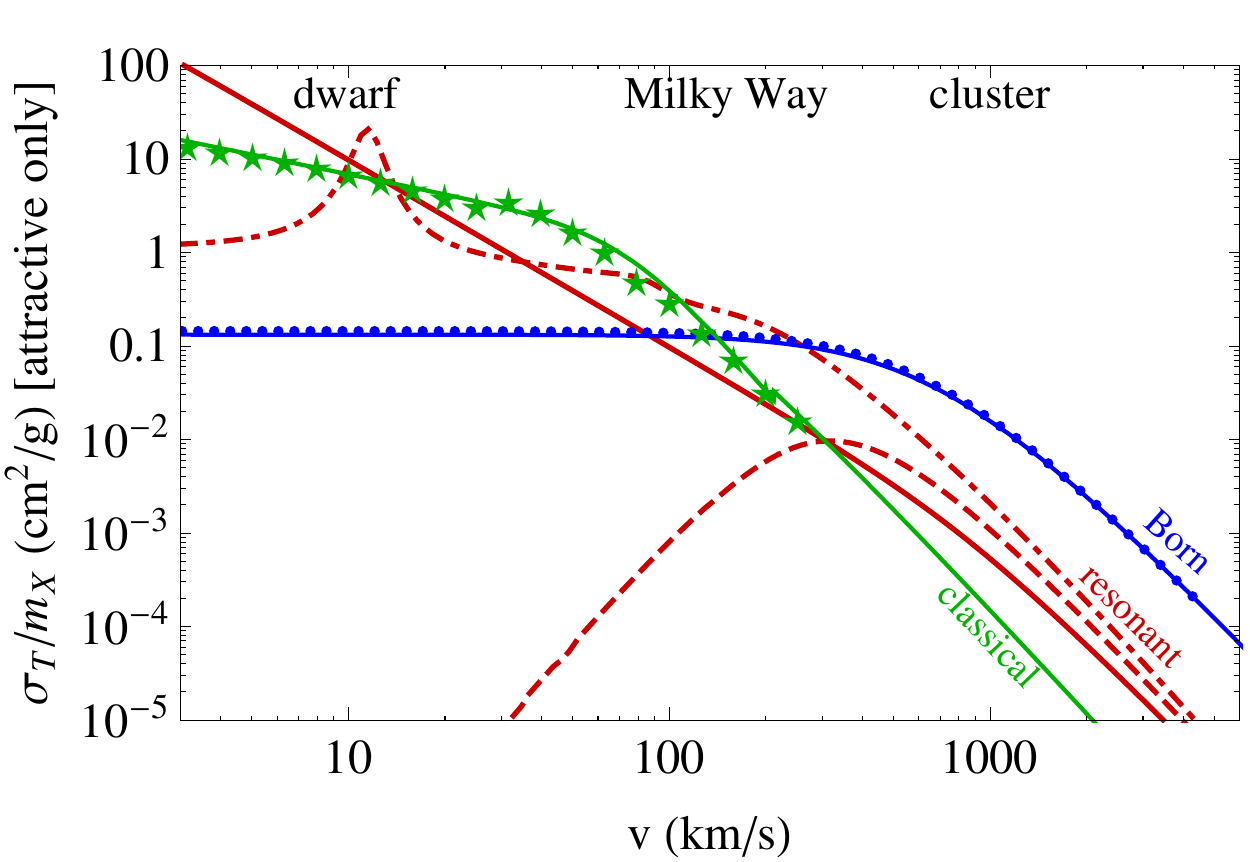}\;\;
\includegraphics[scale=0.63]{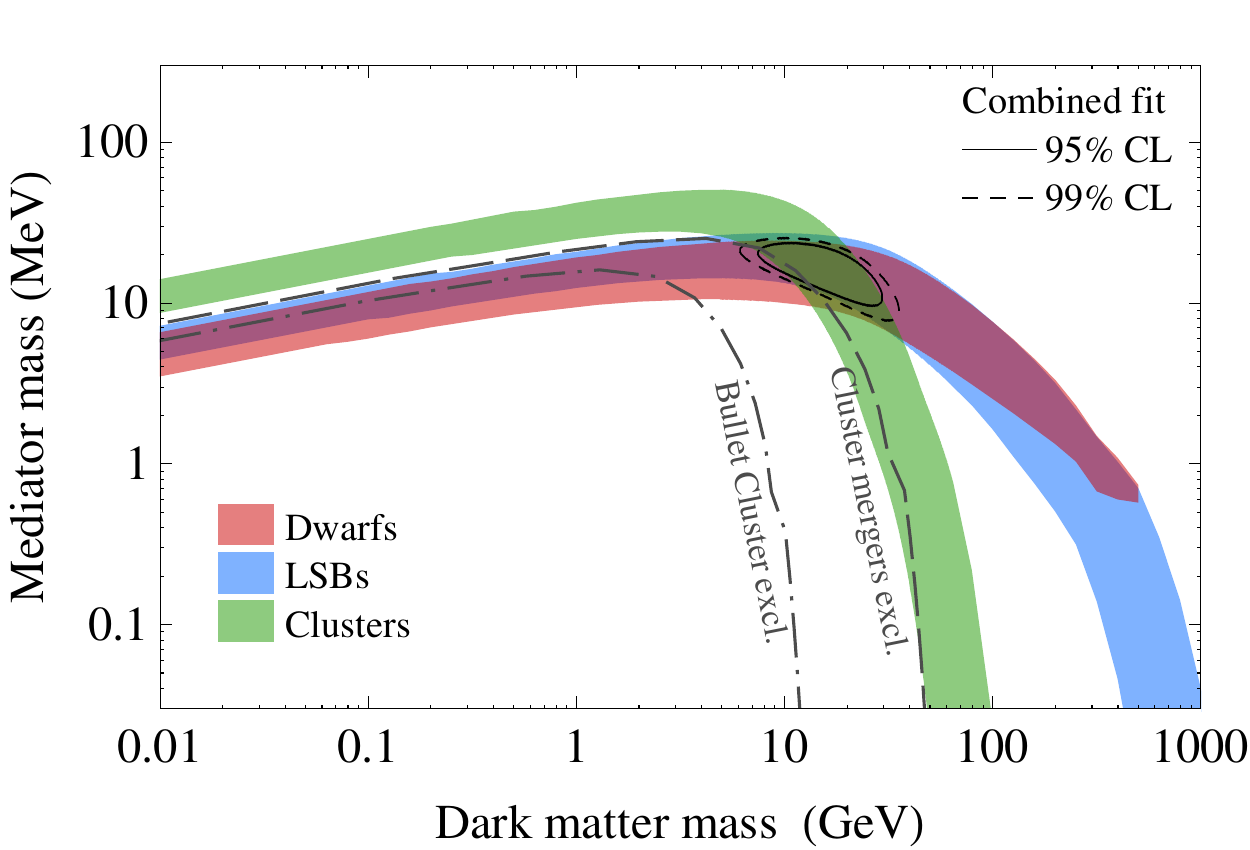}
\caption{ \it Left: Velocity-dependence of the DM self-scattering transfer cross section with an attractive Yukawa potential. Parameters are chosen to illustrate the dependence in different regimes, including weakly-coupled Born limit (blue), strongly-coupled classical limit (green), $s$-wave quantum resonance (red solid), $p$-wave resonance (red dot-dashed), and $s$-wave antiresonance (red dashed). Reprinted from Ref.~\cite{Tulin:2012wi}. Right: Parameter space for repulsive Yukawa model of DM self-interactions ($\alpha_\chi\approx1/137$), preferred by dwarf galaxies (red), LSB galaxies (blue), and galaxy clusters (green), at $95\%$ CL. Combined region at $95\%$ ($99\%$) CL is shown by the solid (dashed) contours. The estimated exclusion regions from the Bullet Cluster (dot-dashed) and the ensemble of merging clusters (long dashed) are also shown. Reprinted from Ref.~\cite{Kaplinghat:2015aga}.
}
\label{fig:yukawa}
\end{figure}

Even this simple SIDM model exhibits many possible velocity dependencies for the scattering cross section.  Fig.~\ref{fig:yukawa} (left) shows the transfer cross section $\sigma_T/m$, as a function of relative velocity.  Each curve represents a different choice of parameters $(\alpha_\chi, m_\chi, m_\phi)$ for an attractive Yukawa potential.  In these cases, $\sigma_T$ is suppressed on clusters scales and larger ($\vrel\gtrsim1500\kms$), corresponding to the Rutherford limit.  However, the dependence becomes much more complicated on galactic scales with the potential for quantum mechanical resonances (and anti-resonances).  For example, for $\ax\mx/m_\phi \approx 1.6 \, n^2$ where $n$ is a positive integer, scattering has an $s$-wave resonance and $\sigma_T/m_\chi$ becomes enhanced at low velocity.  

Since the velocity dependence of $\sigma_T/m$ is quite sensitive to the choice of the particle physics parameters, astrophysical observations on different scales can constrain or even discover these parameters.  Kaplinghat, Tulin \& Yu~\cite{Kaplinghat:2015aga} showed that SIDM with a repulsive Yukawa interaction could yield a suitable cross section inferred from both galaxy and cluster data; see Fig.~\ref{fig:jeans}~(left).  Fig.~\ref{fig:yukawa} (right) illustrates the favored range of ($\mx,~\mphi$) for fixed coupling, chosen to be $\ax\approx1/137$.  Colored bands denote the regions preferred by dwarf galaxies (red), LSB galaxies (blue), and galaxy clusters (green). Remarkably, there exists a closed common region (black contours) that points to DM mass of $\sim15~{\rm GeV}$ and mediator mass of $\sim17~{\rm MeV}$. The result is {\it independent} of whether the dark and visible sectors are coupled via interactions beyond gravity.

An interesting question is whether the mediator $\phi$ can be massless~\cite{Feng:2009mn,Ackerman:2008gi,Agrawal:2016quu}. In the absence of recombination into bound states, scattering is described by the Coulomb potential.  Since the scattering cross section scales as $1/v_{\rm rel}^4$, fixing $0.1~{\rm cm^2/g}$ in galaxy clusters~\cite{Kaplinghat:2015aga} leads to an enormous cross section $\sim10^6~{\rm cm^2/g}$ on dwarf scales.  Such large cross sections are expected to lead to gravothermal collapse of a dwarf halo.  However, Agrawal et al.~\cite{Agrawal:2016quu} argue that core formation and collapse are inhibited for large cross sections since heat conduction is suppressed by the small mean free path, suggesting an approximate duality between strongly and weakly self-interacting regimes.  Ref.~\cite{Ahn:2002vx} predicted that a large cross section value $\sigma/m \sim 10^4 \; {\rm cm^2/g}$ leads to similar cored profiles as a smaller cross section $\sigma/m \sim 1 \; {\rm cm^2/g}$.  On the other hand, cosmological simulations that had been performed in the fluid limit for SIDM yield singular isothermal halos that are steeper than CDM halos~\cite{Moore:2000fp,Yoshida:2000bx}, disfavoring this picture.  Nevertheless, further simulations are required to make these statements more quantitative.  Lastly, another complication is that collective plasma effects due to dark electromagnetic fields may be important in affecting the halo~\cite{Ackerman:2008gi,Heikinheimo:2015kra,Sepp:2016tfs}.

In the early Universe, light mediator states can play an important role in setting the DM relic abundance through $\chi\bar{\chi}\rightarrow\phi\phi$ annihilation.  For symmetric DM, the required annihilation cross section is $\langle \sigma_{\rm ann} v_{\rm rel} \rangle \approx 5\times10^{-26}{\rm~cm^3/s}$ assuming the dark and SM sectors are thermalized.  For asymmetric DM, the relic density is determined by a primordial DM asymmetry and the annihilation cross section has to be larger than this value to deplete the symmetric density (see Refs.~\cite{Davoudiasl:2012uw,Petraki:2013wwa,Zurek:2013wia} and references therein).  These considerations imply $\ax\gtrsim4\times10^{-5} \, ({\mx/{\rm GeV}})$ for the vector mediator case.  On the other hand, smaller couplings are viable if the two sectors are thermally decoupled and the SM has a higher temperature~\cite{Feng:2008mu}.  In addition, the SIDM abundance could also be set by non-thermal production mechanisms~\cite{Bernal:2015ova,Reece:2015lch}.

\subsection{Strongly interacting dark matter}

SIDM may be a composite state of a confining non-Abelian gauge theory in the dark sector~\cite{Carlson:1992fn,Faraggi:2000pv,Frandsen:2011kt,Cline:2013zca,Boddy:2014yra,Boddy:2014qxa,Soni:2016gzf,Forestell:2016qhc,Ko:2016fcd,Prilepina:2016rlq}.
Since such theories are already known to exist in nature---namely, quantum chromodynamics (QCD)---it is appealing that similar physics may exist for DM.  QCD enforces that the proton is long-lived (due to an accidental symmetry) and accounts for most of the mass of visible matter in the Universe.  Similarly, DM may be a dark hadron whose mass arises through nonperturbative physics and whose stability is imposed automatically by symmetry.  Proposed candidates typically fall into the categories of dark baryons, dark mesons, or dark glueballs.  However, the gauge group and properties of the dark quarks (number of flavors, masses, representations) are unknown {\it a priori} and many possible theories exist (see Ref.~\cite{Kribs:2016cew} for a review).

Due to the nonperturbative nature of the theory, it is nontrivial to compute the self-interaction cross section and mass spectrum.  In the low-energy limit, the self-interaction cross section can be expressed as $\sigma = 4\pi a^2$ where $a$ is the scattering length.  On dimensional grounds, one estimates $a \sim \Lambda_{\rm DM}^{-1}$, where $\Lambda_{\rm DM}$ is the dark confinement scale (in analogy with $\Lambda_{\rm QCD} \approx 300$ MeV).  The DM mass $m$ depends in some detail on the nature of the DM state.  For instance, for $SU(N)$ gauge theory, dark baryons have mass $m \sim N \Lambda_{\rm DM}$ if composed of effectively massless constituents (nucleon-like), while $m$ can be larger if one or more constituents are massive (like a heavy flavor baryon).  If DM is a meson-like state, $m$ is somewhat arbitrary and DM can be lighter than $\Lambda_{\rm DM}$ (pion-like) or heavier (like a heavy flavor meson).  For dark glueballs, the mass of the lightest state typically scales as $m \sim {\rm few} \times \Lambda_{\rm DM}$~\cite{Ochs:2013gi}.  

The cross section per unit mass for self-interactions can be expressed as follows:
\beq
\label{eq:strong}
\sigma/m\sim3 \; {\rm cm^2/g} \times \left(\frac{\Lambda_{\rm DM}}{m}\right) \left(\frac{\Lambda_{\rm DM}}{a^{-1}}\right)^2 \left(\frac{100~{\rm MeV}}{\Lambda_{\rm DM}}\right)^3.
\eeq
If all dimensionful scales are set by $\Lambda_{\rm DM}$, then taking $\Lambda_{\rm DM} \approx \mathcal{O}(0.2-0.6) \times \Lambda_{\rm QCD}$ gives a cross section in the right ballpark for addressing small scale issues on dwarf scales.  However, in this simple picture, the nonrelativistic cross section is constant with velocity, which is disfavored by cluster bounds.  

If self-interactions are the solution to small scale issues, then a velocity-dependent cross section is required.  The cross section is expected to fall with velocity once the de Broglie wavelength becomes smaller than the scattering length, i.e., when $m v_{\rm rel} \gtrsim a^{-1}$.  Clearly if both $m,a^{-1} \sim \Lambda_{\rm DM}$, this condition can never be satisfied for nonrelativistic scattering.  On the other hand, if 
\beq
m a \gg 1 \, , \label{eq:ma1}
\eeq
it may be possible to achieve the desired cross section on both galactic and cluster scales.  

There are two simple ways for a composite DM model to satisfy Eq.~\eqref{eq:ma1}.  First, the constituents (e.g., dark quarks) may be very heavy such that $m \gg \Lambda_{\rm DM}$.  As an example along these lines, Boddy et al.~\cite{Boddy:2014yra} proposed a model consisting of heavy fermions in the adjoint representation of a confining $SU(N)$ gauge theory.  DM particles consist of single fermions---with mass $\sim 10$ TeV to yield the correct relic density---plus dark gluons (``glueballinos'').  The range of the interaction is set by $a^{-1} \sim \Lambda_{\rm DM} \sim 100$ MeV to give an acceptable self-scattering cross section.  

Second, self-interactions may have a larger-than-expected scattering length $a \gg \Lambda_{\rm DM}^{-1}$.  Precisely this effect occurs in QCD.  Neutron-proton scattering has a resonant enhancement at low energy due to the weakly bound deuteron, such that $a^{-1} \approx 15 \; {\rm MeV} \ll \Lambda_{\rm QCD}$.  As shown in Fig.~\ref{fig:npscat}, the self-scattering cross section falls with velocity for $v_{\rm rel} \gtrsim (m_p a)^{-1}c \approx 5000\; {\rm km/s}$, while it is constant at low velocity (neglecting electromagnetism).  The same physics may provide a viable SIDM candidate~\cite{Cline:2013zca}.  However, to satisfy constraints on cluster scales ($v_{\rm rel} \sim 1500 \; {\rm km/s}$), larger values of $ma$ are required so that the turn-over in the cross section is at lower velocities, ${\cal O}(100\;{\rm km/s})$.  Other models with long range forces mediated by pseudo-Goldstone bosons may also be viable.

The observed abundance of composite SIDM particles can be realized in different ways. Since the gauge coupling is large, the usual freeze-out mechanism only works if $m \sim {\rm TeV}$.  For example, in Ref.~\cite{Boddy:2014yra}, massive dark quarks annihilate into dark gluons before the dark confinement phase transition.  On the other hand, if $m \ll {\rm TeV}$, other mechanisms must be utilized since the DM abundance would be too small in the usual freeze-out scenario.  For DM carrying a conserved charge---analogous to baryon number---it is natural to assume that its abundance is set by a primordial asymmetry similar to the baryon asymmetry of the Universe (see, e.g.,~\cite{Davoudiasl:2012uw,Petraki:2013wwa,Zurek:2013wia} and references therein).  Here, a large annihilation cross section is important to deplete the symmetric density efficiently.  One needs make sure that there are additional light degrees of freedom into which the symmetric component can annihilate, such as dark pions, which must subsequently get depleted.  Dark pions may decay into SM fermions through massive mediators~\cite{Cline:2013zca}.  Alternatively, if dark quarks are massless, dark pions are Goldstone bosons and get redshifted away as dark radiation, which remains allowed if the dark sector is slightly colder than the visible sector~\cite{Feng:2008mu,Cline:2013zca}.  In the case of glueball DM, $3\rightarrow2$ cannibalization can achieve the correct DM abundance~\cite{Carlson:1992fn,Hochberg:2014kqa,Soni:2016gzf} and excited states may be important~\cite{Forestell:2016qhc}.  A strongly-coupled SIDM sector could also be realized in the dynamical DM framework~\cite{Dienes:2011ja,Dienes:2016vei}.  In this scenario, the DM density comprises an ensemble of different hadron-like states and the DM relic abundance evolves with time as some states decay away.

\subsection{Dark atoms}

Composite SIDM candidates may arise in weakly coupled theories in the form of atomic (or molecular) bound states~\cite{Mohapatra:2000qx,Mohapatra:2001sx,Foot:2001hc,Khlopov:2005ew,Kaplan:2009de,Behbahani:2010xa,CyrRacine:2012fz,Cline:2012is,Fan:2013yva,Fan:2013tia,Cline:2013pca,Foot:2014mia,Boddy:2016bbu}. 
The simplest atomic DM model consists of a hydrogen-like bound state composed of two distinct species with opposite charges under a massless $U(1)$ gauge symmetry.  Following the same nomenclature as for normal hydrogen, there is a dark electron $e$ and dark proton $p$, with masses $m_{e,p}$, and a dark fine structure constant $\alpha$.  The DM particle is denoted $H$, with mass $m_H \approx m_e + m_p$.  The effective potential between hydrogen atoms is known in the atomic physics literature and depends on the total electronic spin configuration (singlet or triplet)~\cite{Silvera:1980abc}.  Refs.~\cite{Cline:2013pca,Boddy:2016bbu} adopted these results to compute self-interactions between dark atoms.  As an example, the potential for the triplet state is
\beq \label{eq:VHtrip}
V(x) \approx {\alpha}^2\mu_{H} \left[e^{0.09678-1.10173 x-0.003945 x^2}-e^{-(10.04/x-1)^2}\left(\frac{6.5}{x^6}+\frac{124}{x^8}+\frac{3285}{x^{10}}\right)\right],
\eeq
where $\mu_{H}=m_{e}m_{p}/m_H$ is the atomic reduced mass and the numerical factors are fitted from experimental atomic data~\cite{Silvera:1980abc,Cline:2013pca}.  The complicated expression in brackets---the first term models repulsive exchange and the second term represents the Van der Waals interaction---is easily adapted to the dark sector by expressing Eq.~\eqref{eq:VHtrip} in terms of the radius $x=r/a_0$ normalized in units of the Bohr radius $a_0=(\alpha \mu_H)^{-1}$.

\begin{figure}
\includegraphics[scale=0.4]{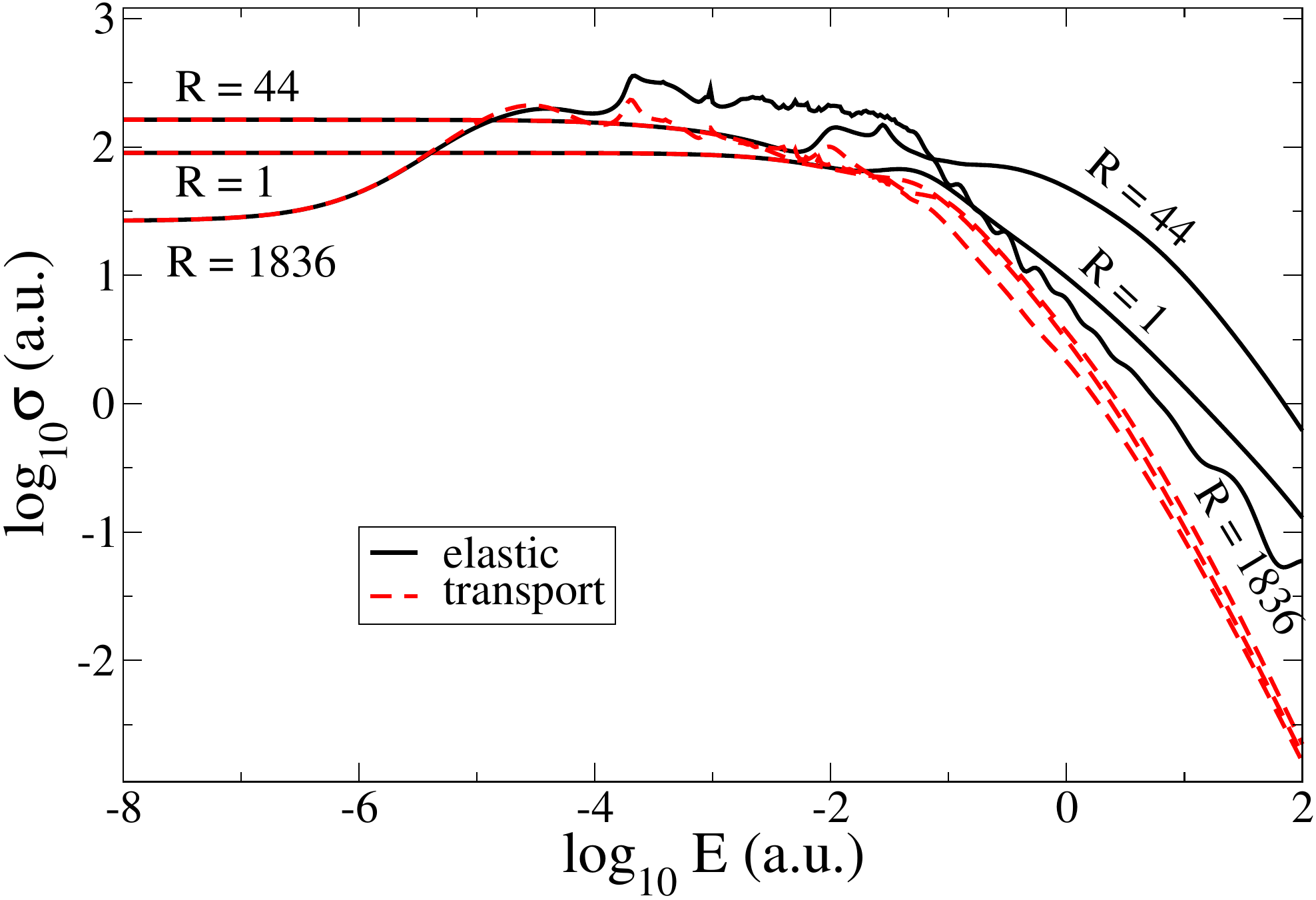}\;\;
\includegraphics[scale=0.4]{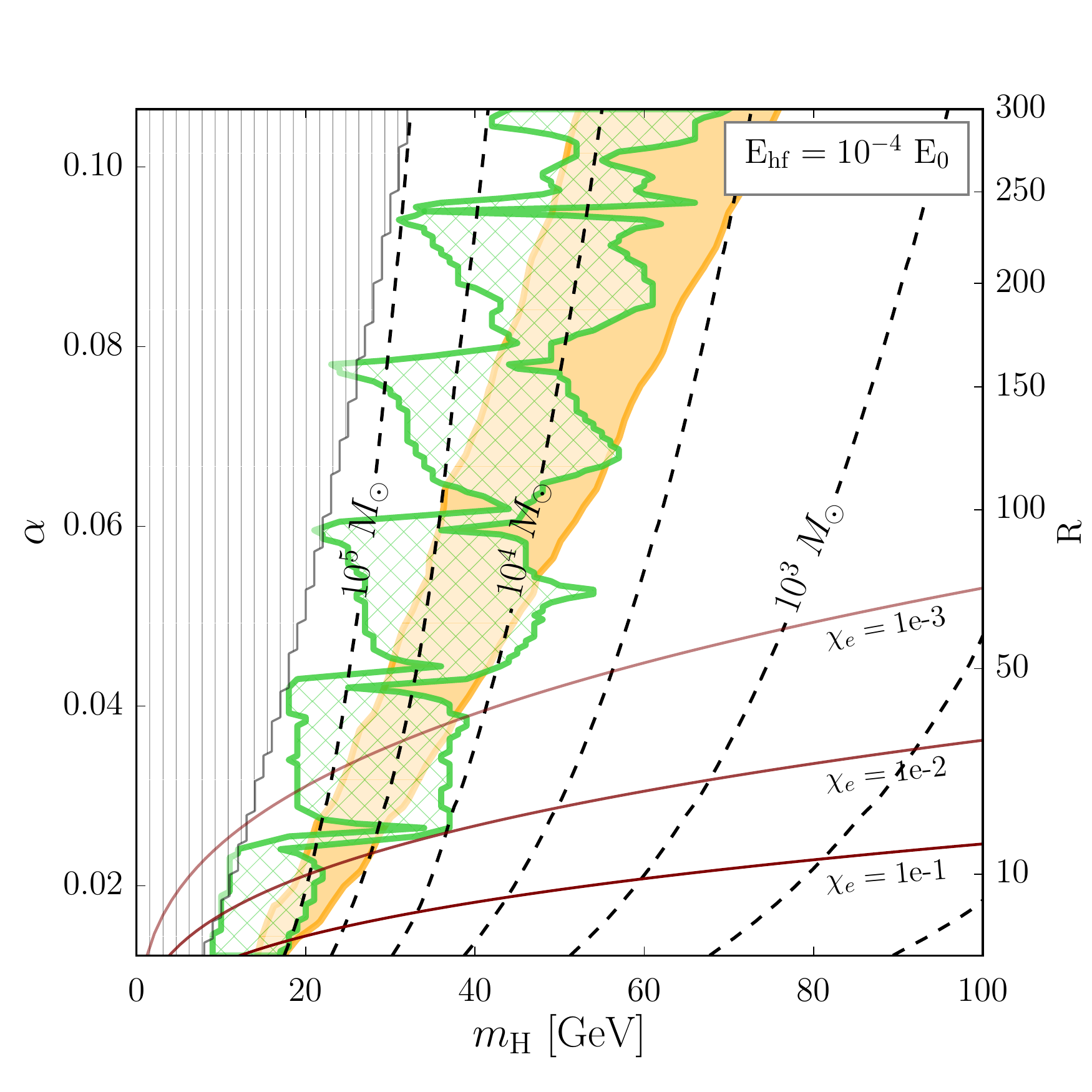}
\caption{ \it Left: Elastic total and viscosity (``transport'') cross sections as a function of collision energy for the atomic SIDM model for different ratios $R = m_p/m_e$.  The axes are normalized to atomic units for energy and length: $\alpha^2\mu_{H}$, and $(\alpha\mu_H)^{-1}$, respectively. Reprinted from Ref.~\cite{Cline:2013pca}. Right: Parameter space for atomic SIDM. The cross-hatched green region satisfies $0.5~{\rm cm^2/g<\sigma/m<5~{\rm cm^2/g}}$ for $\vrel\approx30\textup{--}100~\kms$, leading to core formation in lower-mass halos. The solid orange area denotes the region where the viscosity cross section can reproduce cluster core sizes, $\sim 0.1~{\rm cm^2/g}$ at $\vrel\approx1500~\kms$. The vertical hatched gray region is disfavored by the halo shape measurement in galaxy clusters, corresponding to $\sigma/m_H >1~{\rm cm^2/g}$ for $\vrel\approx1000~\kms$. The dashed (solid) lines show contours of constant minimal halo mass (ionization fraction) predicted in the model, where the hyperfine splitting energy is $10^{-4}$ times the atomic energy, $E_0$. Reprinted from Ref.~\cite{Boddy:2016bbu}.
}
\label{fig:atom}
\end{figure}

To calculate the atomic self-scattering cross section, Cline et al.~\cite{Cline:2013pca} used a partial wave analysis.  Boddy et al.~\cite{Boddy:2016bbu} further included inelastic scattering due to hyperfine transitions. At low energy, $s$-wave scattering dominates and the cross section scales with the geometric size of the dark atom, $\sigma\approx100 \, a^2_0$~\cite{Cline:2013pca}.  
In the limit $m_{p}\gg m_{e}$, the cross section is numerically given by
\beq
\sigma/m_H \approx1~{\rm cm^2/g} \times \left(\frac{0.1}{\alpha}\right)^2\left(\frac{8~{\rm GeV}}{m_{p}}\right)^3\left(\frac{m_{p}/m_{e}}{15}\right)^2 \, .
\eeq

Dark atom scattering can produce a velocity-dependent scattering cross section, which is preferred for viable SIDM models.  Fig.~\ref{fig:atom} (left) shows that the cross section decreases once the collision energy exceeds about one-tenth of the atomic binding energy, corresponding to the de Broglie wavelength becoming smaller than $a_0$~\cite{Cline:2013pca,Boddy:2016bbu}.  In this region, the total (``elastic'') cross section is larger than the viscosity (``transport'') cross section due to the spurious enhancement from forward and backward scattering, even although they are normalized to be identical in the low-energy limit. Boddy et al.~\cite{Boddy:2016bbu} showed that atomic scattering can have the right velocity dependence to explain density cores inferred in dwarf galaxies while being consistent with observations on cluster scales.  Fig.~\ref{fig:atom} (right) maps out the parameter space favored by these observations, which imposes $\alpha>0.01$ and the DM mass to be in the $10\textup{--}100~{\rm GeV}$ range.

There are several realizations of atomic DM. In the mirror DM model, the dark sector is a mirror copy of the SM sector~\cite{Mohapatra:2000qx,Mohapatra:2001sx,foot:2014uba}. In the double disk DM model~\cite{Fan:2013yva}, dark atoms only compose a very small fraction of DM, and they may form a dark disk in DM halos through dissipative processes. For atomic DM, the relic density should be set by a primordial asymmetry. Since $\alpha$ is typically large for atomic bound states to form in the early Universe, the annihilation cross section of baryons and anti-baryons to massless dark photons is significant enough to deplete the symmetric component. One important constraint on atomic DM is that the energy dissipation rate should be small enough such that DM halos can form~\cite{Mohapatra:2001sx,CyrRacine:2012fz,Boddy:2016bbu}, although Foot \& Vagnozzi~\cite{foot:2014uba,Foot:2016wvj} argued that the energy injection from supernova explosions in the visible sector could compensate the energy loss due to the dissipative cooling process if the two sectors are connected with a kinetic mixing term, and both the core-cusp and the missing satellites problems could be resolved.

\subsection{SIDM with an excited state}
\label{subsec:excited}

\begin{figure}
\includegraphics[scale=0.4]{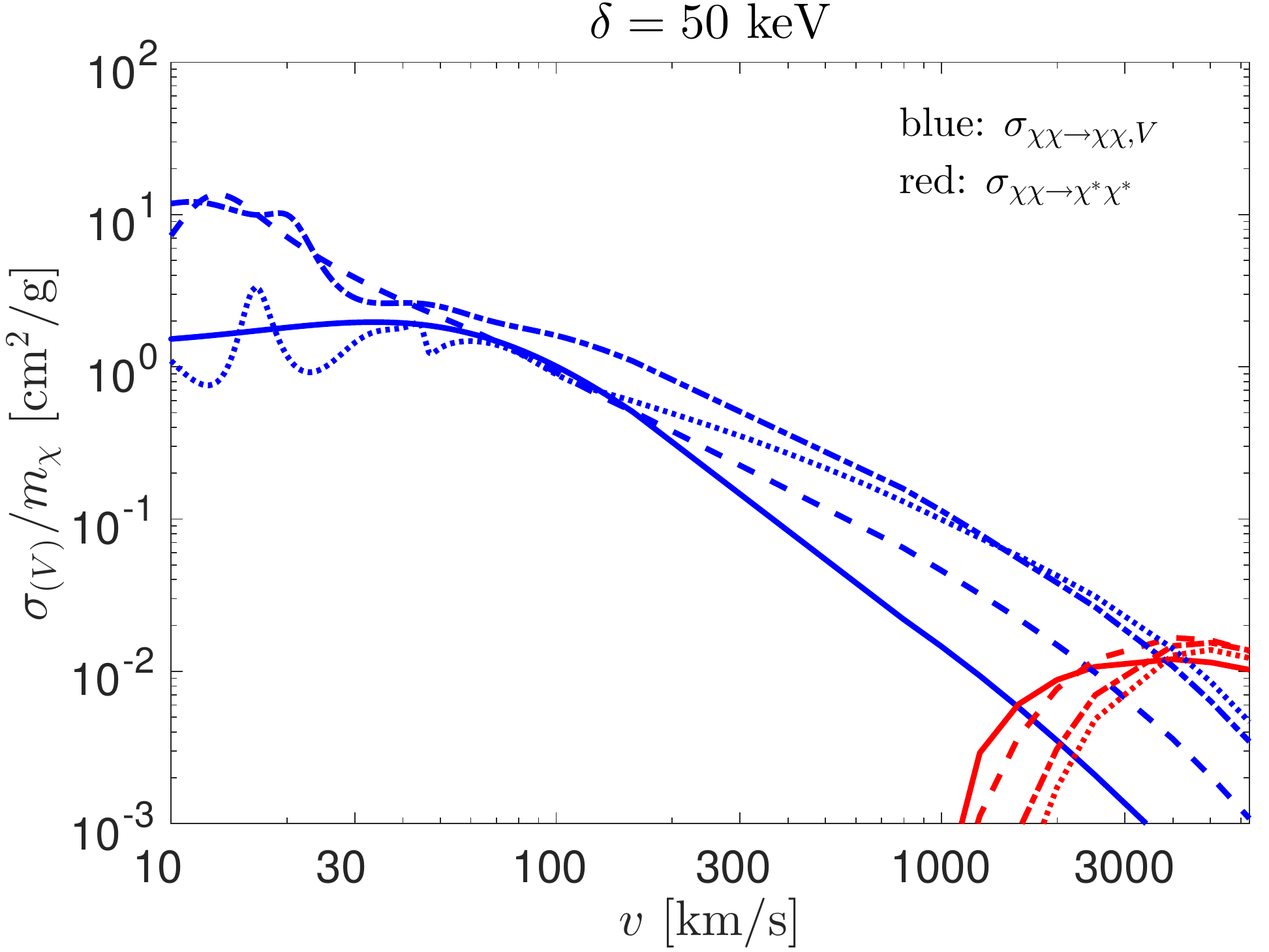}\;\;
\includegraphics[scale=0.4]{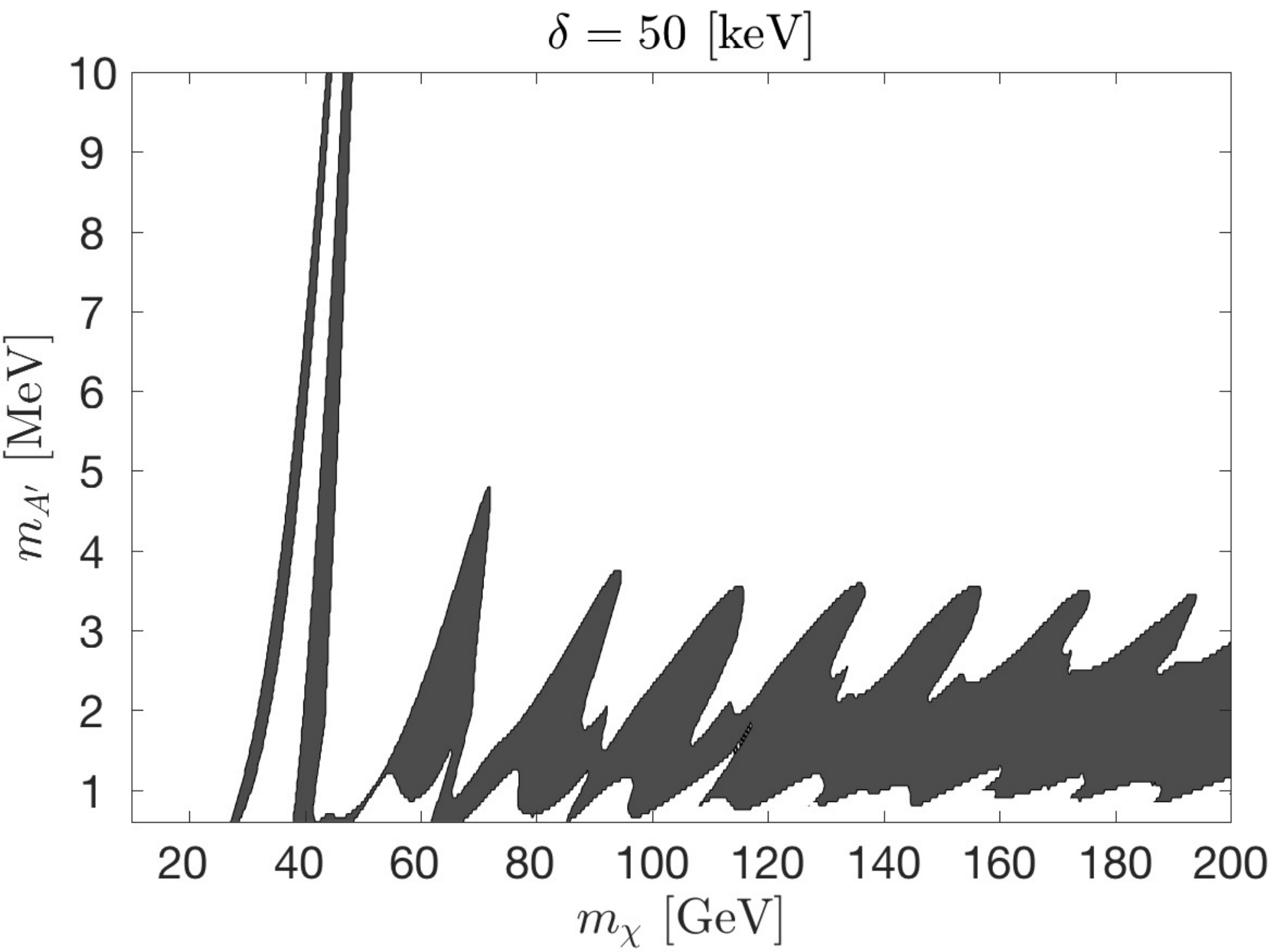}
\caption{ \it Left: Elastic (blue) and inelastic (red) viscosity cross sections versus scattering velocity for inelastic SIDM benchmark models, where the mass splitting is $50~{\rm keV}$. Different line styles represent different DM and mediator masses in the range of $40-160$ GeV and $1-2$ MeV, respectively.  Right: SIDM parameter space (shaded), mediator mass ($m_{A^\prime}$) versus DM particle mass, where the elastic scattering cross section is in the range of $0.5\textup{--}5~{\rm cm^2/g}$ for $v=30\textup{--}100~{\kms}$. Both figures reprinted from Ref.~\cite{Blennow:2016gde}; see therein for further details.
}
\label{fig:inelastic}
\end{figure}

Self-interactions need not be a purely elastic process.  The dark sector may include a spectrum of closely lying states and self-interactions can mediate transitions between them if allowed kinematically.  In the literature, DM models with inelastic transitions have been motived for other phenomenological studies, such as direct detection~\cite{Han:1997wn,TuckerSmith:2001hy,Batell:2009vb,An:2011uq,Bramante:2016rdh}, indirect detection~\cite{Finkbeiner:2007kk,ArkaniHamed:2008qn,Tulin:2012uq,Finkbeiner:2014sja}, and collider signals~\cite{Baumgart:2009tn,Bai:2011jg}.  For SIDM, such a spectrum of states may arise as energy levels of atomic DM~\cite{Cline:2014eaa,Boddy:2016bbu} or strongly interacting composite DM~\cite{Boddy:2014qxa}.

The minimal model along these lines consists of two states, the DM ground state $\chi$ and an excited state $\chi^*$, with an interaction that mediates transitions between them.  If DM particles are charged under a $U(1)$ gauge symmetry, symmetry breaking can generate a mass term that splits the charged state into two real states $\chi$ and $\chi^*$ with a coupling to the gauge boson $\phi$ that is purely off-diagonal.  In the non-relativistic limit, the potential is
\begin{eqnarray}
V(r)=\begin{pmatrix}
    0 & \frac{-\ax}{r}e^{-\mphi r} \\
    \frac{-\ax}{r}e^{-\mphi r} & 2\delta
  \end{pmatrix}
\end{eqnarray}
in the basis of ($\chi\chi$, $\chi^*\chi^*$) two-body scattering states, where $\delta$ is the mass splitting~\cite{Slatyer:2009vg,Hisano:2004ds}.  In general, all possible combinations of initial and final states need to be considered, i.e., 
\beq
\chi \chi, \chi^* \chi^* \to \chi \chi, \chi^* \chi^* \, , \quad \chi \chi^* \to \chi \chi^* \, ,
\eeq
since multiple interactions are important outside the Born regime.  Schutz \& Slatyer~\cite{Schutz:2014nka} derived analytic expressions for these cross sections including $s$-wave contributions only.  Zhang~\cite{Zhang:2016dck} used an adiabatic approximation to investigate parameters with a large mass splitting and small coupling, while Blennow, Clementz \& Herrero-Garcia~\cite{Blennow:2016gde} obtained numerical solutions covering the full parameter range.

Fig.~\ref{fig:inelastic} (left) shows the elastic (blue) and inelastic (red) scattering cross sections as a function of the DM velocity, for the benchmark models presented in Ref.~\cite{Blennow:2016gde}, where the coupling constant is fixed as $\ax=0.01 \, (\mx/270~{\rm GeV})$ from relic density. In all cases, the viscosity cross sections for $\chi\chi\rightarrow\chi\chi$ are larger than $\sim1~{\rm cm^2/g}$ in dwarf halos, but decrease below $\sim0.1~{\rm cm^2/g}$ in clusters. Inelastic scattering $\chi\chi\rightarrow\chi^*\chi^*$ only occurs in cluster halos, where the DM velocity is large enough to overcome the mass gap $2\delta$. Fig.~\ref{fig:inelastic} (right) illustrates the parameter regions, where the DM self-interactions create cores in dwarf galaxies. Compared to the elastic case, the inelastic model requires a smaller mediator mass ($m_{A^\prime}$) to compensate for the mass splitting.  

Down-scattering $\chi^*\chi^*\rightarrow\chi\chi$ may also play an interesting role in heating a halo by converting excitation energy into kinetic energy.  Typically, however, the abundance of $\chi^*$ is negligible in DM halos even if it is stable because down-scattering is in chemical equilibrium in the early Universe, which depletes its abundance~\cite{Finkbeiner:2009mi,Batell:2009vb,Blennow:2016gde}.


\section{Complementary searches}
\label{sec:comp}

The unifying feature of {\it all} SIDM models is the requirement of a new mass scale below the GeV-scale.  For example, in weakly coupled theories, self-interactions arise through dark sector mediators whose mass must be sufficiently light to give a large enough cross section.  In this section, we consider how complementary searches for light dark sectors can shed light on the physics of SIDM.  Our discussion focuses on two general themes: {\it (1)} interactions between DM and SM particles through sub-GeV mediators and {\it (2)} cosmological probes of decoupled dark sectors that do not interact with the SM.


\begin{figure}
\begin{tabular}{c}
\includegraphics[scale=1]{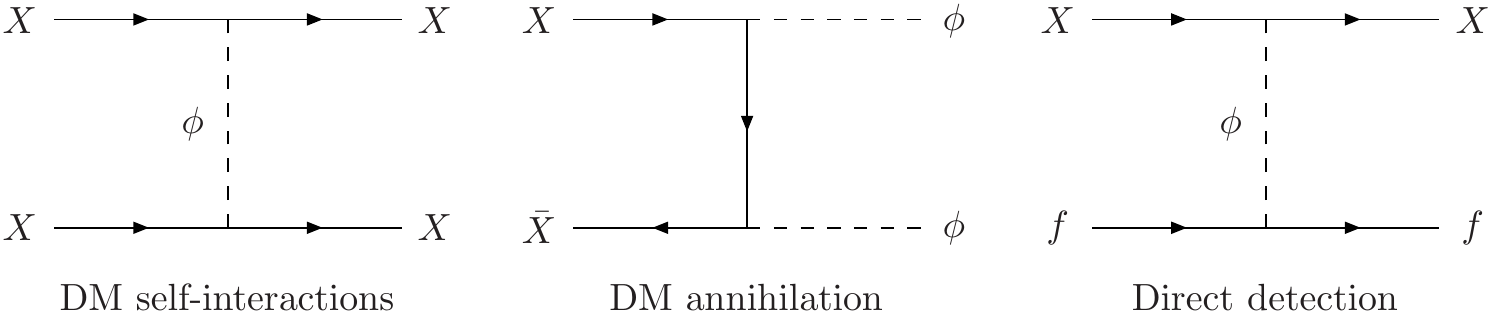}
\end{tabular}
\caption{\it Feynman diagrams arising from a DM particle $\chi$ coupled to a dark force mediator $\phi$. Self-interactions modify the DM halo properties, while annihilation can give rise to the observed DM abundance. Direct detection experiments are highly sensitive to potential interactions of the mediator to SM fermions ($f$), which allow decays of $\phi\rightarrow f\bar{f}$ in the early Universe. Reprinted from Ref.~\cite{Kaplinghat:2013yxa}.}
\label{fig:feyman}
\end{figure}

\subsection{Portals for light mediators}

Self-interactions through the Yukawa potential provide one of the simplest viable frameworks for SIDM.  Typically, the mediator $\phi$ has mass in the range of $\sim1-100$ MeV to yield the correct self-scattering cross section.  If $\phi$ was thermally produced in the early Universe, it must decay, otherwise its abundance would dominate the total energy density, leading to a cosmic overclosure problem~\cite{Kaplinghat:2013yxa,Boddy:2014yra}. Kaplinghat, Tulin \& Yu~\cite{Kaplinghat:2013yxa} suggested that the mediator may decay to SM particles if it couples to the SM sector.  The same coupling, therefore, may lead to signals in DM direct and indirect detection searches, as sketched in Fig.~\ref{fig:feyman}, as well as collider searches. 

There are several ways that $\phi$ may couple to the SM.  Depending on its spin, $\phi$ can mix with photon or $Z$ boson if it is a vector, or it can mix with the Higgs boson if it is a scalar.  This mixing allows relic $\phi$ particles to decay, typically into $e^+e^-$ and $\nu \bar \nu$ pairs. Here, we follow Ref.~\cite{Kaplinghat:2013yxa} and discuss these examples.

For the vector mediator case, mixing with photon or $Z$ boson is described by the Lagrangian
\beq \label{Lmix1}
{\cal L}_{\rm mixing} = \frac{\varepsilon_\gamma}{2}\,  \phi_{\mu \nu}  F^{\mu\nu} + \delta m^2 \, \phi_\mu Z^{\mu} 
\eeq
where $\phi_\mu$ is the vector mediator field and $\phi_{\mu\nu} \equiv \partial_\mu \phi_\nu - \partial_\nu \phi_\mu$ is its field strength, and $F_{\mu\nu}$ is the photon field strength. The first term corresponds to photon kinetic mixing~\cite{Holdom:1985ag}, parametrized by $\varepsilon_\gamma$, which has been widely studied as a DM portal, e.g.,~\cite{Foot:2004pa,Feldman:2006wd,Pospelov:2008jd, ArkaniHamed:2008qn}. The second term corresponds to mass mixing with the $Z$~\cite{Babu:1997st,Davoudiasl:2012ag,Davoudiasl:2013aya}, and it can be parametrized by $\varepsilon_Z \equiv \delta m^2/m_Z^2$, where $m_Z$ is the $Z$ boson mass.  We assume the mixing parameters are small: $\varepsilon_{\gamma,Z} \ll 1$.

The mixing terms in Eq.~\eqref{Lmix1} induce a coupling of $\phi$ to SM fermions
\beq \label{LDD1}
{\cal L}_{\rm int} = \left(  \varepsilon_\gamma e J_{\rm em}^\mu + \varepsilon_Z \frac{g_2}{\cos\theta_W}  J_{\rm NC}^\mu \right) \phi_\mu \, ,
\eeq
where $J_{\rm em}^\mu = \sum_f Q_f \bar f \gamma^\mu f$ and $J_{\rm NC}^\mu = \sum_f  \bar f \gamma^\mu (T_{3f} P_L - Q_f \sin^2\theta_W) f$ are the usual electromagnetic and weak neutral currents, respectively.  Additionally, $g_2$ is the SU($2$)$_L$ coupling, $\theta_W$ is the weak mixing angle, $P_L$ is the left-handed projection operator, and $Q_f$ and $T_{3f}$ denote the charge and weak isospin for SM fermion $f$, respectively.

If the dark mediator $\phi$ is a scalar, the leading renormalizable couplings to SM particles arise through the Higgs portal~\cite{Patt:2006fw,MarchRussell:2008yu,Ahlers:2008qc,Andreas:2008xy,Arina:2010wv,Chu:2011be,Djouadi:2011aa,Bhattacherjee:2013jca,Greljo:2013wja,Bian:2013wna,Choi:2013qra,Kouvaris:2014uoa,Kahlhoefer:2017umn}. Assuming $\phi$ is a real scalar singlet, the relevant terms in the scalar potential are
\beq
V(H,\phi) \supset (a \phi + b \phi^2) |H|^2
\eeq
where $H$ is the Higgs doublet and $a,b$ are coupling constants.  After electroweak symmetry breaking, mixing arises between $\phi$ and the physical Higgs boson $h$ due to the Higgs vacuum expectation value (vev) $v\approx 246$ GeV.  In the limit $a,m_\phi \ll v, m_h$, this mixing angle is $\varepsilon_h \approx a v /m_h^2$.  This generates an effective $\phi$ coupling to SM fermions
\beq
{\cal L}_{\rm int} = - \frac{m_f \varepsilon_h }{v} \bar f f \phi \, .
\eeq

In the early Universe, $\phi$ decays must not affect Big Bang Nucleosynthesis (BBN) and dilute the baryon density, which puts a constraint on the mixing parameter~\cite{Kaplinghat:2013yxa}. For example, in the kinetic mixing case ($\phi$ is known as a ``dark photon''), the final states are $e^+ e^-$ and the decay rate is $\Gamma_\phi = \alpha_{\rm em} m_\phi \varepsilon_\gamma^2 /3$, resulting in a lifetime
\beq
\tau_\phi \approx {\rm 3 \; seconds} \times \left( \frac{\varepsilon_\gamma}{10^{-10}} \right)^{-2} \left( \frac{m_\phi}{\rm 10 \; MeV} \right)^{-1} \; .
\eeq
Kaplinghat, Tulin \& Yu~\cite{Kaplinghat:2013yxa} estimated a lower bound on the mixing parameter, $\epsilon_\gamma\gtrsim10^{-10}$, by assuming that the mediator decays before weak freeze-out, $\sim1~{\rm s}$. Fradette et al.~\cite{Fradette:2014sza} studied BBN and CMB constraints on the $(\mphi,\varepsilon_\gamma)$ plane with the assumption that the cosmic abundance of the mediator particle is via the freeze-in process, i.e., the inverse decay process $f\bar{f}\rightarrow\phi$. Since the mixing parameter controls both production and decay rates in this case, the constrained parameter regions are closed. The injected photon energy released from $\phi$ decays could disassociate the light elements ($\sim10^4\textup{--}10^6~{\rm s}$) and reduce their abundances. The BBN constraint is sensitive to the mixing parameter in the range of $10^{-10}-10^{-14}$ for $\mphi\gtrsim 10~{\rm MeV}$~\cite{Fradette:2014sza}. If the mediator has a longer lifetime and decays after cosmic recombination ($\sim10^{13}~{\rm s}$), it could leave an imprint on the CMB power spectrum~\cite{Adams:1998nr,Chen:2003gz,Zhang:2007zzh,Galli:2009zc,Slatyer:2009yq}. The Planck experiment has excluded $\epsilon_\gamma\sim10^{-16}\textup{--}10^{-17}$ for the mediator mass in the range of $1\textup{--}300~{\rm MeV}$~\cite{Fradette:2014sza}. Proposed CMB missions, such as PIXIE~\cite{Kogut:2011xw}, could be sensitive to $\epsilon_\gamma\sim10^{-13}\textup{--}10^{-17}$ for a similar mediator mass range~\cite{Berger:2016vxi}. We expect these constraints will be stronger if the mediator abundance is also produced thermally, in addition to the freeze-in process.


\subsection{Direct detection}
\label{sec:directdetection}

\begin{figure}[t]
\centering
\begin{tabular}{@{}cc@{}}
\includegraphics[scale=0.9]{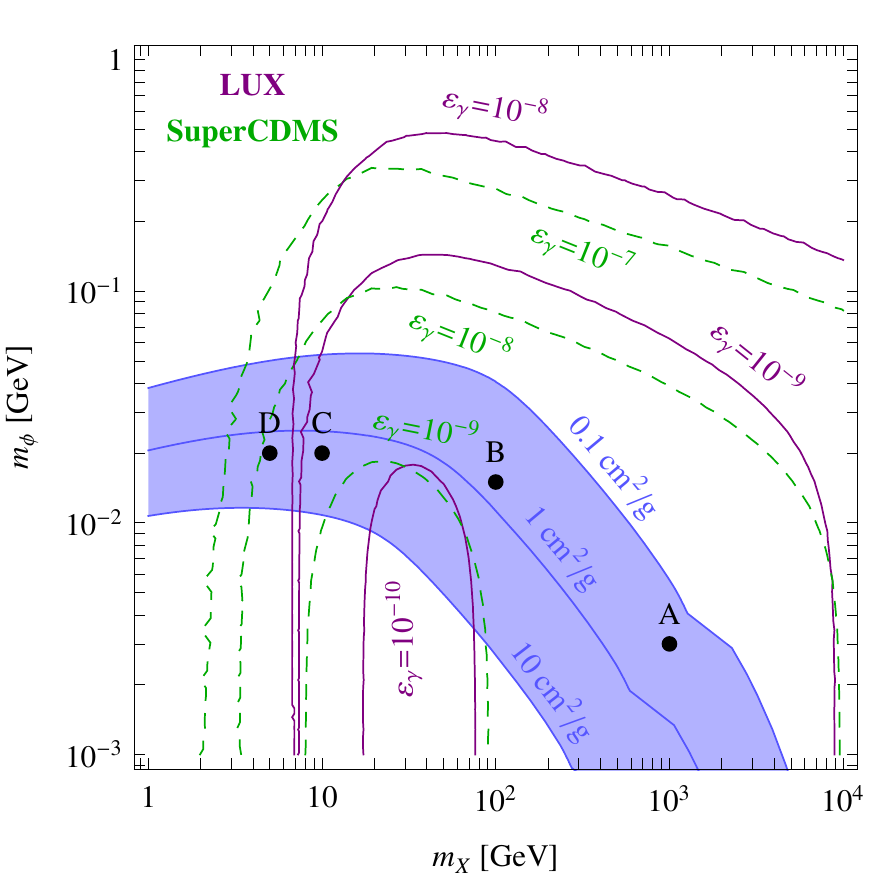}
\end{tabular}
\caption{\label{fig:direct} {\it Constraints on the $(\mx,\mphi)$ plane for the elastic Yukawa SIDM model, for different values of the kinetic mixing parameter $\epsilon_\gamma$, from LUX (purple lines) and SuperCDMS (dashed green lines) direct detection experiments. The region below each curve is excluded at 90\% CL. The shaded band is where DM self-interactions could affect the density profile of DM halos on dwarf scales. Four SIDM benchmark points discussed in~\cite{DelNobile:2015uua} are also shown (black dots). The dark fine structure constant is assumed to be $\ax=0.01$. Reprinted from~\cite{DelNobile:2015uua}.}}
\end{figure}

\begin{figure}
\centering
\begin{tabular}{@{}cc@{}}
\includegraphics[scale=0.9]{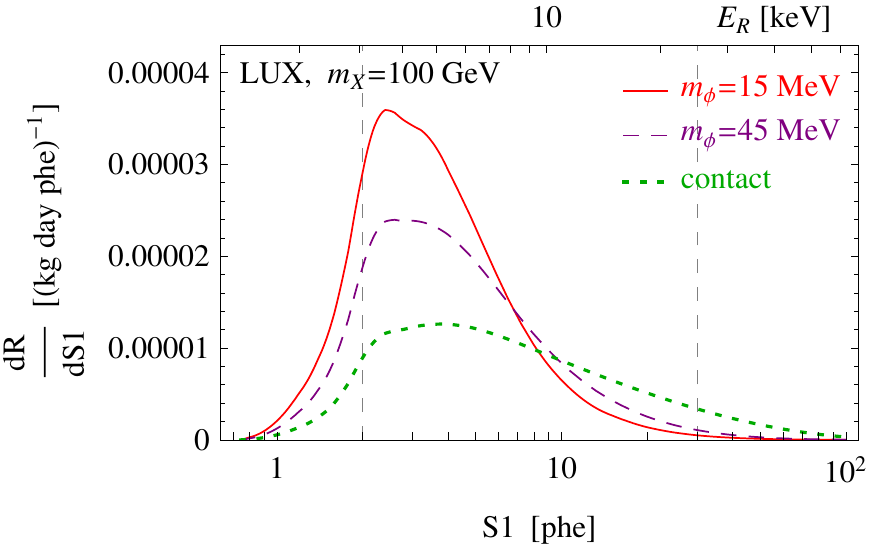}&
\includegraphics[scale=0.9]{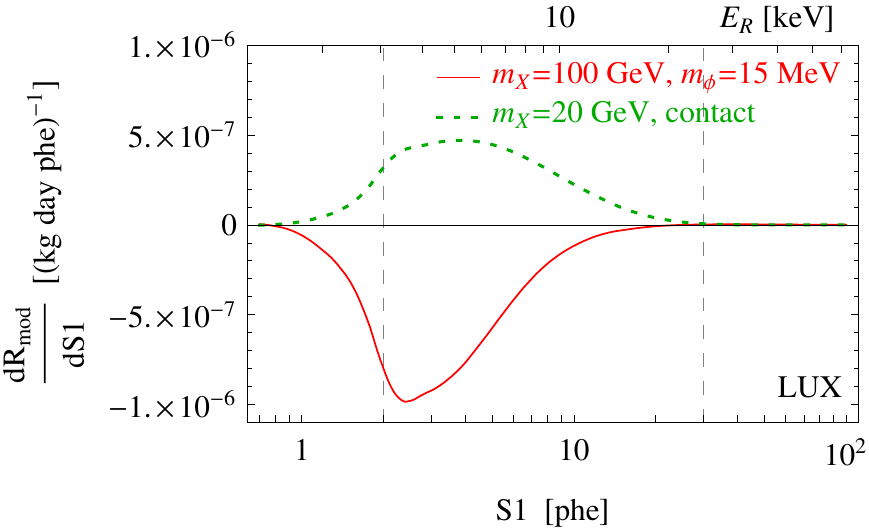}
\end{tabular}
\caption{\label{fig:spectrum} {\it Left: Unmodulated differential scattering rates at LUX for SIDM $(\mx=100~{\rm GeV},\mphi=15~{\rm MeV})$ (solid red line). For comparison, a model with three times the mediator mass (dashed purple line) and a model with a spin-independent contact interaction (dotted green line) are also presented. The spectra are normalized to have the same area within the signal range, enclosed by the two vertical dashed lines. On the top axis, the average recoil energy $E_R$ corresponding to the detected signal $S1$ in photoelectrons. Right: Modulated rates at LUX for a SIDM benchmark model (solid red line) and a $20~{\rm GeV}$ DM particle with contact interactions. Their unmodulated spectra are similar at LUX, but their modulated ones are very different. Reprinted from~\cite{DelNobile:2015uua}.}}
\end{figure}

Direct detection experiments search for nuclear recoils due to particle collisions with DM in the local halo.  Such interactions necessarily arise if mediator states interact with SM fermions as well as DM, as illustrated in the right panel of Fig.~\ref{fig:feyman}. For typical weakly interacting massive particles (WIMPs), direct detection cross sections are suppressed by the mass scale of heavy mediators (such as the Higgs boson).  However, since the SIDM framework requires mediators to be light, the cross section could be strongly enhanced.  Therefore, current null results place strong constraints on the mixing parameter connecting the two sectors~\cite{Fornengo:2011sz,Kaplinghat:2013yxa}. In addition, since DM self-scattering erases the non-thermalized features seen in CDM simulations, a Maxwell-Boltzmann DM velocity distribution, often used in calculating the direct detection rate, is more justified in SIDM than CDM~\cite{Vogelsberger:2012sa}.

The low mediator mass scale makes an interesting prediction for direct detection.  In contrast to WIMPs, for which scattering is a contact interaction, the mediator mass for SIDM is comparable to the typical momentum transfer $q$ in DM-nucleus scattering.  Thus, SIDM may interact with nuclei through a long-range force, which yields an event spectrum that is more peaked toward lower recoil energies~\cite{Fornengo:2011sz,Li:2014vza,DelNobile:2015uua}. This feature may help distinguish between SIDM and WIMP recoil spectra in experiments~\cite{DelNobile:2015uua}. If the SIDM candidate is a bound state, the form factor of the bound state and the possibility of breakup of the bound state could produce new signatures in the recoil spectrum~\cite{Laha:2013gva,Laha:2015yoa}.

Del Nobile, Kaplinghat \& Yu~\cite{DelNobile:2015uua} recast the LUX~\cite{Akerib:2013tjd} and SuperCDMS~\cite{Agnese:2014aze} constraints on the WIMP-nucleon cross section to set a limit on SIDM parameter space, as shown in Fig.~\ref{fig:direct}. The model considered here consists of DM particle $\chi$ with a vector mediator $\phi$ that couples to the SM by kinetic mixing (dark photon).  The blue band denotes the parameter space where $\sigma/m = 0.1-10 \; {\rm cm^2/g}$, as relevant for solving small scale anomalies, for $\alpha_\chi = 0.01$.  At present, LUX excludes this entire parameter space for $\varepsilon_\gamma\gtrsim10^{-9}$ and $\mx\gtrsim 7~{\rm GeV}$.  The SuperCDMS limit is about an order of magnitude weaker, but it is more sensitive to low mass SIDM models because of its lower energy threshold and lighter target nucleus (germanium compared to xenon). 

In the event of a positive detection, detailed studies of the signal spectrum and annual modulation can distinguish between SIDM and WIMP models.  Fig.~\ref{fig:spectrum} (left) shows the signal spectrum for SIDM in LUX, for fixed DM mass, taking into account realistic efficiency, acceptance, and energy resolution of the detector.  The three spectra indicate models with different mediator masses and have been normalized to give the same rate within the signal range (area enclosed by the vertical dashed lines).   SIDM models with lower mediator masses have spectra peaked toward lower nuclear recoil energy, due to the long-range nature of the interaction, compared to a contact interaction, e.g., in WIMP models. 

However, in some cases, there is a degeneracy in the signal spectrum between mass and the nature of the interaction.  For example, a $20~{\rm GeV}$ WIMP with contact interactions can yield the same spectrum at LUX compared to SIDM with $\mx=100~{\rm GeV}$ and $m_\phi=15~{\rm MeV}$.  One way to break the degeneracy is to compare signals between experiments employing different target nuclei, such as SuperCDMS.  In addition, these two models have very different annual modulation spectra, as shown in Fig.~\ref{fig:spectrum} (right).  Remarkably, they have {\it opposite phase}: light WIMPs are peaked when the Earth moves into the galactic DM wind, since the number collisions above threshold is larger, while the SIDM model is peaked moving with the wind, since the cross section is enhanced for smaller velocities.

Direct detection experiments have already placed stringent constraints on possible couplings between SIDM and SM particles.  For the minimal models considered here, upcoming searches have the opportunity to cover SIDM parameter space down to the BBN floor around $\varepsilon_\gamma \sim 10^{-10}$~\cite{Kaplinghat:2013yxa}.  However, these limits may be avoided if SIDM is light ($m_\chi \lesssim {\rm GeV}$) or if the couplings between SM particles and mediators are different than those considered here.  

Inelastic self-interactions provide another possibility for avoiding stringent direct detection constraints on the couplings between DM and the visible sector.~\cite{Zhang:2016dck,Blennow:2016gde}.  These limits may be evaded completely if scattering from nuclei requires an inelastic transition that is kinematically forbidden.  For example, for $\mx=10~{\rm GeV}$, endothermic DM-nucleon scattering is kinematically forbidden in both xenon- and germanium-based detectors when $\delta\gtrsim30~{\rm keV}$~\cite{Blennow:2016gde}.  The required minimal DM velocity is $v_{\rm min}\approx\sqrt{2\delta/\mu_{\chi A}}$, where $\mu_{\chi A}$ is the DM-nucleus reduced mass, which is larger than the escape velocity of the MW halo, $\sim750~{\rm km/s}$.  For the same reason, up-scattering $\chi\chi\rightarrow\chi^*\chi^*$ is suppressed or forbidden in galactic halos.  However, elastic self-scattering $\chi\chi\rightarrow\chi\chi$, with $\chi^*$ as an intermediate state, is {\it not} suppressed due to nonperturbative effects~\cite{Schutz:2014nka}.


\subsection{Indirect detection}

A compelling feature of WIMP models is that the same annihilation processes that set the relic abundance can also give rise to indirect detection signals in DM halos at present. The signal strength is determined by the requirement that the WIMP relic density should be consistent with the observed DM abundance.  For SIDM models with a light mediator $\phi$, the expectation is similar.  First, DM will annihilate into $\phi\phi$ pairs, shown in the middle panel of Fig.~\ref{fig:feyman}, each of which in turn will decay to produce a detectable signal, e.g., $\phi \to e^+ e^-$.  The signal strength is largely determined by dark sector couplings {\it only} and do not turn off in the limit that the portal couplings to the SM become tiny (but nonzero).  In addition, there are a number of characteristic features of SIDM indirect detection, which we now discuss.

\begin{figure}[t]
\centering
\begin{tabular}{@{}cc@{}}
\includegraphics[trim=0 -1cm -1cm 0, clip, scale=1.2]{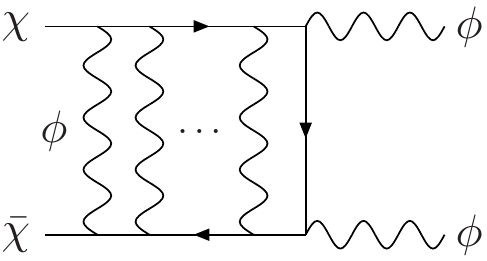}\;&
\includegraphics[scale=0.88]{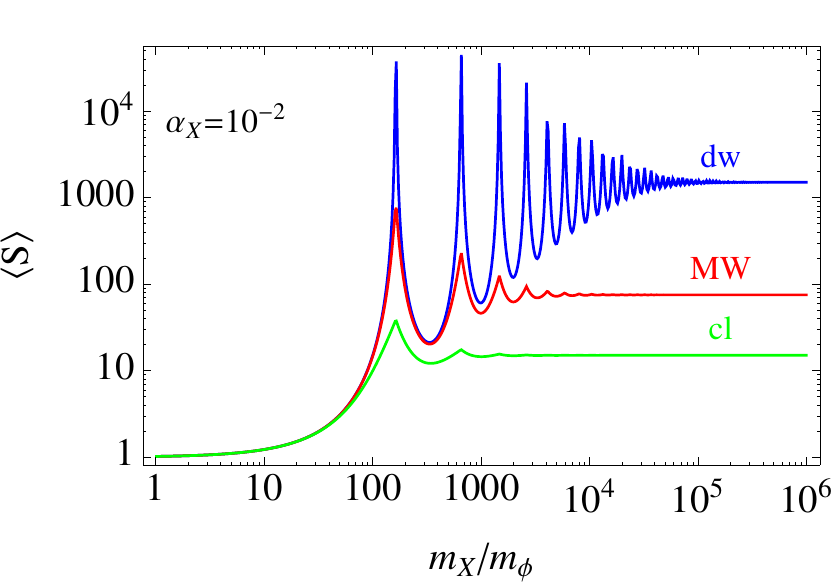}
\end{tabular}
\caption{\label{fig:sommerfeld} {\it Left: DM annihilation cross section can be boosted by the non-perturbative Sommerfeld enhancement effect.  Right: Thermally-averaged enhancement factor in dwarf galaxies (blue), the MW (red), and galaxy clusters (green). Reprinted from~\cite{Tulin:2013teo}.}}
\end{figure}

Light mediators can boost the annihilation cross section in low-velocity environments due to the Sommerfeld enhancement~\cite{Sommerfeld:1931}.  Attractive self-interactions distort the plane wave of the incoming DM particles and increase the probability of annihilation, illustrated diagramatically in Fig.~\ref{fig:sommerfeld} (left).  The DM annihilation cross section can be factorized as 
\beq
\sigma_{\rm ann} \vrel= (\sigma_{\rm ann}\vrel)_0\times S(\vrel),
\eeq
where $(\sigma_{\rm ann}v_{\rm rel})_0$ is the velocity-weighted Born annihilation cross section and $S(\vrel)$ represents the nonperturbative Sommerfeld enhancement factor.  It can be computed by numerically solving the non-relativistic Schr\"{o}dinger equation (see, e.g., Refs.~\cite{Hisano:2004ds,ArkaniHamed:2008qn}) or analytically by using the Hulth\`{e}n potential as an approximation for the true Yukawa potential~\cite{Cassel:2009wt,Slatyer:2009vg,Feng:2010zp,Blum:2016nrz}. 

This effect has important implications for indirect detection~\cite{Baer:1998pg,Hisano:2002fk,Hisano:2004ds,Hisano:2005ec,Cirelli:2007xd,MarchRussell:2008yu, Cirelli:2008pk,ArkaniHamed:2008qn,Lattanzi:2008qa,Feng:2010zp}.  During freeze-out, the DM velocity is $v_{\rm rel} \sim 0.3 \, c$, while at later times---in DM halos today and during the recombination epoch---the velocity is far smaller.  For light mediator models, this implies that the annihilation signal can be enhanced compared to the value needed to fix the relic density.\footnote{To accurately calculate the relic density in DM models with a light mediator, one needs to consider several subtle effects~\cite{Feng:2009mn,Dent:2009bv,Zavala:2009mi,Feng:2010zp,vandenAarssen:2012ag,vonHarling:2014kha}, such as enhanced annihilation after freeze-out, chemical recoupling, and bound state formation. If the two sectors were not in thermal equilibrium during DM freeze-out, the relic abundance also depends on the temperature ratio of the two sectors, see, e.g.,~\cite{Feng:2008mu}.}  Fig.~\ref{fig:sommerfeld} (right) shows that the thermally-averaged Sommerfeld enhancement can provide a large boost to annihilation on dwarf (10 km/s), MW (200 km/s), and cluster (1000 km/s) scales compared to the Born cross section.  The effect becomes important for $\alpha_\chi m_\chi/m_\phi \gtrsim 1$.  The peaks are resonant enhancements due to quasi-bound state formation, which is the same mechanism that leads to the resonant DM self-scattering discussed in \S\ref{subsec:yukawa}.  When the mass ratio $m_\chi/m_\phi$ is larger than $\sim10^3\textup{--}10^5$, the mediator is effectively massless and the enhancement factor is simply Coulomb-like, given by $S\approx\pi\ax/\vrel$.  On the other hand, the case with $\mx/\mphi\sim1$ corresponds to the usual WIMP and the enhancement is typically negligible.

Models along these lines were originally proposed by Arkani-Hamed et al.~\cite{ArkaniHamed:2008qn} and Pospelov \& Ritz~\cite{Pospelov:2008jd} to explain the increasing positron fraction in high-energy cosmic rays reported by the PAMELA collaboration~\cite{Adriani:2008zr}, later confirmed by the Fermi-LAT~\cite{FermiLAT:2011ab} and AMS-02~\cite{Aguilar:2013qda} collaborations. The signal can be fitted if the DM mass is $\sim{\rm TeV}$ and the annihilation cross section is a factor of ${\cal O}(100\textup{--}1000)$ larger than the standard WIMP cross section due to the Sommerfeld enhancement~\cite{Cirelli:2008jk}.  However, this explanation remains in tension with other observations. For instance, energy injection from DM annihilation may increase the ionization fraction during recombination, resulting in a distorted CMB spectrum~\cite{Chen:2003gz,Kamionkowski:2008gj,Galli:2009zc,Slatyer:2009yq,Hisano:2011dc,Hutsi:2011vx,Madhavacheril:2013cna,Lopez-Honorez:2013cua}. Since the DM velocity is very low during the recombination epoch, $v_{\rm rel} \sim10^{-8}c$, the predicted Sommerfeld effect is very large at this epoch~\cite{Feng:2010zp,Ade:2015xua}. 

More recently, Fermi-LAT observations have provided a purported excess in GeV $\gamma$-rays coming from the Galactic Center (GC) with an energy spectrum and spatial morphology consistent with WIMP annihilation~\cite{Goodenough:2009gk,Hooper:2010mq,Abazajian:2012pn,TheFermi-LAT:2015kwa}.  Kaplinghat, Linden \& Yu~\cite{Kaplinghat:2015gha} proposed that SIDM annihilation, shown in Fig.~\ref{fig:feyman} (center), may explain this excess as well.  In this model, DM annihilates into mediator particles, which then decay into energetic $e^+ e^-$ pairs, which in turn yield GeV $\gamma$-rays by Compton upscattering starlight in the GC.  In contrast to WIMPs, which produce $\gamma$-rays more directly in decay cascades, $\gamma$-ray signals from SIDM are negligible in dwarf galaxies due to their small stellar density.  On the other hand, the clustering of GC photons suggests that unresolved point sources are a more likely explanation of this excess~\cite{Bartels:2015aea,Lee:2015fea}.

Recently, Bringmann et al.~\cite{Bringmann:2016din} and Cirelli et al.~\cite{Cirelli:2016rnw} have found that SIDM annihilation is excluded over a wide range of parameters, $100 \; {\rm MeV} < m_\chi < 100\; {\rm TeV}$ and $1\; {\rm MeV} < m_\phi < 1\; {\rm GeV}$, based on indirect detection bounds from Fermi, AMS-02, and Planck.  However, there are number of caveats to keep in mind: 
\begin{itemize}
\item The bound can be relaxed if the dark and visible sectors have different temperatures during the freeze-out. In fact, given present constraints from direct detection, the couplings between the two sectors are likely too small to achieve thermalization~\cite{Kaplinghat:2013yxa,DelNobile:2015uua}.  To fit the GC excess, the dark-to-visible temperature ratio is $\sim 0.1$~\cite{Kaplinghat:2015gha}.  
\item It is assumed that $\phi$ is a vector mediator that decays via kinetic mixing.  If $\phi$ decays via $Z$ mixing, then a sizable fraction of decays produce neutrinos, which are far less constrained.  Alternatively, if $\phi$ is a scalar, annihilation is a $p$-wave process and thus the cross section is velocity-suppressed. However, CMB bounds may still be relevant for heavy SIDM due to bound state formation~\cite{An:2016kie}.
\item Indirect detection constraints are avoided if SIDM is asymmetric.  (Late-time particle-antiparticle oscillations can in principle repopulate the symmetric density~\cite{Cohen:2009fz,Buckley:2011ye,Cirelli:2011ac,Tulin:2012re}.)
\end{itemize}

Inelastic self-interactions can give rise to a different class of indirect detection signals.  If DM $\chi$ up-scatters to an excited state $\chi^*$, it may subsequently de-excite through a decay, e.g., $\chi^* \to \chi \gamma$, producing $\gamma$-ray signatures. Finkbeiner \& Weiner~\cite{Finkbeiner:2014sja} suggested that the recently claimed 3.5 keV line signal in M31 and several clusters~\cite{Bulbul:2014sua,Boyarsky:2014jta} could arise from this effect.  Since up-scattering is restricted by the kinematics of the mass gap, this scenario explains why the signal is seen in larger systems but not in dwarf galaxies.  Boddy et al.~\cite{Boddy:2014qxa} showed that composite SIDM can produce the same effect, where the mass gap arises as a hyperfine splitting that is naturally in the right energy range.

We may also expect indirect detection signals from the centers of large astrophysical bodies such as the Sun and the Earth, which may capture ambient DM particles~\cite{Gould:1987ir}. Annihilation of those captured DM particles may lead to some striking signals. For example, WIMP annihilation in the Sun could produce high-energy neutrinos detectable at the IceCube~\cite{Aartsen:2016zhm} and SuperK~\cite{Choi:2015ara} experiments, and annihilation in the Earth may produce anomalous heat~\cite{Mack:2007xj}. Compared to WIMPs, DM models with a light mediator have several interesting features in generating such signals. First, mediators may be so long-lived that their decays occur outside of the astrophysical object, allowing for the detection of charged final states~\cite{Batell:2009zp,Schuster:2009fc}. Second, the capture rate can be enhanced in the Sun due to the self-scattering~\cite{Zentner:2009is,Albuquerque:2013xna,Chen:2014oaa,Chen:2015bwa}. Third, the Sommerfeld effect can increase the annihilation cross section and shorten the equilibrium timescale of DM particles captured in the Earth~\cite{Delaunay:2008pc,Feng:2015hja}, resulting in a maximal signal flux. Fourth, the capture cross section is momentum dependent~\cite{Feng:2015hja,Feng:2016ijc,Kouvaris:2016ltf,Smolinsky:2017fvb}, in contrast a contact interaction for WIMPs.

Lastly, even if annihilation signals are absent (e.g., in the asymmetric case), DM accumulation in astrophysical objects may still cause observable effects. For example, captured asymmetric SIDM in stars can affect their evolution~\cite{Frandsen:2010yj,Zentner:2011wx, Bramante:2013nma,Kouvaris:2011gb, Eby:2015hsq}. Interestingly, a Rutherford-like DM-baryon interaction can significantly enhance the energy transport and produce an impact on the helioseisomology data, improving the agreement between the best solar model and the helioseismic data~\cite{Lopes:2014aoa}.


\subsection{Collider searches}

DM particles produced in colliders, although not detected directly, are usually inferred through a net momentum imbalance when they escape the detector.   The typical WIMP signal consists of initial-state radiation recoiling against missing energy~\cite{Birkedal:2004xn,Feng:2005gj,Beltran:2010ww,Goodman:2010yf,Bai:2010hh,Goodman:2010ku,Bai:2012xg,Zhou:2013fla,Carpenter:2013xra,Hochberg:2015vrg}, which is being actively searched for at the Large Hadron Collider (LHC)~\cite{Chatrchyan:2012tea,ATLAS:2012ky}. However, if the dark and visible sectors are coupled through sub-GeV mediators {\it only}, the LHC is not the best environment to test these models (compared to direct detection and other types of searches discussed) since the couplings must be small and the light mediator cannot produce DM through an on-shell decay.

\begin{figure}
\centering
\includegraphics[scale=0.5]{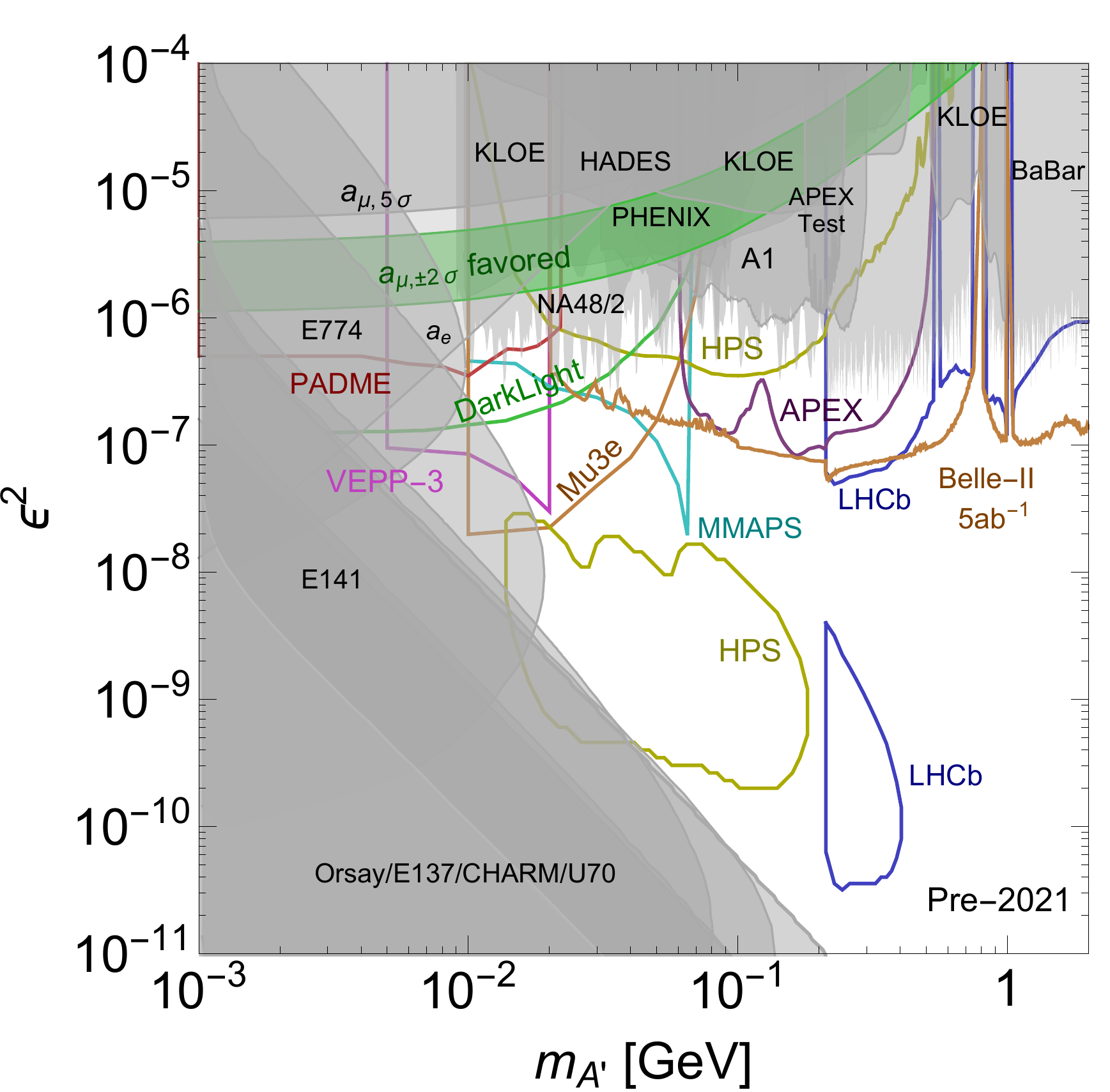}
\caption{\label{fig:beamdump} \it Collider constraints on a light dark photon $A^\prime$ with kinetic mixing parameter $\varepsilon$ and mass $m_{A^\prime}$.  Shown are existing bounds (shaded regions) and projected bounds from future experiments (open regions). Green band shows $2\sigma$ favored region in which the dark photon model can explain the discrepancy between the measured and calculated value of muon anomalous magnetic moment. Reprinted from~\cite{Alexander:2016aln}; see therein for further details.}
\end{figure}

On the other hand, production can be enhanced at high energy colliders if there exist other types of interactions distinct from what mediates self-interactions.  For example, if the light mediator $\phi$ couples to the SM through Higgs or $Z$ mixing, then on-shell decays of the Higgs or $Z$ boson can produce DM particles if kinematically allowed, contributing to their invisible widths~\cite{Kouvaris:2014uoa}.  Alternatively, Tsai et al.~\cite{Tsai:2015ugz} proposed that SIDM may couple to quarks through non-renormalizable contact operators.  In particular, for operators that are poorly constrained by direct detection, these interactions may be large enough to allow sizable production of SIDM at the LHC, yielding mono-jet signatures similar to WIMP searches.

Additionally, SIDM may have unique signals at the LHC due to the light mediator $\phi$.  First, since $m_\phi$ is much smaller than the typical parton energy, highly boosted DM particles will emit collinear $\phi$ radiation, analogous to collinear photon emission in QED~\cite{Buschmann:2015awa}.  Since these $\phi$ particles will decay to produce leptons, the signature of this effect is pairs of collimated lepton-jets.  If $\phi$ is a dark photon, this type of search is complementary to other dark photon probes (discussed below), albeit with an additional model dependence for production.  Second, if two SIDM particles are produced near the threshold, they may form a bound state due to strong self-interactions mediated by $\phi$, provided the interaction range is larger than the Bohr radius~\cite{Shepherd:2009sa,An:2015pva,Tsai:2015ugz,Bi:2016gca}. The lifetime of the bound state is short and it subsequently decays back to mediators or SM particles inside the detector.  This process does not have a large SM background and provides a direct way for testing the self-interacting nature of DM at particle colliders~\cite{An:2015pva,Tsai:2015ugz}. 


Recently, there has been considerable progress in searches for light weakly-coupled dark sectors at high-luminosity experiments, such as $e^+ e^-$ colliders and fixed target experiments~\cite{Batell:2009yf,Essig:2009nc,Bjorken:2009mm,Batell:2009di,Essig:2010xa}
 (see, e.g., Ref.~\cite{Alexander:2016aln} and references therein).  These studies may be able to discover the mediator $\phi$ for self-interactions purely through its coupling to the SM.  For the case that $\phi$ is a dark photon (denoted $A^\prime$), Fig.~\ref{fig:beamdump} shows current and prospective constraints on the parameter space of kinetic mixing parameter $\varepsilon$ and mass $m_{A^\prime}$.  In these studies, it is assumed that $A^\prime$ will decay into SM fermions, resulting in distinctive dilepton resonances.  Alternatively, if DM is sufficiently light, low-mass mediators may decay predominantly into DM particles.  This scenario may be detected as missing energy signatures at fixed target experiments~\cite{Izaguirre:2014bca,Izaguirre:2015yja}.

\begin{figure}
\centering
\includegraphics[scale=0.32]{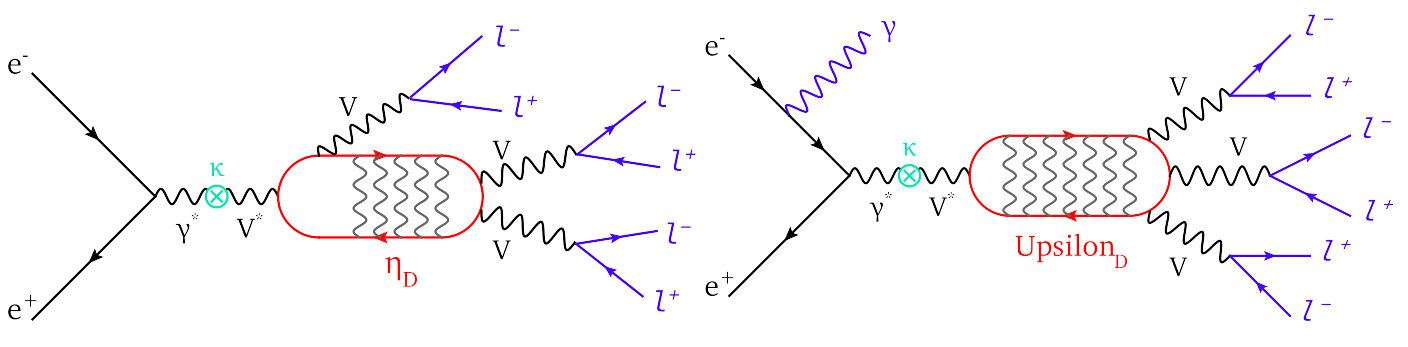}
\caption{\label{fig:boundstate} \it Production and decay of the $^1S_0$ (left) and $^3S_1$ (right) SIDM bound states. Since the $^1S_0$ ($^3S_1$) state is even (odd) under charge conjugation while the dark photon (denoted $V$) is odd, it connects with even (odd) numbers of dark photons in its production and decay.  Reprinted from~\cite{An:2015pva}.}
\end{figure}

An et al.~\cite{An:2015pva} showed that SIDM bound states may form at low-energy $e^+e^-$ colliders if the DM mass is sufficiently light and explored constraints from B-factories.  Fig.~\ref{fig:boundstate} illustrates the production and decay processes. When bound states form, they subsequently decay to charged leptons, resulting in six-lepton final states with low SM backgrounds.  Compared to direct $A^\prime$ searches at $e^+ e^-$ colliders, these bound state signatures improve the sensitivity to SIDM by $\mathcal{O}(10)$ for very large DM self-couplings ($\alpha_\chi = 0.5$)~\cite{An:2015pva}.

\subsection{Cosmological probes}

If the dark sector is completely decoupled from the SM, all of the complementary signals of SIDM discussed above do not apply.  However, in this case, many SIDM models predict a suppression of cosmological structure on small scales, which can be tested observationally.  

Most SIDM studies follow the assumption that the linear matter power spectrum remains unchanged compared to CDM.  However, several SIDM models---especially those that are decoupled from the SM---require additional model ingredients that can change this picture.  For example, in light mediator models, the mediator $\phi$ must decay to avoid overclosure.  Hence, there must exist new light states for $\phi$ to decay into if it is decoupled from the SM.  Similarly, for atomic SIDM models, massless mediators responsible for atomic binding have a thermal density in the early Universe.  Tight coupling between DM and a dark radiation component leads to acoustic oscillations and diffusion damping for DM, analogous to the baryon-photon fluid, modifying the linear matter power spectrum~\cite{Feng:2009mn,CyrRacine:2012fz,Aarssen:2012fx,Cyr-Racine:2013fsa,Buckley:2014hja,Boddy:2016bbu}.   Massless particles may also contribute to the effective degrees of freedom in the early Universe, potentially detectable in CMB observations (see, e.g., Refs.~\cite{Baumann:2015rya,Chacko:2015noa,Chacko:2016kgg,Abazajian:2016yjj,Ko:2016fcd,Ko:2016uft,Banerjee:2016suz,Tang:2016yt,Brust:2017nmv}).  Alternatively, in cannibalization models, DM particles cool more slowly during freeze-out if entropy is conserved in the dark sector, which can also suppress small scale power~\cite{Carlson:1992fn,deLaix:1995vi}.  

Damping caused by interactions between DM and radiation can lead to observable cosmological signals, as first noticed by Boehm et al.~\cite{Boehm:2000gq,Boehm:2001hm,Boehm:2004th}.\footnote{These works considered interactions between DM and SM radiation, i.e., photons or neutrinos.  Subsequent works have shown that both the missing satellites and TBTF problems can be alleviated due to damping if the DM-radiation scattering cross section is $2\times10^{-9}\mx/{\rm GeV}$ times the Thomson cross section~\cite{Boehm:2014vja,Schewtschenko:2015rno}.  Furthermore, van der Aarssen, Bringmann \& Pfrommer~\cite{Aarssen:2012fx} suggested that a light mediator coupled to both DM and neutrinos will generate DM self-interactions and DM-neutrino interactions to solve these issues {\it plus} the core-cusp problem.  However, couplings between SM neutrinos and the dark sector are subject to various experimental constraints~\cite{Ahlgren:2013wba,Laha:2013xua,Bertoni:2014mva}.}  The damping scale is set by the horizon size when kinetic decoupling occurs, which is of order $10 \; {\rm pc} \times (T_{\rm kd}/{\rm 10 \; MeV})$ for kinetic decoupling temperature $T_{\rm kd}$~\cite{Chen:2001jz,Hofmann:2001bi,Loeb:2005pm,Profumo:2006bv,Bertschinger:2006nq,Bringmann:2009vf}.  To estimate $T_{\rm kd}$, one equates the Hubble rate to the momentum transfer rate for scattering 
\beq
\Gamma =  n_r \left<\sigma_{\chi r} v_{\rm rel} \right>\frac{T}{\mx},
\eeq
where $n_r\sim T^3$ is the radiation number density, $\left<\sigma_{\chi r} v\right>$ is the velocity-weighted scattering cross section between DM and radiation, and $T$ is the temperature of the thermal bath.  Parametrizing the cross section to be $\left<\sigma_{\chi r} v\right> =  T^2/m_\phi^4$, where $m_\phi$ represents the mediator mass scale for DM-radiation interactions, yields\footnote{If the two sectors have different temperatures, $T_{\rm kd}$ also depends on the temperature ratio, see, e.g.~\cite{Feng:2009mn,Cyr-Racine:2015ihg}.}
\beq
T_{\rm kd} \sim 10 \; {\rm MeV} \; \left(\frac{m_\phi}{\rm 100 \;  GeV}\right) \left( \frac{ m_\chi}{\rm 100 \; GeV} \right)^{1/4} \, .
\eeq
For WIMPs, the mediator $\phi$ represents weak-scale degrees of freedom that couple DM $\chi$ to SM radiation, with all masses $m_\chi, m_\phi$ set by the weak-scale. Thus, the damping scale is $\mathcal{O}(10 \; {\rm pc})$, much smaller than current observational limits.  

On the other hand, SIDM may couple to a {\it dark} radiation component, such as light dark photons or sterile neutrinos.  In this case, the relevant mediator mass scale is likely $m_\phi \sim 10 \; {\rm MeV}$, which significantly delays DM kinetic decoupling from the thermal bath~\cite{Hooper:2007tu,Aarssen:2012fx,Dasgupta:2013zpn,Bringmann:2013vra,Ko:2014bka,Chu:2014lja,Cherry:2014xra,Bertoni:2014mva,Binder:2016pnr,Bringmann:2016ilk}. In this case, even for a weak-scale DM mass, one has $T_{\rm kd}\sim1~{\rm keV}$ and damping scales of $\mathcal{O}(\sim0.1 \; {\rm Mpc})$.  Through this enhanced damping effect, the missing satellites problem may be addressed within the SIDM framework~\cite{Aarssen:2012fx}.

\begin{figure}[t]
\centering
\begin{tabular}{@{}cc@{}}
\includegraphics[scale=0.3]{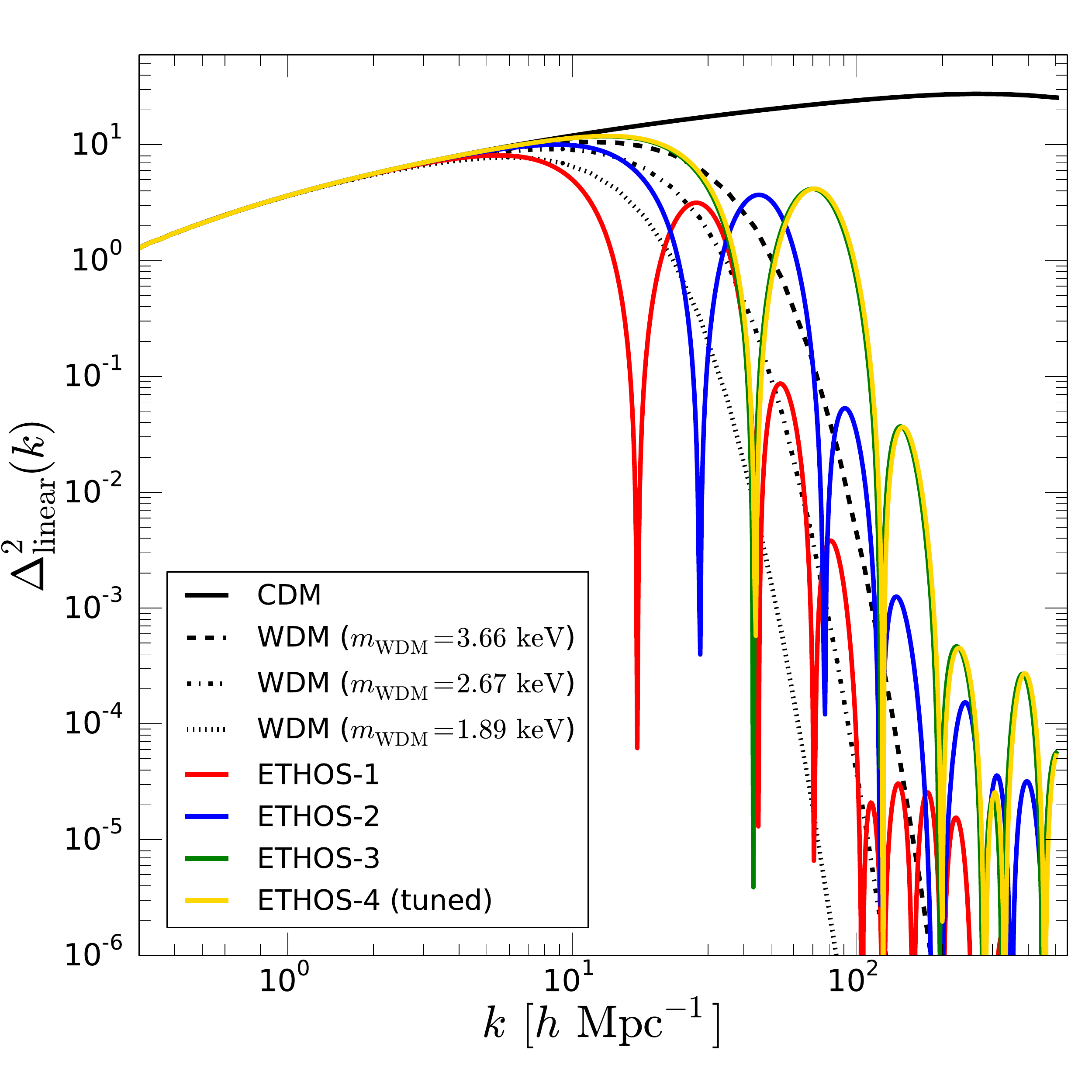}&
\includegraphics[scale=0.3]{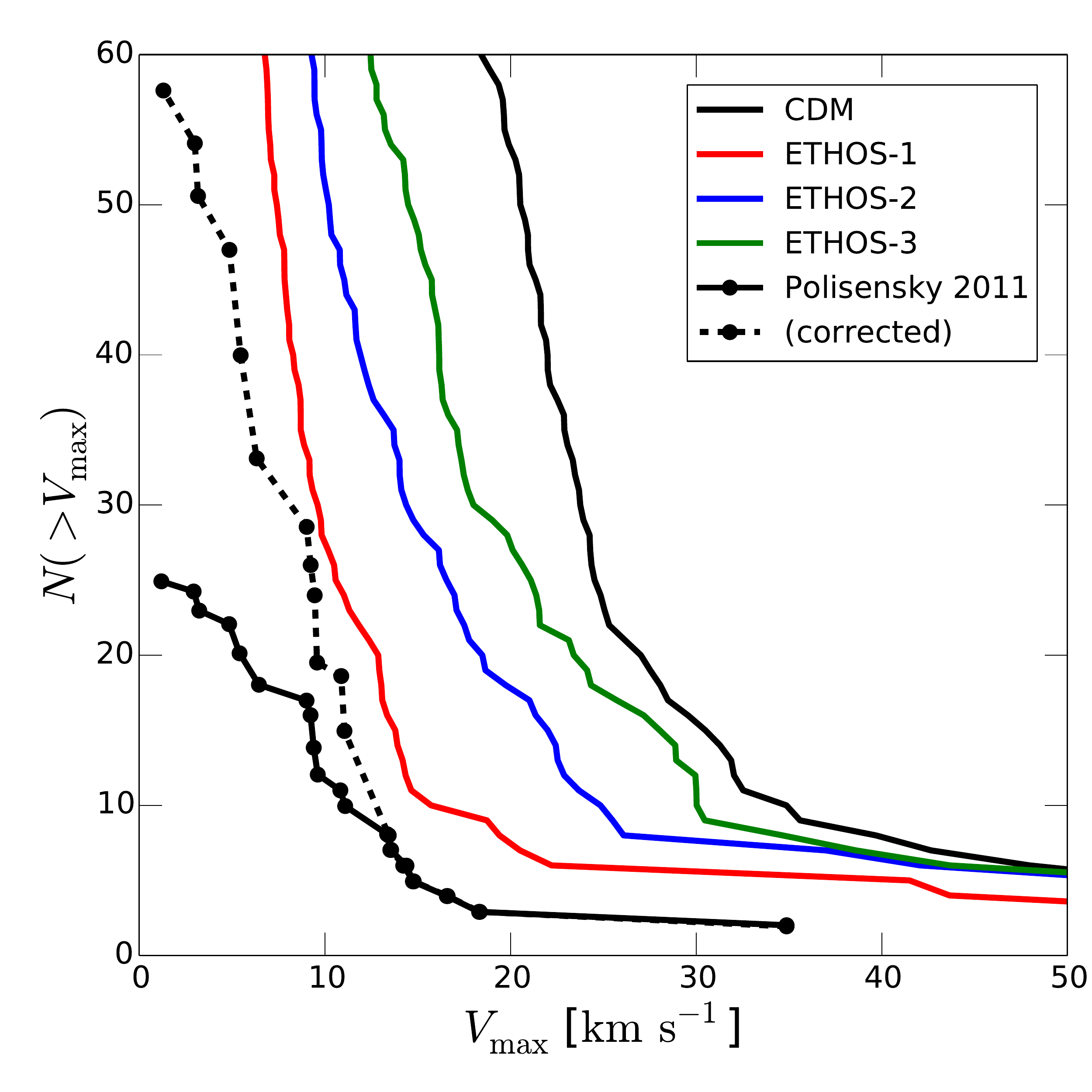}
\end{tabular}
\caption{\label{fig:ethos}\it Left: Linear matter power spectra for four ETHOS models that differ in the DM coupling strength to dark radiation, resulting in different damping and acoustic oscillation scales. ETHOS-1, 2, 3 have $\sigma/\mx\gtrsim5~{\rm cm^2/g}$ on dwarf scales, while ETHOS-4 has $\sigma/\mx\sim0.1~{\rm cm^2/g}$. CDM and thermal-relic warm DM models are also shown. Right: The number of subhalos as a function of their maximal circular velocity for four different ETHOS models, compared to observed satellites of the MW with a sky coverage correction~\cite{Polisensky:2010rw}. DM acoustic damping can reduce the tension between the number of observed satellites and simulated CDM subhalos. Reprinted from~\cite{Vogelsberger:2015gpr}.}
\end{figure}
   
The ETHOS collaboration performed cosmological N-body simulations of SIDM with damping due to dark radiation~\cite{Vogelsberger:2015gpr}. Their study considers four SIDM models (ETHOS-1--4) with different self-scattering cross sections and damping scales.  Their simulations assume DM couples to a massless hidden fermion via the same mediator $\phi$ that generates self-interactions, however they also explore a more general parametrization of damping within a broader class of models~\cite{Cyr-Racine:2015ihg}.  Fig.~\ref{fig:ethos} (left) shows the linear DM power spectra for the ETHOS models, compared to those for warm DM and CDM. For each ETHOS model, the power spectrum and the self-scattering cross section are calculated in a self-consistent way for a given set of model parameters. As expected, the power spectra of the ETHOS models exhibit acoustic oscillations, which are different from free-streaming damping for warm DM.  Fig.~\ref{fig:ethos} (right) shows that the number of subhalos in a MW-sized halo is significantly reduced in ETHOS models. These models agree with observational results better than CDM, thus potentially alleviating the missing satellites problem.  

It is of great interest to reassess all small scale issues in SIDM scenarios that include dark acoustic damping.  From the model building perspective, dark radiation (or other modifications to the power spectrum) is a generic feature of hidden sectors that are completely decoupled from the visible Universe.  Even in this ``nightmare" scenario, SIDM may still be testable.  In fact, the ETHOS studies have made clear that all small structure issues are highly interconnected: the combined effect of DM self-scattering and acoustic damping may provide too large of an effect on dwarf scales to solve all issues simultaneously.  Future work remains to find the optimal model to solve all issues.

\section{Conclusions}

\subsection{Summary}

Studies of astrophysical structure, combining both observations and N-body simulations, are confronting the paradigm of cold, collisionless DM.  While this paradigm works exceedingly well to explain data on large scales, the situation is less clear on smaller scales.  The inner densities of dwarf and LSB galaxies, as well as galaxy clusters, exhibit a mass deficit in their inner halos compared to expectations from CDM-only simulations.  While this may signal the importance of baryon dynamics, or other limitations in interpreting these observations, the particle dynamics of self-interacting DM can provide a viable explanation.  The figure of merit for self-interactions is $\sigma/m$, the cross section per unit DM mass.  Provided that $\sigma/m$ is nonzero, but not too large, heat is transferred inward to the center of the halo, inducing a mass deficit as DM particles are energized and move outward to larger orbits.  

The past few years have seen a renaissance for SIDM.  On the simulation side, there have been several important developments:
\begin{itemize}
\item Halo shape observations of clusters~\cite{MiraldaEscude:2000qt} and elliptical galaxies~\cite{Feng:2009hw,Feng:2009mn,Lin:2011gj} had previously provided stringent upper limits on $\sigma/m$, at the level of $\sim 0.01 \; {\rm cm^2/g}$.  In light of higher resolution simulations with greater halo statistics, these limits have been reassessed and weakened to $\sim 1 \; {\rm cm^2/g}$~\cite{Peter:2012jh}.
\item Simulations have begun to explore the importance of baryons in SIDM halos.  However, when it comes to stellar feedback on dwarf scales, there is little consensus.  For a bursty star formation history, CDM and SIDM halos both have large density cores and are virtually indistinguishable~\cite{Fry:2015rta}, while for a smoother history, CDM and SIDM halos are very different since baryonic effects are insufficient to generate cores~\cite{Vogelsberger:2014pda}.
\item Aside from feedback, the static gravitational potential of baryons can greatly impact halo profiles 	compared to SIDM-only expectations, especially in regions that are baryon-dominated~\cite{Kaplinghat:2013xca}.  This effect, recently verified through simulations, can result in SIDM profiles as cuspy as their CDM counterparts and leads to a predictive correlation between baryonic and SIDM density profiles~\cite{Elbert:2016dbb,Creasey:2016jaq}.  The scatter in rotation curves for a fixed halo mass is increased compared to CDM halos, lessening the diversity problem~\cite{Kamada:2016euw,Creasey:2016jaq}.
\item Recent cluster merger simulations have downgraded the magnitude of DM-stellar offsets induced by self-interactions during core passage~\cite{Robertson:2016xjh,Kim:2016ujt}.  Without $\mathcal{O}(10 \; {\rm kpc})$ resolution on the DM and galactic centroids of the subclusters, it may not be feasible to improve sensitivities much beyond simple scattering depth arguments.  On the other hand, post-merger misalignment of brightest central galaxies may provide greater sensitivities~\cite{Kim:2016ujt}.
\item Simulations have moved beyond the minimal SIDM model with a constant cross section, albeit for only a few benchmark models.  The scenarios considered include both velocity-dependent~\cite{Vogelsberger:2012ku} and angular-dependent~\cite{Robertson:2016qef} cross sections, as well as models that include a dark radiation component as an additional source of damping on small scales~\cite{Buckley:2014hja,Cyr-Racine:2015ihg,Vogelsberger:2015gpr}.  However, the model-space of potential dependencies remains large~\cite{Tulin:2013teo}.
\end{itemize}
In parallel, semi-analytic methods based on the Jeans equation have been developed for relaxed halos, providing a complementary path for confronting SIDM against observations that does not rely on computationally expensive simulations~\cite{Kaplinghat:2013xca}.  These methods have been tested against and agree with results from SIDM simulations, both without and with baryons~\cite{Kaplinghat:2015aga,Elbert:2016dbb,Creasey:2016jaq}.

As summarized in Table \ref{tab:constraints}, dwarf and LSB galaxies prefer a cross section of at least $\sim1 \; {\rm cm^2/g}$ to generate a density core consistent with observations. The strongest constraint on SIDM comes from cluster scales.  For fixed $\sigma/m$, the self-scattering rate in clusters is boosted compared to dwarf galaxies by a larger collisional velocity, providing an effect out to larger physical radii.  The absence of $\mathcal{O}(100 \; {\rm kpc})$ DM cores in massive clusters disfavors a constant cross section of $1 \; {\rm cm^2/g}$.  Interestingly, a recent study of seven massive relaxed clusters, including data from lensing and stellar kinematics, supports the presence of $\mathcal{O}(10 \; {\rm kpc})$ cores~\cite{Newman:2012nw,Newman:2012nv}.  These observations are consistent with $0.1 \; {\rm cm^2/g}$, favoring velocity-dependent self-interactions as a unified explanation across all scales~\cite{Kaplinghat:2015aga,Elbert:2016dbb}.  Earlier cluster studies~\cite{MiraldaEscude:2000qt,Meneghetti:2000gm}, which constrained $\sigma/m < 0.1 \; {\rm cm/g}$ based on the inner halo ellipticity and mass density of cluster MS2137-23, were overestimated for reasons discussed in Sec.~\ref{sec:sidmhalos} and additionally were based on lensing {\it only}, which lacks sensitivity to small radii compared to stellar kinematics data~\cite{Sand:2002cz}.  Although the Bullet Cluster famously confirms the collisionless nature of DM, null results in this and other merging systems are fully consistent with $0.1 \; {\rm cm^2/g}$ on cluster scales.  However, Abell 3827 and Abell 520 are exceptions, exhibiting significant DM-stellar offsets.  If these offsets are explained by SIDM, larger cross sections are required, in tension with cluster core data.

Self-interactions are a general feature of hidden sector models for DM.  The particle physics of self-interactions is not particularly exotic.  As proof of principle, the self-scattering cross section for {\it nucleons} has a similar order of magnitude to what is required for SIDM.  In fact, many SIDM models are quite reminiscent of the SM, such as QED- and QCD-like theories for DM involving dark photons or other hidden gauge forces.  Although the mass of DM itself can take a range of values depending on the model, the hidden sector is typically characterized by additional mass scales that govern the strength of self-interactions.  These scales, which represent the mass of mediator particles or binding/confinement energies in composite DM models, must lie below the GeV-scale in order to achieve $\sigma/m \sim 1 \; {\rm cm^2/g}$.  In contrast, WIMP DM cannot generate a large enough self-interaction cross section since {\it all} mass scales are assumed to lie at or above the weak scale.

Where it comes to experimental signatures, the SIDM paradigm is no less rich than the WIMP paradigm.  Mediators for self-interactions may couple to SM particles through renormalizable interactions, such as kinetic mixing~\cite{Holdom:1985ag} or Higgs portal~\cite{Patt:2006fw} operators.  This has motivated, in part, experimental efforts to discover new sub-GeV states in high-luminosity colliders~\cite{Alekhin:2015byh,Alexander:2016aln}.  

SIDM signals may also arise in traditional WIMP searches.  For DM mass $m \gtrsim 10 \; {\rm GeV}$, DM-nucleon scattering via light mediators is strongly constrained by direct detection searches, requiring very feeble couplings to the visible sector~\cite{Fornengo:2011sz,Kaplinghat:2013yxa}.  Interestingly, the phase of the annual modulation signal can be reversed compared to WIMP DM, due to the momentum-dependence of the interaction, which provides an avenue for model discrimination in the event of a positive signal~\cite{DelNobile:2015uua}.  Alternatively, lower mass SIDM can be detected via electron recoils~\cite{Essig:2011nj,Essig:2012yx,Essig:2017kqs}. 

For indirect detection, SIDM annihilation can be impacted by nonperturbative effects from the light mediator, including Sommerfeld enhancement and bound state formation~\cite{An:2016gad}.   $s$-wave annihilation of SIDM into mediator states, which subsequently decay into SM states, is constrained by null results from AMS-02, Fermi-LAT, and Planck~\cite{Bringmann:2016din}.  However, annihilation limits may be evaded in a variety of ways: e.g., if the dark sector is thermally decoupled from the visible sector (and colder) during freeze-out, a smaller annihilation cross section can produce the observed relic density.  Self-interactions can also enhance the capture rate of DM in the sun, relevant for neutrino signals~\cite{Zentner:2009is}, as well as induce collapse of neutron stars~\cite{Kouvaris:2011gb,Guver:2012ba}.

On the other hand, self-interactions may be completely sequestered from the visible sector, in which case the observational probes discussed above do not apply.  However, we argue that this is {\it not} a ``nightmare'' scenario for SIDM.  Getting the correct relic density in SIDM requires additional model-building ingredients that may be tested cosmologically.  Well-known scenarios involve either a dark radiation component (e.g., massless dark photons in atomic DM models~\cite{CyrRacine:2012fz}) or additional heating due to $3 \to 2$ cannibalization (e.g., glueball DM models)~\cite{Carlson:1992fn,deLaix:1995vi}, both of which affect the matter power spectrum.  For atomic DM, the effect of damping and dark acoustic oscillations at small scales may also address the dearth and diversity of MW satellites~\cite{Buckley:2014hja,Vogelsberger:2015gpr}.

If SIDM makes up only a subleading fraction of the total DM density, additional possibilities may arise.  For example, strongly self-interacting DM ($\sigma/m \gg 1$) can seed the initial growth of supermassive black holes~\cite{Pollack:2014rja} or undergo dissipative collapse to form compact dark disks~\cite{Fan:2013yva}.

\subsection{Outlook}

We conclude this report with a glimpse toward the horizon for astroparticle physics, small scale structure, and DM self-interactions.  On the astrophysics side, this is already a data-rich field and will become even more so once planned instruments come online in the near future.  On the particle physics side, a variety of experimental and observational techniques have the potential to access SIDM, including both traditional WIMP searches and more novel directions.  We highlight several of the latter approaches below.  

\vspace{2mm}

\underline{Future astronomical surveys:} Over the next decade, upcoming instruments will boost the statistics and resolution for all observations relevant for SIDM.  The Square Kilometer Array (SKA) radio telescope has the potential to observe hundreds of nearby spiral galaxies at resolutions below 100 pc, providing a deep look into their gas kinematics~\cite{deBlok:2015gja}.  This will provide a large and detailed sample of rotation curves, clarifying statistical arguments for the consistency of feedback-generated cores with $\Lambda$CDM cosmology~\cite{Katz:2016hyb,Pace:2016oim}, with an eye on potential outliers with large core sizes, $\gtrsim$ few kpc, that appear inconsistent with feedback (see Fig.~\ref{fig:baryon}).  SKA will also provide new insights into the cooling and collapse of H gas---an important consideration for the conversion of gas into stars in numerical simulations~\cite{deBlok:2015gja}.  In addition, the Giant Magellan Telescope (GMT)~\cite{2001aanm.book......} and the European Extremely Large Telescope (E-ELT)~\cite{2005scee.book.....H} can provide improved measurements of stellar kinematics of nearby dwarf galaxies.

Optical/near-infrared surveys with the Large Synoptic Survey Telescope (LSST)~\cite{Abell:2009aa}, Euclid~\cite{Amendola:2012ys} and Wide Field Infrared Survey Telescope (WFIRST)~\cite{Spergel:2013tha} will observe thousands of strong lensing systems on group and cluster scales.  Both space-based (Euclid, WFIRST) and ground-based (LSST) instruments play a complementary role~\cite{Jain:2015cpa}.  For probing self-interactions, statistical tests of halo ellipticity, as advocated in Ref.~\cite{Peter:2012jh}, will improve.  In addition, the prospects for testing SIDM using minor mergers (substructure infalling into cluster halos) will be improved as well, with increases both in cluster statistics and in background source densities used for weak lensing studies~\cite{Massey:2010nd,Harvey:2013tfa}.  For major cluster mergers, mass maps for strong and weak lensing will become more accurate than before and moreover will benefit from a wider field of view compared to the current state-of-the-art observations from the Hubble Space Telescope~\cite{Spergel:2013tha}.

\vspace{2mm}

\underline{Substructure:} Certain classes of SIDM models (e.g., coupling to a radiation component~\cite{Aarssen:2012fx,Bringmann:2013vra,Buckley:2014hja,Cyr-Racine:2015ihg,Boddy:2016bbu}) may damp the matter power spectrum on small scales.  Lyman-$\alpha$ forest data exclude significant damping for halos above $3\times 10^8 \Msun$~\cite{Viel:2013apy}.  Smaller halos predicted by $\Lambda$CDM would be devoid of baryons and must be discovered gravitationally, either through lensing~\cite{Mao:1997ek,Chiba:2001wk,Dalal:2001fq} or perturbations to dynamically cold structures in the MW, such as tidal streams~\cite{Johnston:1997fv,Ibata:2001iv,Johnston:2001wh,SiegalGaskins:2007kq}.  Recently, studies of lensed dusty star-forming galaxies have shown great promise to probe substructure~\cite{Hezaveh:2012ai}.  Observations from the Atacama Large Millimeter/submillimeter Array (ALMA) can exploit individual star-forming clumps as sources to probe the lensing potential on small angular scales.  The first studies based on ALMA science verification data have already found evidence for a $10^9 {\Msun}$ subhalo in one elliptical galaxy, with a potential to reach down to $10^7 \Msun$ in the future~\cite{Hezaveh:2016ltk} (see also~\cite{Inoue:2015lma}).  
Another method for finding substructure uses tidal streams from globular clusters.  These low-dispersion stellar systems are prone to disruption from subhalo fly-bys, leaving gaps in the stream.  While observed tidal stream gaps are qualitatively consistent with the rate for subhalo encounters expected from $\Lambda$CDM~\cite{Yoon:2010iy,Carlberg:2012ur}, improved stream modeling, coupled with future observations~\cite{Erkal:2015kqa}, offer the potential to probe down to $10^5  \Msun$ subhalos~\cite{Bovy:2016irg}.

\vspace{2mm}

\underline{Testing bursty feedback models:} In principle, observations can test whether star formation histories in dwarf and LSB galaxies are bursty enough to transform cusps into cores.  One method is to compare tracers of the star formation rate that have sensitivity to different time scales: e.g., H$\alpha$/H$\beta$ and far-ultraviolet (FUV) emission, which probe the mean rate over the past 10 Myr and 200 Myr, respectively~\cite{2012ApJ...744...44W} (see also Ref.~\cite{Kauffmann:2014cda}).  Nearby dwarf galaxies exhibit considerable scatter in their H$\alpha$/FUV ratios, which is interpretted as evidence for burstiness~\cite{2012ApJ...744...44W}.  Recent FIRE simulations~\cite{2017MNRAS.466...88S} have been able to reproduce these observations for the most part (except for a non-negligible fraction of low-mass galaxies that have very small H$\alpha$/FUV ratios, which may indicate that the simulations are overestimating somewhat the peak-to-trough variation in the star formation rate).  Additionally, if outflows from starbursts drive rapid fluctuations in the gravitational potential, the stellar distribution is heated gravitationally alongside the DM~\cite{2016ApJ...820..131E}.  This yields a testable correlation between star formation rate and stellar velocity dispersion~\cite{2017ApJ...835..193E}.

LSBs have quiescent star formation today~\cite{deBlok:1997zlw}, although bursty feedback may have occurred in the past to form a core. It is important, therefore, to understand the redshift dependence of these processes.  Studies of H$\beta$/FUV ratios in low-mass galaxies at $0.4 < z < 1$ find evidence that star formation is burstier at higher redshifts~\cite{2016ApJ...833...37G}.  However, applying these probes to faint DM-dominated galaxies at distant redshifts is challenging.  
An alternative method is to study the radial distributions of stellar ages and chemical compositions in nearby galaxies.  Here, feedback can imprint itself dramatically: since feedback causes dynamical heating of the stellar (as well as DM) density, stars undergo outward radial migration~\cite{2016ApJ...820..131E}.  Since this effect builds up the most for the oldest (metal-poor) stars, it can potentially invert the age and metallicity gradients in galaxies compared to their intrinsic profiles.  This process is expected to be most pronounced for DM halos with the largest cores if both arise through feedback.  On the observational side, Young et al.~\cite{2015MNRAS.452.2973Y} have initiated a program of MUltiwavelength observations of the Structure, Chemistry, and Evolution of LSB galaxies (MUSCEL) with promise to shed light on these issues.

\vspace{2mm}

\underline{Experimental detection of light dark matter:} Searches for DM at colliders and in direct detection experiments have focused primarily on the weak scale, motivated by theoretical expectations.  With the absence of any definitive signals thus far, theorists have turned toward new directions and the idea of light dark matter (LDM), in the keV--GeV mass range, has garnered much attention recently~\cite{Alekhin:2015byh,Alexander:2016aln}.\footnote{If DM is a thermal relic, it cannot be below the keV scale based on Lyman-$\alpha$ constraints on warm dark matter (see, e.g., Ref.~\cite{Viel:2013apy}).  Even if DM is a thermal relic of a hidden sector that is much colder than the visible sector, its mass cannot be below 1.5 keV due to free streaming and phase space constraints~\cite{Das:2010ts}.}  Models along these lines typically require light mediators to be viable, and hence these scenarios are a natural framework for SIDM~\cite{Lin:2011gj}.  There are many proposals to detect LDM experimentally, offering the potential to study the physics of DM self-interactions in the laboratory.  Accelerator-based proposals include missing-energy signatures at GeV-scale colliders~\cite{Essig:2013vha,Izaguirre:2014bca}, as well as the production and detection of dark matter beams produced in fixed target and beam dump experiments~\cite{Batell:2009di,Izaguirre:2013uxa,Kahn:2014sra,Alekhin:2015byh,Battaglieri:2016ggd}.  The MiniBoone collaboration recently performed a dedicated search along these lines~\cite{Aguilar-Arevalo:2017mqx}.  Alternatively, there are many ideas for the direct detection of LDM as well, such as WIMP search constraints repurposed in terms of electron recoils~\cite{Essig:2011nj,Essig:2012yx,Essig:2017kqs}, as well as new detector technologies with ultra-low thresholds~\cite{Graham:2012su,Hochberg:2015pha,Essig:2015cda,Schutz:2016tid,Derenzo:2016fse}.

\vspace{2mm}

\underline{Theoretical modeling and numerical simulations:} On the theory side, it is important to understand the role of DM self-interactions in galaxy formation together with baryon physics.  Some issues can be studied using a {\it quasi-equilibrium} analysis, as we discussed in \S\ref{sec:jeans}.  One example is the tight connection predicted between DM and baryon distributions~\cite{Kaplinghat:2013xca,Elbert:2016dbb,Creasey:2016jaq}, which may explain the long-mysterious halo-disk conspiracy~\cite{1986RSPTA.320..447V,Kamada:2016euw}.  On the other hand, there are {\it dynamical} questions of great interest as well that must be addressed using hydrodynamical SIDM simulations (see Ref.~\cite{DiCintio:2017zdz}).  For example, SIDM inner halos are expected to be more stable than for CDM in response to violent baryonic outflows since self-interactions rapidly redistribute energy deposited from feedback.  Additionally, the star formation rate may be more suppressed in dwarf galaxies with SIDM halos because they have shallower gravitational potential wells from which UV heating can remove gas more effectively.  This suggests that the missing satellites problem and the TBTF problem in the field may be more easily solved in SIDM than CDM with baryon feedback.

\vspace{0.5cm}

{\it Acknowledgments}: We are indebted to James Bullock, Peter Creasey, Jonathan Feng, Ran Huo, Manoj Kaplinghat, Rachel Kuzio de Naray, David Morrissey, Laura Sales, Kristine Spekkens, Philip Tanedo, and Kathryn Zurek for helpful discussions and collaborations on this subject. 
This work is supported by the Natural Science and Engineering Research Council of Canada
(ST), the U.S. Department of Energy under Grant No. de-sc0008541 (HBY), and the Hellman Fellows Fund (HBY). 

\bibliography{review_Haibo,review_Sean}

\end{document}